%% file: MagicNumber_last.tex
\documentclass[twoside,12pt]{article}
\usepackage{epsfig,color}

\begin{document}

\title{ \vspace{1cm} Nuclear magic numbers: new features far from 
stability}
\author{O.\ Sorlin$^1$, M.-G.\ Porquet$^2$ \\
\\
$^1${\small Grand Acc\'el\'erateur National d'Ions Lours (GANIL),}\\
{\small CEA/DSM - CNRS/IN2P3, B.P. 55027, F-14076 Caen Cedex 5, France}\\
$^2${\small Centre de Spectrom\'etrie Nucl\'eaire et de Spectrom\'etrie 
de Masse (CSNSM),}\\ 
{\small CNRS/IN2P3 - Universit\'e Paris-Sud, B\^at 104-108, 
F-91405 Orsay, France}\\
}
\maketitle
\input{texteAbstract_9avril}

\footnotesize
\tableofcontents
\normalsize
\input{texteIntroduction_9avril}

\input{texteInteraction_10avril}
\input{texteNZ8_10avril}
\input{texteNZ20_9avril}
\input{texteNZ28_9avril}
\input{texteNZ50_9avril}

\input{texteNZ82_10avril}
\input{texteN126_25fev}

\input{texteLastZ_9avril}
\input{texteConclusions_10avril}
\input{texteAnnex_9avril}

\section*{Acknowlegments}
This review owes much to dicussions with many collegues. Related
to the theoretical parts, F. Nowacki, A. Poves, D. Lacroix, S.
Typel, T. Otsuka and B.A. Brown are greatly acknowkedged. M.-G.P. 
thanks M. Girod and S. P\'eru for providing her with the HFB-D1S code 
and for many helpful discussions. Concerning the experimental
topics, discussions and comments from close collegues as F.
Azaiez, Z. Dombr\`adi, L. Gaudefroy, S. Gr\'evy, K.-L. Kratz, A. Navin, and 
J.C. Thomas, to name a few, were very enlightening. O.S. also 
wishes to thank P. Chomaz, who trusted us to write this review 
article.

\end{document}

%% file: texteAbstract_9avril.tex
\begin{abstract}

The main purpose of the present manuscript is to review the
structural evolution along the isotonic and isotopic chains around
the "traditional" magic numbers $8, 20, 28, 50, 82$ and $126$. 
The exotic regions of the chart of
nuclides have been explored during the three last decades.
Then the postulate of permanent magic numbers was
definitely abandoned and the reason for these structural
mutations has been in turn searched for. General trends in the
evolution of shell closures are discussed using complementary
experimental information, such as the binding energies
of the orbits bounding the shell gaps, the trends of the first
collective states of the even-even semi-magic nuclei, and the
behavior of certain single-nucleon states. Each section is devoted to
a particular magic number. It describes the underlying physics of
the shell evolution which is not yet fully understood and indicates
future experimental and theoretical challenges. 
The nuclear mean field embodies various facets of
the Nucleon-Nucleon interaction, among which the spin-orbit 
and tensor terms play decisive roles in the shell evolutions. 
The present review intends to provide experimental constraints 
to be used for the refinement of theoretical models aiming at a
good description of the existing atomic nuclei and at more 
accurate predictions of hitherto unreachable systems.

\end{abstract}

%% file: texteIntroduction_9avril.tex
\section{Introduction}
As early as 1934, W.~Elsasser noticed the existence of
"special numbers" of neutrons and protons which confer to the
corresponding nuclei a particularly stable
configuration~\cite{Elsa34}. In analogy with atomic 
electrons, he 
correlated these numbers with closed shells in a model of
non-interacting nucleons occupying energy levels generated by
a potential well. This hypothesis was not further pursued at that
time both because of the apparent paradox that strong inter-nucleon
forces would average out in such a simple way, and 
the paucity of experimental data in favor of a single-particle
description.

More than a decade later, the study of shell structure 
regained interest through the review of M.~Goeppert-Mayer which,
including a large number of precise experimental data~\cite{Goep48},
pointed out to the existence of closed shells at numbers 8, 20, 50, 82 and 126.
Unfortunately only the first two, 8 and 20, could be
explained from solutions of simple potential wells. Rather\emph{
ad-hoc} rearrangements of the level ordering were required to find 
the higher closed shell numbers. This
strongly casted doubt on this view of the nucleus in which
independent particles were moving in an average field.

Success was finally achieved in 1949 by Mayer, Haxel, Suess and
Jensen who independently showed that the inclusion of a spin-orbit
potential could give rise to the observed gaps between the
shells~\cite{Goep49,Haxe49}. Simultaneously all these special
numbers - renamed "magic numbers" - as well as a vast amount of nuclear properties on
nuclei reachable at that time, e.g. spins, magnetic moments,
isomeric states, and $\beta$-decay systematics could then be explained. This discovery
opened the path to great progress in the understanding of nuclear
structure and these magic numbers were the cornerstones for future
theoretical developments in nuclear physics. The permanence of
these nuclear magic numbers remained a dogma for several decades.

With the technical possibility of exploring nuclei with relatively
large N/Z ratios, the persistence of these magic shells for nuclei far
from stability was first examined for $N=20$. Experimental results 
on mass, nuclear
radius and spectroscopy obtained in the three last decades have shown
that the $N=20$ gap in energy was quite fragile in nuclei far from
stability~\cite{Thib75,Hube78,Detr79,Guil84}. These combined results 
provided the
first pieces of evidence that magic numbers were not immutable,
and triggered a large number of experimental and theoretical
work devoted to study the fate of the magic numbers far from the
valley of stability.

Since then, many radioactive ion beam facilities have emerged
worldwide and it is widely believed that magic numbers may
evolve when extreme proton-to-neutron ratios are explored. The
detailed location and magnitude of shell gaps in the neighborhood
of the valley of stability have served to develop satisfactory
mean-field models. However, these models diverge quickly far from
stability, implying that some unkown degrees of freedom are required 
to describe the low-energy properties of the atomic nuclei.

The main purpose of the present manuscript is to review the major
structural features along the isotonic and isotopic chains around
the\emph{ spherical} magic numbers 8, 20, 28, 50, 82, and 126. Attempts
to prove the existence of new magic numbers will also be commented 
on. A compilation of experimental results is presented 
on the following properties : (i) evolution of the
binding energies of the two orbits bounding the spherical shell gaps, (ii)
trends of first collective states of even-even semi-magic
nuclei, (iii) characterization of single-nucleon states. Each
section is devoted to one magic number, which potentially
provides suitable constraints on the spin-orbit and
tensor terms contained in the effective interactions used to
describe the nuclei. The present compilation intends to lead to 
a better understanding of the action of nuclear forces in the 
nucleus and to provide experimental constraints to refine
theoretical treatments. Henceforth better predictive power could 
be achieved throughout the whole nuclear chart,
and in particular for the most neutron-rich isotopes which cannot
be studied experimentally. In parallel, some of the major
remaining experimental challenges arising while exploring the
evolution of shell closures will be emphasized.\\

\noindent{\bf Foreword}\\
All the experimental information on the excited states discussed
in this review can be found in the ENSDF database~\cite{NNDC}, the
references of the original work are not systematically given.
The literature has been covered till March 2008.
The adopted values of the reduced electric
quadrupole transition probabilities B(E2) are those tabulated in
Ref.~\cite{Rama01},  but when a new value  has been obtained recently,
we refer explicitely to the last measurement.
The experimental values of the atomic masses used to compute the
nucleon separation energies come from the last atomic mass
evaluation~\cite{Audi03}, except for a few cases
explicitely reported in the
text. The use of binding energy of the last 
nucleon as a single-particle energy is discussed in Sect.~\ref{annex},
which also contains the limitations of such a method.

%% file: texteInteraction_10avril.tex
\section{Effective interactions between nucleons}

Many nuclear structure models exist to describe the properties of
the nucleus. One can ask why is there so many models rather than a
unique one ? Reasons are both practical and phenomenological. From
the practical point of view, the most fundamental approach, which
would treat the nucleus as N interacting fermions by means of
short-range forces, is often untractable. On the phenomelogical
side, the nucleus explores different degrees of freedom, ranging
from the spherical magic, the deformed (axial, triaxial,
pear-like) or cluster shapes, the one or two nucleon haloes, to
the particle-unbound systems. It is an extremely changeable
system, which is also making the field of nuclear structure so
diverse and fascinating. But it is tempting and often wise to use
models which are suitable to describe one or few of these observed
phenomena.

As the present review is addressing the evolution of spherical
shell closures, it implies that we shall focus on nuclei which are
sitting at or in the vicinity of shell closures. Their presence is
rooted in the approximation of the single-particle (SP) motion of
nucleons in a mean-field potential and by the non-uniform
distribution of single SP levels. But what makes shell closures to
change ? Several reasons can be invoked. First, while nucleons
occupy various orbits, the interactions between them are changed
according to the radial distribution, angular momentum and spin
orientation of their orbits. The nuclear mean field and the SP
level distribution are modified accordingly. In this spirit, the
present review aims at identifying which properties of the nuclear
forces can account for observed changes of shell closures. If done
properly, better extrapolations to unknown regions of the chart of
the nuclides can be proposed. Second, correlations of pairing or
quadrupole type can weaken the robustness of spherical shell gaps.
These many-body correlations drive the nucleus to deform, acting
on top of SP energy changes. Third, for weakly bound nuclei,
interactions between bound states, resonances and scattering
states~\cite{Doba07} can modify the spectroscopic properties of
nuclei located close to drip-lines. Their effects could jeopardize
extrapolations of nuclear properties far from stability (such as
for the study of halo nuclei or for the astrophysical r-process),
if neglected. Accents will be however put on the two first changes
for which experimental data exist, i.e. the force-driven change of
SP energies will constitute the body of the present manuscript and
the mechanism to develop correlations will be outlined when
encountered.

Three kinds of models could address the evolution of shell
closures from nucleon-nucleon interactions (see for instance
Fig.~1 of Ref.~\cite{Dean07}). The most fundamental approach, the
so-called\emph{ ab-initio} calculations, use\emph{ realistic}
nucleon-nucleon ($NN$) forces derived from the results of the free
nucleon-nucleon scatterings (such as Argonne $v_{18}$
\cite{Wiri95}, CD-Bonn~\cite{Mach01} and the recent soft
interaction $V_{low k}$ \cite{Bogn03}). As these calculations are
very computationally demanding, they have kept pace with recent
technological progresses. Despite this, the actual precision in
describing nuclei with such two-body forces is still not fully
satisfactory, pointing out to the need of three-body
forces~\cite{Piep01}. Also technical problems of convergence
remain for nuclei with $A > 20 $ (see for
instance~\cite{Roth07,Dean07b}). Therefore, even if very
promising, other approaches should be used at present in order to
describe the evolution of shell structure over a wide range of
atomic masses.

An alternative manner is to use a large-scale shell-model approach
(for recent reviews, see Refs.~\cite{Brow01,Graw04,Caur05}).
Starting from a core nucleus with experimental SP 
energies,\emph{ effective} two-body interactions are used to describe the
properties of nuclei in a given valence space. Note that in this
model the $NN$ interactions are effective, being different from
the bare (or free) ones. This owes to the fact that, as the core
is never completely inert, it contains some of the interactions.
In addition the nucleons cannot scatter freely to states of the
core which are already occupied by nucleons. These effective
two-body forces are often derived from G-matrix calculations using
realistic microscopic forces, with some empirical readjustments to
experimental data (see for instance Fig.~7 of Ref.~\cite{Brow01}).
They can also be derived from experimental multiplet of states
and/or from the evolution of binding energies. The two-body matrix
elements (TBME) contain all the ingredients, such as the monopole
and multipole energies, which affect the SP energies and the
development of correlations, respectively. The monopoles can also
be broken into central, spin-orbit and tensor forces, as for the
bare $NN$ forces. Doing so one can  eventually derive from which
of these terms the evolution of SP energy comes from. Even though
a systematic comparison between the properties of the realistic and
effective forces has not been done, it is seen in
Ref.~\cite{Schw08} that\emph{ some} of the monopole interactions
obtained from $V_{low k}$ and the phenomenological ones used in
the $fp$ shell exhibit comparable strength and properties
\footnote{A notable exception arises for the neutron-neutron
$V_{\ell j \ell j}$ matrix elements which, in the $V_{low k}$
approach, are much too small as compared to effective forces. This
feature calls for the need of three-body forces, which has already
been shown~\cite{Zuke03} when using realistic interactions such as the
Kuo-Brown ones.}.

To model the evolution of nuclear structure in the whole chart of
nuclides, the self-consistent mean-field calculations are more
often employed. They use effective energy density functionals, the
parameters of which are adjusted phenomenologically to reproduce
many average nuclear properties like the total energy per nucleon,
the size of the nuclei, the nuclear incompressibility... The most
common non-relativistic ones are the Skyrme and the Gogny
interactions. For recent reviews on non-relativistic and
relativistic models, see
Refs.~\cite{Bend03,Ston07,Vret05,Meng06}). Despite the overall
good description of the nuclear properties, two 'problems' remain.
As many parametrizations are existing, this method is not yet
universal. There is need to agree on parameters to be constrained
to chosen experimental data.
In addition, the link between\emph{ ab initio} calculations using
realistic and effective  nucleon-nucleon interactions is not yet
reached. Therefore the sets of  parameters of the effective
forces, determined from the properties of known nuclei, will
probably not be predictive anymore when applied to regions in
which new components of the forces are involved. This explains the
large divergence of predictions for unknown atomic masses using
various forces. A last remark concern the tensor force. Even if it
is intrinsically present in realistic interactions it was, until
recently~\cite{Otsu06,Brow06,Colo07,Lesi07,Zale08}, not considered
in the effective forces due to the complexity to implement it and
the lack of experimental data to constrain it.

This section will introduce first the basic properties of the
nuclear force that are required to understand the following parts
which review all magic numbers from $8$ to $126$. Two
major theoretical approaches to solve the nuclear many body
problem, the mean field and the shell model, will be subsequently
presented. The two-body matrix elements are the cornerstones of
the shell model approach, which will be extensively used in this
review. They govern the evolution of shells through their monopole
part, whereas the onset of deformation is driven by the quadrupole
correlations. Thus means to derive the monopole matrix elements
from the experimental multiplet of states will be shown, followed
by a description on how they modify the SP energies. Finally these
monopoles are broken into central, spin-orbit and tensor terms.
Their effects on the nuclear structure are presented in a
qualitative manner for some examples.

\subsection{\it Properties of two-body forces in nuclear matter}
\subsubsection{\it
Features of the nucleon-nucleon forces\label{general}}

A brief summary of the most important features of the
nucleon-nucleon interaction in nuclear matter are presented below.
More details can be found in many textbooks, such as Refs.
\cite{Bohr69,Ring80,Heyd94}.
\begin{itemize}
\item The nuclear interaction is of short range, of the order of
1~fm. As the size of the nucleus scales with
$R=r_0A^{\frac{1}{3}}$, with r$_0$ $\simeq$ 1.1-1.3 fm, it follows
a saturation of the following nuclear properties~: the binding
energy per nucleon for nuclei with $A>10$ is almost constant and
barely exceeds  8~MeV, the nuclear density remains roughly
constant at 0.16 (nucleon) fm$^{-3}$. 
\item The nearly identical
spectra for mirror nuclei suggest that the nuclear force is
'charge symmetric' (i.e. the proton-proton and neutron-neutron
interactions are equal). 
\item The spectra of nuclei forming
isobar triplets indicate that the nuclear force is 'charge
independent': It is the same for any pair of nucleons,
irrespective of their charges ($\pi\pi$, $\nu\nu$, and $\pi\nu$),
provided that they are in the same configuration. 
\item  Because
of the Pauli principle, two like nucleons 
in the same orbit cannot have
identical quantum numbers, so two-neutron or two-proton
configurations only exist in the $S=0$ total spin state, as a
$S=1$ state would imply that both nucleons have the same spin
projection. Hence the 'charge independence' means that the nuclear
force is the same for any pair of nucleons in the $S=0$ total spin
state. On the other hand, proton-neutron configurations can have
$S=1$. Moreover the nuclear force binds this $S=1$ state to form
the ground state of the deuteron, whereas the three other
configurations with $S=0$, $\pi\pi$, $\nu\nu$, and $\pi\nu$, are
unbound. This feature implies that the nuclear interaction is
'spin dependent'. 
\item Other properties of the deuteron give
important clues on the nuclear interaction. The values of the
angular momentum and of the magnetic moment of its bound state
indicate that, to a first approximation, the intrinsic spins of
the two nucleons are parallel, without contribution from relative
orbital angular momentum (i.e. $L=0$, $S=1$). However the non-zero
quadrupole moment of the deuteron indicates that its ground state
configuration should contain an $L=2$, $S=1$ admixture which
amounts to about 5\%. The fact that $L=0$ is not a good quantum
number for the deuteron implies that the nuclear interaction
cannot be a pure\emph{ central} force, which depends only on the
distance $r$ between the nucleons, but also contains a\emph{ tensor} part, which depends on the relative orientation of the
distance $\overrightarrow{r}$ between the nucleons and their
intrinsic spins. 
\item The nuclear interaction has a\emph{ spin-orbit} component. A decisive step in the development of the
nuclear shell model was the recognition that the assumption of a
relatively strong spin-orbit interaction in the nucleonic motion
leads to a naturel explanation of the major shell closures at 28,
50, 82, and 126 \cite{Goep49,Haxe49}. This component provokes a
large splitting in energy between any two levels having the same
orbital momentum $\ell$ with aligned or anti-aligned intrinsic
spins. The 'aligned' configuration, $j \uparrow = (\ell +
\frac{1}{2})$, is energetically favored, whereas the anti-aligned
one, $j \downarrow (= \ell - \frac{1}{2})$, is at a higher
energy. Later, the occurrence of polarization phenomena measured
in the nucleon-nucleus scattering process has been a direct proof
for the existence of the spin-orbit interaction, 
ascertaining both its sign and order of magnitude
\footnote{Part of the splitting in energy between spin-orbit partners originates
in other terms of the bare $N-N$ interaction, this will be discussed in 
following sections}.
\end{itemize}
A general formulation of\emph{ two-body effective $NN$
interactions} in terms of 'central' and 'non-central' (spin-orbit
and tensor) forces is given in Sect.~\ref{NNcomponent}.

\subsubsection{\it Some theoretical approaches used to solve the nuclear
many-body problem}

Soon after the discovery of the main 
spectroscopic properties, it was realized that a one-body
mean-field potential, representing the average interaction of all
the nucleons, can be used to model the atomic nucleus. One of the
first proposed potential was that of an harmonic oscillator,
complemented by a spin-orbit potential and an $\ell^2$ term which
correct the relatively steep behavior of the harmonic oscillator 
at the surface~\cite{Nils95}. Nowadays self-consistent mean-field 
potentials provide a better description of the stable and far-off 
stability (or exotic) nuclei. The basic concept of transforming 
the many-body problem into a one-body one is briefly depicted below.

Taking only two-body forces into account, the Hamiltonian for $A$
nucleons in interaction is a sum over the kinetic-energy terms
$T(k)$ and the two-body nucleon-nucleon interaction $V(k,\ell)$,
\begin{equation}
H=\sum\limits_{k=1}^A T(k) + \sum\limits_{1=k<\ell}^A V(k,\ell)
\end{equation}

An exact solution of such a many-body problem can rarely be
obtained, except for the lightest masses~\cite{Piep01}. Therefore 
the first step
towards an approximate solution is to introduce a single-particle
potential $U(k)$ by writing the Hamiltonian as :
\begin{equation}
H=\sum\limits_{k=1}^A [T(k)+U(k)] + [\sum\limits_{1=k<\ell}^A
V(k,\ell) - \sum\limits_{k=1}^A U(k)] =H^{(0)}+H^{(1)}
\label{decompHamil}
\end{equation}
where $H^{(0)}$ describes an ensemble of independent particles
moving in an effective average potential $U(k)$ and $H^{(1)}$
corresponds to the residual interactions. 
The very notion of a mean field is fulfilled when $H^{(1)}$ is small.\\

\textbf{The mean field approach}\\
Practically, calculations using a\emph{ realistic} two-body force for
$V(k,\ell)$ in Eq.~\ref{decompHamil} are untractable. This was
attributed to the hard-core part of the $NN$
interaction~\cite{Ring80}.  Therefore the Hartree-Fock (HF) theory
starts with a parameterized\emph{ effective} two-particle
interaction $V(k,\ell)$ to derive the corresponding mean-field
potential which can at best reproduce selected properties of the
atomic nucleus. Results obtained with some effective forces widely
used today, the Gogny and the Skyrme interactions, have been
recently reviewed in Refs.~\cite{Bend03,Ston07}. Similar 
Relativistic Mean Field (RMF) models are based on forces 
originating from the exchange of effective mesons between 
the nucleons~\cite{Vret05,Meng06}.

\begin{figure}[h!]
\begin{center}
\epsfig{file=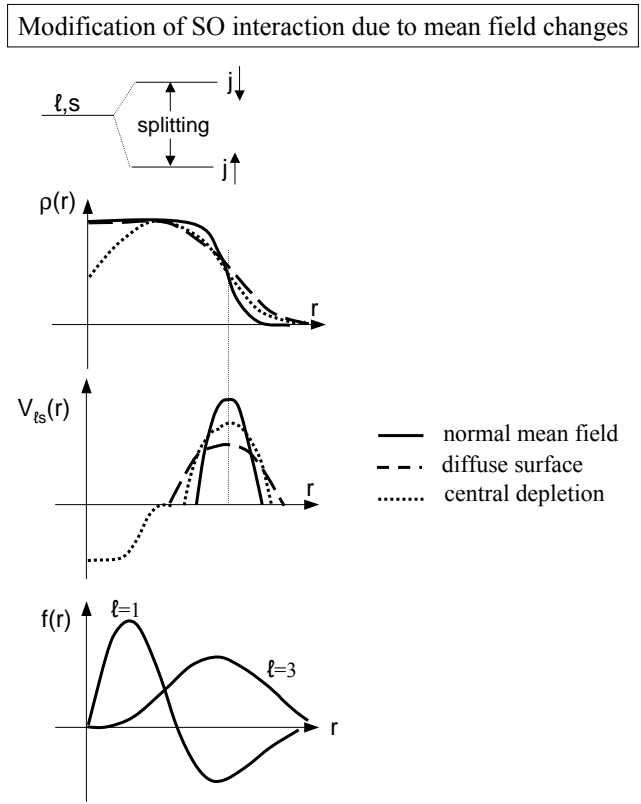,scale=0.5}
\begin{minipage}[t]{16.5 cm}
\caption{Schematic illustration showing the effect of a modification
of the nucleon density $\rho (r)$ on the SO potential $V_{\ell s}
(r)$. The graph on the first row shows various density
distributions for nuclei exhibiting a standard value (full line),
a diffuse surface (dashed line), and a central density depletion
(dotted line). On the second row, the corresponding $V_{\ell s}
(r)$ profiles are shown. The last row shows examples of 
$\ell=1$ and
$\ell=3$ radial wave functions in order to judge how they could be
influenced by surface or central density changes. \label{SOMF}}
\end{minipage}
\end{center}
\end{figure}

Although mean-field theories start from effective interactions
which are not directly connected to 'bare' $NN$ interactions, they
are very helpful to understand some phenomenological aspects of
the nuclear structure. In the following we consider the SO part
of the interaction. Starting from a Skyrme interaction, it can be
shown that the two-body SO interaction leads to one-body SO
potentials for neutrons and protons, $V_{ls}^n(r)$ and
$V_{ls}^p(r)$, respectively, which scale with the derivative of
the neutron $\rho_n(r)$ and proton $\rho_p(r)$ densities 
as~\cite{Vaut72}:
\begin{equation}\label{SOMF1}
V_{\ell s}^n(r)\propto \frac{1}{r} \; \frac{\delta}{\delta r} \;
[2\rho_n(r)+ \rho_p(r)] \; \overrightarrow{\ell} \cdot
\overrightarrow{s},
\end{equation}
\begin{equation}\label{SOMF2}
V_{\ell s}^p(r)\propto \frac{1}{r} \; \frac{\delta}{\delta r} \;
[\rho_n(r)+ 2\rho_p(r)] \; \overrightarrow{\ell} \cdot
\overrightarrow{s},
\end{equation}
These definitions contain a weighting factor in front 
of the proton or neutron densities which is twice as
large when applied to like particles. In particular, 
it follows that any change of neutron density
influences more the neutron SO splitting than the proton density
does. For some parametrisations of Skyrme interaction as well as
for relativistic mean field descriptions, equal weightings between
like and unlike particles are used. This in particular
relates to the iso-vector part of the SO interaction, the
intensity of which is very small in the RMF approaches.
Experimental data are required to justify the choice of these weight
factors. Anyhow, the SO interaction provokes a large splitting in
energy between any two levels having the same orbital momentum
$\ell$ with aligned or anti-aligned intrinsic spins, as shown in
the upper part of Fig.~\ref{SOMF}. As the SO interaction scales
with the derivative of the nucleon densities, it is believed
to be peaked at the surface where this derivative is
maximum. It naturally follows that, with the advent of new
radioactive ion beams facilities, searches for possible reductions
of the SO interaction have been oriented to very neutron-rich
atomic nuclei where the surface is more diffuse and the
derivative of the nuclear density is weaker~\cite{Doba94}. So far
these attempts were not conclusive enough, as the effect is
substantial only for large N/Z nuclei.

It should be noted that a depletion of the density at the center of the
nucleus should also induce a reduction of the SO splitting for the
nucleons in vicinity. The derivative of the density in
the interior of the nucleus is of the opposite sign to the one
obtained at the surface. It follows a global reduction of the SO
interaction for nucleons located in the center. As the centrifugal
barrier prevents nucleons of high orbital momentum $\ell$ to
reside there, such a reduction of the SO splitting would be seen
for the lowest $\ell$-values, i.e. mainly for $\ell$=1 ($p$
states). The reduction of the $\nu p_{1/2} - \nu p_{3/2}$ SO
splitting as $s_{1/2}$ protons are removed from the interior of
the nucleus is taken as an illustrative example of this phenomenon
in Sect.~\ref{nuf7f5}.

Pictorial descriptions of the aforementioned mean-field effects
which modify the strength of the SO interaction are given in
Fig.~\ref{SOMF}. Such changes appear in a self-consistent manner
in the mean-field description, as soon as the densities are 
modified. As will be discussed in Sect.~\ref{decompmonopol},
additional modifications of the SO\emph{splitting} are also
expected from the tensor 
interaction\footnote{Several attempts have been recently made to 
include tensor terms in mean-field
approaches~\cite{Otsu06,Brow06,Colo07,Lesi07,Zale08}}. This
points to the fact that, if a reduced SO\emph{splitting}
is found, this does not necessarily imply a change of the
SO one-body potential in the terms depicted here.

\textbf{The Shell-model approach}\\
In the interacting shell-model (SM) approaches (see for
instance~\cite{Brow01,Graw04,Caur05}), the spherical mean field
produced by the closed core -the $H^{(0)}$ part- is a standard
phenomenological single-particle potential given by an harmonic
oscillator. Configuration mixings between all valence
nucleons outside the core -interacting through the $H^{(1)}$ part-
are taken into account for describing nuclei in a given model
space. For the large-scale SM calculations, the effective
interactions can be derived from the free $NN$ interaction (various
ways can be found in~\cite{Dean04}), i.e. the so-called\emph{ realistic
interactions}. They can also be
obtained in a more phenomenological manner, combining results from
the free $NN$ interaction and from fits on selected experimental
data~\cite{Brow01,Graw04,Caur05}. 

The numerical values of some two-body matrix elements (TBME) of 
the\emph{ effective $NN$ forces} can be directly determined from the
experimental data, as described in the following section. The sets
of TBME can be subsequently decomposed into multipole expansion,
monopole, dipole, quadrupole, ... (the formula are given in
Ref.~\cite{Moin69}). The monopole part implicitly
contains central, spin-orbit and tensor terms, the intensity of
which are discussed in Sect.~\ref{decompmonopol}.

\subsection{\it Two-Body Matrix Elements and determination of 
monopole interaction}

The Hamiltonian of a nucleus formed by an inert core and two
orbiting nucleons in the valence shells can be split into two
parts, $H=H_{core}+H_{12}$. Using the same notations as those of
Eq.~\ref{decompHamil},
\begin{equation}
H_{core}=\sum\limits_{k=3}^A [T(k)+U(k)] +
[\sum\limits_{3=k<\ell}^A V(k,\ell) - \sum\limits_{k=3}^A U(k)],
\end{equation}
\begin{equation}
H_{12}=\sum\limits_{k=1}^2 [T(k)+U(k)] + [\sum\limits_{k=1}^2
\sum\limits_{\ell=3}^A V(k,\ell) + V(1,2) - \sum\limits_{k=1}^2
U(k)],
\end{equation}
$H_{core}$ refers to the interactions between the core particles
(labeled by $k=3, ..., A$). $H_{12}$ describes the contribution
from the two additional particles. It can be written in the
following way, $H_{12}=H_{12}^{(0)}+H_{12}^{(1)}$ where
$H_{12}^{(0)}=[T(1)+U(1)]+[T(2)+U(2)]$ comprises the
single-particle kinetic and potential energies felt by particles 1
and 2 while $H_{12}^{(1)}$ denoted the residual interaction,
\begin{equation}
H_{12}^{(1)}= [\sum\limits_{\ell=3}^A V(1,\ell)-U(1)]+
 [\sum\limits_{\ell=3}^A V(2,\ell)-U(2)] + V(1,2). \label{H12}
\end{equation}
These equations hold for any average field $U(k)$. By choosing
$U(k)=\sum\limits_{\ell=3}^A V(k,\ell)$, it follows that
\begin{equation}
H_{12}^{(1)} = V(1,2). \label{H12_V12}
\end{equation}
since only the two-particle term $V(1,2)$ remains in
Eq.~\ref{H12}. In other words, information on the properties of
the $NN$ interaction, $V(k,\ell)$, can in principle be obtained by
analyzing the contribution of the two particles to the total
energy of nuclei composed by an\emph{ inert core 
plus 2 nucleons}.
The same arguments can be applied to other nuclei, provided that
their corresponding cores are inert enough to isolate the
contribution of the two valence particles. In the following section
typical examples showing how to extract these TBME are given.

\subsubsection{\it TBME from multiplet of states in odd-odd nuclei
and monopole interaction\label{expmultiplet}}

As an example for determining $\pi-\nu$ matrix elements, we 
consider the ($Z+1,N+1$) odd-odd nucleus formed by the ($Z,N$)
core and two extra nucleons located in the $j_\pi$ and $j_\nu$
orbits respectively~\footnote{Another example using a ($Z-1,N+1$)
odd-odd nucleus will be given later. This procedure
can be generalized to any system comprising a core $\pm$ one
neutron $\pm$ one proton.}. An empirical eigenvalue of the
truncated Hamiltonian, $H_{core}+H_{12}^{(0)}$, (i.e. without
residual interaction) is given by $M_{free} c^2$, with
\begin{equation}
M_{free}(Z+1,N+1)=M(Z,N)+M_\pi+M_\nu
\end{equation}
\noindent The contributions of the two extra nucleons, $M_\pi$ and
$M_\nu$, are added free of any interaction to the measured mass
$M(Z,N)$ of the even-even core. The $M_\pi$ and $M_\nu$ terms are
determined from the odd-A neighboring nuclei as,
\begin{equation}
M_\pi=M(Z+1,N)-M(Z,N),~~ M_\nu=M(Z,N+1)-M(Z,N).
\end{equation}
The empirical value $M_{free}(Z+1,N+1)$ differs from the measured
mass of the odd-odd nucleus $M_{exp}(Z+1,N+1)$ because of the
residual interaction provided by $H_{12}^{(1)}$ (or V(1,2), see
Eq.~\ref{H12_V12}) between the two added particles.

\begin{figure}[h!]
\begin{minipage}{9.cm}
\begin{center}
\epsfig{file=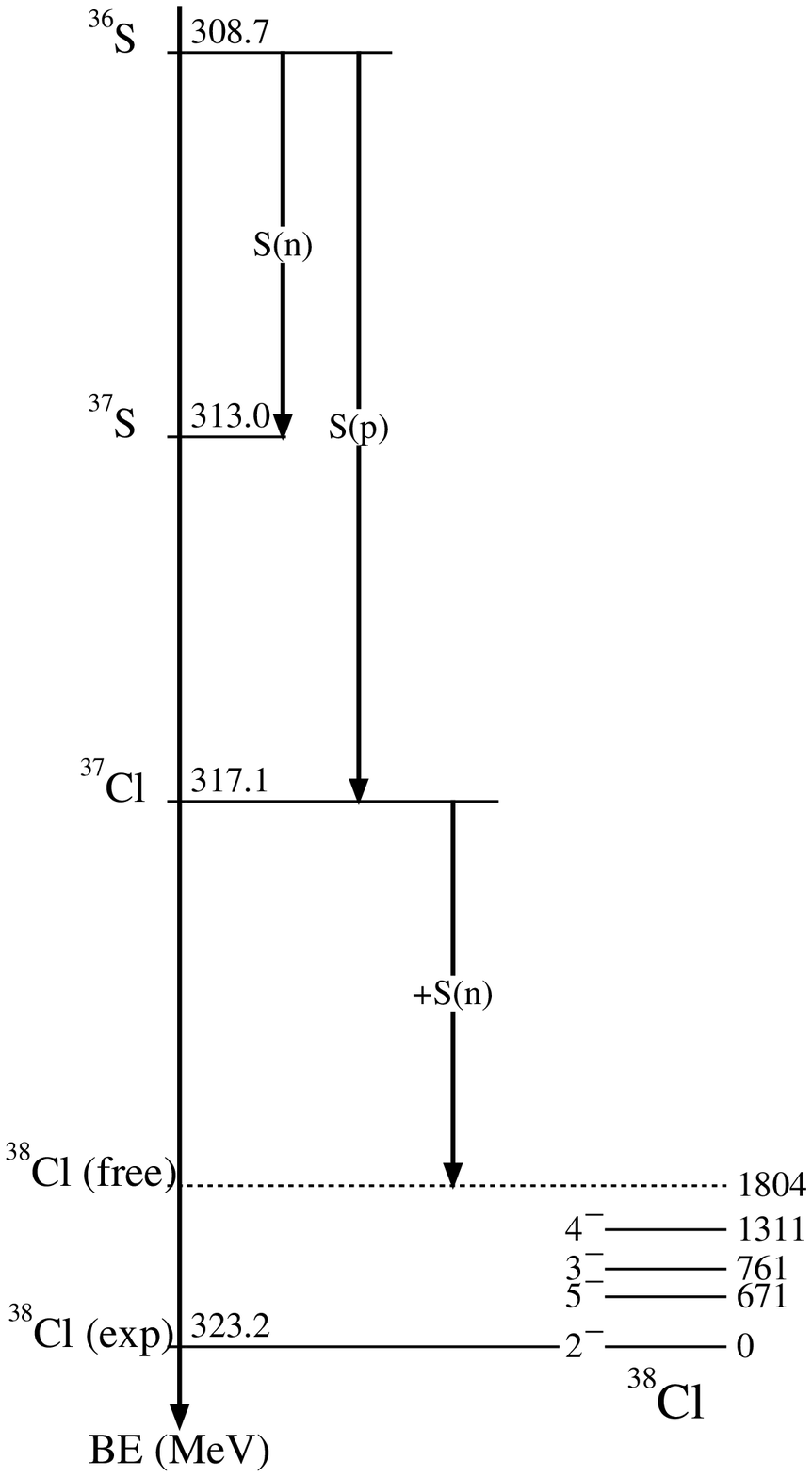,scale=0.55}
\end{center}
\end{minipage}\hfill
\begin{minipage}{9.cm}
\begin{center}
\epsfig{file=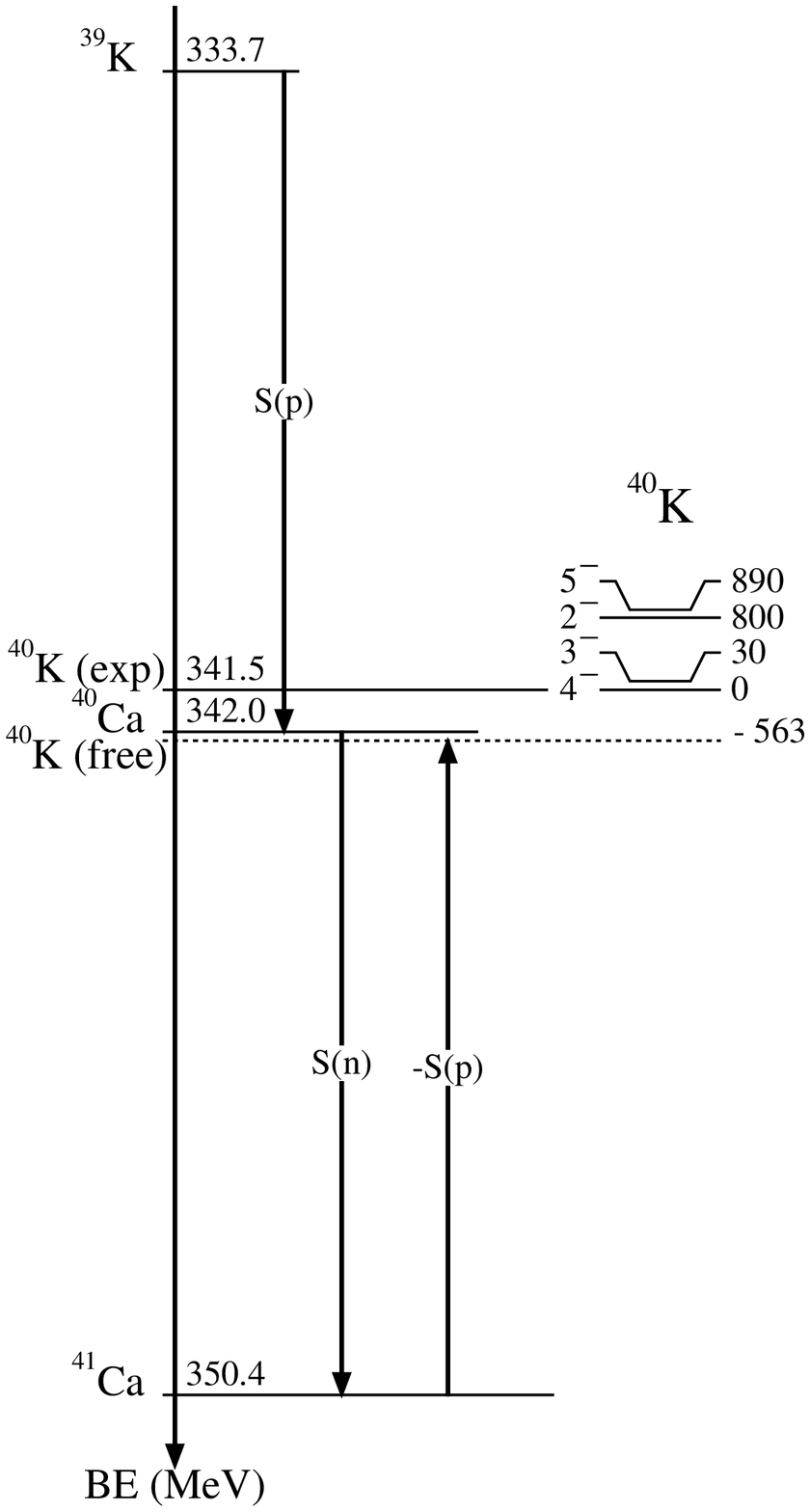,scale=0.55}
\end{center}
\end{minipage}
\begin{center}
\begin{minipage}[t]{16.5 cm}
\caption{Pictorial illustration of the empirical $\pi-\nu$
interaction energies in two cases, the $(\pi d_{3/2})^{+1}(\nu
f_{7/2})^{+1}$ configuration in $^{38}_{17}$Cl$_{21}$ starting
from the $^{36}_{16}$S$_{20}$ closed-shell core (left) and the
$(\pi d_{3/2})^{-1}(\nu f_{7/2})^{+1}$ configuration in
$^{40}_{19}$K$_{21}$ from the $^{40}_{20}$Ca$_{20}$ closed-shell
core (right). \label{IntResEmpiric}}
\end{minipage}
\end{center}
\end{figure}

The left part of Fig.~\ref{IntResEmpiric} illustrates how 
the TBME arising from the $\pi d_{3/2} - \nu f_{7/2}$
interaction in the case of $^{38}_{17}$Cl$_{21}$ are extracted.
Starting from the binding energy of the closed-shell core,
$^{36}_{16}$S$_{20}$, one adds successively contributions of the
odd $\pi d_{3/2}$ and $\nu f_{7/2}$ particles, the binding
energies of which are derived from the $^{37}$Cl and $^{37}$S
nuclei, respectively. When applying this method we obtain a value
denoted $^{38}$Cl(free). It is less bound by 1804~keV than the
measured ground state, $^{38}$Cl(exp). The experimental
$^{38}$Cl spectrum contain several states, with spin values
ranging from $J=|j_\pi - j_\nu|=2$ (the ground state) to $J=j_\pi
+ j_\nu=5$, originating from the $\pi d_{3/2} - \nu f_{7/2}$
configuration. Hence the $\pi-\nu$ residual interaction lifts the
degeneracy in energy of the multiplet of $J$ states, in addition
to provide a\emph{ global} gain of binding energy.

\begin{figure}[h!]
\begin{center}
\epsfig{file=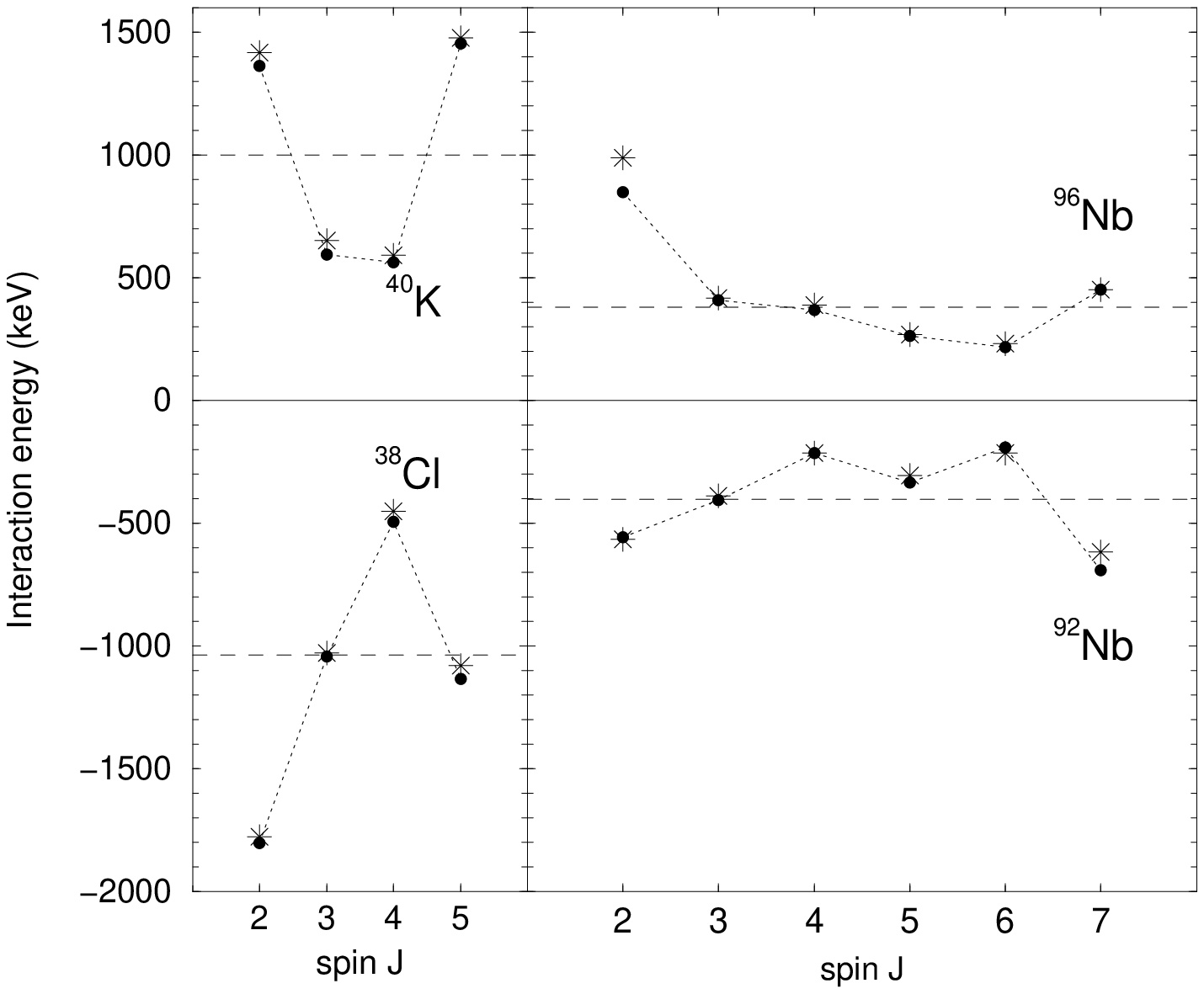,scale=0.6}
\begin{minipage}[t]{16.5 cm}
\caption{Empirical interaction energy in two configurations as a
function of angular momentum: $\pi d_{3/2} \nu f_{7/2}$ extracted
from the level schemes of $^{38}_{17}$Cl$_{21}$ ($pp$ excitation)
and $^{40}_{19}$K$_{21}$ ($ph$ excitation), $\pi g_{9/2} \nu
d_{5/2}$ from $^{92}_{41}$Nb$_{51}$ ($pp$) and
$^{96}_{41}$Nb$_{55}$ ($ph$). The results of the Pandya
transformation of the four experimental sets (filled circles) are
drawn with stars. The dashed lines indicate the values of the
monopole contribution to the interaction energies of each
configuration. \label{IntRes_examples1}}
\end{minipage}
\end{center}
\end{figure}

The values of the TBME, $E_J(j_\pi,j_\nu)$, for $^{38}$Cl are
reported as a function of $J$ in the left part of
Fig.~\ref{IntRes_examples1}. In this case of two  particles 
($pp$ interaction\footnote{To avoid possible confusion of 
particle-particle interaction with proton-proton interaction, 
we use the following notations, $p$ stands for particle and 
$\pi$ for proton.}), the
interaction is attractive and the TBME values are negative. The
largest amounts to -1804~keV for 2$^-$ state, whereas the smallest
is -493~keV for the 4$^-$ one. The dashed line drawn in this
figure shows the mean energy of the two-particle interaction,
$V^{pn}_{j_\pi j_\nu}$, averaged over the relative orientations,
$J$, of the orbits. In this procedure, each $E_J$ is weighted by
its $(2J+1)$ degeneracy, that is the number of magnetic sub-states:
\begin{equation}\label{monopole_val}
V^{pn}_{j_\pi j_\nu} = \frac{\sum (2J+1)\times E_J(j_\pi,j_\nu)}
{\sum (2J+1)}.
\end{equation}
This mean energy corresponds to a certain proton-neutron 
interaction, the so-called\emph{ monopole} term. It does not
depend on the $J$ value anymore, and can be written as
$V^{pn}_{d_{3/2} f_{7/2}}$ in the present case.

The monopole term corresponds to the mean energy brought to the
nucleus by the addition of two interacting nucleons, irrespective 
of the
relative orientation of their orbits. It contains the fundamental
properties of the $NN$ interactions which influence the
evolution of shells closures~\cite{Dufo96}, as described in
Sect.~\ref{monopole}. Noteworthy is the fact that for the two
extreme values of $J$, corresponding to states in which the two
nucleon orbits are coplanar (here, $J=5$ when $j_\pi$ and $j_\nu$
have the same direction and $J=2$ when $j_\pi$ and $j_\nu$ have
opposite directions), the gain in energy exceeds that of the sole
monopole. Conversely, when nucleons are orbiting orthogonal to
each other, their TBME is the weakest. This $J$-dependent part of
the two-body matrix elements is contained in the multipole
components (dipole, quadrupole, ....). The method of
decomposition is given in Ref.~\cite{Moin69}. It is important to note
that the quadrupole term usually has, after the monopole, the
highest amplitude and it is responsible for the parabolic behavior of
$E_J(j_\pi,j_\nu)$ as a function of $J$.

A similar procedure can be applied to determine the same monopole
interaction from a ($Z-1,N+1$) odd-odd nucleus formed by a ($Z,N$)
core, one proton hole in the $j_\pi$ orbit and one neutron
particle in the $j_\nu$ orbit. The right part of
Fig.~\ref{IntResEmpiric} shows the case of $^{40}_{19}$K$_{21}$.
Starting from the binding energy of the closed-shell core
$^{40}_{20}$Ca$_{20}$, adding the contribution of the $\nu
f_{7/2}$ particle neutron from $^{41}$Ca and subtracting the one
of the $\pi d_{3/2}^{-1}$ hole proton from $^{39}$K, we obtain the
value denoted $^{40}$K(free). Owing to the $\pi-\nu$ interaction,
the $^{40}$K states are found to be less bound by a quantity
varying from 563 keV for the 4$^-$ state to 1443 keV (=563 + 890)
for the 5$^-$ state. Here, the residual interaction between one
particle and one hole ($ph$) is mainly repulsive. The values of
the two-body matrix elements, $E_J(j_\pi,j_\nu)$, are positive for
the $ph$ configuration (as shown in Fig.~\ref{IntRes_examples1}).

In summary, the TBME of the $\pi d_{3/2} - \nu f_{7/2}$
interaction were determined from the $pp$ and $ph$ configurations
in the $^{38}_{17}$Cl$_{21}$ and $^{40}_{19}$K$_{21}$ nuclei. 
It is interesting to compare the values derived in both cases.
Using the so-called Pandya transformation \cite{Moin69,Heyd94} the
$pp$ matrix elements can be converted into $ph$ ones, and
vice-versa. The result of these transformations are indicated by
stars in Fig.~\ref{IntRes_examples1}, whereas the experimental
points are shown as the filled circles. The very good agreement
between the TBME extracted from the $pp$ or $ph$ configurations
ascertains the reliability of the procedure, i.e. both $^{36}$S
and $^{40}$Ca can be considered as good cores.

The same procedure has been applied to the $\pi g_{9/2} - \nu
d_{5/2}$ interaction, using experimental results of
$^{92}_{41}$Nb$_{51}$ ($pp$ configuration) and
$^{96}_{41}$Nb$_{55}$ ($ph$ configuration). Similar agreement is
obtained using the Pandya transformation.  The only
exception is for the 2$^+$ state of $^{96}$Nb whose energy would
be expected 150~keV higher than that measured. This state is likely to
be mixed to nearby 2$^+$ state of another configurations,
leading to a decrease of its excitation energy.

\subsubsection{\it Monopole interaction from double difference of
binding energies\label{expmonopole}}

The method depicted in the previous paragraph  is extremely
powerful to determine both the monopole and the other multipole
terms. However it requires that all the $J$ values of
$E_J(j_\pi,j_\nu)$ are experimentally known, which is seldom the
case. An alternative way to determine the monopole term is to
use differences in binding energies between neighboring nuclei.

Let us consider an inert core of closed shells, the $(Z,N)$
nucleus, and its three neighbors, the $(Z+1,N)$, $(Z,N+2)$, and
$(Z+1,N+2)$ nuclei. The difference in the binding energies of the
$(Z+1,N)$ nucleus and the core is due to the interaction of the
added proton with the core. In the same manner, the difference in
the binding energies of the $(Z,N+2)$ nucleus and the core is due
to the interaction of the two added neutrons with the core.
Lastly, the difference in the binding energies of the $(Z+1,N+2)$
nucleus and the core is due to three terms, the interactions of
the extra proton with the core, of the two extra neutrons with the
core, and of the extra proton with the two extra neutrons.
Therefore the interaction of the extra proton and two neutrons,
$V^{1p-2n}$, is given by the differences,
\begin{equation}\label{doublediff}
V^{1p-2n}=[BE(Z+1,N+2)-BE(Z,N)]-[BE(Z+1,N)-BE(Z,N)]-[BE(Z,N+2)-BE(Z,N)]
\end{equation}
\begin{equation}
= [BE(Z+1,N+2)-BE(Z,N+2)]-[BE(Z+1,N)-BE(Z,N)]
\end{equation}
Assuming that the proton and the two neutrons occupy the $j_\pi$
and $j_\nu$ single-particle orbits, respectively, we obtain
formally
\begin{equation}
V^{1p-2n}= 2 \; V^{pn}_{j_\pi j_\nu},
\end{equation}
where $V^{pn}_{j_\pi j_\nu}$ is the same as in
Eq.~\ref{monopole_val}. Indeed all the nuclei involved in
Eq.~\ref{doublediff} have an even number of neutrons. Therefore
the single proton interacts with paired neutrons without any
privileged axis, meaning that the $V^{1p-2n}$ value is implicitly
averaged on all the possible orientations.

\begin{figure}[h!]
\begin{center}
\epsfig{file=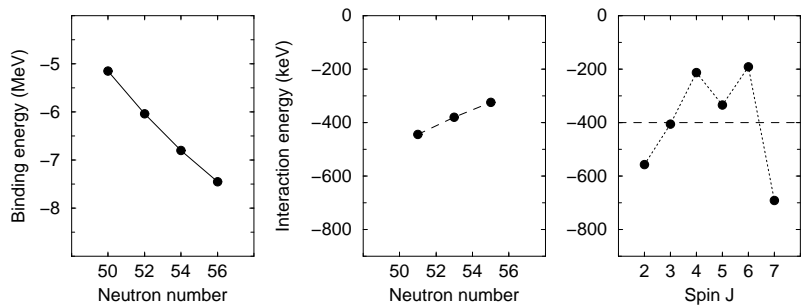,scale=0.6}
\begin{minipage}[t]{16.5 cm}
\caption{\textbf{Left} : Experimental binding energies of the $\pi
g_{9/2}$ orbit in the odd-A $_{41}$Nb isotopes as a function of
neutron number. \textbf{Middle} : Mean proton-neutron interaction
extracted from the slopes of the proton binding energies.
\textbf{Right} : Monopole interaction $V^{pn}_{g_{9/2} d_{5/2}}$ 
(dashed line) from
the multiplet of states in $^{92}$Nb (see caption of
Fig.~\ref{IntRes_examples1} for details). \label{Sp_Nb}}
\end{minipage}
\end{center}
\end{figure}

This method is applied in the case of the odd-Z $_{41}$Nb isotopes
for $N=50-56$. The experimental binding energies of the last
proton, located in the $\pi g_{9/2}$ orbit, display an almost
constant decrease as a function of the number of neutrons filling
the $\nu d_{5/2}$ orbit (see the left part of Fig.~\ref{Sp_Nb}).
The values of the proton-neutron interaction extracted from the
slopes of these binding energies, which amount to $\sim$ -400~keV,
are given in the middle part of Fig.~\ref{Sp_Nb}. As previously
mentioned, the multiplet of the $\pi g_{9/2} \nu d_{5/2}$ states
observed in the odd-odd $^{92}$Nb provided a monopole value as
well, and is given in the right part of Fig.~\ref{Sp_Nb}. The two
methods compare reasonably well.

This example illustrates the fact that the monopole term
$V^{pn}_{g_{9/2} d_{5/2}}$ can be obtained from the shift of the
'experimental' single-proton energy, provided that it displays a
quasi-linear behavior for the whole interval of the $\nu d_{5/2}$
filling, from $N=50$ to $N=56$. Deviations to the linear trend are
seen by the slight variation of the monopole values drawn in the
middle part of Fig.~\ref{Sp_Nb}. It is seen from this example that the
evolution of single-particle levels and subsequently of shell gaps
are, to a first order, governed by monopole interactions. This
feature is discussed in more details in the next sections.

\subsection{\it Properties of the monopole interaction\label{monopole}}
\subsubsection{\it Effects of the monopole 
interaction\label{monopoleffect}}

{\bf Case of unlike nucleons}\\
The forthcoming discussion illustrates the
effects of monopole proton-neutron interactions to modify 
two neutron binding energies, $\epsilon_{n1}$ and 
$\epsilon_{n2}$, due to the
addition of protons in two successive shells.
In this example, 
the value of the monopole $V^{pn}_{j_{p1}j_{n2}}$ is chosen to be 
significantly more attractive than the three others, 
$V^{pn}_{j_{p1}j_{n1}} \simeq
V^{pn}_{j_{p2}j_{n1}} \simeq V^{pn}_{j_{p2}j_{n2}}$.

\begin{figure}[h!]
\begin{center}
\epsfig{file=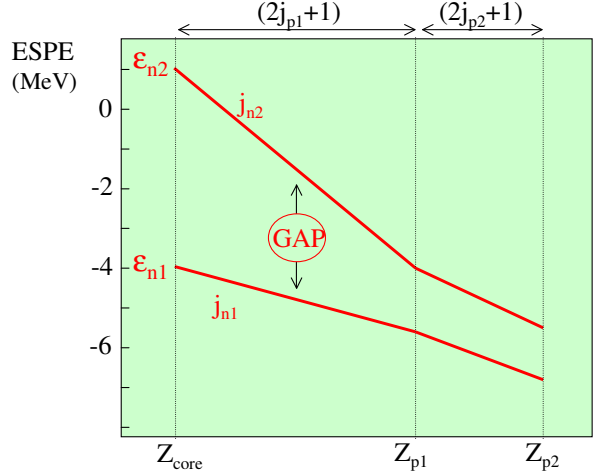,scale=0.5}
\begin{minipage}[t]{16.5 cm}
\caption{Evolution of effective spherical single-particle energy
(ESPE) of two neutron states $n1$ and $n2$ due to the addition of
protons in the $j_{p1}$ and $j_{p2}$ shells.
Starting from energies $\epsilon_{n1}$ and $\epsilon_{n2}$ of the
core nucleus $Z_{core}$, the monopole interactions involved while
filling the $j_{p1}$ orbit give rise to a steep increase
of binding energy of $\epsilon_{n2}$ and a large reduction of the
neutron gap 'GAP'. \label{ESPE}}
\end{minipage}
\end{center}
\end{figure}

Within the Shell-Model approach, the monopole interaction induces
a shift of the effective\emph{ spherical} single-particle energy
(ESPE), $\epsilon_{ni}$, of the neutron orbits
${ni,j_{ni},\ell_{ni}}$ by the mean field generated by protons
added to an inert core $Z_{core}$. Starting from the energy
$\epsilon_{n1}$ of a potentially occupied state
$j_{n1}$, the variation of ESPE is linear with the
monopole proton-neutron interaction $V^{pn}_{j_{p1}j_{n1}}$ when
$x$ protons occupy the $j_{p1}$ orbit ($0 < x < 2j_{p1} + 1$) :
\begin{equation}\label{shiftE}
\Delta \epsilon_{n1} = x \; V^{pn}_{j_{p1}j_{n1}}
\end{equation}
The same relation applies to the variation of $\epsilon_{n1}$ during the filling
of the $j_{p2}$ orbit, as well as the variation of $\epsilon_{n2}$
as protons are added in the $j_{p1}$ and $j_{p2}$ orbits 
(see the schematic view in
Fig.~\ref{ESPE}). From the
trends of ESPE, the variation of the neutron gap between
$Z_{core}$ and $Z_{p1}$ can be expressed as the difference of the
involved monopoles:
\begin{equation}\label{GAP}
\Delta (GAP) =  (2j_{p1} + 1)\; (V^{pn}_{j_{p1}j_{n2}} -
V^{pn}_{j_{p1}j_{n1}})
\end {equation}
As exemplified in Fig.~\ref{ESPE}, the neutron gap is
weakened between $Z_{core}$ and $Z_{p1}$ and remains constant
later. These features arise from the fact that the
monopole $V^{pn}_{j_{p1}j_{n2}}$ has been chosen to be significantly 
more attractive
than the three others. The presence
of one dominating monopole term compared to many others with a small
intensity, is widespread throughout the chart of nuclides. It
arises in particular in the $N=20$ isotones, where
$V^{pn}_{d_{3/2} d_{5/2}}$ is significantly larger than any other
monopole in the $sd$ valence space (this is discussed in 
Sect.~\ref{disappearN20}).

When the 'experimental' binding energies (BE) display a linear
behavior for a wide interval, the corresponding monopole matrix
elements can be derived from the slopes of the experimental BE as
shown in Fig.~\ref{Sp_Nb}. The analysis of the binding
energies of the last proton (neutron) in semi-magic nuclei will be
largely used in this paper to pin down the role of the
residual interactions, particularly those dealing with certain
specific proton-neutron configurations only available in
exotic nuclei. However we should bear in mind the following points:
\begin{itemize}
\item  Any observed curvature in the experimental BE trend
indicates that either the core nucleus cannot be considered as
inert anymore or/and that mixing between added nucleons is
present. In particular when several orbits are close in
energy, they are filled together because of the pairing
correlations between them. Then the evolution of ESPE arises from
an averaged monopole interactions of the degenerate orbits,
weighted by their occupation number. 
\item This representation in
term of ESPE provides an oversimplified view of the global
evolution of nuclear shells due to $NN$ interactions. In the
relativistic and non-relativistic mean-field (MF) approaches a
progressive re-arrangement of the mean field potential occurs.
Therefore direct comparison between ESPE derived from the shell
model approach and the trends on binding energy obtained with MF
calculations cannot be made in a straightforward way.
\end{itemize}

{\bf Case of like nucleons}\\
Before closing this section we need to add a few remarks on the
strength of monopoles between\emph{ like} nucleons, and on how
they modify the shell gaps. The major difference with the interaction
between\emph{ unlike} nucleons is a consequence of the Pauli 
principle and the antisymmetrisation of the two-nucleon wave
function, which requires the space-spin part of the wave function to be 
antisymmetric.

\begin{figure}[h!]
\begin{minipage}{9cm}
\begin{center}
\epsfig{file=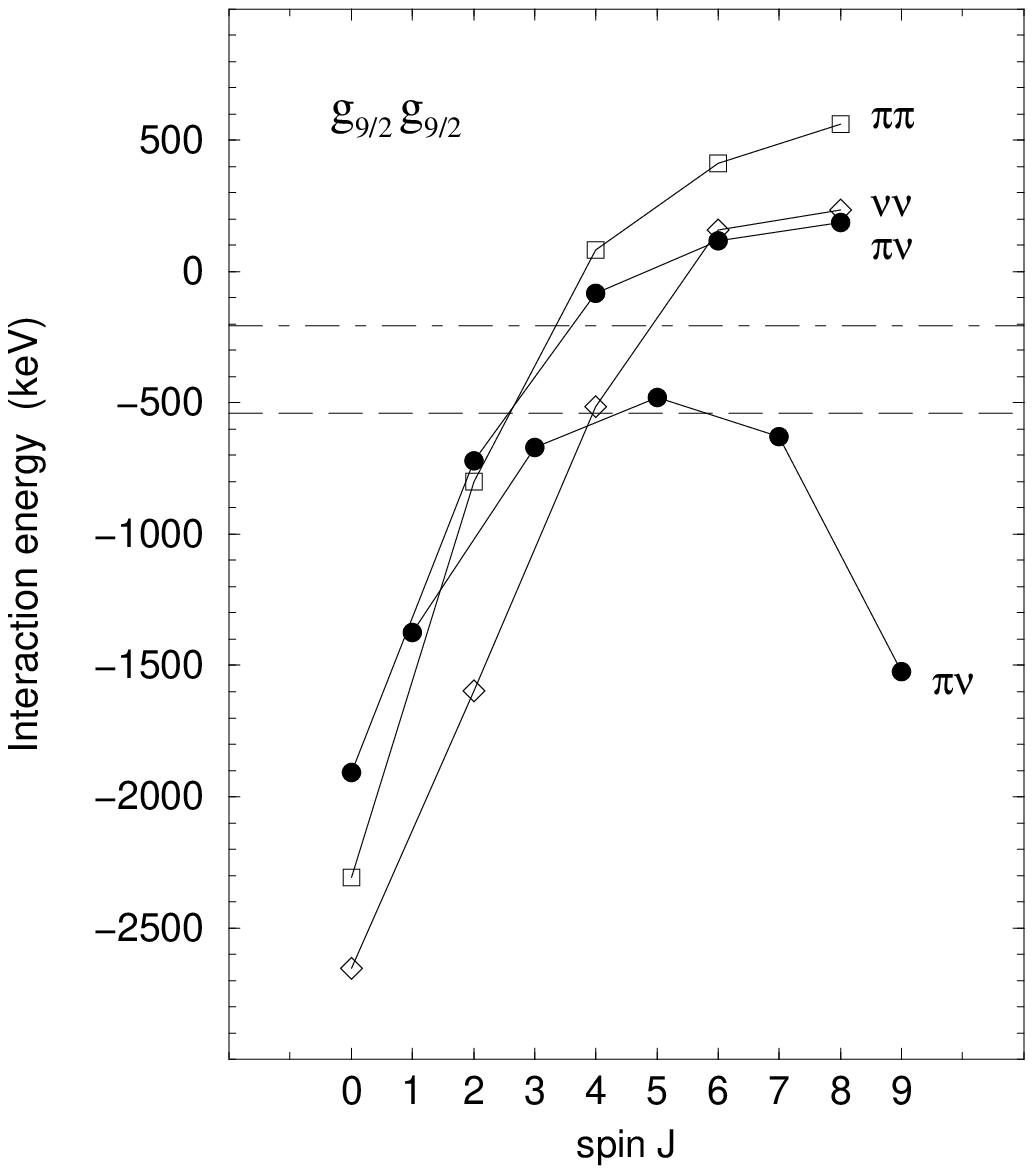,scale=0.55}
\end{center}
\end{minipage}\hfill
\begin{minipage}{9cm}
\begin{center}
\epsfig{file=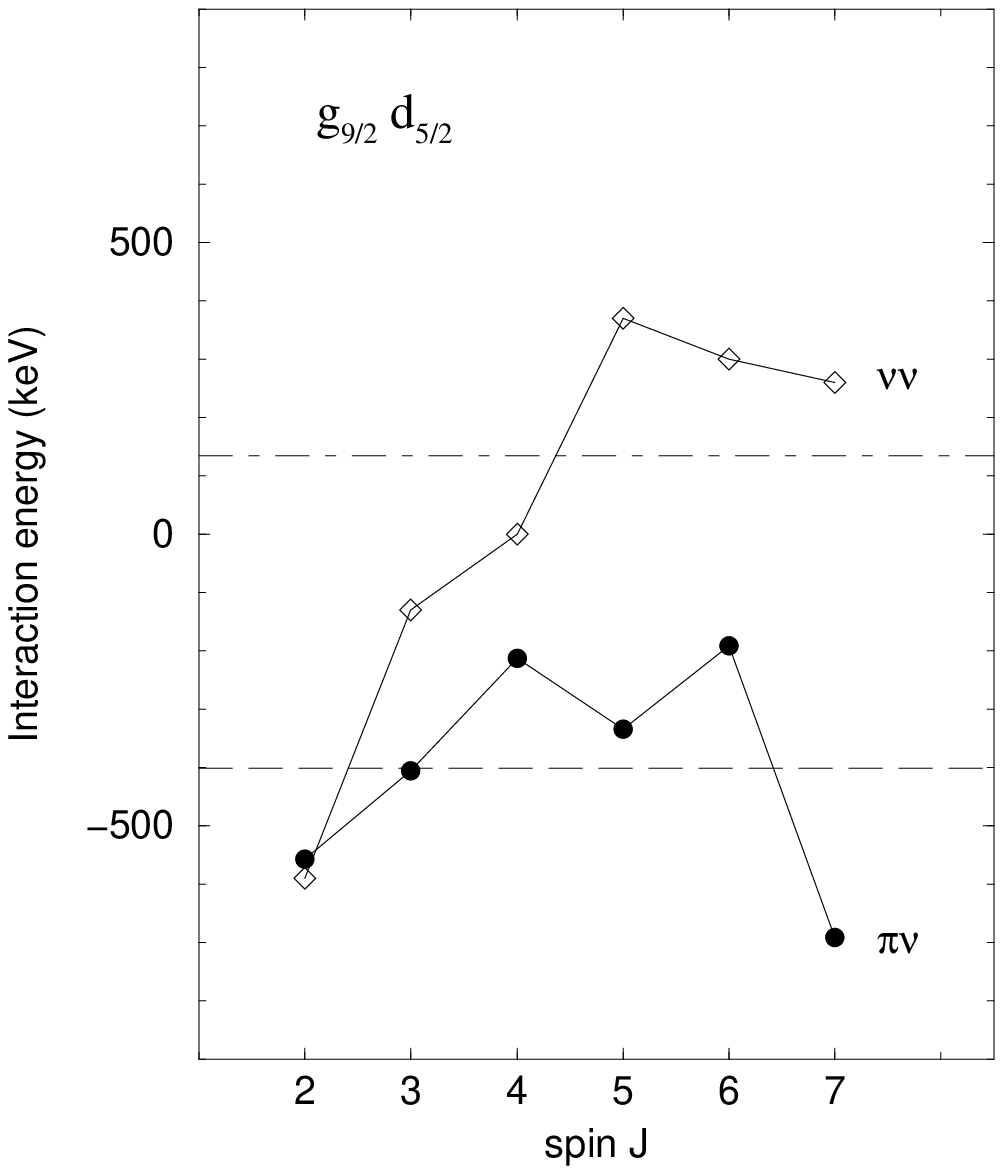,scale=0.55}
\end{center}
\end{minipage}
\begin{center}
\begin{minipage}[t]{16.5 cm}
\caption{Empirical interaction energy between two nucleons in the
same orbit (left) and in two different orbits (right).
For the $g_{9/2}^2$ case, results for the three configurations are presented, $\pi \pi$
from $^{92}_{42}$Mo, $\nu \nu$ from $^{88}$Zr$_{48}$ and 
$\pi \nu$ from $^{90}_{41}$Nb$_{49}$. For the $g_{9/2} d_{5/2}$ case, the $\pi \nu$ results
come from $^{92}_{41}$Nb$_{51}$ and the $\nu \nu$ ones from $^{90}$Zr$_{50}$.
The dashed line indicate the
monopole energy obtained for the\emph{ unlike} nucleons and 
the dashed-dotted line, the one for\emph{ like} nucleons.
\label{pn_nn}}
\end{minipage}
\end{center}
\end{figure}

When the two\emph{ like} nucleons occupy the same orbit $j$, the 
only allowed $J$ states are those with even values, 
$J=0, 2, 4, ... (2j-1)$. 
Conversely the wave functions with odd values of $J$ 
(1, 3, 5, ... 2$j$), which are allowed for two\emph{ unlike} nucleons, 
have a symmetric space-spin part. As an
example, the case of two nucleons in the $g_{9/2}$ orbit is shown 
in the left part of Fig.~\ref{pn_nn}. 
Their interaction energies as a function of $J$
have been extracted from $^{90}_{41}$Nb$_{49}$, $^{92}_{42}$Mo 
and $^{88}$Zr$_{48}$, using the relations given in Sect.~\ref{expmultiplet}.    
The interaction energies obtained for the even $J$ values are very 
similar irrespective of the nucleons concerned, 
$\pi \pi$, $\nu \nu$ or $\pi \nu$. 
In particular a very favored configuration corresponds to the 
pairing situation $J_{min}=0$: Here, nucleons move in opposite 
directions in coplanar orbits. Another very favored configuration 
corresponds to $J_{max}=9$, such a total $J$ value can be only 
obtained when the two nucleons have exactly the same quantum 
numbers (classically, they move in the same direction in coplanar 
orbits). As mentioned above, such a situation is only allowed 
for\emph{ unlike} nucleons. 

As shown in the left part of Fig.~\ref{pn_nn}, there is a large difference in the
interaction energies for the high-$J$ values, the odd ones
displaying a larger attraction. This is also a consequence of 
the Pauli principle. To attain a high-$J$ value, the 
two $j$ momenta have to be close in direction, this implies nearly 
coplanar orbits in which the two nucleons orbit in the same 
direction. The condition of an antisymmetric space-spin wave function 
for\emph{ even} $J$ can be viewed classically as a rotation out of 
phase, so the two nucleons always remain at large distance and 
have a very weak interaction~\cite{Cast90}.  

The major consequence of the Pauli principle in this
context is that the monopole interaction between\emph{ like} particles
is about two times weaker than those between\emph{ unlike} particles,
as shown in the case of two interacting nucleons in
$g_{9/2}$ orbit (dashed and the dashed-dotted lines 
in the left part of Fig.~\ref{pn_nn}).

The right part of Fig.~\ref{pn_nn} displays
another case, the interaction energies between two particles
occupying two different orbits, $g_{9/2}$ and 
$d_{5/2}$. The interaction energies for $\pi \nu$ configuration
come from $^{92}_{41}$Nb$_{51}$ and those for the $\nu \nu$ 
configuration from excited states of
$^{90}$Zr$_{50}$~\cite{Daeh83}. Whereas the interaction between the
two\emph{ unlike} particles is attractive for all spin values,
resulting in a monopole value of -400 keV 
(see the dashed line in the right part of Fig.~\ref{pn_nn}), the 
other one is mainly repulsive with a positive monopole value 
of 130 keV (see the dashed-dotted line). 

\begin{figure}[h!]
\begin{minipage}{9cm}
\begin{center}
\epsfig{file=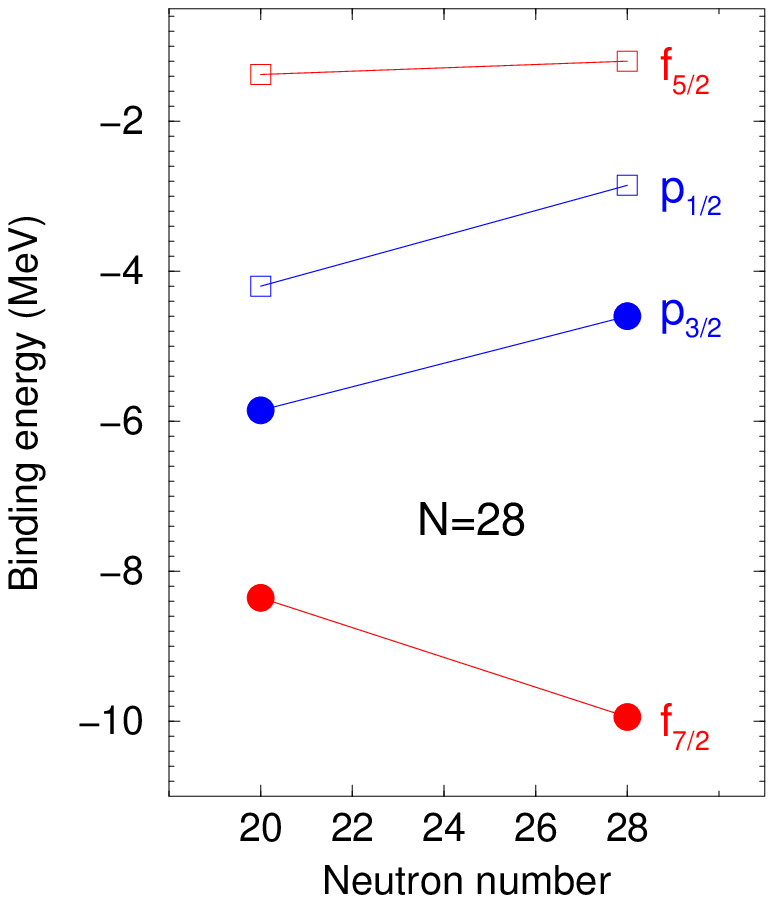,scale=0.55}
\end{center}
\end{minipage}\hfill
\begin{minipage}{9cm}
\begin{center}
\epsfig{file=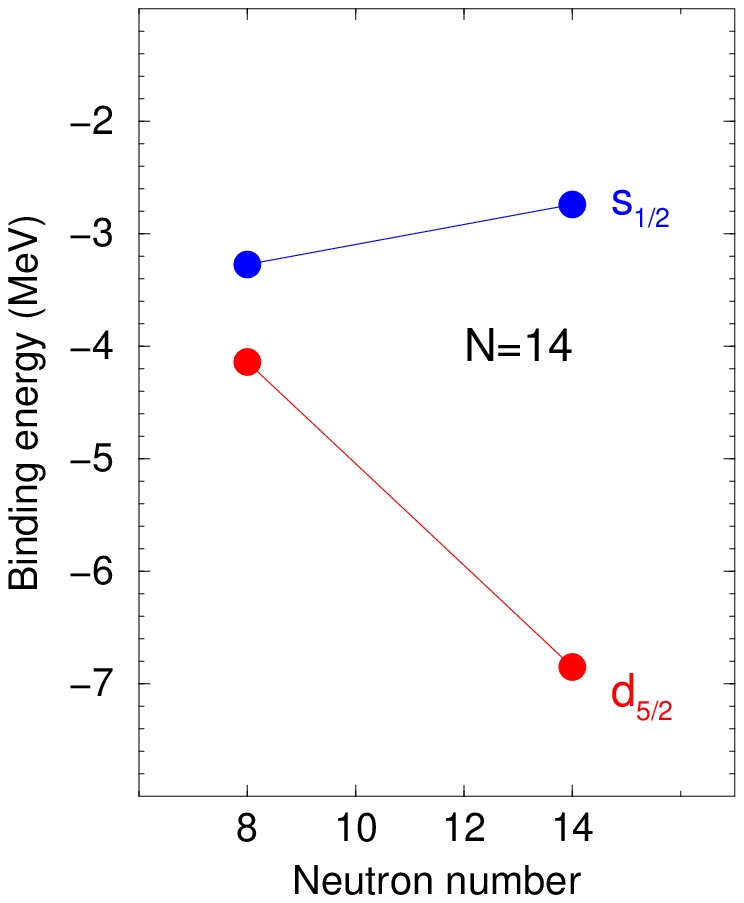,scale=0.55}
\end{center}
\end{minipage}
\begin{center}
\begin{minipage}[t]{16.5 cm}
\caption{{\bf Left}: Experimental binding energies of the neutron $fp$ 
orbits in Ca isotopes, at the beginning and at the end of the 
$\nu f_{7/2}$ shell filling. The straight lines display the 
evolution expected from the effect of the monopole interactions
(as depicted in Fig.~\ref{ESPE}).
{\bf Right}: Experimental binding energies of the $\nu d_{5/2}$ 
and $\nu s_{1/2}$ orbits in O isotopes, at the
beginning and at the end of the $\nu d_{5/2}$ filling.
\label{gapnf7p3}}
\end{minipage}
\end{center}
\end{figure}

Even if weaker in strength, the monopole terms 
between\emph{ like nucleons}  may play an important role to modify 
the shell gaps. To give a 
first example, the gap between the $\nu f_{7/2}$ and
$\nu p_{3/2}$ orbits strongly increases from $N=20$ to $N=28$, i.e. 
during the filling of the $\nu f_{7/2}$ shell.
The single-particle states of $^{41,49}$Ca have been thoroughly
studied using (\overrightarrow{d}, p) reactions 
~\cite{Uozu94a,Uozu94b}, with the observation of the whole strength 
of the $fp$ shells. The obtained results are shown in the left part of 
Fig.~\ref{gapnf7p3}. 
The behaviors of the four single-particle states 
during the filling of the $\nu f_{7/2}$ shell are those expected from the
effect of the monopole interactions, which are 
attractive between two\emph{ like} nucleons in the same orbit 
and repulsive for  two\emph{ like} nucleons in different orbits,  
as mentioned above. Using the experimental
binding energies displayed in the left part of Fig.~\ref{gapnf7p3},
one obtains $V^{nn}_{f_{7/2}f_{7/2}} \sim$ -230~keV, 
$V^{nn}_{f_{7/2}p_{3/2}} \sim$ +160~keV,
$V^{nn}_{f_{7/2}p_{1/2}} \sim$ +170~keV, and
$V^{nn}_{f_{7/2}f_{5/2}} \sim$ +20~keV, values 
close to the ones extracted from the effective TBME used in 
the $fp$ valence space. 
As a consequence, the $N=28$ gap increases by $\sim$~3Mev, making $^{48}$Ca 
a doubly-magic nucleus. It also leads to a large
change in the SO splitting of the $\nu f$ orbits from $N=20$ to
$N=28$.

The appearance of the $N=14$ gap between the $\nu d_{5/2}$ and 
$\nu s_{1/2}$  orbits in $^{22}$O$_{14}$ has a similar origin. 
While they are very close in energy at $N=8$ (see the
right part of Fig.~\ref{gapnf7p3}), the 
addition of six neutrons in the $\nu d_{5/2}$ orbit leads to a
new shell gap of about 4.4~MeV (this will be discussed in
Sects.~\ref{2plusoxygene} and ~\ref{carboxygene}). Using the experimental
binding energies displayed in the right part of Fig.~\ref{gapnf7p3},
one obtains $V^{nn}_{d_{5/2}d_{5/2}} \sim$
-540~keV and $V^{nn}_{d_{5/2}s_{1/2}} \sim$ +90~keV. 

Finally it is important to note that SM calculations using two-body 
{\it realistic} interactions derived from the free $NN$ force fail to 
reproduce some shell closures. As was already mentioned for 
the\emph{ ab-initio} calculations (see Sect.~\ref{general}), the three-body forces 
have to be taken into account. Thus many of the previously observed
discrepancies are now solved~\cite{Zuke03,Caur05}. 

\subsubsection{\it Strength of the monopole interaction}

\begin{figure}[h!]
\begin{center}
\epsfig{file=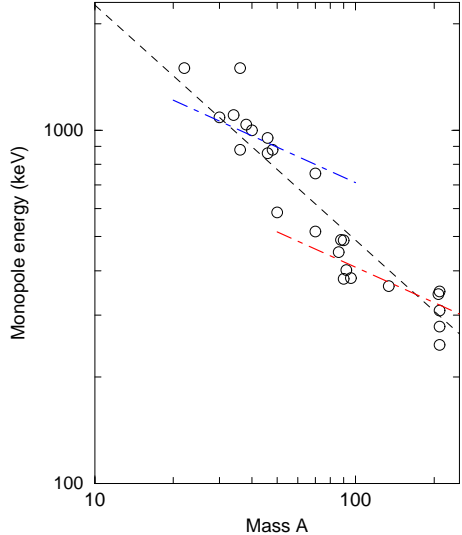,scale=0.6}
\begin{minipage}[t]{16.5 cm}
\caption{Absolute values of the proton-neutron monopole energies
calculated from many multiplets identified in odd-odd nuclei, as a
function of the mass number \cite{Porq08}. The dashed line is a fit
performed within the total mass range, $V^{pn}_{j_1 j_2} \propto
A^{-2/3}$, the two other ones (dashed-dotted lines) being  $\propto
A^{-1/3}$. \label{monop_A}}
\end{minipage}
\end{center}
\end{figure}

Many TBME values were determined throughout the periodic table
thirty years ago from the experimental data available at that
time~\cite{Schi76}. Using the latest experimental results in 
odd-odd exotic nuclei, and the AME2003 atomic mass
evaluation~\cite{Audi03}, we propose new monopole values of
various\emph{  proton-neutron} configurations~\cite{Porq08}. These are 
displayed as a function of mass number in Fig.~\ref{monop_A}. 
The most obvious feature is that the monopole energy of the two-body
proton-neutron interaction strongly decreases with the mass number,
from $\sim$ -1 MeV for $A=30$ to $\sim$ -300 keV for $A=210$.
Therefore the effects of the monopole interactions to modify shell
gaps, if any, are expected to be more sudden in the light-mass
nuclei.

In many Shell-Model calculations, the TBME values are assumed to
scale with $A^{-1/3}$ (the two dashed-dotted lines in
Fig.~\ref{monop_A}). Such a behavior is relevant when the mass range
is not too large, which is the case in all the SM calculations which
apply to restricted valence spaces. When examining a
wider mass range, an $A^{-2/3}$ trend for the 
$V^{pn}_{j_1 j_2}$ is
more appropriate, as shown by the dashed line in
Fig.~\ref{monop_A}. This latter scaling factor indicates that the
strength of residual interactions varies inversely to the surface
of the nucleus, consistent with the fact that the valence 
nucleons are mainly located at the nuclear surface.

In addition to this global trend, the empirical monopoles
displayed in Fig.~\ref{monop_A} clearly exhibit large spreading in
strength for a given mass number. As seen around masses $A \simeq
30, 90$ and $200$, it amounts to about $\pm 30\%$ of the mean
value. This could be accounted for by multiple properties of the
specific proton and neutron configurations involved in the
monopoles. This is borne out by the properties of the monopole 
interaction, which explore some specific parts of the $NN$
interactions, discussed in the following section.

The case of weakly bound systems or more generally of extreme
isospin values is also to be discussed both for the like- and
unlike-nucleon cases. While comparing $V^{pp}$
and $V^{nn}$ derived for weakly bound systems involving similar
orbits, the Coulomb field should lead to a significant breaking of
the charge independence of the $NN$ interaction, giving rise to $V^{pp} \neq
V^{nn}$. Generally speaking, the interaction with states in the 
continuum may
change the value of the derived monopoles for weakly bound systems
as they potentially contain interaction with unbound states of
similar configurations. The monopole interaction $V^{nn}$
extracted in weakly-bound neutron-rich nuclei should drop more
steeply than the $A^{-2/3}$ factor. This is ascribed to
the fact that the radius of the nucleus no longer scales with
$A^{1/3}$, as a wide neutron cloud develops as the neutron binding
energy decreases. Finally at extreme isospin values, protons and
neutrons have considerably different binding energies, which
reduces the overlap of their wave functions. This effect could
lead to weaker $V^{pn}$ monopole matrix elements.

\subsection{\it Decomposition of the monopole interaction into
central, spin-orbit and tensor parts\label{decompmonopol}}

For most of the descriptions of the nucleus (except for
the\emph{ ab-initio} calculations), two-body\emph{ phenomenological
effective forces} are introduced (i) in MF approaches to
describe the interaction of two nucleons in the presence of 
other nucleons, (ii) in the SM to describe the interaction of two nucleons
added to a core nucleus from TBME. In both cases, the absolute
values of the two-body effective interactions significantly differ
from the 'bare' $NN$ values. However reminiscent properties of the
$NN$ forces -such as those of the central, spin-orbit and
tensor terms- prevail in the atomic nucleus, as witnessed by
significant modifications of shell structure.

The present section first reviews the main general components of
an effective $NN$ force. Some
TBME derived from the study of atomic nuclei are split into these
components, triggering comments on the their respective amplitudes
and their spin-dependence. Finally some examples on how 
nuclear shell gaps and SO splitting are affected are discussed 
qualitatively. 

\subsubsection{\it Formal description of the various components 
of the NN interaction\label{NNcomponent}}

The more general potentials between two nucleons depend on three
vector coordinates of the radius, spin and isospin as follows :
\begin{equation}\label{Vgeneral}
V(1,2)=V(\vec{r_1}, \vec{\sigma_1}, \vec{\tau_1}; \; \vec{r_2},
\vec{\sigma_2}, \vec{\tau_2}).
\end{equation}
Following the general prescription derived in Sect.~\ref{general},
the nucleon-nucleon interaction can be expanded in two main parts
which are 'central' and 'non-central'. The various components are
written as a function of the coordinate combinations, provided that
they fulfill the standard required symmetries.

Using the spin-isospin representation, the\emph{ central} interaction
is decomposed in four terms,
\begin{equation}
V_c(1,2)=V_0(r) + V_{\sigma}(r) \: \vec{\sigma_1} \cdot
\vec{\sigma_2} + V_{\tau}(r) \: \vec{\tau_1} \cdot \vec{\tau_2} +
V_{\sigma\tau}(r) \: \vec{\sigma_1} \cdot \vec{\sigma_2}  \;
\vec{\tau_1} \cdot \vec{\tau_2} \label{eqcentral}\end{equation}
where $r=\vert \vec{r_1}-\vec{r_2}\vert$ is the distance between the
two nucleons.

The\emph{ non-central} interaction
contains two terms,\\
(i) the two-body\emph{ spin-orbit} interaction given by
\begin{equation}
V_{LS}(1,2)= \Big( V^{is}_{LS}(r) + V^{iv}_{LS}(r) \; \vec{\tau_1} \cdot
\vec{\tau_2} \Big) \; \vec{L} \cdot \vec{S},
\label{eqspinorbit}\end{equation} where $\vec{L}$ is the relative
orbital momentum between the two interacting nucleons,$\vec{S}$ is
their total intrinsic spin,
$\vec{S}=\frac{1}{2}(\vec{\sigma_1}+\vec{\sigma_2})$, and the
superscript $is$ and $iv$ hold for iso-scalar and iso-vector part of the potential.\\
(ii) the\emph{ tensor} part defined by
\begin{equation}
V_{T}(1,2) \: = \: \Big( V^{is}_{T}(r) + V^{iv}_{T}(r) \; \vec{\tau_1} \cdot
\vec{\tau_2} \Big) \: S_{12}(r), \label{eqtensor}
\end{equation} 
where
\begin{equation}
S_{12}(r)=\frac{3}{r^2}
(\vec{\sigma_1}\cdot\vec{r})(\vec{\sigma_2}\cdot\vec{r}) -
\vec{\sigma_1}\cdot\vec{\sigma_2}
\end{equation}

An example of a realistic $NN$ interaction comprising such 
components is the Argonne V18 interaction~\cite{Wiri95}.
The eight functions of $r$, i.e. $V_0(r)$, $V_{\sigma}(r)$,
$V_{\tau}(r)$, $V_{\sigma\tau}(r)$, $V^{is,iv}_{LS}(r)$ and
$V^{is,iv}_{T}(r)$, are chosen according to the specific problem
which is handled. Various radial functions can be used, such as
$\delta(r)$ for contact interaction (zero range) or Gauss forms for
finite-range interactions. All of them depend on a certain
number of parameters which are adjusted to reproduce selected
experimental data. 

\subsubsection{\it Selected examples of decomposition of Two Body 
Matrix Elements}

The present part aims at a decomposition of the monopole interaction
into the various components mentioned above, i.e. the\emph{
central}, \emph{ spin-orbit} and\emph{ tensor} parts in order to assess
their possible effect on the shell evolution.  

For this purpose a transformation from the $jj$-coupled 
representation, which holds for a direct comparison to experimental 
data where $J = |\vec{j_\pi} + \vec{ j_\nu}|$, to the $LS$-coupled 
scheme has to be achieved.
In order to perform this decomposition, it is necessary to have 
a\emph{ whole set} of TBME available, i.e. all the $J$ states of 
the\emph{ four} multiplet states  dealing with the spin-orbit pairs
($j_\pi=\ell_\pi \pm \frac{1}{2}$ and $j_\nu=\ell_\nu \pm
\frac{1}{2}$) have to be known~\footnote{All the relations used to
obtain the three components (central, spin-orbit and tensor terms)
of the TBME can be found in Ref.~\cite{Brow88}.}. This is by far
not an easy task as it often requires the spectroscopy of very
exotic nuclei.  This is especially true for determining TBME in
which a deeply bound proton ($j_\pi \uparrow$) interacts with a 
weakly bound neutron ($j_\nu \downarrow$).

Such decompositions have been achieved in constructing semi-empirical 
effective interactions in the $1s0d$ shell~\cite{Brow88} and more 
recently, in the $0p$~\cite{Umey04} or $0p-1s0d$ shells~\cite{Umey06}, 
and in the $sd-fp$ shells~\cite{Mill06}. Typical results obtained 
from these sets\footnote{It has to be noted that some terms of these
decompositions may be modified in the future if better
experimental determinations of TBME arise from experimental data
far from the valley of stability. Therefore this
discussion may to some extent be not fully quantitative.} are
gathered in Table~\ref{decomp_mono}. They corroborate the general
features of the proton-neutron interaction given in
Sects.~\ref{general} and \ref{NNcomponent}.
\begin{table}[h!]
\begin{center}
\begin{minipage}{16.5 cm}
\caption{Decomposition of monopole interactions (in MeV) into central,
spin-orbit and tensor parts for some configurations involving
spin-orbit partners. \label{decomp_mono}}
\end{minipage}
\vskip 0.3cm
\begin{tabular}{|c|c|c|c  c c c c|}
\hline
& &Total&Central&      &     &Spin-Orbit & Tensor \\
& &     & tot   & $S=0$ & $S=1$&           &            \\
\hline
\textbf{case A }             &$\pi 0p_{3/2} \nu 0p_{1/2}$  & -3.59 &  -3.47 & (-0.10 &  -3.37)       & +0.14 & -0.26 \\
Table IV of \cite{Umey04}&$\pi 0p_{1/2} \nu 0p_{1/2}$  & -3.21 &  -4.30 & (-0.33 &  -3.97)& +0.57 & +0.53 \\
               &variation                         & +0.38 &  -0.83 &        &        & +0.43 & +0.79   \\
& & &&& &  & \\
\textbf{B}              &$\pi 0p_{3/2} \nu 1s_{1/2}$ & -1.59& -1.52 & (-0.03 &  -1.49)& -0.07 &  0.0 \\
Ref.~\cite{Umey06} &$\pi 0p_{1/2} \nu 1s_{1/2}$ & -1.38& -1.52 & (-0.03 &  -1.49)& +0.14 &  0.0 \\
                 &variation                & +0.21&  0.0  &        &        & +0.21 &    \\
 && &&& &  & \\
\textbf{C }             &  $\pi 0d_{3/2} \nu 0d_{5/2}$ &-1.88 & -1.79 & (-0.20 & -1.59 )& +0.01 & -0.19  \\
Ref.~\cite{Umey06} & $\pi 0d_{3/2} \nu 0d_{3/2}$ & -1.68 & -2.03 & (-0.18 & -1.85 )& +0.07 & +0.29 \\
                & variation                &+0.20 & -0.24  &        &        & +0.06 & +0.48  \\
 && &&& &  & \\
\textbf{D}              &$\pi 0d_{3/2} \nu 0f_{7/2}$ & -1.27& -1.27 &(-0.23  &  -1.04)& +0.22 & -0.22 \\
Ref.~\cite{Mill06}  &$\pi 0d_{3/2} \nu 0f_{5/2}$ & -0.99& -1.23 & (-0.08 &  -1.15)& -0.05 & +0.29 \\
                 &variation                & +0.28&  +0.04 &        &         & -0.27 & +0.51   \\
 && &&& &  & \\
\textbf{E }              &$\pi 0d_{3/2} \nu 1p_{3/2}$ &-0.93&  -1.03 & (-0.13 &  -0.90)& +0.16 & -0.06 \\
Ref.~\cite{Mill06} &$\pi 0d_{3/2} \nu 1p_{1/2}$ &-0.91&  -1.00 & (-0.06 &  -0.94)& -0.03 & +0.12 \\
                 &variation                &+0.02&  +0.03 &        &        & -0.19 & +0.18   \\
\hline
\end{tabular}
\end{center}
\end{table}
\begin{itemize}
\item Whatever the orbits, the strongest term of the monopoles 
comes from the\emph{ central} part , particularly the one corresponding
to $S=1$, the term responsible for the bound state of the
deuteron. Moreover for configurations dealing with the same
$\ell_\pi$ and $\ell_\nu$ values (such as cases A and C), the
central term is more attractive for  $j_\pi = j_\nu$ than for 
$j_\pi = j_\nu \pm 1$.
\item Compared to
the central part, the absolute values of the\emph{ spin-orbit} term are
weaker. Nevertheless the differential value between spin-orbit
partners is sometimes larger than the one obtained from the
central part, e.g. cases B and E. 
\item \emph{Tensor terms} are present
both when identical as well as different $\ell_\pi$ and $\ell_\nu$
values are involved. They amount to about 20\% of the total
monopole force in case D. In case B no tensor effect is present as
no preferred spin-orientation could arise when a
$s_{1/2}$ nucleon is involved.\\ 
The tensor term is repulsive (positive energies in the table) 
when the intrinsic spins are both aligned or both anti-aligned 
with the orbital momenta, and attractive (negative energies) 
otherwise. In particular, this counteracts the effect of the central
part when $\ell_\pi = \ell_\nu$. For instance, while the central term
of $V_{d_{3/2}d_{3/2}}$ is more attractive than the one of 
$V_{d_{3/2}d_{5/2}}$, its total monopole becomes less attractive
because of the tensor term.

Within two $\pi\nu$ configurations 
dealing with spin-orbit partners -such as those presented in the
table- the following identity (see
Ref.~\cite{Otsu05}) 
\begin{equation}\label{tensor_sum}
(2j_\uparrow +1) \; V_T(j',j_\uparrow) + 
(2j_\downarrow +1) \; V_T(j',j_\downarrow) = 0
\end{equation}
can be derived from angular momenta algebra by summing all spin 
and orbital magnetic substates for a given $\ell$. 
This sum rule is fulfilled ,
e.g. applied to the $d_{3/2}-p_{3/2}$ versus $d_{3/2}-p_{1/2}$
monopoles, it writes as -0.06 x 4 + 0.12 x 2 = 0.
\end{itemize}

\subsubsection{\it Qualitative effects of some parts of the interaction}

As shown in the previous section, the values of the monopole
interaction strongly depend on the $\pi\nu$ configurations.  
Particularly those involving spin-orbit partners display large
differences, which may modify the shell structure and the gap 
values. Some typical examples are given now.  

\begin{figure}[h!]
\begin{center}
\epsfig{file=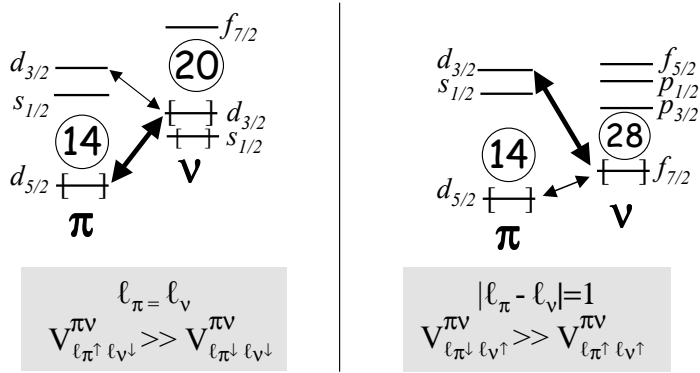,scale=0.6}
\begin{minipage}[t]{16.5 cm}
\caption{Variation of the $\pi d_{5/2} - \pi d_{3/2}$  SO splitting 
induced by the properties of the $NN$ interaction. 
Left: Case of $\pi d - \nu d$ from a $\Delta \ell = 0$ tensor term. 
Right: Case of $\pi d - \nu f$ from a $\Delta \ell = 1$ tensor term.
\label{SONN}}
\end{minipage}
\end{center}
\end{figure}

The first example, dealing with nucleons having the\emph{ same} orbital
momentum $\ell$, is given in the left part of Fig.~\ref{SONN}.
The strongly attractive $\pi
d_{5/2} - \nu d_{3/2}$ force generates large $Z=14$ and $N=20$
gaps in $^{34}_{14}$Si$_{20}$, which consequently behaves as a
doubly magic nucleus. As soon as protons are removed from the $\pi
d_{5/2}$ orbit, the $\nu d_{3/2}$ becomes much less bound (and
actually particle unbound in the neutron-rich O isotopes).
 This
closes the $N=20$ shell gap to the benefit of a new growing $N=16$
one, formed between the neutron $s_{1/2}$ and $d_{3/2}$ shells.
This re-arragement of shell gaps
explains also why $^{24}_{~8}$O$_{16}$ has the properties of a
doubly magic nucleus (with $Z=8$ and $N=16$ shell gaps), and why
$^{28}_{~8}$O$_{16}$ is unbound with respect to neutron emission
(see the discussions and references included in 
Sect.~\ref{disappearN20}). 
As this interaction acts differentially on spin-orbit partners, a 
reduction of the proton $d_{5/2}-d_{3/2}$ SO splitting is
expected while removing neutrons from the $d_{3/2}$ orbit, i.e.
from $^{34}$Si$_{20}$ to $^{30}$Si$_{16}$.

This particular effect acting in a $\pi \nu$ configuration with 
$\Delta \ell =0$ was ascribed in Ref.~\cite{Otsu01} to the 
the fourth term of the central potential derived in
Eq.~\ref{eqcentral}, $V_{\sigma\tau}(r) \: \vec{\sigma_1} \cdot 
\vec{\sigma_2} \: \vec{\tau_1} \cdot \vec{\tau_2}$. However in a 
later publication~\cite{Otsu07}, the same authors have claimed that 
such a term is not strong enough, so this interaction 
between spin-flip proton-neutron partners has to come from the tensor 
term of the $NN$ interaction (see Eq.~\ref{eqtensor}). As derived in the
decomposition of the monopole interaction in the
previous section (see Table~\ref{decomp_mono}), 
the tensor interaction is attractive when the intrinsic spins of the 
neutron and the proton are anti-parallel and repulsive when they are
parallel and is strong enough to counteract the difference in the 
central parts when the two
nucleons have the same orbital momentum $\ell$.  

The second example, dealing with nucleons having 
$\Delta \ell =1$, is given in the right part of Fig.~\ref{SONN}. 
The $V^{pn}_{d_{3/2}f_{7/2}}$
monopole contains an attractive tensor term 
(see Table~\ref{decomp_mono}), which acts in favor
of a large $N=28$ shell gap in $^{48}$Ca where the $\pi d_{3/2}$
orbit is filled completely. 
The removal of all $d_{3/2}$ protons
provokes a reduction of the neutron $f_{7/2}-f_{5/2}$ SO
splitting, and a reduction of the $N=28$ gap. Important is to note
however that the reduction of the $N=28$ gap does not come solely
from this effect (see Sect.~\ref{N28} for details).

In a more general manner, the proton-neutron tensor force felt by
a given nucleon orbiting in $j'$ globally cancels when the two
orbits $j_\uparrow$ and $j_\downarrow$ are filled, as given by 
the tensor sum rule of Eq.~\ref{tensor_sum}. 
When only the $j_\uparrow$ orbits are filled for both
protons and neutrons, the tensor forces that prevail in the
nucleus are maximized. Applied to the case of $^{42}_{14}$Si$_{28}$ 
(see the discussion and references in Sect.~\ref{trendN28}), 
the last occupied proton and neutron shells are
$\pi d_{5/2}$ ($\ell=2,j_\uparrow$)  and $\nu f_{f/2}$
($\ell=3,j_\uparrow$) (shown in the right part of Fig.~\ref{SONN}). 
The proton-induced tensor force 
$V_{\pi d_{5/2} \nu f_{\uparrow, \downarrow}}$ reduces the neutron
$f_{7/2} - f_{5/2}$ SO splitting as compared to the situation in
which both the $d_{5/2}$ and $d_{3/2}$ orbits are filled, as in
$^{48}$Ca. In a symmetric manner, the neutron-induced tensor
$V_{\pi d_{\uparrow, \downarrow} \nu f_{7/2}}$ force weakens the
proton $d_{5/2} - d_{3/2}$ SO splitting as compared to $^{34}$Si,
in which no neutron occupies the $\nu f_{f/2}$ orbit. Added
together, these mutual proton and neutron induced tensor forces
give rise to significant reductions of\emph{ both} proton and neutron SO
splittings and hereby of the $Z=14$ and $N=28$ shell gaps which
were generated by the SO force. In the case of $^{42}$Si, in which
only the proton and neutron $j_{\uparrow}$ configurations are
filled, the tensor force acts against the SO force.

To summarize this part, the tensor interactions which strongly 
act in $\pi \nu$ configurations either with $\Delta \ell =0$ or 
with $\Delta \ell =1$, 
can significantly modify the SO\emph{ splittings}, and hereby the
shell gap for magic numbers originating from the SO interaction.
  
\subsection{\it Summary and Outlooks}

The present section has provided an introduction to the properties
and effects of the effective nuclear forces in atomic nuclei.
Though they differ in definitions, effective forces are widely
used in various theoretical approaches such as mean field
calculations (Skyrme, Gogny and RMF forces) and shell models
(through Two-Body Matrix Elements). This use is due to
the fact that the solving of the N-body problem, from the 'bare'
$NN$ forces, is untractable already for medium mass nuclei. Even
though the effective forces differ with the 'bare' ones (at least in
strength), they do contain some reminiscent properties of the
bare interaction, such as the existence of SO and
tensor forces. These forces are expected to play key roles
in the evolution of nuclear structure.
 
In the SM approach these forces are implicitely contained into the
Two Body Matrix Elements (TBME) of the interaction, among which
the so called monopole term - the average value of TBME over all
directions - is one of the fundamental pillars. Indeed, the
evolution of the spherical shell structure is guided by the changes of this
monopole term, illustrated through the variations of
Effective Single Particle Energies (ESPE). The TBME
used in the SM approach were decomposed into various components, e.g.
central, SO and tensor forces. Quantitative and qualitative effects of
these terms have been discussed. By gathering all these pieces
of information, one notices that new nuclear degrees of freedom -
such as those provided by the action of the
tensor and change in the nuclear density - are expected to modify the
apparent SO splitting and hence the nuclear magic shells. If some
of these features were already invoked more than thirty years ago
(see Ref.~\cite{Sche76}), renewed interest is
growing nowadays from the experimental possibility to explore
these new facets of the in-medium $NN$ interaction. Being aware of
the possible importance of these effects to model the atomic
nucleus, new parameters are being implemented into the effective
forces, such as that of the tensor forces.
 
One fundamental question will therefore pervade the whole document: 
to what extent will the properties of the $NN$ interaction be
seen in the nucleus or will they be diluted among other degrees of fredom such
as configuration mixings ? After having reviewed the properties of
all major-shell closures, this will be addressed in the conclusions.

%% file: texteNZ8_10avril.tex
\section{The magic number 8}

The magic number $8$ naturally comes from any phenomenological
mean-field description of the nucleus. There are so large
distances in energy between the first eigenvalues of the
Hamiltonian with a central potential, whatever its radial form
(square well, harmonic oscillator, Woods-saxon, ...) that the $0p$
states are well separated from the $1s0d$ ones. The $N,Z=8$ shell
gaps are thus formed between occupied states having negative
parity and valence ones having positive parity.  In this section,
we limit our discussion to the evolution of the $N,Z=8$ shell
closures in terms of single-particle energies, trying to derive
which components of the $NN$ force have predominant roles. This
description can be applied up to a certain limit, as the very
concept of mean field is often in trouble for light nuclei close
to the drip-lines. It intends to pave the way for other
theoretical approaches taking into account correlations inherent
to cluster~\cite{Oert03,Free03,Oert06} or halo 
systems~\cite{Tani03,Orr03,Riis06}.

This section is divided in two parts. The first one reviews the
evolution of the $N=8$ shell closure whereas the second describes
the evolution of shell structure along the $Z=8$ isotopic chain.
The neighboring isotopic chains of F ($Z=9$) and C ($Z=6$) will be
discussed as well.

\subsection{\it Evolution of the $N=8$ shell closure\label{evolgapN8}}

\subsubsection{\it Binding energies and the $N=8$ shell
gap\label{BEN8}} 
\begin{figure}[h!]
\begin{minipage}{8cm}
\begin{center}
\epsfig{file=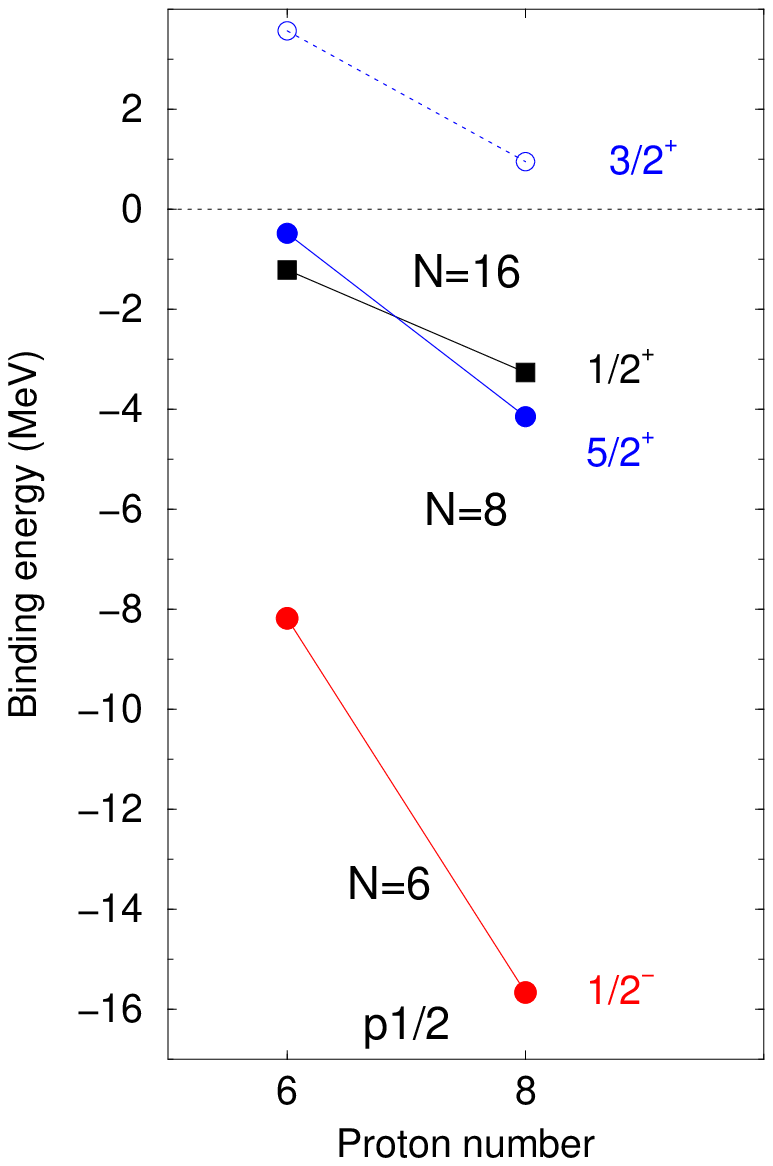,scale=0.65}
\end{center}
\end{minipage}\hfill
\begin{minipage}{8cm}
\epsfig{file=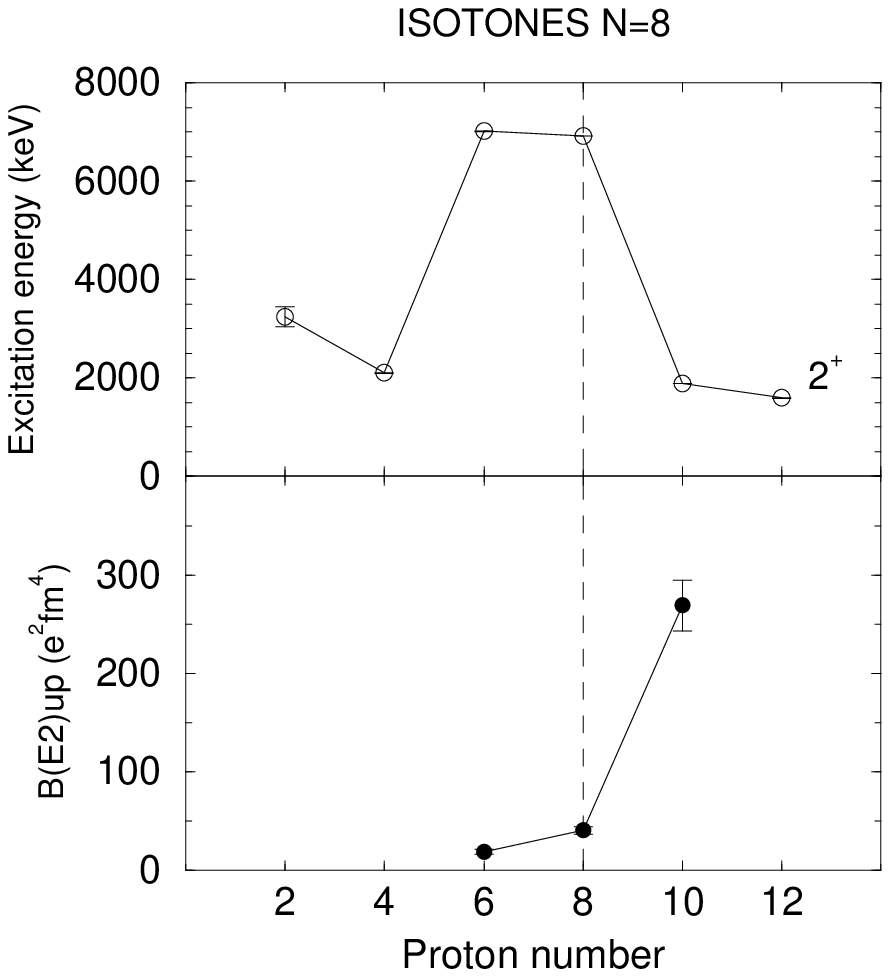,scale=0.65}
\end{minipage}
\begin{center}
\begin{minipage}[t]{16.5cm}
\caption{{\bf  Left}: Binding energies of the $1/2^-$
($5/2^+,1/2^+,3/2^+$) states located below (above) the $N=8$ magic
number (see Sect.~\ref{annex}). The proton $p_{1/2}$ orbit is
getting filled between $Z=6$ and $Z=8$. {\bf Right}: Experimental
$E(2^+)$ and $B(E2; 0^+ \rightarrow 2^+)$ values in the $N=8$
isotopes.} \label{neutronBEZ8}
\end{minipage}
\end{center}
\end{figure}

The neutron $N=8$ shell gap is formed between the
occupied $p_{1/2}$ and the valence $d_{5/2}$ and $s_{1/2}$ orbits.
The binding energies of the corresponding states $1/2^-$, $5/2^+$
and $1/2^+$ are shown in the left part of Fig.~\ref{neutronBEZ8}.
Taking the binding energy values at $Z=6$, it is found that the
size of $N=8$ shell gap amounts to about 6~MeV. The $1/2^-$ state
at $Z=8$ contains additional binding energy due to the Wigner
term, the $^{16}$O nucleus being a self conjugate $N=Z$ nucleus.
Taking the prescription of Chasman~\cite{Chas07}, the Wigner
energy  to substract in order to determine the size of the gap is
5~MeV in the case of $^{16}$O. Doing so the $N=8$ gap at $Z=8$
amounts to 7~MeV. The unbound $3/2^+$ states are reported also in
Fig.~\ref{neutronBEZ8}, showing the existence of a sizeable $N=16$
subshell gap of about 3~MeV. These data points at $Z=6$ and $Z=8$
correspond to the first $3/2^+$ states which carry a large -but
for $^{15}$C not the full- spectroscopic strength observed in the
$^{14}$C(d,p)$^{15}$C and $^{16}$O(d,p)$^{17}$O stripping
reactions, respectively.
A crossing between the $5/2^+$ and $1/2^+$ states occurs between
$Z=6$ and $Z=8$, while the proton $p_{1/2}$ orbit is getting
filled. These states $5/2^+$ and $1/2^+$ arise from the $d_{5/2}$
and $s_{1/2}$ orbits, respectively. As early mentioned by Talmi
and Unna~\cite{Talm60} "the order of the filling of the neutron
shells ($d_{5/2}$ and $s_{1/2}$) may depend on the proton
configuration". In the present case the addition of 2 protons in
the $p_{1/2}$ shell induces a differential change of the neutrons
binding energies of the $d_{5/2}$ and $s_{1/2}$ orbits and provoke
their inversion. The $V^{pn}_{p_{1/2}d_{5/2}}$ has larger absolute
intensity than the $V^{nn}_{p_{1/2}s_{1/2}}$ one. The former
matrix element contains an attractive tensor term, whereas the
second contains a repulsive two-body spin-orbit term (see
Table~\ref{decomp_mono}). Taking the experimental energies of the
first $1/2^+$ and $5/2^+$ in $^{15}$C$_{9}$ and $^{17}$O$_{9}$,
the monopole energy difference  2($V^{pn}_{p_{1/2}d_{5/2}} -
V^{pn}_{p_{1/2}s_{1/2}})$ amounts to approximately -1.6~MeV. This
value should be taken with caution as the full spectroscopic
strength of these $5/2^+$ and $1/2^+$ states is not found in
$^{15}$C.

One would also be tempted to comment on the variation of the
neutron $d_{5/2}- d_{3/2}$ spin-orbit splitting between $Z=6$ and
$Z=8$ by taking the energy difference between the $5/2^+$ and
$3/2^+$ states. Doing so, one finds a reduction of about 1~MeV,
taking the mean energy values of the resonant $3/2^+$ states in
$^{15}$C and $^{17}$O. Taken at face value, this indicates that a
differential change of binding energy is occuring between the
$d_{5/2}$ and $d_{3/2}$ orbits, as two protons are added to the
$p_{1/2}$ orbit. In the present case, the
$V^{pn}_{p_{1/2}d_{5/2}}$ monopole would globally be more
attractive than the $V^{pn}_{p_{1/2}d_{3/2}}$ one, such as
2($V^{pn}_{p_{1/2}d_{5/2}} - V^{pn}_{p_{1/2}d_{3/2}}) \simeq$
-1~MeV. The action of tensor forces could account for this
variation. However, as said earlier, this conclusion should be
ascertained by further theoretical calculations to derive the
single-particle energies from the experimental states which
inevitably contain some amount of correlation.

\subsubsection{\it Trends of E(2$^+$) and B(E2) values in the $N=8$
isotones\label{2plusN8}}

The top right part of Fig.~\ref{neutronBEZ8} shows the systematics
of the $2^+$ energies in the $N=8$ isotones. From the left to the
right hand side of the figure, $2^+$ energies of $^{10}$He,
$^{12}$Be, $^{14}$C, $^{16}$O, $^{18}$Ne, and
$^{20}$Mg~\cite{Gade07} are reported. Large $2^+$ energies are
found for the two doubly magic nuclei $^{14}$C and $^{16}$O, while
they drop by a factor of two to three on each side of the valley
of stability. This effect does not necessarily document for a
sudden erosion of the $N=8$ gap. Indeed it arises principally from
the fact that $2^+$ states can be created by proton excitations,
which are hindered in the $^{14}$C and $^{16}$O nuclei due to the
existence of large $Z=6$ and $Z=8$ shell gaps (see the next
Sections).

The trend of the B(E2) values confirms the statements given in the
previous paragraph. The low B(E2) values²²² of $^{14}$C and
$^{16}$O owe to the fact that they are doubly-magic nuclei. The
B(E2) of $^{16}$O is larger by a factor of 2 as, being a
self-conjugate nucleus, its $2^+_1$ state contains an almost equal
amount of neutron and proton excitations. As protons carry the
major fraction of the effective charge, this enhancement of proton
excitation accounts for the relatively large B(E2) value for
$^{16}$O. On the other hand, the $2^+$ state of $^{14}$C has a
dominant wave function built with neutron excitation. At $Z=10$
the B(E2) value suddenly rises to a large value~\cite{Macd76}.
This coincides with the filling of the proton $sd$ shells, in
which $2^+$ excitations can be easily generated. The large value,
B(E2)=18(2) W.u., indicates that $^{18}$Ne is rather collective.

\subsubsection{\it Trends of E(1$^-_1$) state in the $N=8$ isotones}

\begin{figure}[h!]
\begin{center}
\begin{minipage}[t]{8 cm}
\epsfig{file=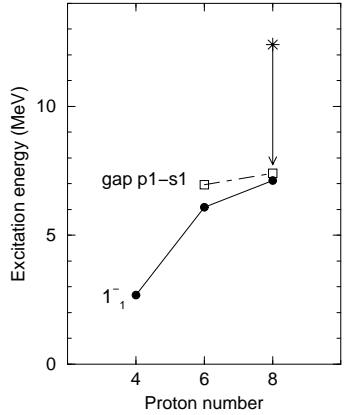,scale=0.6}
\end{minipage}
\begin{minipage}[t]{16 cm}
\caption{Experimental E(1$^-_1$) values in the $N=8$ isotones,
$^{12}_{~4}$Be, $^{14}_{~6}$C and $^{16}_{~8}$O (filled circles).
The values of the $N=8$ gap are shown with empty squares. The data
point for $^{16}$O (indicated with a star) contained the Wigner
energy, which has been substracted to give the real size of the
gap (see Sect.~\ref{BEN8} for details).} \label{E1minus}
\end{minipage}
\end{center}
\end{figure}

To get a further insight into the evolution of the $N=8$ shell
closure as a function of the proton number, one can look at the
evolution of the energy of cross shell excitations. The $N=8$
shell gap is formed between the $p_{1/2}$ and $s_{1/2}$ (or
$d_{5/2}$ in $^{16}$O) orbits. Therefore particle-hole excitation
across this gap give rise to states having total spin values
$J=0^-,1^-$. Such states should be found at approximately the
energy of the $N=8$ gap, shifted by the residual energies of the
$(\nu p_{1/2})^{-1}(\nu s_{1/2})^{+1}_{J=0,1}$ configurations.
Energies of the first $1^-$ state in three lightest isotones are
shown in Fig.~\ref{E1minus}, as compared to the size of the $\nu
p_{1/2}-\nu s_{1/2}$ gap extracted from Fig.~\ref{neutronBEZ8}.
The data point on $^{12}_{~4}$Be has been obtained by means of the
inelastic scattering technique at intermediate energy at the RIKEN
facility~\cite{Iwas00}. In this experiment two targets (Pb and C)
were used to ascertain the multipolarity of the transition and the
1$^-$ assignment of the observed state. The strong decrease of
$E(1^-_1)$ in $^{12}$Be has been ascribed to a drastic narrowing
of the $N=8$ shell gap there. The authors argued that the large
measured B(E1) value is consistent\footnote{For weakly bound
nuclei exhibiting a well developed neutron halo, the transition
between $p$ and $s$ states occurs with a sudden increase of
neutron radius. As the E1 operator scales with the nuclear radius,
it follows that enhanced B(E1) transition is a generic property of
halo nuclei.} with a quasi-degeneracy between the $p_{1/2}$ and
$s_{1/2}$ states, a feature which is also present in
$^{11}$Be~\cite{Mill83} . Additionally knockout reaction of
$^{12}$Be measured at the NSCL/MSU facility has also shown a
strong indication of shell melting in its the ground
state~\cite{Navi00}. Where a pure $\ell=1$ wave function would
have been expected for an $N=8$ closed shell, the ground-state
wave function contains two-thirds of $\ell=0,2$ admixtures.

This sudden reduction of the $N=8$ shell gap for $Z<6$ could be
interpreted in term of proton-neutron interactions, at least on a
semi-quantitative manner. Below $Z=6$, protons are removed from
the $p_{3/2}$ orbit. The reduction of the $\nu p_{1/2}- \nu
s_{1/2}$ gap could be ascribed to the monopole matrix element
difference ($V^{pn}_{p_{3/2}p_{1/2}}$- $V^{pn}_{p_{3/2}s_{1/2}}$)
per proton removed from the $p_{3/2}$ orbit. This linear
extrapolation was early made in Ref.~\cite{Talm60} to account for
the inversion between the $\nu p_{1/2}$ and $\nu s_{1/2}$ state in
$^{11}$Be. As the spin-flip proton-neutron matrix element
$V^{pn}_{p_{3/2}p_{1/2}}$ is strongly attractive, the $\nu
p_{1/2}$ orbit becomes suddenly much less bound while the partner
$\pi p_{3/2}$ shell empties. This effect weakens the $N=8$ gap.
Following the trend of the $1^-$ state (see Fig.~\ref{E1minus}),
the inversion between the $p_{1/2}$ and $s_{1/2}$ states will be
found at $N=8$ in the drip-line $^{10}$He nuclei.

This description in term of single particle energy evolution,
making use of the 'monopole-driven' reduction of shell gaps,
should be taken as qualitative only as states in halo or cluster
nuclei are no of longer single particle nature.

\subsubsection{\it Conclusion}

The $N=8$ shell gap is formed between the $p_{1/2}$ and $d_{5/2}$
and the nearby $s_{1/2}$ orbits.  In $^{16}$O the protons
occupying the $p_{3/2}$ and $p_{1/2}$ orbits bind the neutron
$p_{1/2}$ as compared to other neutron orbits. While the $p$
protons are progressively removed, the $\nu p_{1/2}$ shell become
less bound. Therefore the $N=8$ shell gap is quickly eroded,
provoking finally an inversion between the $p_{1/2}$ and $s_{1/2}$
states. This reduction is infered from the ground-state components
of $^{12}$Be, the low energy of its first $1^-$ state as well as
the large value of its reduced transition probability B(E1).
Incidentally, the inversion of levels in addition to the presence
of a nearby $s$ orbit available, give birth to halo nuclei.

\subsection{\it Evolution of the $Z=8$ shell closures\label{evolgapZ8}}

\subsubsection{\it Binding energies and the $Z=8$ shell gap}

\begin{figure}[h!]
\begin{minipage}{8cm}
\begin{center}
\epsfig{file=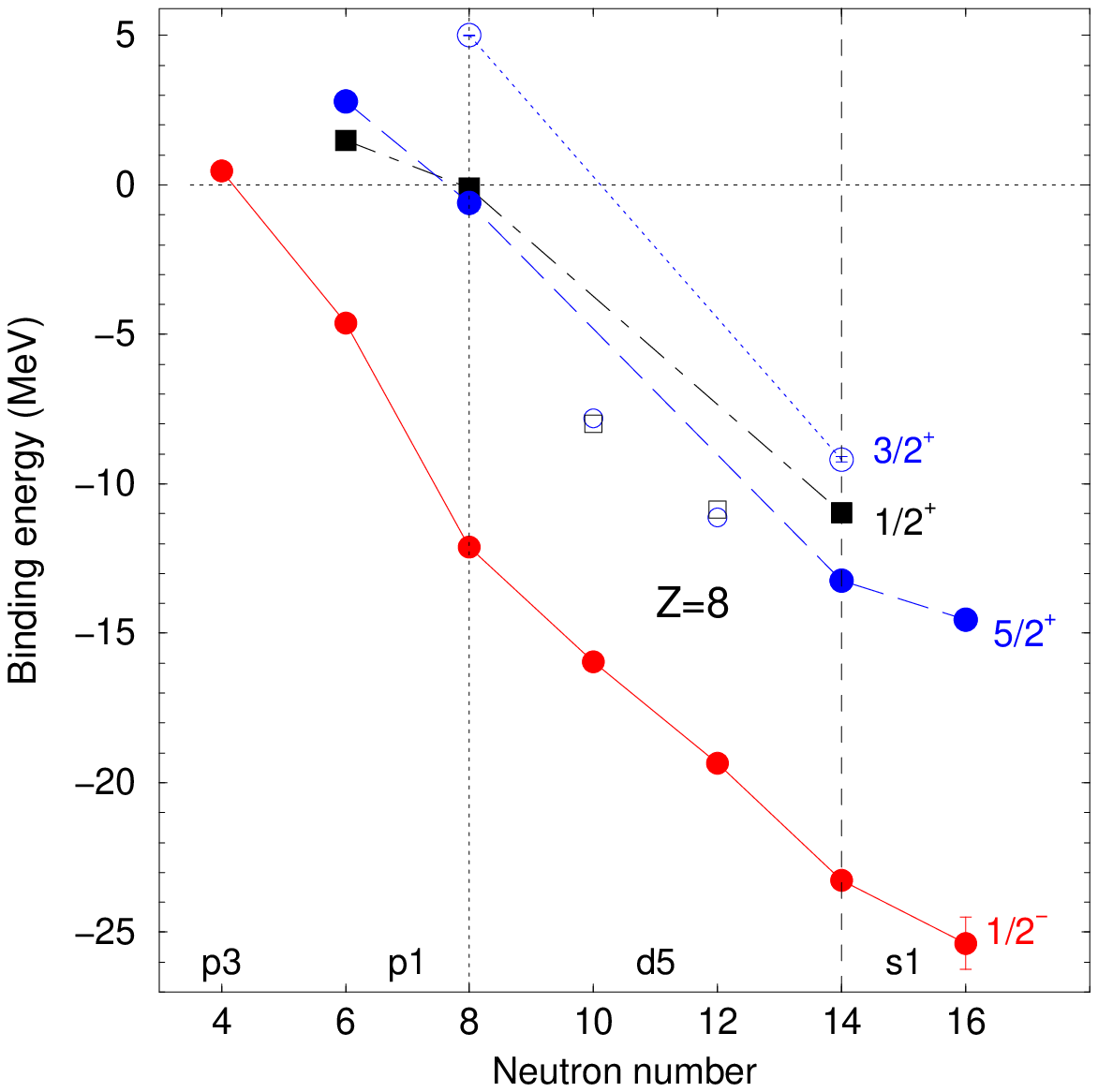,scale=0.6}
\end{center}
\end{minipage}\hfill
\begin{minipage}{8cm}
\epsfig{file=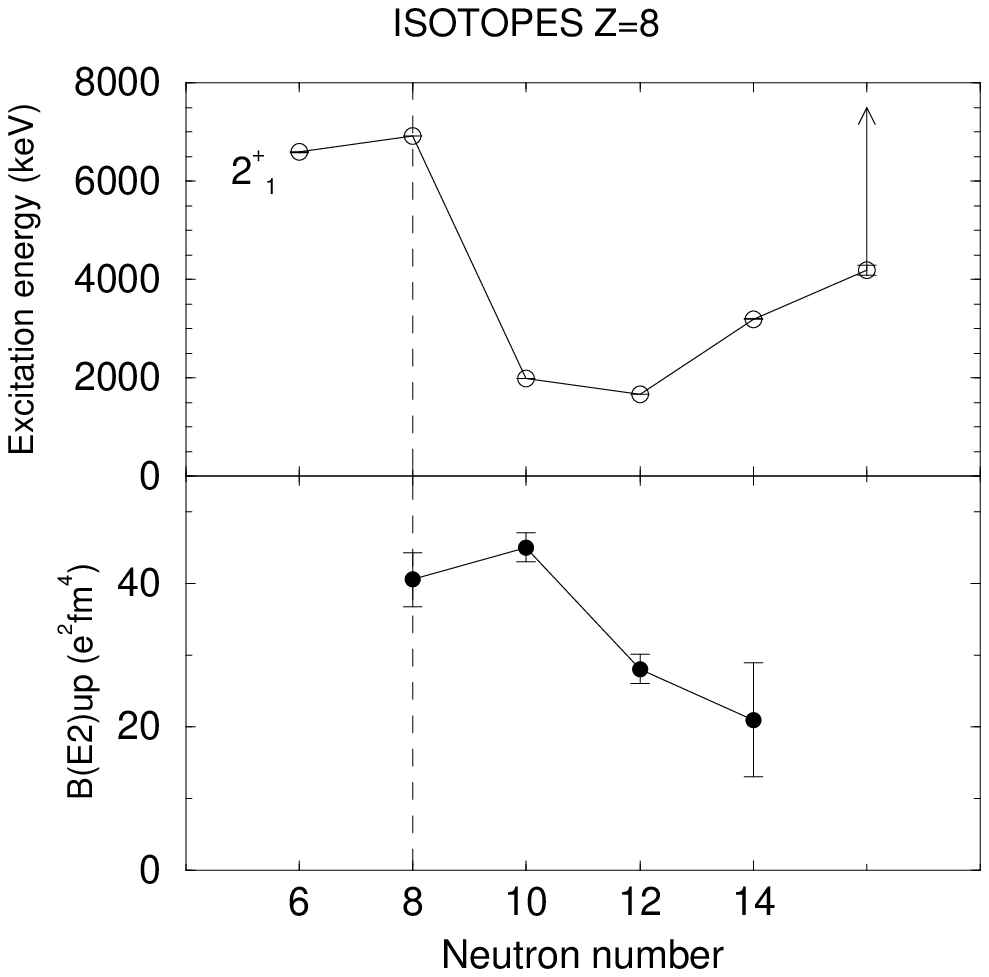,scale=0.7}
\end{minipage}
\begin{center}
\begin{minipage}[t]{16.5cm}
\caption{{\bf  Left}: Binding energies of the $1/2^-$
($5/2^+,1/2^+$) states located below (above) the $Z=8$ magic
number. The neutron orbitals which are getting filled as a
function of increasing $Z$, $p_{3/2}$, $p_{1/2}$, $d_{5/2}$ and
$s_{1/2}$, are indicated above the neutron number axis. {\bf
Right}: Experimental $E(2^+)$ and $B(E2; 0^+ \rightarrow 2^+)$
values in the oxygen isotopes. The $E(2^+)$ value of
$^{24}$O$_{16}$ is greater than the neutron emission threshold
(see text).} \label{protonBEZ8}
\end{minipage}
\end{center}
\end{figure}

The proton $Z=8$ shell gap is formed between the occupied
$p_{1/2}$ and the valence $d_{5/2}$ and $s_{1/2}$ orbits. The
binding energies of the corresponding states $1/2^-$, $5/2^+$ and
$1/2^+$ are shown in the left part of Fig.~\ref{protonBEZ8}, the
proton-unbound states of the $^{12}$O and $^{15}$F nuclei being
also included.
The very last measured mass values of the $^{24}$O and $^{23}$N
neutron-rich nuclei, by B. Jurado et al.~\cite{Jura07}, have been
used for the neutron-rich side of the figure. As regards the
$5/2^+$ state, the filled symbols are used for nuclei which are
thought to be spherical. Conversely, empty symbols at $N=10,12$
correspond to binding energies involving deformed $_9$F isotopes.
These data points are located aside of the interpolation lines, as
they correspond to nuclei containing additional correlation energy
due to their deformation. The binding energy values for the
$^{16}$O nucleus contains the Wigner term as being an $N=Z$
nucleus. Taking all these remarks into account, it is found that
the $Z=8$ shell gap equals to about 10~MeV. This value is large
enough to preserve a spherical shape for all $_8$O nuclei.

The drip line occurs already at $N=16$ in the O chain, while one
would have expected the doubly magic nucleus $^{28}_{~8}$O$_{20}$
to be bound. By looking at the evolution of the calculated ESPE in
Fig.~\ref{E2SPE_CO}, one sees that the $d_{3/2}$ orbit is only
bound by about 1~MeV there. The pairing correlations in $^{26}$O
increase the binding energy of this nucleus, thus making $^{28}$O
unbound relative to $^{26}$O. Interesting is the fact that the
addition of one proton makes the F isotopes bound up to $N=20$ and
even beyond (the $^{31}$F$_{22}$ isotope has been
observed~\cite{Saku99}). This feature is discussed in detail in
Sect.~\ref{evolgapN20} which describes the behavior of the $N=20$
shell closure. But from Fig.~\ref{E2SPE_CO}, one already
understands the mechanism which is responsible from this sudden
change of structure between O and F isotopes. The addition of a
single proton in the $\pi d_{5/2}$ orbit makes its spin-flip
partner $\nu d_{3/2}$ orbit more bound by about the value of the
monopole involved, i.e. by $V^{pn}_{d_{5/2}d_{3/2}} \simeq$ 2~MeV
(see Table~\ref{decomp_mono}). This large value is enough to bind
all the F isotopes. In addition to this monopole energy, the F
isotopes located between the $N=8$ and $N=14$ shell closures gain
quadrupole energy, as discussed in the previous paragraph. One
notes that, as the $Z=8$ gap is large, a similar gain of
quadrupole energy is hindered for the O isotopes.

The slope of the binding energy curves could provide information
on the strength of the monopole proton-neutron interactions at
play\footnote{Doing so, we assume that most of the spectroscopic
factors are contained in these states, which is reasonable only
for spherical nuclei. Additional word of caution concern the use
of the $N=8$ data point, the energy of which is shifted by the
Wigner energy term of the $N=Z$, $^{16}$O nucleus.}. In
particular, the steepest slopes correspond to the largest monopole
terms, as for instance the ones involving the same angular momenta
and number of nodes in their proton and neutron wave functions.
Ranked by decreasing intensity, one finds
$V^{pn}_{p_{1/2}p_{3/2}}$, $V^{pn}_{p_{1/2}p_{1/2}}$ and
$V^{pn}_{d_{5/2}d_{5/2}}$. The $V^{pn}_{p_{1/2}d_{5/2}}$ monopole
looks important as well, possibly due to the fact that it contains
an attractive tensor part. After having passed the $N=14$ neutron
number, smoother slopes for the $5/2^+$, and to a weaker extent
for the $1/2^-$ binding energy curves are found. This is due to
the fact that the monopoles  $V^{pn}_{d_{5/2}s_{1/2}}$ and
$V^{pn}_{p_{1/2}s_{1/2}}$ are weaker than the ones involved before
$N=14$. As these latter monopoles link orbits separated by at
least one unit of angular momentum, the overlap of valence proton
and neutron wave functions is small.

The structure of the $^{23}_{~9}$F$_{14}$ nucleus has been
recently studied in order to identify the proton single-particle
states lying above the $Z=8$ gap~\cite{Mich06}, i.e. $d_{5/2}$,
$s_{1/2}$ and $d_{3/2}$ orbits. The situation is expected to be
simple as its core, $^{22}$O, behaves as a doubly-magic nucleus
(as discussed in Sect.~\ref{carboxygene}). In this experiment,
done at the secondary-beam line in the RIKEN Accelerator Research
Facility, the identification of the proton single-particle states
has been done from the comparison of the population strengths of
the excited states in $^{23}$F from four direct reactions: the
one-proton transfer reaction onto $^{22}$O, the $\alpha$ inelastic
scattering of $^{23}$F, the neutron knock-out from $^{24}$F, and
the two-nucleon knock-out from $^{25}$Ne. Such a comparison is
appropriate for identifying single-particle states among the many
excited states. The states at 2268 and 4059~keV are
good candidates, having large cross-sections in the proton transfer
reaction while their populations are weak in the neutron knock-out
reaction. They are assigned to be the $\pi s_{1/2}$ and $\pi
d_{3/2}$ states respectively, by analysis of the population
strengths and the angular distributions of the outgoing $^{23}$F
nuclei. The authors conclude that the energy gap between the two
spin-orbit partners, $\pi d_{5/2}$ - $\pi d_{3/2}$, amounts to
4.06~MeV, since the ground state of $^{23}$F is
5/2$^+$~\cite{Sauv00}. The decrease of the $\pi d_{5/2} - \pi
d_{3/2}$ splitting by about 1~MeV, from $\sim$5.0~MeV in $^{17}$F
to 4.06~MeV in $^{23}$F (as seen in the left part of
fig.~\ref{protonBEZ8}), occurs simultaneously to the addition of 6
neutrons into the $\nu d_{5/2}$ orbit. Therefore this reduction of
the proton $d$ SO splitting was attributed in~\cite{Sugi07} to the
differential action of the proton-neutron tensor forces, viz $\pi
d_{5/2}- \nu d_{5/2}the $ versus $\pi d_{5/2}- \nu d_{3/2}$
forces.

Complementary to these works, Signoracci and Brown~\cite{Sign07}
have compared the same data on $^{23}$F and other existing data on
$^{19}$F to shell model calculations using various versions of the
USD interactions (USD~\cite{Wild84}, USDA and
USDB~\cite{Brow06b}). By taking into account the full
single-particle strength, the size of the $\pi d_{5/2}$ - $\pi
d_{3/2}$ splitting is increased by about 50\% as compared to the
value computed from the lowest state energies. The USDB
interaction gives remarkable agreement with the energies obtained
for the first $3/2^+$ and $5/2^+$ states in $^{23}$F. Such an
excellent agreement is not found in the case of $^{19}$F, which
has very fragmented single-particle strengths, as being a rather
deformed nucleus. The authors could unfortunately not use
experimental data on $^{17}$F, which are rather yet incomplete. By
comparing  calculated and experimental level schemes and
spectroscopic factors, the authors of Ref.~\cite{Sign07} deduced
that the proton $d$ SO splitting is 6.2(4) MeV in $^{23}$F and
stays almost constant as a function of neutron number, from the
proton-drip line, 6.0~MeV in $^{17}$F$_8$, to the neutron drip
line, 6.9~MeV in $^{29}$F$_{20}$, within the present uncertainties
inherent to the calculations.

To conclude, it seems that a more detailed study of
proton transfer reactions on $^{16,22}$O is required to determine
more precisely the proton single-particle energies in $^{17,23}$F.
This remark is especially true for the $^{17}$F case. This is
required to clarify the present situation concerning the evolution
of the $\pi d_{5/2} - \pi d_{3/2}$ SO splitting along the F
isotopic chain, while the $\nu d_{5/2}$ orbit is filled. This is
undoubtedly a crucial test with respect to the action and strength
of tensor forces in atomic nuclei.

\subsubsection{\it Trends of E(2$^+$) and B(E2) values in the Oxygen
isotopes\label{2plusoxygene}}

In addition to the doubly magic $N=Z$ nucleus, $^{16}$O, the
Oxygen isotopic chain contains three quasi doubly-magic nuclei
$^{14}$O, $^{22}$O and $^{24}$O. This owes to the $N=6$, $N=14$
and $N=16$ large neutron subshell gaps between the $\nu p_{3/2} -
\nu p_{1/2}$, $\nu d_{5/2} - \nu s_{1/2}$ and $\nu s_{1/2} - \nu d_{3/2}$ orbits,
respectively (see fig.~\ref{neutronBEZ8}). The presence of the $N=6$ subshell closure leads to
a large $2^+$ energy at 6.5~MeV as well as a $1^-$ state at
5.17~MeV in $^{14}$O$_6$. The latter state corresponds to a
cross-shell
excitation $\nu p_{3/2}^{-1} \otimes \nu p_{1/2}^{+1}$ configuration.
Concerning the neutron-rich isotopes, the $2^+$ energy rises up to
3.2~MeV at $^{22}$O$_{14}$~\cite{Stan03}, and at more than
4.19(10)~MeV at $^{24}$O$_{16}$. As no $\gamma$-ray was observed
during the de-excitation of $^{24}$O, it was surmised
in Ref.~\cite{Stan03} that all excited states, including the $2^+_1$
one, lie about the neutron emission threshold\footnote{The value reported
in the figure, $S_n$=4.19(10)~MeV, is computed from the new atomic
masses of Ref.~\cite{Jura07}}. Hence this value is a lower limit of the $2^+$
energy in $^{24}$O (drawn with an arrow in the
top right part of Fig.~\ref{protonBEZ8}).

The trend of the B(E2) values corroborates the existence of an
$N=14$ subshell closure (see the bottom right part of
Fig.~\ref{protonBEZ8}). Within the present error bars, the B(E2)
value of $^{22}$O is the smallest of the isotopic
chain~\cite{Thir00}. As the $Z=8$ shell gap is rather large,
the configuration of the $2^+$ in most of the oxygen isotopes is
\emph{mainly} of neutron origin~\cite{Bech06}. This statement does
not hold for the $N=Z$ nucleus, the $2^+$ state of which is made
of equally mixed protons and neutrons configurations, as already
mentioned in Sect.~\ref{2plusN8}.

\subsubsection{\it Comparison between the C and O isotopic chains
and the $N=14,16$ subshell closures\label{carboxygene}}

\begin{figure}[h!]
\begin{center}
\epsfig{file=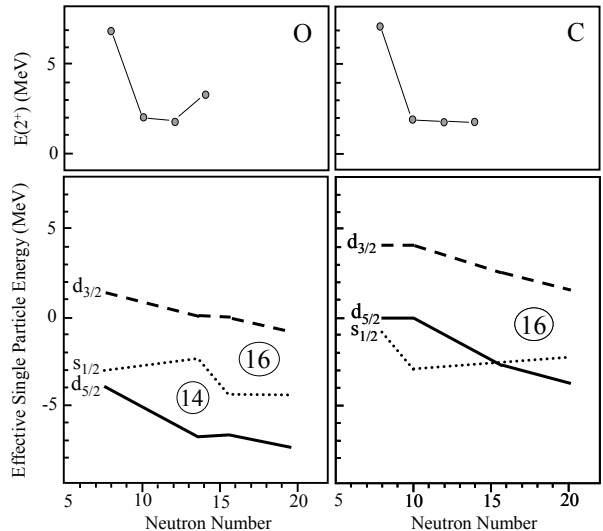,scale=0.5}
\begin{minipage}[t]{16 cm}
\caption{ \textbf{Top} : $2^+$ energies in the O (left) and C
(right) isotopic chains. \textbf{Bottom} Neutron Effective Single
Particle Energies (ESPE) as a function of the neutron number in
the O (left) and C (right) isotopic chains. The variation of the
ESPE have been derived by using the monopole matrix elements of
the USDB interaction~\cite{Brow06b,Brow08}.}
\label{E2SPE_CO}
\end{minipage}
\end{center}
\end{figure}

The $^{14}$C nucleus has the remarkable properties of a doubly
magic nucleus, such as a high $2^+$ energy value at 7.012~MeV and
a weak B(E2) value of 18.7(2)~$e^2fm^4$. These properties arise
from the combination of large $Z=6$ and $N=8$ shell gaps. As the
size of the $Z=6$ gap is almost equivalent to that of the $Z=8$
one, a comparison between the $_6$C and $_8$O isotopic chains towards
the neutron drip line is therefore interesting. The subshell
closures at $N=14$ and $N=16$ in the oxygen
isotopes have been presented in the previous sub-section.
By comparing the $2^+$ energy trends of the O and C
nuclei represented in Fig.~\ref{E2SPE_CO}, one sees a similar
behavior up to $N=12$ and a drastic change at $N=14$. Instead of a
rise at $N=14$, the $2^+$ energy remains constant at
$^{20}$C$_{14}$. This demonstrates that the $N=14$ subshell
closure no longer exists in the C isotopes, making it effective
only in the O chain. What is the reason for this difference of
behavior at $N=14$ between the O and C isotopic chain ?

To answer this question, one should first understand the mechanism
leading to the formation of the $N=14$ gap in oxygen, and second
brings reasons for its non-existence in the carbon isotopes. The
evolution of Effective Single Particle Energies (ESPE) of the $\nu
d_{5/2}$ and $\nu s_{1/2}$ orbits is shown for the O isotopes in
the bottom left part of Fig.~\ref{E2SPE_CO}. These values are
obtained when using the USDB interaction~\cite{Brow06b}, the
matrix elements of which were derived by adjustments to  many
experimental data in this mass region. From  Fig.~\ref{E2SPE_CO}
one sees that the $N=14$ gap is created by the filling of 6
neutrons in the $d_{5/2}$ shell between $N=8$ and $N=14$. As the
monopole matrix element $V^{nn}_{d_{5/2}d_{5/2}}$ is globally
attractive, the ESPE of the $d_{5/2}$ orbit gains binding
energy\footnote{Interesting here to remark is that this mechanism
is generic. It creates other magic numbers such as the N=28 and
N=50 ones by the filling of 8 and 10 neutrons in the $f_{7/2}$ and
$g_{9/2}$ orbits, respectively (see the end of
Sect.~\ref{monopoleffect}).}. Meanwhile the ESPE of the $s_{1/2}$
moves upwards, due to the fact that the $V^{nn}_{d_{5/2}s_{1/2}}$
monopole is slightly repulsive. As the neutrons start to fill the
$s_{1/2}$ orbit, the ESPE of the $\nu s_{1/2}$ orbit suddenly
drops down to large binding energy by virtue of the large
$V^{nn}_{s_{1/2}s_{1/2}}$ matrix element. It is seen that a large
$N=16$ gap is preserved up all along the O isotopic chain. The
$^{24}$O$_{16}$ behaves as a doubly magic nucleus, as being
surrounded by two neutrons gaps, $N=14$ and $N=16$. This Figure
should however be taken in a qualitative manner as few monopoles
involved far off stability, such as $V^{nn}_{d_{3/2}d_{3/2}}$, are
not yet strongly constrained by experimental data. The major
constraint existing so far is to get the neutron $d_{3/2}$ unbound
with respect to 2 neutron emission at $N=20$ to account for the
particle instability of $^{28}$O$_{20}$ (this is also shown in the
right part of fig.~\ref{ESPE_N20} in Sect.~\ref{disappearN20}, for
another set of TBME).

Why is the $N=14$ gap not existing in the C chain? The difference
of structural behavior between the C and O nuclei originates from
the inversion of the 1/2$^+$ and 5/2$^+$ levels between the
$^{17}$O and $^{15}$C isotones (shown in the left part of
Fig.~\ref{neutronBEZ8}). Taking this early change into account,
the structural evolution in the neutron-rich C isotopes can be
straightforwardly deduced using the same $V^{nn}$ terms as those
involved in the spectroscopy of the O nuclei~\footnote{This
assumption assumes that the neutron-neutron monopoles in the O and
C chains could be considered to be similar. As stated in
Refs.~\cite{Stan07,Camp06} the monopoles of the C chain are lower
by about 20\% as compared to those of the O chain. But the global
picture described here still holds.}. Soon after $^{14}$C$_{8}$,
the $\nu s_{1/2}$ orbit is filled first, making the study of
$^{15}$C of particular interest for the appearance of a
one-neutron halo, as the $s_{1/2}$ orbit is weakly bound
(S$_n$=1.22~MeV). This has been searched for in one-neutron
break-up reactions~\cite{Bazi98}, but the results were not
conclusive enough. After $^{16}$C the $\nu d_{5/2}$ orbit starts
to be filled. Its binding energy increases gradually, to
eventually cross the $\nu s_{1/2}$ orbit at $N \sim 15$. This
change of ESPE has three noticeable consequences : the $N=14$
subshell gap is not formed, large admixtures are present between
the nearby $\nu d_{5/2}$ and $\nu s_{1/2}$ orbits, and halo
nucleus shows up at $N \sim 15$. This latter feature was derived
in the one-neutron break-up of
$^{19}$C~\cite{Bazi95,Marq96,Bazi98}. Moreover a greatly enhanced 
$E1$ strength
at low excitation energy has been measured in the Coulomb
dissociation of $^{19}$C, in agreement with the halo
structure~\cite{Naka99}.

It is also seen from Fig.~\ref{E2SPE_CO} that the $N=16$ subshell
gap is expected to be a large, making $^{22}$C$_{16}$ a good
candidate for a new doubly-magic nucleus. Even more interestingly,
$^{22}$C would be a two-neutron halo nucleus, with a large $(\nu
s_{1/2})^2$ component, weakly bound to the $^{20}$C core.

\subsubsection{\it Conclusion}

The $Z=6$ and $Z=8$ shell closures have been studied in this
chapter. The existence of a large $Z=8$ shell gap together with
large neutron subshell gaps allows the existence of four almost
doubly-magic nuclei, $^{14,16,22,24}$O$_{6,8,14,16}$.
The new magic numbers at $N=14$ and $N=16$ come from large gaps in
energy at subshell closures.
These nuclei serve as good cores to derive the role of
nucleon-nucleon interactions at play in this mass region and
explain the spectroscopy of neighboring nuclei. In particular, the
strongly attractive spin-flip $\pi d_{5/2} - \nu d_{3/2}$
interaction explains why the the neutron drip line extends up to
$N=20$ and beyond in the $_{9}$F isotopes, whereas it is located at
$N=16$ in O. Another example is the role of the $\nu d_{5/2} - \nu
d_{5/2}$ interaction to create the $N=14$ gap. In the C isotopic
chain, this subshell gap no longer exists. This was ascribed to
other specific proton-neutron interactions, involving the
$\pi p_{1/2}$ orbit, and leading to the swapping of the
$\nu s_{1/2}$ and $\nu d_{5/2}$ shells between $^{17}$O and $^{15}$C.
Note that in these light nuclei, the values of the monopole interactions
are rather large ($\sim$ 2~MeV for many of them). Therefore sudden 
changes in
nuclear structure quickly occur while adding of removing few
nucleons. In addition to local magic numbers, haloes or cluster
shapes are omnipresent in this mass region when approaching the
proton or neutron drip-lines.

%% file: texteNZ20_9avril.tex
\section{The magic number 20}
The magic number $20$ arises naturally when using a harmonic
oscillator potential well to simulate the mean field interactions
in fermionic systems. The shell gaps $N,Z=20$ are formed between
the $d_{3/2}$ and $f_{7/2}$ orbits originating from the $N=2$ and
$N=3$ major oscillator shells, respectively. These orbits have
opposite parities, which is essential to explain the behavior of
this magic number against collective nucleonic excitations. The
following sections treat the evolution of the $N=20$ and
$Z=20$ shell closures at the two extremes of the valley of
stability. Historically, the first indication of a reduction of a
shell closure among those identified nowadays in the nuclear chart
was obtained for the magic number $N=20$. Contrary to what has
been and is still often taught in the textbooks, the magic numbers
can evolve far from the valley of stability due for instance to
specific proton-neutron interactions. This discovery has triggered
the physics of "exotic nuclei", a term which was inspired by the
discoverers of exotic lands during their journey all 
around the
world.

\subsection{\it Evolution of the $N=20$ shell closure\label{evolgapN20}}

The first indications of a vanishing of a shell closure in the chart
of nuclides were revealed around the neutron magic number
$N=20$. It was found that $^{31}_{11}$Na and $^{32}_{12}$Mg
exhibit anomalies in their binding energies~\cite{Thib75},
mean-square radii~\cite{Hube78} and nuclear
spectra~\cite{Detr79,Guil84}. The extra binding energy of these
nuclei, with respect to their neighbors, was attributed to their
deformation which was quickly associated with particle-hole
excitations across the $N=20$ shell
gap~\cite{Camp75,Pove87,Warb90}. The spherical $N=20$ gap is thus
not large enough to prevent excitations and correlations to
develop to the $fp$ shells in the $_{12}$Mg isotopic chain. As
these orbits lie outside the normal model space description of the
$sd$ shells, they are often referred to as \emph{ intruder states}. For some
nuclei, the intruder and normal configurations are inverted in
energy. They belong to the so-called "Island of
Inversion"~\cite{Warb90}. Soon after this finding many
experimental and theoretical efforts have been devoted to
determine the boundaries of this Island, and to understand the
underlying physics driving this sudden change between
pure $sd$ and full $sdfp$ configurations. The gain of binding
energy through deformation was invoked to explain why the drip
line extends further in the F and Ne than in the O isotopic chain.
Experiments performed at GANIL~\cite{Guil90,Tara97,Luky02},
RIKEN~\cite{Saku99,Nota02} and NSCL~\cite{Faue96} facilities have
shown that $^{26}$O and $^{28}$O are unbound against particle decay,
whereas the heaviest F and Ne isotopes contain 6 and 8 more
neutrons, respectively. It is fascinating to understand how a
single proton added to the $^{24}$O nucleus can bind 6 additional
neutrons to form the $^{31}$F nucleus. Also why the drip line is
so close to stability in the O chain ? These were burning
questions raised amongnuclear physicists that lasted the end
of the previous century and the beginning of the present one. It
was surmised that the onset of deformation at $N=20$ originates
from the reduction of the $N=20$ spherical shell gap, although 
there was no direct
experimental proof of this assumption. Additionally,
while the $d_{3/2}$ single particle moves closer in energy to the
$fp$ orbits to reduce the $N=20$ shell gap, a new shell gap at
$N=16$ should be formed below, between the $s_{1/2}$ and the
$d_{3/2}$ orbits. Also there was no proof for the emergence of
this new $N=16$ shell gap.

Thanks to the recent advent of radioactive ion-beam facilities worldwide,
the evolution of the $N=20$ shell closure and the appearance of a
new magic number at $N=16$ far from the valley of stability are
now well established and much better understood.
In the following sections we will use different experimental information
to investigate the evolution of the $N=20$, starting with the
systematics of the binding energies, followed by trends in $2^+$
energies and reduced transition probabilities B(E2; $0^+
\rightarrow 2^+$), beta-decay studies and transfer reactions.
Finally the underlying physics parameters which reduce the size of
the $N=20$ gap and increase that of the $N=16$ ones will be
discussed.

\subsubsection{\it Binding energies}

\begin{figure}[h!]
\begin{minipage}{8cm}
\begin{center}
\epsfig{file=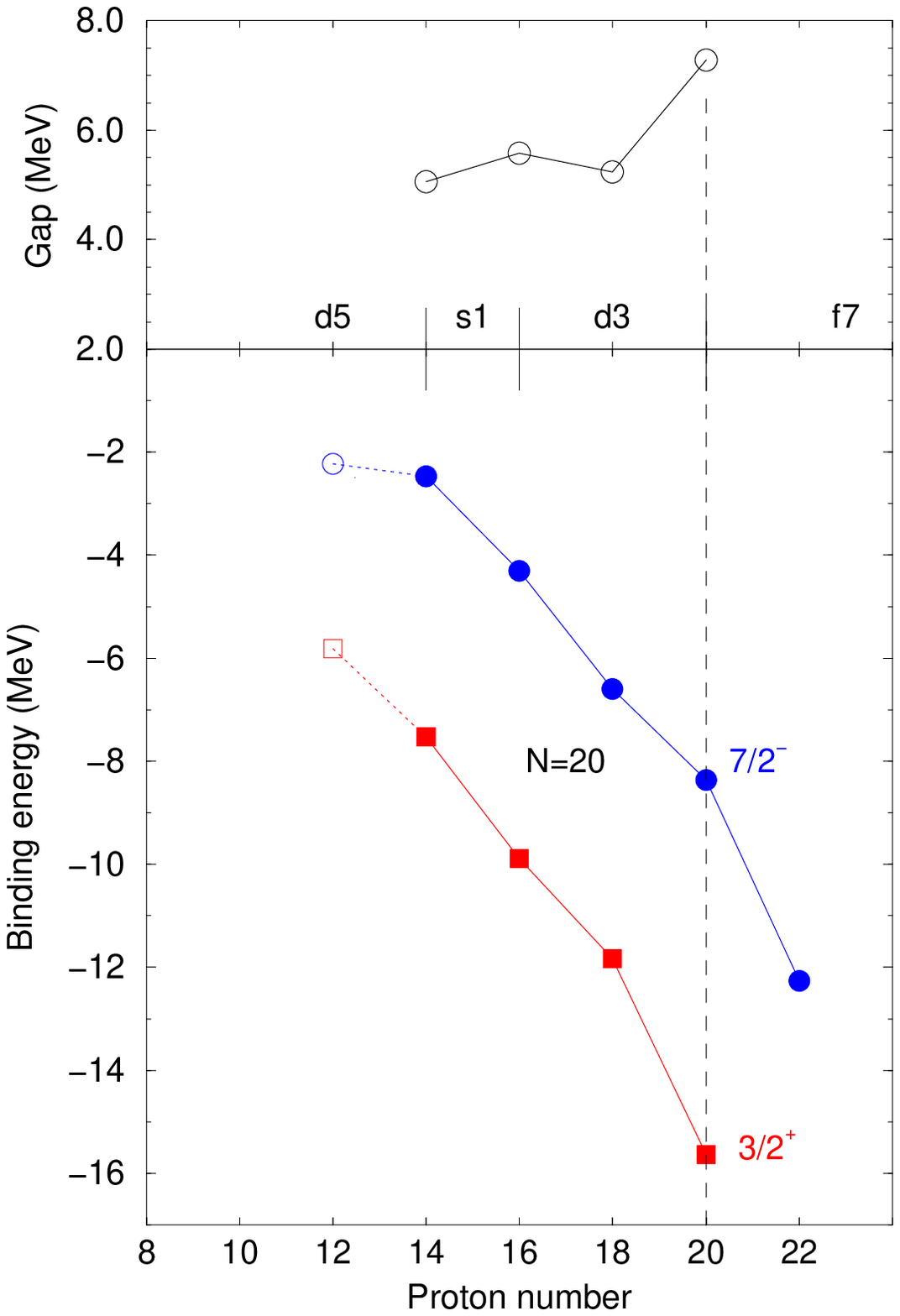,scale=0.55}
\end{center}
\end{minipage}\hfill
\begin{minipage}{8cm}
\epsfig{file=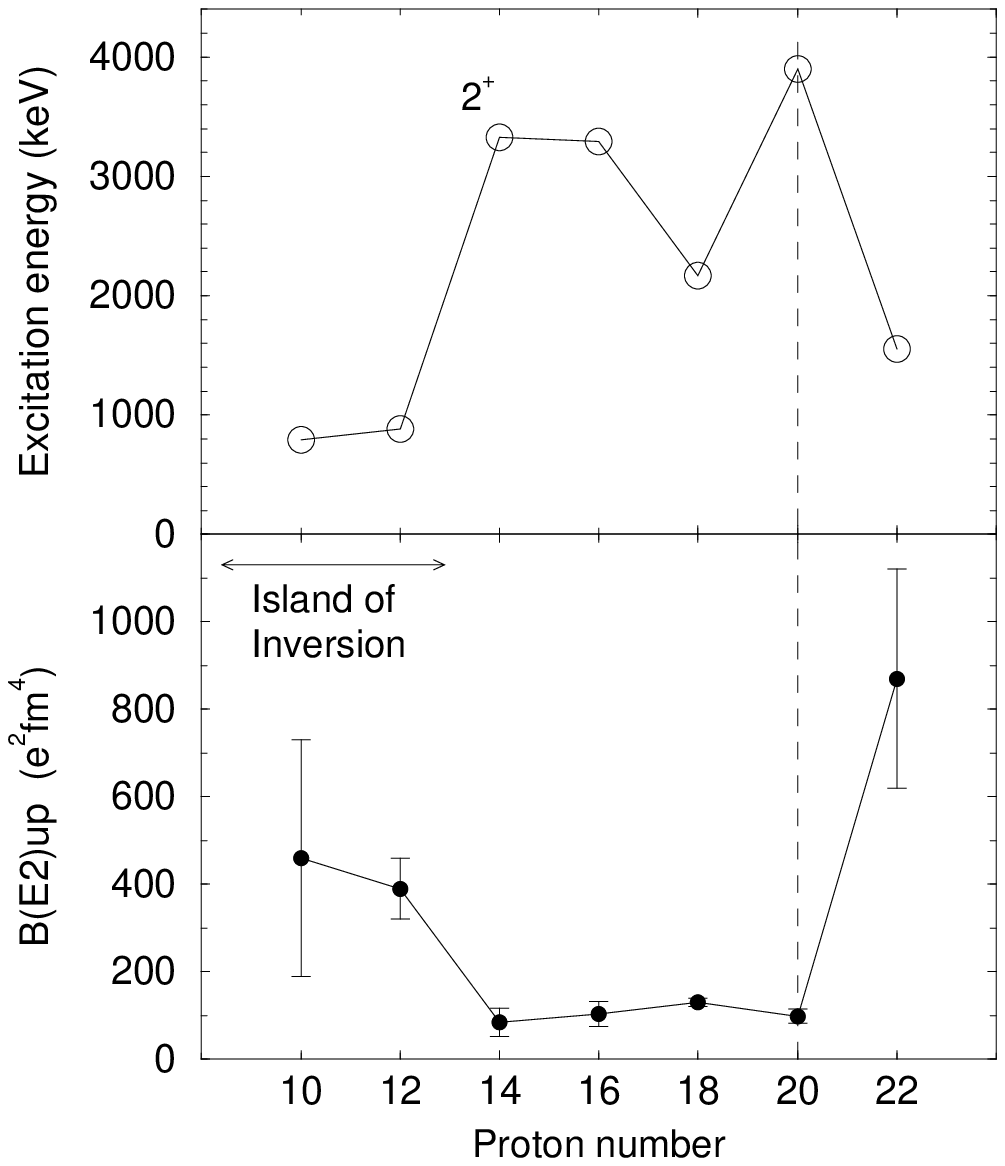,scale=0.65}
\end{minipage}
\begin{center}
\begin{minipage}[t]{16.5cm}
\caption{{\bf  Left}: Binding energies of the states $3/2^+$
($7/2^-$) located below (above) the $N=20$ magic number and
difference of the binding energies of these two states
surrounding the gap at $N=20$ (see Sect.~\ref{annex}). The proton
orbitals which are getting filled as a function of increasing
$Z$ are given in the middle of the figure.
{\bf  Right}: Experimental $E(2^+)$ and $B(E2; 0^+ \rightarrow
2^+)$ values in the $N=20$ isotones.} \label{gapN20}
\end{minipage}
\end{center}
\end{figure}

The binding energies of the last neutron in the $N=21$ and $20$
isotones are drawn in the left part of Fig.~\ref{gapN20}.
The energy difference between the two binding-energy curves - or
the size of the $N=20$ shell gap - amounts to about 5 MeV. This
curve displays a maximum at $Z=20$, due to the extra binding
energy of $^{40}$Ca which is  self-conjugate (see
Sect.~\ref{annex}). The $N=20$ shell gap remains large and almost
constant until 6 protons are removed from the $d_{3/2}$ and
$s_{1/2}$ shells. Below $Z=14$, the size of the $N=20$
\emph{spherical} gap can no longer be obtained due to the onset of
deformation that is occurring there. The corresponding data points
for $Z=12$ are also shown in Fig.~\ref{gapN20}. As compared
to a spherical case, they contain additional binding energy due to
quadrupole correlations. Interestingly the
$Z=8$ proton shell closure should in turn lead 
$^{28}$O$_{20}$ nucleus to be bound and spherical in shape. As we 
know that it is 
unbound by
a certain amount, the slope of the neutron binding energy curve of
the $3/2^+$ state below $Z=14$ (as protons are removed from the
$d_{5/2}$ orbit) should be steeper than above $Z=14$ (as protons
occupy the $d_{3/2}$ and $s_{1/2}$ orbits). This means that the
strength of the proton-neutron interactions is changing at $Z=14$,
while entering the island of inversion.

\subsubsection{\it Trends of E(2$^+$) and B(E2) values and the onset 
of collectivity}

The disappearance of the $N=20$ spherical shell closure far from
the valley of stability is now a well established phenomenon.
It is however interesting to remember that the $N=20$ shell closure
remains quite strong from $^{40}_{20}$Ca to $^{34}_{14}$Si
after the removal of 6 protons, successively from the $d_{3/2}$
and $s_{1/2}$ orbitals. There are a plethora of experimental data
which strengthen this statement, in particular derived from atomic
masses, $2^+$ excitation energies, and reduced transition
probabilities B(E2; $0^+ \rightarrow 2^+$). From atomic masses, it
is found that the size of the $N=20$ shell gap remains almost
constant and large, as shown in the left part of Fig.
\ref{gapN20}. The $2^+$ energies amount
to about 4~MeV for $^{40}$Ca and 3.5~MeV for the $^{36}$S and
$^{34}$Si isotones, (see the right part of Fig. \ref{gapN20}).
In $^{40}$Ca, a minimum of $2p-2h$ excitations is 
required to create a
$J=2$ positive-parity state across the gap formed by orbits of
different parities ($sd$ ($fp$) orbits have positive (negative)
parities). As a consequence the B(E2; $0^+ \rightarrow 2^+$)
value (drawn in the right part of Fig. \ref{gapN20}) is low 
as quadrupole excitations to the
$fp$ shells can proceed only via small matrix elements provided by
the overlap of the $sd$ and $fp$ wave functions. While decreasing
the number of protons from $^{40}_{20}$Ca$_{20}$, the vacant
proton holes created within the $sd$ shells should \emph{a priori}
increase the possibility to generate $p-h$ excitations to reach a
maximum value of the B(E2) at the mid proton $sd$ shell. As the 
full $sd$ shell contains a total of 12 protons, we would have 
expected a maximum collectivity for 6 protons in the $sd$ 
shells, i.e. for the $^{34}_{14}$Si$_{20}$ isotope. Experimental 
B(E2) values~\cite{Rama01,Ibbo98} clearly deviate from this simple 
argument, which assumed that the $sd$ states were 
quasi-degenerate in energy. It is the presence of significant 
sub-shell closures at $Z=14$ ($\sim$ 4.3~MeV)  and $Z=16$ 
($\sim$ 2.5~MeV) and the persistence of a large $N=20$ shell 
gap that keeps the B(E2) at small values and the $2^+$ state at 
high energy in the $^{36}$S and $^{34}$Si nuclei~\cite{Baum89}.

For isotopes below $Z=14$ the picture changes suddenly. The $2^+$ 
energy drops
down to 885~keV~\cite{Guil84} in $^{32}_{12}$Mg$_{20}$, whereas
the B(E2) value increases by a factor of about 4. The value
reported in the right part of Fig. \ref{gapN20} is obtained from
the weighted averaged results of the measurements obtained by
Coulomb excitation at intermediate energy at
RIKEN~\cite{Moto95,Iwas01} and NSCL~\cite{Prit99,Chur05}. The
energy of the first $2^+$ state in $^{30}$Ne has been tentatively
determined to be 791(26)keV from proton inelastic scattering 
measurement~\cite{Yana03} at RIKEN. A liquid hydrogen target was
bombarded by a radioactive beam of $^{30}_{10}$Ne$_{20}$ at
intermediate energy with a mean intensity of 0.2 ions per second.
$\gamma$-rays emitted in flight were detected by an array of
Na(Tl) scintillator detectors placed around the target. As 
mentioned by the
authors, the identification of the scattered particles based on
the energy loss and total energy method was difficult. Despite
this problem, a B(E2) value of 460(270)$e^2fm^4$ has been derived
for $^{30}_{10}$Ne. This value is, within error bars, similar
to that of $^{32}_{12}$Mg. The sudden increase of the B(E2) values
at $Z=12$ indicate that neutron are occupying not only the $sd$
but also the $fp$ shells. Otherwise the B(E2) values would have
been much lower~\cite{Moto95}.

The large values of B(E2) in the Mg chain can arise only if the
valence neutrons include the $fp$ orbits. Within the
$N_p N_n$ scheme, the B(E2) values are expressed as a function of
the product of the number of valence protons $N_p$ and neutrons
$N_n$~\cite{Cast90}. If the $N=20$ shell gap is large, the twelve
neutrons above the $^{16}$O core are blocked inside the $sd$
shells. Thus $N_n \simeq 0$ and hence the value of B(E2) value 
is small.
Conversely, if the $N=20$ shell gap vanishes, neutrons can move in
a wider space which extends to the $fp$ shells, leading to
$N_n$=12. Shell model calculations restricted to the $sd$ shells
cannot account for the sudden increase of B(E2) from $^{34}$Si to
$^{32}$Mg (see the right part of Fig. \ref{gapN20}). It is sufficient
to
invoke $2p-2h$ neutron excitation from the $sd$ to the $fp$ shells to
increase the B(E2) value. For the nuclei which reside in the
Island of Inversion, the ground states are dominated by $2p-2h$
configurations. There, the loss of energy due to the promotion of
2 neutrons across the $N=20$ shell gap ($2 E_{gap}$) is largely
compensated by the gain of correlation energy obtained from
$2p-2h$ excitations~\cite{Caur98}. As
described by Heyde et al.~\cite{Heyd91} this correlation energy
comprises proton-neutron and neutron-neutron monopole and
quadrupole terms, the sum of which is larger than $2 E_{gap}$
in $^{32}$Mg. Naturally the dominance of $2p-2h$ over the normal
$sd$ configuration will also be favored if the single-particle
energy difference between the $d_{3/2}$ to the $fp$ orbits,
$E_{gap}$, is lowered. Between $Z=20$ and $Z=14$, no such reduction is
found as shown in Fig.~\ref{gapN20}. But at $Z=12$, the nucleus
suddenly becomes deformed. This sudden change of structure
triggers several questions that the following section intends to
address. Why does the removal of six protons leave the nuclei in a
spherical shape, whereas the removal of two additional ones leads
to prompt deformation~? In other words, what is the major
deformation-driving mechanism below $Z=14$ and what makes it so
efficient~? Fundamental reasons are taken from properties of the
nucleon-nucleon interactions which are at play in these nuclei.

\subsubsection {\it Evolution of neutron SPE's\label{disappearN20}}

The $N=20$ shell gap is formed between the $\nu d_{3/2}$ and the
$\nu f_{7/2}$ (or $\nu p_{3/2}$) orbit. To simplify the discussion
we do not consider the $\nu p_{3/2}$ orbit, to which similar
arguments could be applied. From $^{40}_{20}$Ca$_{20}$ to
$^{36}_{16}$S$_{20}$, 4 protons are removed from the $d_{3/2}$
orbit. Therefore the size of the $N=20$ gap scales with the
difference of monopole interactions (see
Sect.~\ref{monopoleffect}), $4 \times (V^{pn}_{d_{3/2} f_{7/2}}$ -
$V^{pn}_{d_{3/2} d_{3/2}}$). The fact that the $N=20$ gap remains
constant comes from the fact that the two monopole interactions,
$V^{pn}_{d_{3/2} f_{7/2}}$ and $V^{pn}_{d_{3/2} d_{3/2}}$, are
almost similar in strength. They are determined to be
$\simeq$ -1~MeV from several experimental binding energy curves
and energies of multiplets in several odd-odd nuclei \cite{Porq08}.

\begin{figure}[t!]
\begin{center}
\epsfig{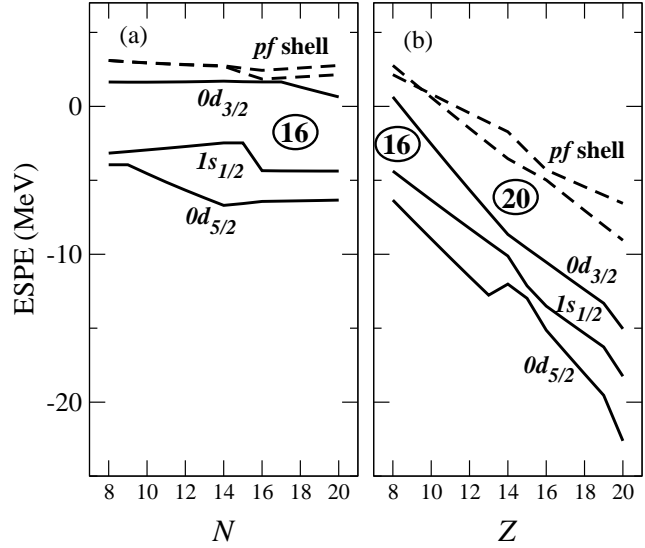}
\begin{minipage}[t]{16 cm}
\caption{ Effective Single Particle Energies (ESPE) of neutrons in
the $_{8}$O isotopic chain (left) and in the $N=20$ isotones with
$8 < Z < 20$ (right). These figures have been adapted from Fig.~5
of Ref.~\cite{Utsu99} and Fig.~1 of Ref.~\cite{Otsu02}. }
\label{ESPE_N20}
\end{minipage}
\end{center}
\end{figure}

As soon as protons are removed from the $d_{5/2}$ orbit, the size
of the $N=20$ gap is changing by virtue of another difference of
monopole interactions, i.e. ($V^{pn}_{d_{5/2} f_{7/2}}$ -
$V^{pn}_{d_{5/2} d_{3/2}}$). As mentioned by Otsuka et
al.~\cite{Otsu01,Otsu07} and shown in Fig.~\ref{ESPE_N20}, the
$V^{pn}_{d_{5/2} d_{3/2}}$ value is expected to be large, due to
the spin-isospin dependence of the nuclear interaction. This
interaction is maximized for the exchange of nucleons and spin
orientations, and when protons and neutrons have the same orbital
momentum. Conversely the overlap of protons and neutrons wave
functions in the $V^{pn}_{d_{5/2} f_{7/2}}$ matrix element is
weaker as their angular momenta are different ($\ell_p =$2,
$\ell_n =$3). As the spins of the protons in $d_{5/2}$ and the
neutrons in $f_{7/2}$ are both aligned with respect to their
angular momenta, the $V^{pn}_{d_{5/2} f_{7/2}}$ interaction
contains repulsive tensor forces. These two effects make the
$V^{pn}_{d_{5/2} f_{7/2}}$ monopole smaller in absolute value than
the $V^{pn}_{d_{5/2} d_{3/2}}$ one, giving rise to a large
reduction of the $N=20$ shell gap. The
monopole values used in Ref.~\cite{Utsu99} and Fig.~\ref{ESPE_N20}
have several origins: the USD interaction for the $sd$-shell part
has been fitted so as to reproduce the structure of stable nuclei,
the $fp$-shell part (Kuo-Brown) has been obtained from the
renormalized G matrix, whereas the cross-shell part is based on
the Millener-Kurath interaction. Several monopole interactions had
to be subsequently modified in order to reproduce the location of
the drip-line in the O isotopic chain. Moreover the variation of 
the ESPE shown in Fig.~\ref{ESPE_N20} is based on some estimated 
monopole values, the intensity of which could be updated if new 
physics ingredients are added in the future.

The present predictions on the evolution of ESPE lead to several
consequences that should be ascertained experimentally. First, as
the $N=20$ shell gap decreases, $2p - 2h$ excitation are increased
and the ground state wave-function of nuclei in the island of
inversion should contain admixtures of $sd$ and $fp$ states.
Second, the 'intruder' $fp$ states should be present at the
proximity of the ground state for lighter, less deformed, nuclei.
Third, the reduction of $N=20$ is associated to the emergence of a
new shell gap at $N=16$. The search for this new shell gap, which
is predicted to be large for the heaviest O isotopes (see left
part of Fig.~\ref{ESPE_N20}) has been carried out by several means
these last years. These three topics are addressed in the next
sections.

\subsubsection {\it Static properties and $\beta$-decay}

Static properties of the nuclei, such as the g-factors and
quadrupole moments~\cite{Neug06}, and $\beta$-decay schemes have been
extensively used to map the island of inversion.
Looking at the Mg isotopic chain~\cite{Kowa08,Neye05,Yord07}, 
these combined approaches are
bringing a wealth of information on the spin-parities of the
ground states and their mixed configurations involving normal and
intruder orbits.

Starting at $^{29}_{12}$Mg$_{17}$, the measured $g$-factor and
quadrupole moments are in good agreement with the ones calculated
with the USD interaction~\cite{Kowa08} which applies to a
valence space limited to the $sd$ orbits. This nucleus is
therefore sitting outside the Island of Inversion. The low-energy
level scheme and the Gamow-Teller distribution of $^{31}$Mg have
defined $N=19$ as the neutron number at which the transition takes
place~\cite{Klot93}. The spin, parity and magnetic moment of its 
ground state have been
determined by Neyens et al.~\cite{Neye05} by combining
hyperfine-structure and $\beta$-NMR measurements with an optically
polarized ion beam at ISOLDE (CERN). A $J=1/2^+$ configuration of
the ground state, with $\mu_{exp}=-0.88355 \mu_N$ can be accounted
for by an almost pure $2p-2h$ configuration. Shell model
calculations using the SDPF-NR interaction reproduce perfectly the
ground state as well as the high level density below 
500~keV~\cite{Mare05}. 
Hence the global physics picture is well understood.

The $\beta$-decay pattern of these nuclei is also very sensitive
to the configurations of the mother and daughter nuclei. In the
neutron-rich Mg isotopes, the $\beta^-$ decay occurs mainly
through the $\nu d_{3/2} \rightarrow \pi d_{5/2}$ Gamow-Teller
transition. If the filling of the neutron  is restricted to the
$sd$ orbits, the $\nu d_{3/2}$ orbit contains two neutrons in
$^{30}$Mg$_{18}$ and four neutrons in $^{32}$Mg$_{20}$. The
$\beta$-decay transitions of the $J^{\pi}=0^+$ ground state of
these even-even nuclei feed mainly $J^{\pi}=1^+$ states in the
daughter odd-odd nuclei via allowed Gamow-Teller transition with
$\Delta \ell=0, \Delta J=1$. In the $^{30}$Mg $\rightarrow$
$^{30}$Al decay, the feeding to the first $1^+$ state at 688 keV
exhausts almost all the $\beta$-decay strength, according to
Ref.~\cite{Piet03}. This shows that only neutrons from the $sd$
shell contribute to this decay pattern. On the other hand, the
$\beta$-decay of $^{32}$Mg proceeds to the $1^+$ ground state of
$^{32}$Al with a $\beta$-feeding $I_\beta$ of about 55\%. Another
large part goes to high lying levels at 2.7 and 3.2 MeV, with
$I_\beta$ of about 25 and 10\%, respectively. The remaining 5\%
goes to neutron-unbound states giving rise to $\beta$-delayed
neutron emission. The $\beta$ feeding writes as a function
of the $\beta$ strength $S_\beta$ and the $Q_\beta$ value as
follows:

\begin{equation}
I_\beta (E^*) = \frac{S_\beta(E^*) \times (Q_{beta} -
E^*)^5}{\sum_{E^*} S_\beta(E^*) \times (Q_{beta} - E^*)^5}
\end{equation}

the $(Q_{beta} - E^*)^5$ factor, originating from the phase space
in the $\beta$-decay, strongly favors the decay to low excited
states, such as the ground state with $E^*=0$. In the case of the
$^{32}$Mg decay, the $Q_{beta}$ value amounts to about 10 MeV.
When assuming a given value of $S_\beta$, a ratio of $(10/7)^5
\simeq 6$ is found between the ground state and a level at 3 MeV.
Taking this phase space factor into account, it is found that the
observed strength to excited states is indeed four times larger
than that to the ground state. Had the neutrons been restricted to
the $sd$ shells, the $\beta$-decay pattern would have been similar
to that in the $^{30}$Mg nucleus, i.e. almost all focused to low
energy states. When assuming that the $^{32}$Mg ground state has a
fairly large $2p-2h$ component, part of the $d_{3/2}$ neutrons
have moved to the $fp$ orbits\footnote{ 
Single-neutron knock-out experiments on $^{30,32}$Mg$_{18,20}$
projectiles have been performed, revealing significant direct 
population of
negative-parity intruder states of $^{29,31}$Mg~\cite{Terr08}.
The total observed spectroscopic strength to these states are
0.60(12) and 1.78(38) respectively, indicating a large increase 
of the $2p-2h$ component from $N=18$ to $N=20$.}.
Thus the $\beta$-decay pattern is
shared in two parts, the one which arises from the remaining
$d_{3/2}$ neutrons occurs to the ground state, the other which
originates from neutrons in the $fp$ orbits is found at higher
energy. This qualitative description is of course over simplified
as the complete mixing of configurations of the mother and
daughter nuclei have to be taken into account.

Finally the discovery of an allowed $\beta$-decay from
$^{33}_{11}$Na$_{22}$ to the ground state of
$^{33}_{12}$Mg$_{21}$, with log($ft$)=5.27(26), also provides
direct evidence of the inversion of the 7/2$^-$ and
3/2$^+$ states~\cite{Numm01b}. With this inversion, the $\nu
d_{3/2} \rightarrow \pi d_{5/2}$ decay becomes possible
between the ground-state levels.  Otherwise the allowed
$\beta$-strength would have been found at higher excitation
energy.

\subsubsection{\it Search for intruder $fp$ states by transfer
and knock-out reactions\label{transfer20}}

As depicted in the previous paragraph, the \emph{ground-state}
composition of a given nucleus could be partly probed from its
$\beta$-decay pattern. Stripping $(d,p)$ or one-nucleon knock-out
reactions are known to be ideal tools to determine the
single-particle energy and the spectroscopic factors (SF) or
occupancies of the valence and occupied orbits, respectively.
These complementary methods have been used to search for the
presence of intruder states, with the aim of determining which
fraction of the nucleons are promoted from the $sd$ to the $fp$ shells.

\begin{figure}[h!]
\begin{center}
\epsfig{file=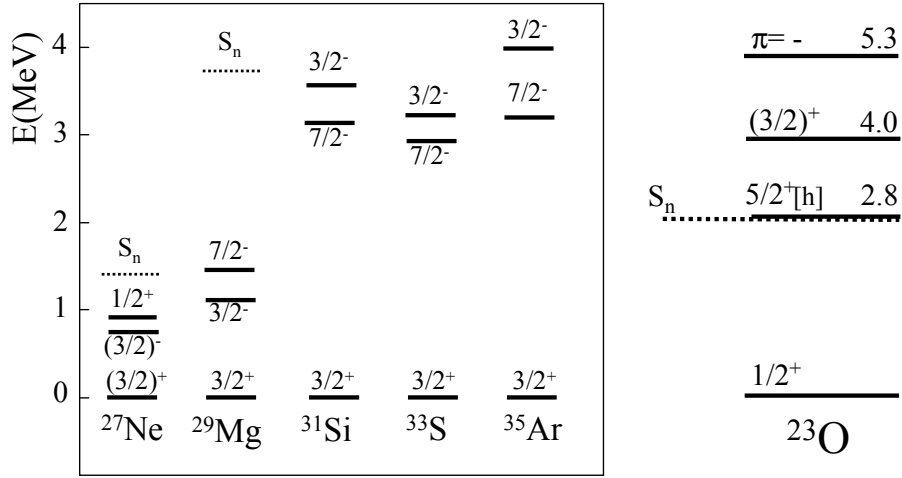,width=12cm}
\begin{minipage}[t]{16 cm}
\caption{ \textbf{Left} : Experimental energies of the excited
states in the $N=17$ isotones, showing the decrease of the
intruder states $3/2^-$ and $7/2^-$ with respect to the ground
state (adapted from the results of~\cite{Ober06, Terr06}).
\textbf{Right}: Experimental energies in $^{23}$O$_{15}$ taken from
Refs.~\cite{Elek07, Schi07}, [h] corresponds to a hole state.
\label{N17isotones}
}
\end {minipage}
\end{center}
\end{figure}

The reduction of the $N=20$ shell gap should be accompanied with
the appearance of negative-parity states (arising from the $fp$
orbits) at low excitation energy. In the two $N=17$ isotones,
$^{31}_{14}$Si and $^{29}_{12}$Mg, the energy of the first
negative-parity state drops by more than 2~MeV with respect to the
$3/2^+$ ground state as shown in the left part of
Fig.~\ref{N17isotones}. This feature is likely to be due to the
reduction of the $N=20$ shell gap or/and to the appearance of
deformation below $Z=14$. In this later case, the wave functions
of these states carry a small fraction of the total
single-particle strength originating from the spherical orbits.
Experimental $(d,p)$ results obtained so far in these two nuclei
do not really permit to determine the evolution of the $N=20$
shell closure as the spectroscopic factors of the states are
either unknown (in $^{29}_{12}$Mg) or incertain (in
$^{31}_{14}$Si). This task is not an easy one as the core nuclei
$^{30}$Si and $^{28}$Mg do not exhibit strong closed-shell
configurations. Therefore, the neutron single-particle states mix
with excited states of the cores. A proper unfolding of
these correlations should be undertaken before attributing this
sudden change of excitation energy to a change of proton-neutron
monopole matrix elements below $Z=14$.
\vspace{0.7cm}

As compared to the $^{29}_{12}$Mg$_{17}$ case, the situation is
expected to be a little simpler in the $^{27}_{10}$Ne$_{17}$
nucleus as its $^{26}$Ne core has a weaker B(E2) value and a $2^+$
state at higher energy as compared to $^{28}$Mg. The ground-state
configuration is due to a single neutron in the $d_{3/2}$ shell.
Shell model calculations within the $sd$ shell predict a single
bound excited state in $^{27}$Ne, $J=1/2^+$, at about 800~keV. Its
configuration corresponds to a hole into the neutron $s_{1/2}$
orbit, a pair of neutron being coupled to $J=0^+$ in the $d_{3/2}$
shell. The presence of intruder states in the $N=17$ isotones
would reveal the reduction of the $N=20$ shell gap between the
$\nu d_{3/2}$ and $\nu f_{7/2}$ or $\nu p_{3/2}$ orbits. By
including the possibility of excitations into the $fp$ shells, and
estimations of the relevant monopoles matrix elements, two
intruder states are predicted below 1~MeV using the SDPF-M
effective interaction~\cite{Utsu99}.

The $^{26}_{10}$Ne$(d,p)$ reaction has been used in inverse 
kinematics by Obertelli et al.~\cite{Ober06} to determine the 
states in $^{27}_{10}$Ne$_{17}$ from the observation of their
$\gamma$-decay into segmented Ge detectors. Two states were fed at
765~keV and 885~keV. The $(d,p)$ cross section for the state at
765~keV leads to a SF value, $C^2S(d,p)=0.6 \pm 0.2$. This is
consistent with the value obtained with the SDPF-M interaction for
a negative parity state which could be $3/2^-$ or $7/2^-$. A
weaker spectroscopic factor, $C^2S (d,p)=0.3 \pm 0.1$ is found for
the state at 885~keV for which a spin assignment of $1/2^+$ was
suggested on the basis of its excitation energy and spectroscopic
factor values. One expects a rather small SF for the $s_{1/2}$
level, which is in principle fully occupied when pair scattering
to the $d_{3/2}$ state is not taken into account. As protons were
stopped into the thick target used to induce the $(d,p)$ reaction,
their angular distribution could not be used to ascertain the spin
assignments of the observed states.

Complementary information on $^{27}$Ne was obtained from the work
of Terry et al.~\cite{Terr06}. The one-neutron knock-out reaction
$-1n$ from a $^{28}$Ne beam was used to study excited states in
$^{27}$Ne. As compared to the $(d,p)$ technique which aimed at
probing neutron valence states, the one-neutron knock-out occurs
from occupied levels. To see any component arising from an
intruder state would mean that the ground state wave function of
$^{28}$Ne already contains some amount of intruder configuration.
A segmented Ge array surrounded the production target for the
detection of $\gamma$-rays in coincidence with the $^{27}$Ne
nuclei in the focal plane of a spectrograph. The level scheme was
found to be similar to the one proposed in Ref.~\cite{Ober06}. In
addition, a \emph{ prompt} $\gamma$ transition between the $1/2^+$
and the intruder state has been observed. This discards a spin
assignment of $7/2^-$ for the intruder state, which would have no
decay to the $1/2^+$ due to their large spin difference. Moreover
momentum distributions of the knock-out residues were used to
determine the spin values and spectroscopic factors of the two
excited states. It was found that the intruder state has an
orbital angular momentum value, $\ell \leq 1$. Combining these
pieces of information, it was inferred that the state at 765~keV
is likely to have a $J=3/2^-$ (from the $p_{3/2}$ orbit) and not a
$7/2^-$ (from the $f_{7/2}$ orbit) configuration. Its
spectroscopic factor $C^2S(-1n)$ of 0.32(4), means that the
intruder $p_{3/2}$ shell is partly occupied in $^{28}$Ne. This
roughly agrees with the one predicted (0.24) by the shell model
calculations using the SDPF-M interaction.

The presence of intruder state was also investigated 
in the $^{23}_{~8}$O$_{15}$ nucleus by Elekes et al.~\cite{Elek07} using
the $^{22}$O$(d,p)$$^{23}$O reaction with a 30~mg/cm$^2$ thick
CD$_2$ target.
Protons were detected in the backward direction by 156~CsI(Tl)
scintillators, while an array of 80~NaI(Tl) crystals was used to
detect de-excited $\gamma$-rays produced during the stripping
reaction. A neutron wall was used additionally to detect neutrons
emitted after the feeding of neutron-unbound states in $^{23}$O.
Their energy was deduced from their time of flight. The excitation
spectrum of $^{23}$O (see the right part of
Fig.~\ref{N17isotones}) was reconstructed from the momentum of the
neutron and the total energy of the $^{22}$O nucleus measured in a
telescope of Si detectors. Two states were discovered above the
neutron separation energy of 2.74~MeV. The one at 4.0~MeV (unbound
by about 1.3~MeV) was assigned to a $3/2^+$ configuration, while a
negative parity was surmised for the state at 5.3~MeV. The energy
difference between these two unbound levels is as small as
$\sim$1.3~MeV. It is related to the considerable shrink of the
$N=20$ shell gap, the energy of which is compatible with that
obtained from the Monte Carlo Shell Model calculations based on
the SDPF-M interaction~\cite{Otsu01b}.

\subsubsection{\it The appearance of a new magic number at $N=16$}

As was discussed in the previous paragraph, experimental evidences
for the reduction of $N=20$ and the proximity of intruder states
have been found. The emergence of a new shell gap at $N=16$ for $Z
< 14$ should naturally result from the reduction of the $N=20$
gap. Apart from the fascinating discovery of a new doubly-magic
nucleus, such as $^{24}$O, this feature would ensure that the
global evolution of nuclear structure is understood in this mass
region.

The existence the $N=16$ shell closure was first inferred
from the systematics of the two neutron separation energies
$S_{2n}$ by Ozawa et al.~\cite{Ozaw00}. There, a break in the
$S_{2n}$ values at $N=16$ was observed for nuclei having a large
isospin number $T_Z=7/2$. In the same paper, sudden increase of
interaction cross sections ($\sigma_I$) were observed in the F, O
and N isotopes at $N \simeq 15$. The $\sigma_I$ values for the
$^{23}$O and $^{25}$F isotopes could be reproduced with a
dominance of the $s_{1/2}$ shell for the valence orbit. This was
corroborated by Sauvan et al.~\cite{Sauv04} who measured
relatively narrow longitudinal momentum distributions from the
one-neutron removal reaction. Confirmation was obtained by
Cortina-Gil et al.~\cite{Cort04} for $^{23}$O using the exclusive
one-neutron knock-out reaction in coincidence with de-exciting
$\gamma$-rays of the core nucleus $^{22}$O. A large occupancy of
the $s_{1/2}$ orbit has been deduced from these works, which
suggests that the $s_{1/2}$ orbit is well separated from the
$d_{5/2}$ and $d_{3/2}$ orbits by the $N=14$ and $N=16$ gaps,
respectively. Had one of these gap been weakened, the pairing
interaction would have provided a large mixing between the $s$ and
$d$ wave functions for the ground state of $^{23}$O.
The size of the $N=14$ shell gap has been first determined from
the spectroscopy of the $^{22}$O nucleus, which was studied by
Stanoiu et al.~\cite{Stan04} by means of the in-beam $\gamma$-ray
spectroscopy in the fragmentation of stable and radioactive beams.
The energy of the first $2^+$ and $3^+$ states have been found to
be 3.2~MeV and 4.58~MeV, respectively. These excited states have a
$1p-1h$ configuration with one particle in the $s_{1/2}$ orbit
coupled to one hole in the $d_{5/2}$ orbit, the energy of the two
states ($2^+$ and $3^+$) being splitted by the residual $1p-1h$
interaction. A $(2J+1)$ weighted average value of 4.0~MeV is
obtained from their measured excitation energies, meaning that the
distance in energy between the two neutron orbits, $\nu d_{5/2}$ and 
$\nu s_{1/2}$, is likely 4.0~MeV in $^{22}$O. This value was predicted 
by the USD interaction~\cite{Brow01}, and has been recently confirmed 
by the work of Schiller et al.~\cite{Schi07}, from the observation of
a resonance in neutron-fragment coincidences. This corresponds to 
the decay of the first excited state of $^{23}$O at 2.8~MeV, 
assigned as the 5/2$^+$ hole state, into the ground state of 
$^{22}$O (see the right part of Fig.~\ref{N17isotones}).

The in-beam technique could in principle be used to determine the
energy of excited states in the $^{23}$O and $^{24}$O nuclei to
evaluate the size of the $N=16$ shell gap. The non observation of
any $\gamma$-ray \cite{Stan04} from these nuclei has suggested
that all the excited states lie above the neutron decay threshold,
which amounts to 3.7~MeV in $^{24}$O. Based on the arguments
derived above from the study of the $^{22}$O nucleus, a lower
limit could be set on the size of the $N=16$ gap to be $\simeq$
3.7~MeV. As already reported in the previous section (see the right
part of Fig.~\ref{N17isotones}), an unbound state at 4.0~MeV above the
$1/2^+$ ground state has been populated in $^{23}$O by means of
the $^{22}$O$(d,p)$ reaction~\cite{Elek07}. This resonance was
assigned to the $d_{3/2}$ level on the basis of the proton angular
distribution and spectroscopic factor value. The size of the
$N=16$ gap therefore amounts to about 4~MeV, in accordance with
the shell model calculations of Refs.~\cite{Utsu99,Brow01}.

\subsubsection {\it Conclusion}

Since the discovery of the major shell closures  and the shell
gaps between them, about 40 years of nuclear physics have
sustained the postulate
of protons and neutrons shell closures which survived
all through the nuclear chart. The $N=20$ shell closure is the
first in which a weakened gap has been invoked to explain the
anomalies in atomic masses, charge radii, and spectroscopy around
the $^{32}$Mg nucleus. At that time already, the underlying
mechanisms which produced this disappearance of closed were
proposed by the shell model and mean field approaches through
particle-hole excitation and the restoration of broken symmetries,
respectively.

These pioneering works have undoubtedly triggered a new era in the
study of nuclear structure, that of the study of the so-called
"exotic nuclei". Thanks to the development of suitable radioactive
ion beams worldwide, further studies of these nuclei became
available. Now the properties of this shell closure as well as the
reasons for its reduction are rather well understood and are
summarized below.

- The N=20 magic number is formed between two shells of the harmonic
oscillator well having opposite parities. Therefore $1p-1h$
excitations cannot form positive parity states, such as $2^+$
excitations. Therefore, quadrupole collectivity develops only with a
minimum of $2p-2h$ excitations across $N=20$. As the size of the
$N=20$ remains as large as 5~MeV between the $_{20}$Ca and $_{14}$Si
isotopic chains such excitations are limited to high-lying states
such as the superdeformed bands in $^{40}$Ca~\cite{Ideg01}.

- The $2p-2h$ excitations may be favored over the $0p-0h$ configuration
as soon as the size of the $N=20$ shell gap is reduced. This
phenomenon occurs suddenly between the Si and Mg isotopic chain by
the action of the proton-neutron interaction $\pi d_{5/2}$-$\nu
d_{3/2}$ which weakens the $N=20$ gap by about 1.2~MeV by the removal
of 2 protons, from an initial value of about 5~MeV.

- As the $V^{pn}_{d_{5/2}d_{3/2}}$ matrix element has by far the
highest value among the other monopoles which intervene in this
region of the chart of nuclides, it rules the evolution of the
nuclear structure as soon as it is involved. Its large intensity
comes from the intrinsic properties of the nucleon-nucleon
interaction.

- When removing further protons to reach the $_8$O chain, the
$N=20$ gap plummets. Intruder states originating from excitations
across the $N=20$ gap to the $fp$ shells have been discovered only
about 1~MeV above the last occupied $d_{3/2}$ shell. 

- A remarkable consequence of the disappearance of the $N=20$
shell closure is the emergence of a new magic number at $N=16$,
formed in $^{24}_{~8}$O$_{16}$ between the occupied $s_{1/2}$ and
the unbound valence $d_{3/2}$ orbit. This gap amounts to about
4~MeV and the $d_{3/2}$ orbit is unbound by about 1.5~MeV. The
neutron-neutron interactions due to the filling of this orbit do
not suffice to bind the $^{26,28}$O$_{18,20}$ nuclei.

Last but not least, the present study can undeniably serve to
explain similar sudden disappearance of the harmonic oscillator
shell closures, $N=8$ and $N=40$, while removing two protons 
from $^{14}_{~6}$C$_{8}$
to $^{12}_{~4}$Be$_{8}$ and from $^{68}_{28}$Ni$_{40}$ to
$^{66}_{26}$Fe$_{40}$ nuclei, respectively. For these two cases
also, the role of the proton-neutron spin-isospin interactions
(such as $\pi p_{3/2}$-$\nu p_{1/2}$ and $\pi f_{7/2}$-$\nu
f_{5/2}$, respectively) are essential to create other islands of
inversion at $N=8$~\cite{Iwas00, Navi00, Pain06} and
$N=40$~\cite{Hann99, Sorl03}.

\subsection{\it Evolution of the $Z=20$ shell closure\label{evolgapZ20}}

The $_{20}$Ca isotopic chain spans over about 22 hitherto
discovered isotopes, among which four ones behave as doubly-magic
nuclei. This is ascribed to the presence of the well-known $N=20$
and $N=28$ major-shell closures and of the newly-observed $N=16$
and $N=32$ sub-shell closures. As the Ca core is hardly polarized
by adding neutrons, the neutron single-particle energies in the Ca
and neighboring isotopes  serve as references to constraint
effective nuclear interactions used for the shell model or mean
field calculations. Also, the evolution of the  $d_{3/2}$,
$s_{1/2}$ and $d_{5/2}$ proton single-particle energies in the
$_{19}$K isotopic chain with the filling of the neutron $f_{7/2}$
orbit brings a wealth of information on the tensor forces acting
between the $\pi d$ and $\nu f$ orbits. These ingredients will in
turn become essential to understand the evolution of other magic
numbers, such as the $N=28$ ones.

\subsubsection{\it Binding energies\label{bindenergyZ20}}

\begin{figure}[h!]
\begin{minipage}{9.cm}
\begin{center}
\epsfig{file=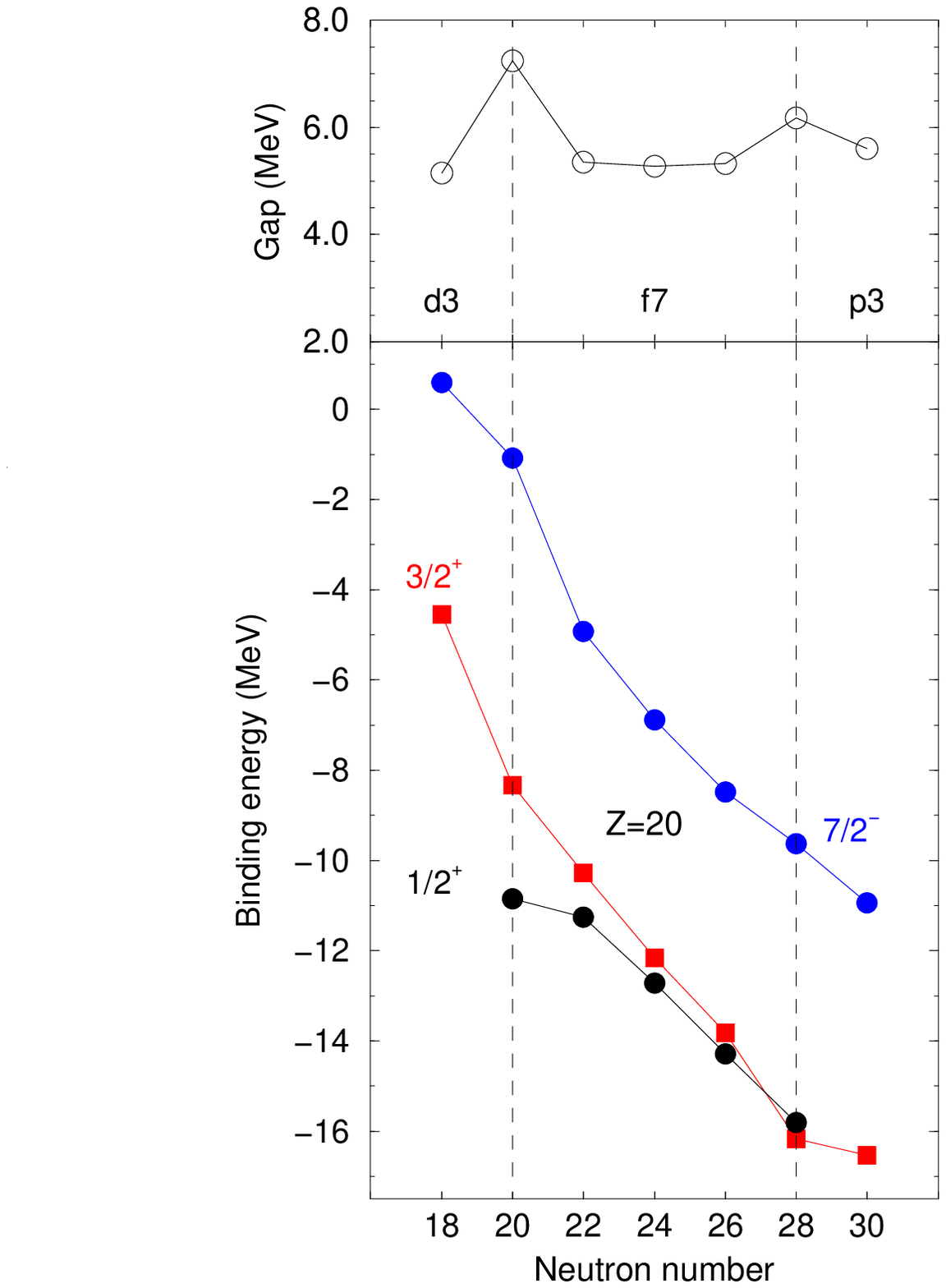,scale=0.55}
\end{center}
\end{minipage}\hfill
\begin{minipage}{9.cm}
\epsfig{file=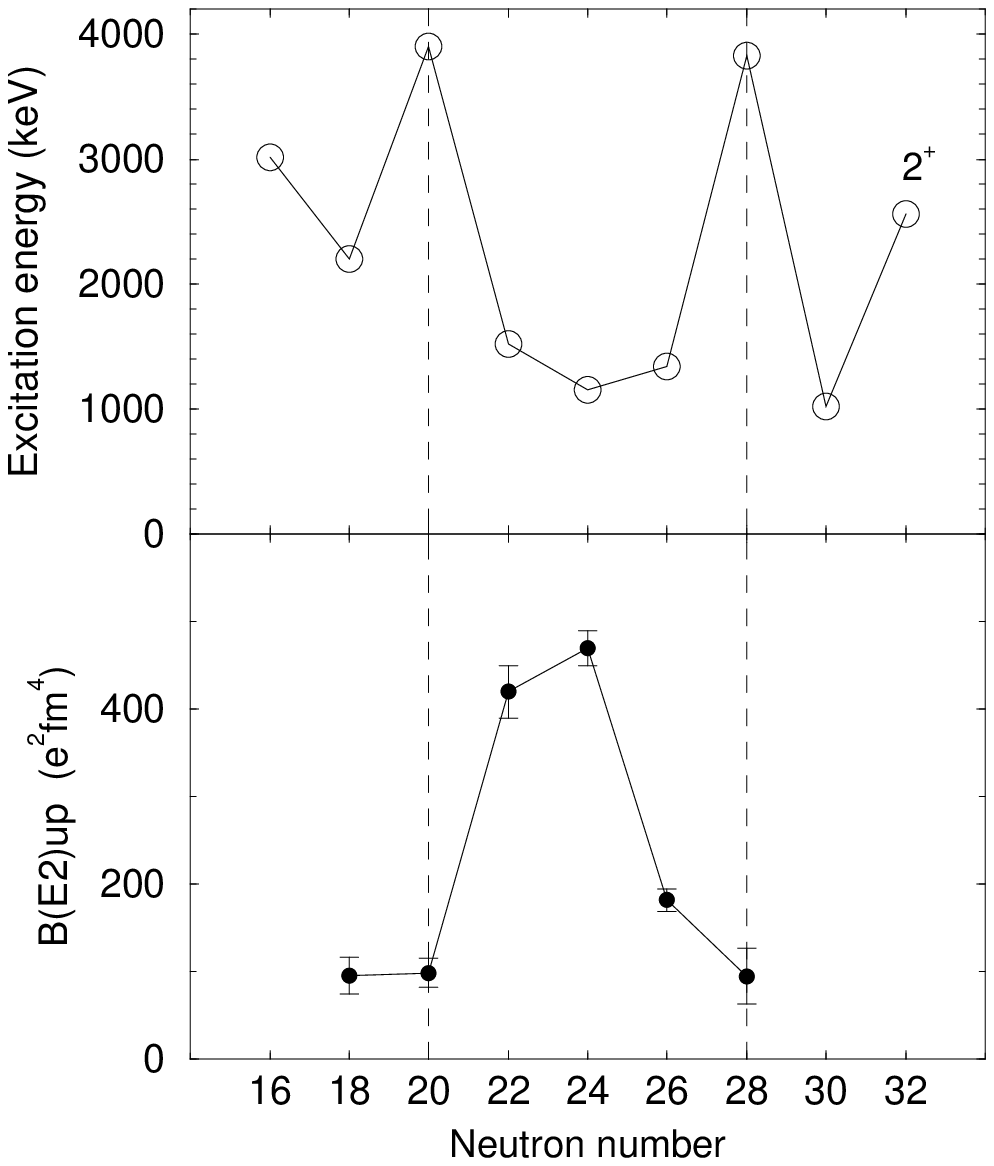,scale=0.55}
\end{minipage}
\begin{center}
\begin{minipage}[t]{16.5 cm}
\caption{{\bf Left}: Binding energies of the states located just above and just
below the $Z=20$ magic number and difference of the binding energies
of the two states surrounding the gap at $Z=20$ (see Sect.~\ref{annex}).
The neutron orbitals which are getting filled as a function of
increasing $N$ are given in the middle of the figure.
{\bf Right}: Experimental E(2$^+_1$) and
B(E2;~0$^+$~$\rightarrow$~2$^+_1$) values in the $_{20}$Ca isotopic
chain.
\label{gapZ20}}
\end{minipage}
\end{center}
\end{figure}

The left part of Fig. \ref{gapZ20} shows the evolution
of the binding energies of the proton orbits (bottom) together 
with that of the
$Z=20$ shell gap (top), for a neutron number N ranging from $N =
18$ to $N=30$. While the ground-state spin value of the $_{21}$Sc
isotopes is 7/2$^-$ for the whole interval, the one of the
$_{19}$K isotopes changes at $N=28$, and is tentatively
established at $N=30$. Apart from the large singularity that shows
up at the $N=20$ neutron magic number, the size of the $Z=20$ gap
is constant, $\simeq$ 5.3~MeV between $N=22$ to $N = 26$, i.e. as
the $\nu f_{7/2}$ is being filled. It seems to be slightly larger
at $N=30$ than at $N=22-26$. However the present knowledge on the
binding energies of the neutron-rich K-Ca-Sc isotopes is not
enough to ascertain this hypothesis. On the proton-rich side, such
an estimation of the gap cannot be obtained below $N=18$ as the
$^{39}$Sc nucleus is already located beyond the proton drip line.
As the size of the $Z=20$ gap remains large and almost constant
along the whole Ca isotopic chain, core excitations of the Ca
isotopes should remain relatively weak at low excitation energy.
This isotopic chain is therefore an ideal test bench for exploring
the spherical neutron shell and sub-shell closures at $N=16$,
$N=20$, $N=28$, $N=32$ and possibly $N=34$ as their existence
would be decoupled from the proton core excitations. This
assertion of a weak core excitation is corroborated by the
behaviors of the 2$^+_1$ energies and the B(E2; 0$^+$
$\rightarrow$ 2$^+$) values in the Ca isotopic chain, discussed in
the following section.

\subsubsection{\it Trends of E(2$^+$) and B(E2) values}
The systematics of the $2^+$ and B(E2) values from $^{40}$Ca to
$^{48}$Ca (see the right part of Fig.~\ref{gapZ20}) is often used in
textbooks to model multi-particle configurations of valence
nucleons in one particular orbit, in the present case the $\nu
f_{7/2}$ one. In this framework, the 2$^+_1$ energies are maximum
at the beginning and end of the shell and almost constant in
between. The B(E2) values follow a bell-shape curve with minima at
the two extremes of the shell.

A strong sub-shell closure is found at $N=16$, through the
observation of a large $2^+$ energy, E($2^+$)=3.015(16)~MeV, in
$^{36}$Ca (see Fig.~\ref{gapZ20}). This state was populated by the
one neutron knock-out reaction from a $^{37}$Ca beam at the GSI
facility~\cite{Door07}. Similar result has been obtained at GANIL
using the same experimental technique~\cite{Burg06}. This $2^+$
state decayed in-flight by $\gamma$-ray emission, even though the
$2^+$ state is located about 400~keV above the proton emission
threshold. This feature is \emph{ a priori} surprising as particle decay
is usually much faster than $\gamma$-decay. The Coulomb barrier
partly delays the particle emission. However the main hindrance
factor against proton decay is due to the weak overlap of the
wave functions of the $2^+$ state and the ground state of
the $^{35}_{19}$K$_{16}$ nucleus. The former is of almost pure
neutron origin due to excitations across the $N=16$ sub-shell,
whereas the second has a proton hole configuration with respect to
$^{36}$Ca.

An $N=32$ sub-shell closure has been evidenced in the $^{52}$Ca
nucleus through the observation of the first $2^+$ state at
2.52~MeV. This state was populated through the $\beta$-decay of
$^{52}$K at the CERN/ISOLDE facility~\cite{Huck85}. This $N=32$
sub-shell is formed between the neutron $p_{3/2}$ and $p_{1/2}$
orbits, the $f_{5/2}$ orbit being located upwards in energy. The
$N=32$ gap is present in the Ca isotopic chain in which the proton
$f_{7/2}$ orbit is empty. As protons are added into this orbit,
the strong $\pi f_{7/2}$-$\nu f_{5/2}$ interaction acts to
progressively move the neutron $f_{5/2}$ orbit between the neutron
$p_{3/2}$ and $p_{1/2}$ orbits, destroying the $N=32$ shell gap.
Therefore the increase of $2^+$ energy at $N=32$ is gradually
smoothened in the $_{22}$Ti~\cite{Lidd04} and $_{24}$Cr
chains~\cite{Pris01}, to eventually vanish in the $_{26}$Fe and
$_{28}$Ni chains. Compared to neighboring isotopes, the B(E2)
values at $N=32$ are slightly reduced in the Ti~\cite{Dinc05} and
Cr~\cite{Burg05} isotopic chain. This feature matches with the
existence of a subshell closure at $N=32$, which induces a local
reduction of the collectivity.

A fairly large $N=34$ sub-shell gap has been predicted by Honma et
al. between the $\nu f_{5/2}$ and $\nu p_{1/2}$ orbits using the
effective GXPF1 interaction~\cite{Honm02}. So far the energy of
the $2^+_1$ state has not been determined in the $^{54}$Ca
isotope. However no significant enhancement of the $N=34$ gap was
found from the study of the excited states in the neutron-rich
$^{50-52}$Ca nuclei~\cite{Rejm07}, which correspond to $f_{5/2}
\otimes p_{1/2}$ excitations. In addition the $2^+$ energy in
$^{56}_{22}$Ti$_{34}$ has been determined to be 1129~keV from the
$\beta$-decay of $^{56}$Sc, which is smaller than that of the
$N=32$ nucleus $^{56}_{22}$Ti$_{32}$ which amounts to
1495~keV~\cite{Lidd04}.

\subsubsection{\it Proton orbits below the $Z=20$ gap:
Levels of $_{19}$K isotopes \label{19K}}

The evolution of the energy of the three single-proton orbits
$d_{3/2}$, $s_{1/2}$, and $d_{5/2}$ (located below the $Z = 20$
gap) during the filling of the $\nu f_{7/2}$ sub-shell, is
essential to derive information on the various components of the
proton-neutron interactions within the $\pi sd$-$\nu f_{7/2}$
configurations. In particular, the strength of tensor forces
between the $\pi d_{3/2}$ and $\pi d_{5/2}$ orbits and the $\nu
f_{7/2}$ one could be determined. So far several
publications~\cite{Otsu05,Gade06,Gaud06,Bast06} are using various
intensities of this force. Therefore all the relevant
experimental data are summarized in the following.\\

\begin{figure}[h!]
\begin{center}
\epsfig{file=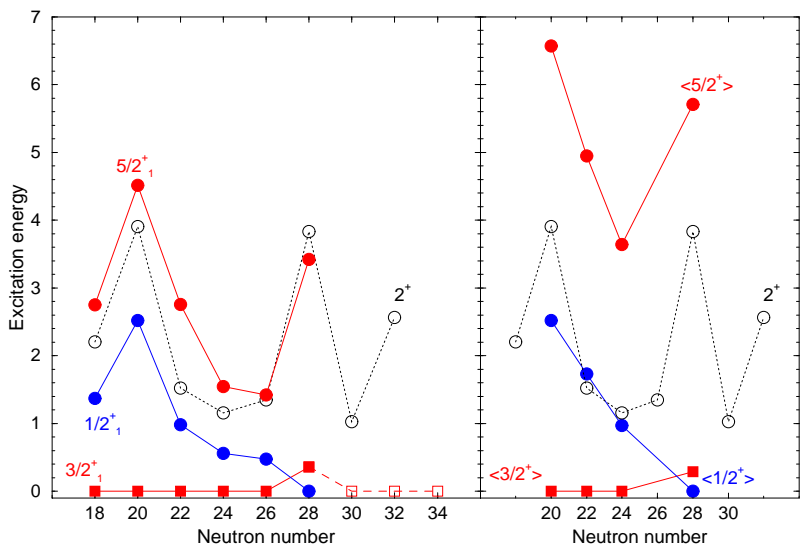,scale=0.6}
\begin{minipage}[t]{16.5 cm}
\caption{{\bf Left}: Energy of the  3/2$^+_1$, 1/2$^+_1$, and 5/2$^+_1$
states of the $_{19}$K isotopes as a function of the neutron number.
{\bf Right}: Average energy of all the 3/2$^+$, 1/2$^+$, and 5/2$^+$
states populated in the $_{20}$Ca($d$,$^3$He) pick-up
reactions \cite{Doll76,Bank85,Kram01} (see Tables \ref{untroisK} and \ref{Epid5K}).
The energy of the first 2$^+$ excitation of the Ca core is also
drawn (empty circles, dotted line).
\label{crossingK}}
\end{minipage}
\end{center}
\end{figure}

{\bf The $\pi d_{3/2}$ and $\pi s_{1/2}$ orbits}\\
From $N = 18$ to $N = 26$, the ground state of the $_{19}$K isotopes
has a 3/2$^+$ spin value, which originates from the
$\pi d_{3/2}$ orbit. The spin value of the first excited state
is 1/2$^+$. Its energy follows that of the 2$^+$ excitation of the
Ca core from $N = 18$ to $N = 24$, as shown in the left part of
Fig.~\ref{crossingK}.
This feature means that, up to $N = 24$, the 1/2$^+_1$ state
possibly arises from the coupling between the 3/2$^+$ state and
the 2$^+$ excitation of the core, and to a weaker extent from the
$\pi s_{1/2}$ single-particle state. For $N > 24$, the excitation
energy of the 1/2$^+_1$ state decreases so rapidly that it becomes
the ground state of $^{47}$K$_{28}$. This indicates, without
ambiguity, that the energy spacing between the two sub-shells,
$\pi d_{3/2}$ and $\pi s_{1/2}$, which defines the $Z = 16$ sub-shell
closure, is strongly decreasing during the
filling of the $\nu f_{7/2}$ orbital.

A more quantitative description of this shell evolution can be
obtained from the study of the ($d$,$^3$He) pick-up reactions on
stable Ca targets, which provide the energies and spectroscopic
factors of the 3/2$^+$ and 1/2$^+$ proton-hole states in the odd-A
$_{19}$K isotopes.  In the work of Doll et al.~\cite{Doll76}, four
targets ($^{40,42,44,48}$Ca) have been used. For each $_{19}$K
isotope, only one state of orbital momentum $\ell$ = 2 has been
assigned to the $\pi d_{3/2}$ sub-shell : the ground state for
$^{39,41,43}$K$_{20,22,24}$ and the first excited state for
$^{47}$K$_{28}$. States of higher energy with orbital momentum
$\ell$ = 2 have been assigned to the $\pi d_{5/2}$ sub-shell, as
reported below. Later, Bank et al.~{\cite{Bank85} have used a
polarized beam of deuterons to distinguish between levels in which
the intrinsic spin is aligned ($j$=$\ell$+1/2) or anti-aligned
($j$=$\ell$-1/2) with the orbital momentum $\ell$. From this
study, a second $\ell$~=~2 state at 3.88 MeV has been assigned to
the $\pi d_{3/2}$ orbit in $^{47}$K. Concerning the $\pi s_{1/2}$
orbit, its strength is split into 2 or 3 states, the first one
having the largest spectroscopic factor. The left part of Table
\ref{untroisK} gives the experimental results of $^{39,41,43}$K
from Ref.~\cite{Doll76} and $^{47}$K from Ref.~{\cite{Bank85}.
\begin{table}[h!]
\begin{center}
\begin{minipage}{16.5 cm}
\caption{The 3/2$^+$ and 1/2$^+$ states observed in the
$^{40,42,44,48}$Ca($d$,$^3$He) pick-up reactions
\cite{Doll76,Bank85} and summary of the properties of the $\pi
d_{3/2}$ and $\pi s_{1/2}$ orbits in the K isotopes}
\label{untroisK}
\end{minipage}
\begin{minipage}[h]{16.5 cm}
\begin{tabular}{|c|cccc|cc|cc|}
\hline
 & E(MeV) & $l$ &sub-shell  & (2J+1)C$^2$S& $\pi d_{3/2}$&    & $\pi s_{1/2}$&\\
 &  & &&& E$_{aver}$ & (2J+1)$\Sigma$C$^2$S & E$_{aver}$ & (2J+1)$\Sigma$C$^2$S\\
\hline
$^{39}$K   & 0   &2     &d3/2&   3.70&0&3.70&&\\
\cite{Doll76}&2.52 &0     &s1/2&   1.65&     &    &   2.52&   1.65\\
\hline
$^{41}$K   & 0   &2     &d3/2&   3.43&0&3.43&&\\
\cite{Doll76}&0.98 &0     &s1/2&   0.77&&&&\\
          &1.57 &0     &s1/2&   0.17&&&&\\
          &2.67 &0     &s1/2&   0.64&      &   &   1.73&   1.58\\
\hline
$^{43}$K    & 0   &2     &d3/2&   3.15&0&3.15&&\\
\cite{Doll76} &0.56 &0     &s1/2&   1.15&&&&\\
           &2.45 &0     &s1/2&   0.32&      &   &   0.97&   1.47\\
\hline
$^{47}$K   & 0   &0     &s1/2&   1.55&&&&\\
\cite{Bank85}&0.36 &2     &d3/2&   4.16&0.87&4.86&&\\
          &3.80 &0     &s1/2&   0.28&   &   &   0.58&   1.83\\
          &3.88 &2     &d3/2&   0.70&    &   &       &   \\
\hline
\end{tabular}
\end{minipage}
\end{center}
\end{table}
The single-particle energies (SPE) of the two shells, $\pi
d_{3/2}$ and $\pi s_{1/2}$, are obtained by averaging the energies
of the states 3/2$^+$ and 1/2$^+$, respectively, weighted by their
spectroscopic factors (see the right parts of Table \ref{untroisK}
and Fig.~\ref{crossingK}). Recently, another experiment has been
performed on $^{47}$K by means of the $^{48}$Ca($e,e'p$)
reaction~\cite{Kram01}. The spectroscopic factors of the populated
states have been determined using the spin values assigned from
the ($\overrightarrow{d}$,$^3$He) reaction. Their values are
systematically smaller than those obtained with the
($\overrightarrow{d}$,$^3$He) reaction, i.e. $(2J+1) \Sigma C^2 S
(\pi d_{3/2})$=2.6 and $(2J+1) \Sigma C^2 S(\pi
s_{1/2})$=1.2~\cite{Kram01}, as compared to 4.86 and
1.83~\cite{Bank85}, respectively. Error bars of about 20-30\%
should be applied to the SF values. The apparent discrepancy
between SF obtained in ($e,e'p$) and ($d$,$^3$He) experiments has
been investigated in Ref.~\cite{Kram01}. It disappears if the
latter ones are reanalyzed with a nonlocal finite-range DWBA
analysis with a bound state wave function that is obtained from
($e,e'p$) experiments. However, as the averaging procedure only
involves the relative SF, the two sets of SPE are close, i.e.
E$_{aver}$($\pi d_{3/2}$)~=~0.81 MeV and E$_{aver}$($\pi
s_{1/2}$)~=~0.52 MeV~\cite{Kram01}, as compared to 0.87 and
0.58~MeV~\cite{Bank85}, respectively. In both cases, the energy
difference between the $s_{1/2}$ and $d_{3/2}$ at $N=28$ is
negative, and amounts to about -0.29 MeV. This indicates that the
$\pi d_{3/2}$ and $\pi s_{1/2}$ shells cross at $N = 28$.
Their spacing decreases from 2.52 MeV at $N = 20$ to -0.29 MeV
at $N=28$, by a total amount 2.81~MeV. As the $f_{7/2}$ orbit
contains up to eight neutrons, the reduction of the $d_{3/2}-
s_{1/2}$ spacing amounts to 2.81/8 $\simeq$ 350~keV per neutron.
This variation can be formally written in terms of the energy
difference between the two proton-neutron monopole interactions
involved in the reduction of the proton $d_{3/2} - s_{1/2}$
spacing,
\begin{equation}\label{Vsdf}
 V^{pn}_{d_{3/2}f_{7/2}}- V^{pn}_{s_{1/2}f_{7/2}} \simeq -350~keV.
\end{equation}

Beyond $N>28$ the physical picture changes as another orbit,
$\nu p_{3/2}$, starts to be filled. Therefore there is no reason
to believe that the $\pi s_{1/2}$-$\pi d_{3/2}$ spacing would
follow the same trend. The spin of the ground state of $^{49}$K
has been assigned tentatively to 3/2$^+$ from its $\beta$-decay
pattern~\cite{NDS49}. Indeed three I$^\pi$~=~5/2$^+$
neutron-unbound states of $^{49}$Ca are strongly fed at 6.55, 6.69
and 7.06 MeV with log($ft$) values of 4.5, 4.5 and 4.6,
respectively. These intensities correspond to Gamow-Teller
transitions between states with $\Delta J$=0,$\pm$1, which holds
true only if the g.s. of $^{49}$K is 3/2$^+$. The ground state
spin values of $^{51,53}$K have been assigned as 3/2$^+$ state
from their $\beta$-decay studies~\cite{Perr04}. Corresponding
states are drawn in the left part of Fig \ref{crossingK} using
empty symbols and dashed line. Firm assignments for the g.s. spin
values of $^{49,51,53}$K and the location of $1/2^+$ excited state
are important to obtain the monopole matrix elements involved
there. For $N>28$, the spacing of the proton  orbits $s_{1/2} -
d_{3/2}$ scales with the difference of the monopole matrix
elements, $V^{pn}_{d_{3/2}p_{3/2}}- V^{pn}_{s_{1/2}p_{3/2}}$.

{\bf The $\pi d_{5/2}$ orbit}\\
The exact location of the $\pi d_{5/2}$ orbit is hard to determine
as it is deeply bound. Consequently its strength is
fragmented into several deeply bound states which carry small
spectroscopic factor each~\cite{Doll76,Bank85}. Table~\ref{Epid5K}
summarizes the number of 5/2$^+$ states, their range of excitation
energy, the sum of the spectroscopic factors (the $(2J+1) C^2S$
sum rule should amount to 6 in the $d_{5/2}$ orbit) as well as the
weighted averaged mean energies. The sum of the spectroscopic
factors barely exceeds 65\% at $N=20$ and $N=28$.

Along the whole K isotopic chain, the energy of the 5/2$^+_1$
state  follows exactly that of the 2$^+$ energy of the core
nucleus as shown in the left part of Fig. \ref{crossingK}. Same is
almost true for the average energy of the $5/2^+$ states, shown in
the right part of the same figure. This feature shed little doubt
on the validity of the location and evolution of the $d_{5/2}$
single particle energy. Moreover, we can refer to the introduction
of Ref.~\cite{Doll76}, from which these data have been extracted :
" ... we shall show that the data are very suggestive of a strong
influence of collective modes, not only on the widths but on the
energies of quasi-holes too",  and to the conclusion "The
quasi-hole energy should not be confused with a centroid
separation energy whose determination would require a knowledge of
the full $1d_{5/2}$ hole strength."
\begin{table}[h!]
\begin{center}
\begin{minipage}{16.5 cm}
\caption{States populated in the pick-up $^A$Ca ($d$,$^3$He)
$^{A-1}$K reactions with orbital angular momentum $\ell$ = 2 and
assigned to the $\pi d_{5/2}$ orbit \cite{Doll76,Bank85}.}
\label{Epid5K}
\end{minipage}
\begin{minipage}[h]{8 cm}
\begin{tabular}{|c|c|c|c|c|}
\hline
     &number  & E$_{min}$-E$_{max}$ & $(2J+1) \Sigma C^2 S $& E$_{aver}$ \\
     &of states& (MeV)      &       &(MeV)\\
\hline
$^{39}$K\cite{Doll76}&    13&         5.27 - 8.90&    4.8&    6.57\\
$^{41}$K\cite{Doll76}&    9&          3.19 - 6.63&    3.8&    4.95\\
$^{43}$K\cite{Doll76}&    12&        1.54 - 5.90&     3.1&    3.64\\
$^{47}$K\cite{Bank85}&    10&         3.32 - 8.02&     3.9&    5.71\\
\hline
\end{tabular}
\end{minipage}
\end{center}
\end{table}

With some words of caution, we could make use of the results on
the two semi-magic isotopes, $^{39}$K and $^{47}$K to derive the
variation between the $\pi d_{3/2}$ and $\pi d_{5/2}$ SPE. The
$d_{3/2} - d_{5/2}$ spacing decreases from 6.57~MeV to 4.84~MeV,
the latter value resulting from the difference of the SPE between
the $\pi d_{5/2}$ energy of 5.71~MeV (reported in
Table~\ref{Epid5K}) and that of the $\pi d_{3/2}$ state at
0.87~MeV (reported in Table~\ref{untroisK}), giving  a variation
of about 1.7~MeV. This is possibly an \emph{ upper bound}, as the
fraction of the identified $\pi d_{5/2}$ hole strength is lower at
$N=28$ than at $N=20$. It is anyhow much weaker than the value of
2.7~MeV used in Ref.~\cite{Gade06}. As the $\nu f_{7/2}$ orbit
contains up to 8 neutrons, the variation of the $\pi d_{3/2} - \pi
d_{5/2}$ splitting amounts to 1.7/8 $\simeq$ 0.210~MeV per
neutron. This could be written formally as a function of the
relevant monopoles:
\begin{equation}
 V^{pn}_{d_{3/2}f_{7/2}}- V^{pn}_{d_{5/2}f_{7/2}} \simeq -210~keV
\end{equation}

A proper unfolding of the correlations  should be performed to
better determine the evolution of the $d_{5/2}$ single-particle
energy between $N=20$ and $N=28$. This requires to consider
proton core-excitations which may not be taken into account so
far.\\

{\bf Theoretical predictions on the evolution of the three orbits}\\
As said in the introduction of this section, the evolution of the
relative energies of the $\pi d_{3/2}$, $\pi s_{1/2}$, and $\pi
d_{5/2}$ shells during the filling of the $\nu f_{7/2}$ orbit,
can provide robust information on the various components of the
proton-neutron interaction.

As a starting point, we can compare the experimental data of the
$_{19}$K isotopic chain to the results of HFB self-consistent
calculations we have obtained using the D1S Gogny force~\cite{Dech80}. 
There,
nuclear interactions are given by the sum of central and
spin-orbit terms, without tensor interaction. The description of
an odd-A nucleus is done using the blocking method, imposing that
the odd particle is located in one particular orbit chosen by its
value of angular momentum and parity. The calculated binding
energy of the nucleus is then analyzed as a function of the
occupied orbital, that gives the ground state and the excited
states. The results of the calculations  on the odd-A K isotopes
are drawn in the right part of Fig.~\ref{crossingK_expHFB}.
\begin{figure}[h!]
\begin{center}
\epsfig{file=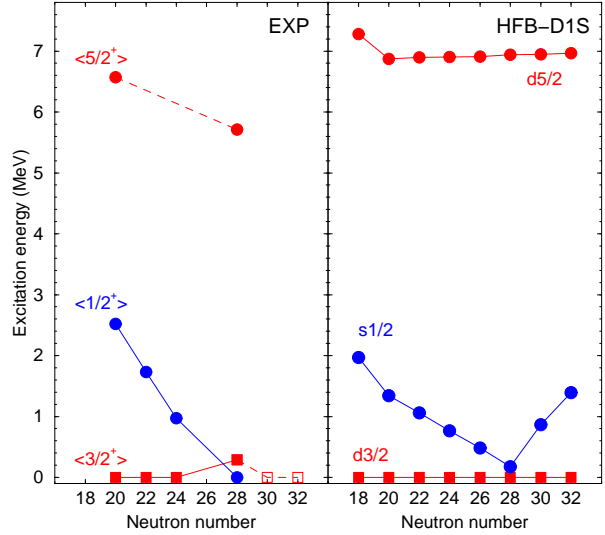,scale=0.6}
\begin{minipage}[t]{16.5 cm}
\caption{{\bf Left}: Average energy of all the 3/2$^+$, 1/2$^+$, and
5/2$^+$ states populated in the $_{20}$Ca(d,$^3$He) pick-up
reactions (see the right part of Fig.~\ref{crossingK}). 
{\bf Right}: Evolution of the
energies of the 3/2$^+$, 1/2$^+$, and 5/2$^+$ spherical states of
the odd-A K isotopes from $N=18$ to $N=32$ predicted in the HFB
self-consistent calculations using the D1S effective force of
Gogny and the method of blocking. \label{crossingK_expHFB}}
\end{minipage}
\end{center}
\end{figure}
The excitation energy of the 1/2$^+$ state decreases during the
filling of the $\nu f_{7/2}$ sub-shell and then increases for
$N>28$. This is in qualitative agreement with the experimental
results, shown for convenience in the left part of
Fig.~\ref{crossingK_expHFB}. Nevertheless, the theoretical slope is
less steep than the experimental one: Between $N=20$ and $N=28$,
the calculated variation of $[E(1/2^+)~-~E(3/2^+)]$ amounts to 1.2
MeV, to be compared to the value of 2.8~MeV derived from
experimental data (see Table~\ref{untroisK}). By adding a
tensor term to the D1S Gogny force, the reduction of
the $1/2^+ - 3/2^+$ splitting would be larger, coming closer to
the experimental values. No variation of the $\pi d_{3/2}-\pi
d_{5/2}$ splitting is predicted during the filling of the neutron
$f_{7/2}$ orbit. There again, a tensor term would reduce this
splitting either.

Shell model calculations have been using constantly improved
effective interactions to describe the structure of the nuclei
involving  the $\pi (sd)$ and $\nu (fp)$  orbits. The $sd-pf$
interaction was originally developed by Retamosa et
al.~\cite{Reta97}. The proton-neutron matrix elements involving
the proton in the $d_{3/2}$, $s_{1/2}$, and $d_{5/2}$ orbits with
the neutron in the $f_{7/2}$ one were constrained by the
systematics of the proton SPE in the K isotopic chain. Later, the
energy differences between the $3/2^-_1$ excited and $7/2^-$
ground states in the $^{37}_{16}$S$_{21}$ and
$^{35}_{14}$Si$_{21}$ nuclei were used to refine the matrix
elements involving the $\pi s_{1/2}$ and the $\nu f_{7/2}$ and
$\nu p_{3/2}$ matrix elements. The energy of the $3/2^-_1$ state
in $^{35}_{14}$Si was obtained from the $\beta$-decay of
$^{35}$Al~\cite{Numm01a}. Further refinements have been done to
account for spectroscopic information of nuclei lying in the
$sd-fp$ shells~\cite{Bast06, Gaud07}. Several recent publications
have discussed about a tensor component of the nucleon-nucleon
interaction (see for instance
Refs.~\cite{Otsu05,Gade06,Gaud06,Bast06}). In the three former
works the tensor term of the $V^{pn}_{{d_{3/2}} f_{7/2}}$ monopole 
interaction, $\tilde{V}^{pn}_{{d_{3/2}} f_{7/2}}$, amounts
to about -~200~keV. In the latter, it is shown than even a weaker
value of the tensor interaction suffice to reduce the $Z=14$
subshell gap and account for the enhanced collectivity of the
$^{42}_{14}$Si$_{28}$ nucleus. This peculiar case will be
discussed in the Sect.~\ref{trendN28}.

\subsubsection{\it Conclusion}
The $Z=20$ isotopic chain is an ideal test bench for nuclear
models, and to derive effective interactions, would they be
neutron-neutron or proton-neutron ones. The reason is that
the gap remains large enough to avoid coupling between single
particle states and collective excitations at low energy. Thus
shell and subshell closures have been studied in the Ca isotopic
chain, such as the $N=16$, $N=20$, $N=28$ and $N=32$ ones.

To induce quadrupole collectivity, a minimum of $2p-2h$ is
required, as the $Z=20$ gap is bound by states having opposite
parities. Hence the B(E2; 0$^+$ $\rightarrow$ 2$^+_1$) values
remain relatively weak. Deformed and superdeformed bands have been
found at high excitation energy \cite{Ideg01}, showing that the
multiparticle-multihole picture is extremely efficient to generate
quadrupole collectivity~\cite{Caur07}.

The energies of the proton-single particle states below the $Z=20$
gap have been determined in the $_{19}$K isotopic chain through
transfer reactions between $N=20$ to $N=28$. A collapse of the
$Z=16$ gap, formed between the proton $d_{3/2}$ and $s_{1/2}$
orbits, is found. Experimental data also seem to show a reduction
of the proton $d_{3/2} - d_{5/2}$ splitting, which can be
accounted for by the $\pi d - \nu f_{7/2}$ tensor forces. This
compression of the proton $d$ orbits echoes that of the $f$ orbits
around the $N=28$ shell closure, which will be addressed in a
following section.

%% file: texteNZ28_9avril.tex
\section{The magic number 28}
The magic number $28$, revealed for instance in the doubly-magic
nucleus $^{48}_{20}$Ca$^{}_{28}$, originates from the spin-orbit coupling in
atomic nuclei. Guided by the existence of such a term in atomic
physics, M.~Goeppert Mayer~\cite{Goep49} and O.~Haxel et al.~\cite{Haxe49}
independently postulated in 1949 the existence of a strong
spin-orbit (SO) force in the nuclear potential to reproduce the
so-called `magic numbers' above 20. The SO interaction is attractive 
for nucleons having
their angular momentum aligned with respect to their spin (denoted
as $\ell _\uparrow$) and repulsive in case of anti-alignment
($\ell _\downarrow$). Its strength was constrained to reproduce
the size of the shell gaps of the various magic numbers formed by
this interaction (28, 50, 82 and 126).
As for the number 28, the effect of such a SO interaction was to lower 
the $f_{7/2}$ orbit into the middle of the gap between the $sd$ and $fp$
oscillator shells, resulting in the magic number between the $f_{7/2}$ 
and $p_{3/2}$ orbits. Nevertheless, the history is not complete since, 
as discussed in the end of Sect.~\ref{monopoleffect} and shown in
Fig.~\ref{gapnf7p3}, we know now that a large part of the $N=28$ gap 
observed at $^{48}$Ca comes from the attractive monopole interaction, 
$V^{nn}_{f_{7/2}f_{7/2}}$.

The sequence of the
orbits around 28 exhibits 2 pairs of spin-orbit
partners $f_{7/2}$, $p_{3/2}$, $p_{1/2}$ and $f_{5/2}$, ranked as
a function of decreasing binding energies. It therefore follows
from this ordering of levels that the existence or vanishing of
this shell closure is possibly linked to the evolution of the SO
force.

Within the mean-field approaches, the SO interaction is maximum at
the surface and vanishes in the interior of the nucleus where the
potential varies slowly for most of the nuclei.  Therefore a wide
consensus still holds on the fact that reductions of the SO
interaction, and henceforth of shell gaps, would occur for
neutron-rich nuclei with increased surface diffuseness (see for
instance Ref.~\cite{Doba94,Lall98}). More recently, a reduction of
SO was suggested when the matter density is strongly depleted in
the center of the nucleus~\cite{Todd04}. 

Beyond the classical mean-field approaches, it was argued that 
the proton-neutron tensor
interaction modifies the energies of the $\ell_\uparrow$ and $\ell
_\downarrow$ components when specific proton and neutron orbits
are filled~\cite{Otsu05}. This effect could modify for instance
the $\nu f_{5/2}$-$\nu f_{7/2}$ splitting as protons are added or
removed into the $\pi d_{3/2}$ orbit 
or the $\pi f_{5/2}$-$\pi f_{7/2}$ 
splitting as neutrons are added to the $\nu g_{9/2}$ orbit.

\subsection{\it Evolution of the $N=28$ shell closure\label{N28}}
The study of the evolution of
$N=28$ magic shell far from stability is an ideal tool for probing
the nuclear forces at play such as the spin-orbit and 
tensor terms for
nuclei with a reduced binding energies or/and with a depletion of
the density in their interior. A lot of experimental studies aimed
at determining whether the $N=28$ shell closure is eroded in very
neutron-rich nuclei through complementary methods, which are
direct or indirect ones. The very first indirect hints of the
vanishing of the $N=28$ gap were obtained by
$\beta$-decay~\cite{Sorl93}, mass measurements~\cite{Sara00} and
Coulomb-excitation ~\cite{Glas97} experiments. This subsequently
triggered other theoretical works, in turn followed by new
experiments. The present understanding of the physics behind the
$N=28$ shell closure results from a subtle interplay between
theory and experiments.

The following sections will start with the systematics of binding
energies and be followed by the evolution of quadrupole properties
of the $N=28$ isotones. A deeper understanding of the nuclear
forces involved around the $N=28$ magic shell will be obtained by
determining the evolution of proton and neutron single particle
energies. These pieces of information will be used to describe the
physics towards the lightest even-$Z$ $N=28$ nuclei, such as
$^{42}_{14}$Si.

\subsubsection{\it Binding energies}

\begin{figure}[h!]
\begin{minipage}{8cm}
\begin{center}
\epsfig{file=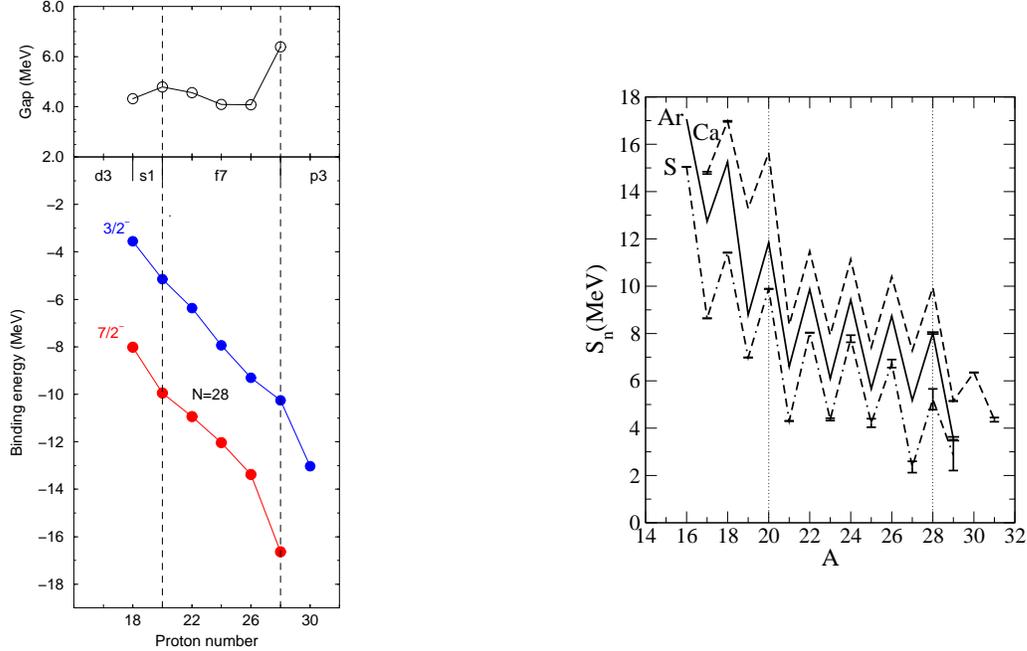,scale=0.5}
\end{center}
\end{minipage}\hfill
\begin{minipage}{8cm}
\epsfig{file=Sn_N28.eps,scale=0.35}
\end{minipage}
\begin{center}
\begin{minipage}[t]{16.5cm}
\caption{{\bf Left}:  Binding energies of the states
located just above and just below the $N=28$ magic number and
difference of the
binding energies of the two states surrounding the gaps at
$N=28$ (see Sect.~\ref{annex}).
The proton orbitals which are getting filled as a function of
increasing $Z$ are given in the middle of the figure.
{\bf Right}: Neutron separation energies $S_n$ as a function of
the atomic mass for the $_{20}$Ca, $_{18}$Ar and $_{16}$S isotopic chains.}\label{SnN28}
\end{minipage}
\end{center}
\end{figure}

The binding energies of the last neutron in the $N=28$ and $N=29$
isotones are drawn in the left part of Fig.~\ref{SnN28}.
These data are restricted to nuclei which are likely to be
spherical. They are taken from the AME2003 atomic mass table
\cite{Audi03}, to which the result for the $^{47}_{18}$Ar
isotope~\cite{Gaud06} has been added. The distance in energy
between the two binding-energy curves is shown in the top part of
Fig.~\ref{SnN28} as a function of the proton number. With the
exception of the maximum at $Z=28$ (see Sect.~\ref{annex}), the
$N=28$ gap amounts to about 4.5~MeV. It is found that
the $N=28$ gap has decreased by 330(80)~keV between $Z=20$ and
$Z=18$. This point is essential to understand the evolution of the
nuclear structure below $Z=20$. This will be discussed more 
extensively later on.

Below $Z=18$ the separation energies of the last neutron in
$^{44,45}_{~~16}$S, recently measured \cite{Jura07},
have not been included in Fig.~\ref{SnN28} as the $^{44}$S ground
state is likely to be deformed. This can be viewed in the right
part of Fig.~\ref{SnN28} which displays the evolution of one-neutron
separation energy $S_n(A,Z)$ as a function of the neutron number
$N$. In the Ca isotopic chain, the decrease of the $S_n$ values is
abrupt after $N=28$ due to the presence of the $N=28$ shell gap
between the $f_{7/2}$ and $p_{3/2}$ orbitals. This drop of the
$S_n$ values after having passed $N=28$ is not observed anymore in
the $_{16}$S isotopic chain. This means that a significant mixing
between these $f$ and $p$ orbits have reduced the magnitude of the
gap, which is a typical feature of a deformed nucleus. The
mass-excess of the nucleus $^{42}_{14}$Si has been determined in
Ref.~\cite{Jura07}. As compared to a liquid drop model, an excess
of binding energy is found. This gain of binding energy could be
due the presence of sub-shell gap or large deformation. The
mass-excess for $^{43}_{14}$Si should be
determined to judge between these two assumptions.

The evolution of the $N=28$ shell gap has been yet viewed from
mass excesses or binding energies. Other complementary data are
required to ascertain the breaking of the spherical $N=28$ gap,
and to see whether this effect is restricted to $^{44}_{16}$S or
extends down to the $^{42}_{14}$Si nucleus. This is addressed in
the following section.

\subsubsection{\it Trends of 2$^+$, 4$^+$ , $0^+_2$ energies and
B(E2) values\label{trendN28}}

\begin{figure}[h!]
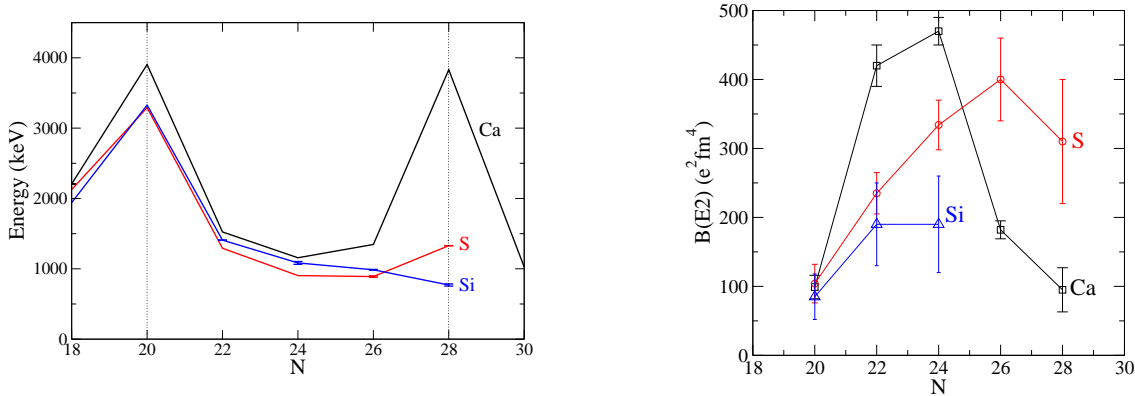

\begin{minipage}{8cm}
\begin{center}
\includegraphics[width=7cm]{E2+N28.eps}
\end{center}
\end{minipage}\hfill
\begin{minipage}{8cm}
\includegraphics[width=6cm]{BE2N28.eps}
\end{minipage}
\begin{center}
\begin{minipage}{16.5cm}
\caption {Experimental E($2^+_1$) energies (left) and B(E2; $0^+_1
\rightarrow 2^+_1$) values (right) in the $_{14}$Si, $_{16}$S
and $_{20}$Ca isotopic
chains as a function of the neutron number $N$. Data are taken
from the compilation of Ref.~\cite{Rama01}, except for
$^{38}$Si~\cite{Lian03}, $^{40}$Si~\cite{Camp06},
$^{42}$Si~\cite{Bast06}, and $^{40}$S~\cite{Wing01}}
\label{E2BE2N2028}
\end{minipage}
\end{center}
\end{figure}

The determination of the 2$^+$, 4$^+$, $0^+_2$ energies and
reduced transition probabilities B(E2) provide complementary
pieces of evidence of significant structural changes in atomic
nuclei, as the onset of deformation. Magic (spherical) nuclei are
commonly characterized by high E(2$^+_1$) and weak B(E2;$0^+_1
\rightarrow 2^+_1$) values. At the mid-shell occupancy the 2$^+_1$
energy drops and the B(E2) value increases. Fig.~\ref{E2BE2N2028}
shows the evolution of the energies of the 2$^+_1$ states and of
their reduced transition probabilities B(E2) as a function of the
neutron number in the $_{20}$Ca, $_{16}$S and $_{14}$Si isotopes.
At $N=20$ the $2^+$ energies are high for the three isotones,
$^{40}_{20}$Ca, $^{36}_{16}$S and $^{34}_{14}$Si, and their B(E2)
values are small and similar. This picture is consistent with a
strong shell closure $N=20$ which persists between $Z=20$ to
$Z=14$ as already reported in the Sect.~\ref{disappearN20}. As
shown in Fig.~\ref{E2BE2N2028}, the correlated increase of the
2$^+$ energies, and reduction of B(E2) does not hold at $N=28$.
There the $2^+$ energies of the $N=28$ isotones, $^{44}$S and
$^{42}$Si, are much weaker than at $N=20$. For $^{44}$S, its
E(2$^+$) and B(E2) values suggest a configuration which is
intermediate between a spherical and deformed nucleus. In
$^{42}$Si, only the $2^+$ energy has been measured. As it is one
of the smallest among nuclei in this mass region, this nucleus is
likely to be very deformed. Further details about the experimental
studies in the S and Si isotopic chains and of the theoretical
interpretations are given in the two following paragraphs.
\vspace{0.5cm}

To understand the evolution of the nuclear structure in the
neutron-rich S isotopes, the first $2^+$ state in the neutron-rich
$^{40,42}_{~~16}$S isotopes has been studied by means of Coulomb
excitation at intermediate energy~\cite{Sche96}. As shown in the
right part of Fig.~\ref{E2BE2N2028}, the $B(E2)$ value of the
$^{42}_{16}$S$_{26}$ nucleus remains large, contrary to the
$^{46}_{20}$Ca$_{26}$ isotone. Complementary pieces of information
on the energy of higher excited states of the S isotopes have been
obtained through in-beam $\gamma$-ray spectroscopy using the
fragmentation reaction of a $^{48}$Ca beam~\cite{Sohl02}. The
level schemes were extended to the $2^+_2$ and $4^+$ states, the
assignments of which were obtained by $\gamma$-ray angular
distributions with respect to the direction of each produced
fragment. From the energy ratio of excited states, such as
$E_{4_1^+}/E_{2_1^+} \sim 3.0$ in $^{42}$S, and the comparison to
HFB predictions using the Gogny D1S interaction~\cite{Sohl02,
Peru00}, it has been deduced that $^{42}$S is a $\gamma$-soft
deformed nucleus. This confirmed the earlier assumption of a
deformed nucleus inferred from its B(E2) value~\cite{Sche96}.

In $^{44}$S$_{28}$ the $2^+_1$ energy and the $B(E2)$ value were
found to be intermediate between a deformed and a spherical
nucleus~\cite{Glas97}. This situation is suggestive of a possible
shape mixing between these two configurations, an assumption which
can be confirmed by the search for a low lying $0^+_2$ state. Such
a state was found to be isomer with a lifetime of 2.3 $\pm$ 0.5
$\mu s$ at an energy of 1362.5(1)keV~\cite{Grev05}. Combining the
conversion-electron and $\gamma$-ray spectroscopy  it was found
that the $0^+_2$ state decays both directly to the ground state
via internal conversion, and to the $2^+_1$ state at 1329~keV.
This experiment has been repeated to obtain more statistics and be
able to determine the ratio between the two reduced transition
probabilities, B(E2; $0^+_1 \rightarrow 2^+_1$) and B(E2; $0^+_2
\rightarrow 2^+_1$). The analysis is in progress~\cite{Grev07}.
Similar and
large B(E2) values would reveal that the 0$^+_1$ and 0$^+_2$
levels have mixed wave functions. A large reduced strength E0
would in turn point to the existence of a spherical and strongly
deformed configurations which co-exist before mixing (see for
instance \cite{Heyd88}).

Excited states in two neutron-rich $_{14}$Si isotopes have been
recently discovered at the NSCL ~\cite{Camp06} and GANIL
laboratories~\cite{Bast06}. These two experiments used nucleon
removal reaction from secondary beams centered around $^{42}$P and
$^{44}$S at intermediate energy to produce the $^{40}$Si and
$^{42}$Si nuclei and study the $\gamma$-rays from their
de-excitation in flight. The detection of the $\gamma$-rays was
achieved by arrays of detectors which surrounded the production
target in which reaction occurred. In the case of
Ref.~\cite{Camp06} a segmented array of Ge detectors was used,
leading to a photo-peak $\gamma$-ray efficiency of about 3\% at
1~MeV. An array composed of 70 BaF$_2$ detectors was used in
Ref.~\cite{Bast06} in order to obtain a higher photo-peak
efficiency of about 30\% at 1~MeV and 20\% at 2~MeV. Campbell et
al.~\cite{Camp06} have determined the energy of the $2^+$ state in
$^{40}_{14}$Si$_{26}$ at 986(5)~keV. While the 2$^+$ energies
increase in Ca isotopes from $N=24$ to reach a maximum around 4
MeV at $N=28$, the 2$^+$ energies in the Si isotopes start to
deviate from those of the Ca isotopes at $N=26$ (see the left part
of Fig.~\ref{E2BE2N2028}) which points to a reduced N=28 shell
gap~\cite{Camp06}. However, as 2$^+$ neutron excitations occurring
before the complete filling of the $\nu f_{7/2}$ shell are mainly
generated inside the shell, they are not that sensitive to the
size of the $N=28$ gap. Conversely, in $^{42}_{14}$Si$_{28}$ the
neutron $f_{7/2}$ shell is\emph{ a priori} filled completely, and
the 2$^+$ state comes mainly from particle-hole excitations across
the $N=28$ gap. Bastin et al.~\cite{Bast06} have established a
$2^+$ state at 770(19)~keV for the $N=28$ nucleus $^{42}$Si. The
dramatic decrease of the 2$^+_1$ energy in $^{42}$Si is a proof of
the disappearance of the spherical $N=28$ shell closure at $Z=14$.
This extremely low energy of 770~keV -actually one of the smallest
among nuclei having a similar atomic mass - cannot be obtained
solely from neutron excitations. Proton core excitations should
play an important role, which could in principle be evidenced by
measuring the B(E2) values in the Si isotopic chain while reaching
$N=28$.

The B(E2) values in the $_{14}$Si isotopic chain~\cite{Ibbo98} seem
to rise after $N=20$, but not as much as in the $_{16}$S ones.
Whether the B(E2) values remain small, steadily increase up to
$N=28$ or follow a parabola is presently not discernable with the
presently large experimental error bars. A reduced $Z=14$ shell gap
would dramatically increase the B(E2) values, as protons are
carrying most of the effective charge in the nucleus. A decrease of
the $N=28$ gap solely would barely change the B(E2) values.

To summarize this part, experimental data on the study of excited
states in the $_{16}$S and $_{14}$Si isotopic chains reveal a
noticeable change of structure between the $N=20$ and the $N=28$
isotones. At $N=20$ the $^{36}$S and $^{34}$Si nuclei have all
characteristics of doubly-magic nuclei, whereas at $N=28$ a rapid
onset of deformation is occurring. As we shall see in the
following, this drastic change originates from a combined
reductions of the proton $Z=14, 16$ and neutron $N=28$ shell gaps.

\subsubsection{\it Evolution of the proton SPE\label{protonN28}}

\begin{figure}[h!]
\begin{center}
\begin{minipage} [c] {8cm}
\includegraphics[width=6.4cm]{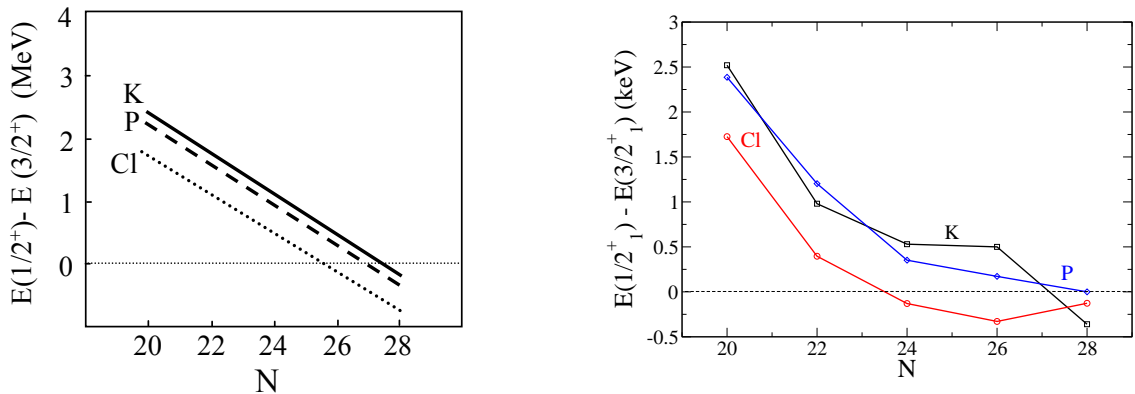}
\end{minipage}
\vspace{0.5cm}
\begin{minipage} [c] {8cm}
\includegraphics[width=7cm]{E12_E32.eps}
\end{minipage}
\begin{minipage}{16.5cm}
\caption{{\bf Left}: Calculated energy difference
$E(1/2^+)-E(3/2^+)$ as a function of the neutron number $N$ when
using Eq.~\ref{eq:Z16monopole} for the $_{19}$K, $_{17}$Cl,
and $_{15}$P isotopic chains and the monopole matrix elements of
Eq.~\ref{Vsdf}. As
the ground state spin value in the P isotopic chain is $1/2^+$,
the sign of $E(1/2^+)-E(3/2^+)$ has been reversed there to put all
data on the same viewgraph. 
{\bf Right}: Experimental energy
difference $E(1/2^+_1)-E(3/2^+_1)$ in the $_{19}$K, $_{17}$Cl, and
$_{15}$P isotopic chains as a function of the neutron number $N$
(see text for references).} \label{ESPEsd}
\end{minipage}
\end{center}
\end{figure}

The change of structural behavior between the $N=20$ and $N=28$
isotones can be partly ascribed to the evolution of the proton
Single Particle Energies~\cite{Cott98}. The ($d$,$^3$He) reactions
from stable $_{20}$Ca target have revealed that the energy spacing
between the two orbits $\pi d_{3/2}$ and $\pi s_{1/2}$ in
$_{19}$K, which defines the $Z=16$ sub-shell closure, has
completely vanished at $N=28$. This was already reported in
Sect.~\ref{19K}. The filling of eight neutrons in the $\nu
f_{7/2}$ orbital induces more binding of the $\pi d_{3/2}$ orbits
compared to the $\pi s_{1/2}$ one, since
$|V{^{pn}_{d_{3/2}f_{7/2}}}| > |V{^{pn}_{s_{1/2}f_{7/2}}}|$.
Taking into account the monopole matrix elements solely, the
evolution of ($E(1/2^+)-E(3/2^+)$) between $N=20$ and $N=28$ would
be linear:
\begin{equation}\label{eq:Z16monopole}
[E(1/2^+)-E(3/2^+)]_{20+x}=[E(1/2^+)-E(3/2^+)]_{20}
+x(V{^{pn}_{d_{3/2}f_{7/2}}}-V{^{pn}_{s_{1/2}f_{7/2}}}),
\end{equation}
\noindent As compared to this linear decrease due to the monopole
interactions, shown in the left part of Fig.~\ref{ESPEsd}, the
experimental evolution between the first 3/2$^+_1$ and 1/2$^+_1$
states, ($E(1/2^+_1)-E(3/2^+_1)$), is steeper as soon as neutrons
are added into the $f_{7/2}$ orbit (compare the left and right
parts of Fig.~\ref{ESPEsd}). This feature is due to the pairing
and quadrupole correlations which already engage at $N=22$, as
soon as the single-particle states $\pi s_{1/2}$ and $\pi d_{3/2}$
come close enough to each others.
\vspace{0.5cm}

One can apply the monopole matrix elements derived in the $_{19}$K
chain to the $_{17}$Cl and $_{15}$P ones, starting from the
experimental values at $N=20$. As shown in the left part of
Fig.~\ref{ESPEsd} the crossing between the proton $3/2^+$ and
$1/2^+$ states is also expected around $N=28$ in the P, and
earlier in the Cl chain. As for the K isotopes, the evolution of
the experimental $3/2^+_1$ and $1/2^+_1$ states shown in the right
part of Fig.~\ref{ESPEsd} in the Cl and P isotopic chains is
distorted by pairing and quadrupole correlations to account for
the mixing between the nearby states~\cite{Lian03,Sorl04,Frid05,
Gade06,Frid06,Bast06}. In particular the steep decrease of
$[E(1/2^+_1)-E(3/2^+_1)]$ in the Cl chain is mainly caused by the
pairing interaction which favors the coupling of proton pairs in
the $d_{3/2}$ shell. This leaves a hole in the $s_{1/2}$ orbit,
which gives rise to a $1/2^+$ spin value. Apart from this
distortion, the reduction of the $Z=16$ gap in the Cl and P chains
is qualitatively accounted for by the monopole part of the nuclear
interaction solely. Shell model calculations using the interaction
developed in Ref.~\cite{Numm01a} show excellent agreement with the
experimental energy difference between the $1/2^+$ and $3/2^+$
states in the $_{19}$K, $_{17}$Cl and $_{15}$P
chains~\cite{Sorl04,Gade06}.

At $N=28$ the $_{17}$Cl and $_{15}$P isotones have almost
degenerate proton $d_{3/2}$ ($\ell=2$) and $s_{1/2}$ ($\ell=0$)
orbits, which are separated by $\Delta \ell$=2. This feature is
essential to develop quadrupole correlations - and henceforth
collectivity- between the $s$ and $d$ nearby states in the
$^{43,45}_{~~~17}$Cl nuclei~\cite{Sorl04}. It also accounts for
increased quadrupole excitations in the neutron-rich $_{16}$S
isotopes which lie in between the $_{15}$P and $_{17}$Cl chains.

The development of collectivity should\emph{ a priori} be
significantly hindered in the $_{14}$Si isotopic chain where
proton excitations should arise across the large $Z=14$ gap, from
the deeply bound $d_{5/2}$ to the $d_{3/2}$ and $s_{1/2}$ orbits.
However this possibility is essential to account for an increased
collectivity in the neutron-rich $^{42}_{14}$Si$_{28}$ nucleus
derived from its low 2$^+$ energy. If the $d_{5/2}$ orbit did not
come closer to the $d_{3/2}$ and $s_{1/2}$ ones, the $^{42}$Si
nucleus would remain spherical. The energy difference between
these orbits is by far not easy to determine as the $d_{5/2}$
orbit is deeply bound. It was estimated in Sect.~\ref{19K} by
looking at the variation of the $d_{5/2}$ single particle energy
between $N=20$ and $N=28$ in the $_{19}$K isotopic chain. However
its single particle energy strength is spread over many states,
the spectroscopic factor sum of which does not exceed 60\%. From
the estimation given in Sect.~\ref{19K} and in Ref.~\cite{Bast06},
the complete filling of the neutron $f_{7/2}$ shell induces the
collapse of the $\pi s_{1/2} - \pi d_{3/2}$ spacing, a reduction
of the $\pi d_{3/2}$-$\pi d_{5/2}$ splitting by about 1.7~MeV and
of the $\pi d_{5/2}$-$\pi s_{1/2}$ splitting by about 400~keV from
an initial value of about 6~MeV. These reductions are large enough
to provide a low energy of the $2^+$ state in
$^{42}$Si~\cite{Bast06}. As mentioned in Sect.~\ref{19K}, larger
reductions of the $\pi d_{3/2}$-$\pi d_{5/2}$ splitting were
predicted in Refs.~\cite{Otsu05,Gade06}.

To summarize this part, the development of collectivity in the
neutron-rich $N=28$ isotones is partly due to the reduction of the
spacing between the protons $d_{5/2}$, s$_{1/2}$ and $d_{3/2}$ as
the neutron $f_{7/2}$ orbit is filled. As these orbits are
separated by 2 units of angular momentum, quadrupole (E2)
collectivity is naturally favored. Added to this, a reduction
of the $N=28$ shell gap would reinforce this tendency to deform.

\subsubsection{\it Evolution of the neutron SPE for $Z <20$\label{neutronN28}}

\begin{figure}[t]
\begin{center}
\includegraphics[width=12cm]{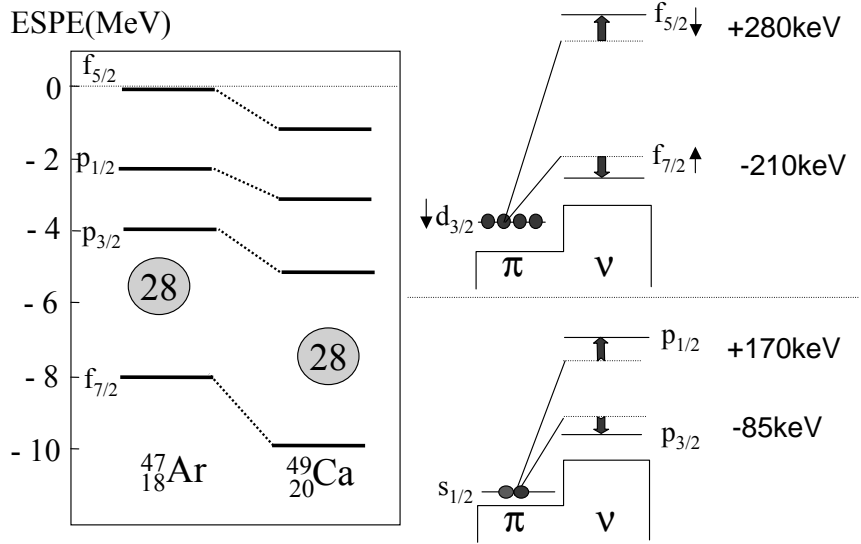}
\end{center}
\begin{center}
\begin{minipage}{16.5cm}
\caption{{\bf Left} : Neutron single
particle energies (SPE) of the $fp$ orbitals for the
$^{47}_{18}$Ar$_{29}$ and $^{49}_{20}$Ca$_{29}$ nuclei.
{\bf Right} : Schematic view of the proton-neutron interactions
involved to change the $f$ (top) and $p$ (bottom) SO splittings
(derived from Refs.~\cite{Gaud06, Gaud07}).} 
\label{spe_N28}
\end{minipage}
\end{center}
\end{figure}

On the basis of experimental data on atomic masses, on energies of
excited states and reduced transition probabilities, it was
inferred that the $N=28$ shell closure is eroded as protons are
removed from the doubly magic $^{48}_{20}$Ca. To ascertain this
assumption, and to derive the reasons for such an erosion, the
$N=28$ shell closure has been investigated via the
$^{46}_{18}$Ar($d,p$)$^{47}$Ar transfer reaction in inverse
kinematics~\cite{Gaud06}. This information will be used to
investigate the change of the $N=28$ gap from $_{20}$Ca to
$_{18}$Ar. A radioactive beam of $^{46}$Ar interacted with a thin
CD$_2$ target at 10.2(1)~A$\cdot$~MeV. The energy and angle of the
reaction protons were measured in eight highly segmented
double-sided Si detectors. From the $Q$ value of the transfer
reaction, the $N=28$ gap was found to be of 4.47(9)~MeV in
$^{46}$Ar, which is 330(90)~keV smaller than in $^{48}$Ca.
Transfer to excited states were used to determine the energies and
spectroscopic factors of the neutron $p_{3/2}$, $p_{1/2}$ and
$f_{5/2}$ states in $^{47}_{18}$Ar. The fact that only part (about
3/4) of the strength for these states has been observed indicates
that some correlation take place in $^{46}$Ar, after the removal
of only 2 protons from the doubly magic $^{48}$Ca nucleus. In
particular it was found in Ref.~\cite{Gaud06} that a fraction
(about 1/4) of neutron already occupies the $p_{3/2}$ orbit
through $1p-1h$ excitation across the $N=28$ gap, from the
$f_{7/2}$ orbit. Monopole terms of the $sdpf$~\cite{Numm01a}
interaction calculations were adjusted to match existing
experimental data in this mass region~\cite{Gaud07}. By applying
this constraint, the full particle strength of the $\nu f_{7/2}$,
$\nu p_{3/2}$ and $\nu p_{1/2}$ and $\nu f_{5/2}$ orbits has been
determined in $^{47}$Ar. The resulting Single Particle Energies
(SPE) are compared to those of the $^{49}_{20}$Ca
isotone~\cite{Abeg78,Uozu94} in Fig.~\ref{spe_N28}. It is found
that the orbits in which the angular momentum is aligned ($\ell
_\uparrow$) with the intrinsic spin, such as $f_{7/2}$ and
$p_{3/2}$, become relatively much less bound than the $f_{5/2}$
and $p_{1/2}$ orbits where the angular momentum and intrinsic spin
are anti-aligned ($\ell_\downarrow$). Bearing in mind that the
$d_{3/2}$ and $s_{1/2}$ orbitals are quasi-degenerate at $N=28$,
(see Sect.~\ref{protonN28}), the removal of  2 protons between 
$^{48}$Ca and $^{46}$Ar is equiprobable from these orbits.
Therefore modifications of neutron SPE arise from proton-neutron
interactions involving these two orbits~\cite{Gaud06}. As regards 
the $f$
states, the experimental trend was mainly ascribed to the fact
that the monopoles contain attractive and repulsive tensor terms
(see Sect.~\ref{decompmonopol}) which amount to 
$\tilde{V}_{d_{3/2}~f_{7/2}}^{pn}$ =-210~keV and
$\tilde{V}_{d_{3/2} f_{5/2}}^{pn}$=+280~keV,
respectively. The
change of the $p$ SO splitting was principally assigned to the
removal of a certain fraction of $s_{1/2}$ protons which deplete
the central density of the nucleus. The corresponding
spin-dependent part of the monopole terms was extracted to be of
+170~keV and -85~keV for $\tilde{V}_{s_{1/2}~p_{1/2}}^{pn}$ and
$\tilde{V}_{s_{1/2}~p_{3/2}}^{pn}$, respectively. These
spin-dependent parts of the monopoles amount to about 20\% of the
cnetral part to which they add. The reduction of the $N=28$ shell
gap arises from a subtle combination of several monopoles involved
while 2 protons are taken equiprobably out of the $d_{3/2}$ and
$s_{1/2}$ orbits~\cite{Gaud06}.

The variation of the single particle energies in term of monopole
interactions can be pursued towards the $^{43}_{14}$Si nucleus in
which, as compared to $^{49}_{20}$Ca, 4 protons have been removed
from the $d_{3/2}$ and 2 from the $s_{1/2}$ orbits. Using the
monopole matrix elements derived in the interaction, a global
shrink between the $fp$ orbits is foreseen, such as reductions of
the (i) $N=28$ gap by $330 \times 3 \simeq 1000$~keV, (ii) $f$ SO
splitting by $ 4(0.28+0.21) \simeq 2000$~keV and (iii) $p$ SO
splitting by $ 2(0.17 + 0.085) \simeq 500$~keV. This global
shrink of SPE is expected to reinforce particle-hole excitations
across $N=28$, which are of quadrupole nature. The existence of
$1p-1h$ excitations in $^{47}$Ar, the discovery of an E0 isomer in
$^{44}_{16}$S$_{28}$ and the weak $2^+$ energy in
$^{42}_{14}$Si$_{28}$ (see Sect.~\ref{trendN28}) weight in favor
of a progressive increase of quadrupole collectivity below $Z=20$.

\subsubsection{\it Evolution of the neutron SPE for $Z > 20$\label{nuf7f5}}

Above $Z=20$ the proton $f_{7/2}$ orbit is getting filled. As
compared to the previous section, the physics involved to account
for the variation of the neutron $f_{7/2}$, $p_{3/2}$, $p_{1/2}$
and $f_{5/2}$ SPE is different. The spin-isospin interaction
between the neutron $f_{5/2}$ and the proton $f_{7/2}$ orbits is
expected to play a major role, as it did between the proton and
neutrons of the $d_{5/2}$ and $d_{3/2}$ shells in the $N=20$
region (see Sect.~\ref{disappearN20}). The strongly attractive part of the
$\pi f_{7/2} - \nu f_{5/2}$ interaction is expected to lower the
neutron $f_{5/2}$ orbit significantly between $Z=20$ and $Z=28$.
In the $^{49}$Ca$_{29}$ nucleus, it lies about 2~MeV above the
$p_{1/2}$ and 4~MeV above the $p_{3/2}$ ground state. The single
particle energies in $^{57}$Ni$_{28}$ were determined at the ATLAS
accelerator at Argonne National Laboratory using the
d($^{56}$Ni,p)$^{57}$Ni reaction at 250~MeV~\cite{Rehm98}. It was
found that the major part of the $f_{5/2}$ single particle energy
is located at 768~keV excitation energy, below that of the
$p_{1/2}$ ones located at 1.13MeV. Though, the authors admitted
that large uncertainties on the spectroscopic factors (up to 50\%)
are possible.

\begin{figure}[h!]
\begin{center}
\epsfig{file=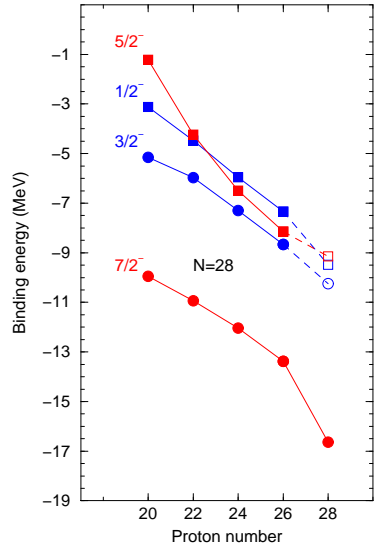,scale=0.55}
\begin{minipage}{16.5cm}
\caption{
Evolution of the binding energies of the states bounding 
the $N=28$ shell closure. The 7/2$^-$ states are the ground states 
of the $N=27$ isotones. The 3/2$^-$, 1/2$^-$  and 5/2$^-$ energies 
are the average of all the states populated in  
$(d,p)$ pick-up reactions on the stable $N=28$ 
isotones \cite{Koch72}. The results on $^{57}$Ni are drawn with 
empty symbols as the spectroscopic factors of its three first 
states are not known.}
\label{BEf7f5p3p1}
\end{minipage}
\end{center}
\end{figure}

The variation of the binding energies of the $f_{7/2}$ and $f_{5/2}$ 
orbits as a function of the proton number are shown in 
Fig.~\ref{BEf7f5p3p1}.
From the mean slopes of the two curves, the monopole matrix 
elements $V_{f_{7/2} f_{7/2}}^{pn}$ and $V_{f_{7/2} f_{5/2}}^{pn}$ 
can be tentatively determined to be $\simeq$ -0.6~MeV 
and -1.0~MeV, respectively. Their large difference 
could be related to the spin-isospin term of the interaction.

The variation of the single- particle states $p_{3/2}$ and
$p_{1/2}$ is also shown in the same figure. The energy spacing
between these two spin-orbit members  does not vary very much with
the addition of eight protons in the $f_{7/2}$ orbit. There, no
tensor force is changing this SO splitting as the protons and 
neutrons orbits are separated by 2 units of orbital angular
momentum. Noteworthy is the fact that this is clearly at variance
with the abrupt change of the $p$ SO splitting with the addition
or removal of protons in the proton $s_{1/2}$ orbit discussed in
the previous section. In such a case a significant part of the
protons and neutrons wave functions were localized in the same
regions of the nucleus.

\subsubsection{\it Conclusion}

The spectroscopy of the neutron-rich $N=28$ isotones has been
investigated by various complementary methods, such as the Coulomb
excitation, in-beam $\gamma$-ray spectroscopy, delayed-electron
spectroscopy and the neutron stripping $(d,p)$ reaction. By
gathering all pieces of information, a coherent description of the
evolution of the $N=28$ shell gap can be proposed for $Z<20$ and
$Z>20$.

For $Z<20$ the shell model description of the $N=20$ and $N=28$
isotones south of Ca isotopes is significantly different. The
$N=20$ shell gap persists with up to the removal of 4 and 6
protons to the quasi-doubly magic $^{36}_{16}$S$_{20}$ and
$^{34}_{14}$Si$_{20}$ nuclei, respectively. For these nuclei the
existence of significant proton sub-shell closures at $Z=16$ and
$Z=14$ is a warrant of their rigidity against quadrupole
excitations. Conversely at $N=28$, the studies of the
proton~\cite{Doll76,Bank85,Cott98} and neutron single particle
energies ~\cite{Abeg78,Uozu94,Gaud06} have revealed significant
differences. The combined compression of the energy of the proton
and neutron orbits, plus the favored quadrupole excitations across
$N=28$, produce a rich variety of behaviors and shapes in the even
$N=28$ isotones; spherical $^{48}$Ca; oblate non-collective
$^{46}$Ar; coexistence in $^{44}$S, and two rotors, oblate
$^{42}$Si and prolate $^{40}$Mg~\cite{Caur04}. This variety of
shapes is also supported by mean-field calculations, relativistic
or non-relativistic~\cite{Wern96,Rein99,Lall99,Peru00,Rodr02}.
Subtle changes in the potential energy surfaces arise from the
choice of the mean field effective interactions, in particular for
the $^{44}$S nucleus. However the global trend to quadrupole
deformation is found as it arises from the fact that
the occupied and valence proton\emph{ and} neutron orbits 
have $\Delta l = 2$.

The reduction of the $f$ and $p$ SO splittings is tentatively
ascribed to new effects that have been put forward these last few
years both from theoretical and experimental point of views. In
intensity these reductions of SO splittings are much larger than
the ones expected by an increased diffuseness of the nuclear
matter far from the valley of stability. They are opening up a new
vista on the various components giving rise to changes of the SO
splittings in hitherto unknown parts of the chart of nuclides.
Noteworthy is the fact that similar physics processes (tensor
force and $E2$ symmetry) are expected to be at play in other
doubly-magic nuclei formed by the SO interaction, such as
$^{78}_{50}$Ni$_{28}$ and $^{100}_{~50}$Sn$_{50}$. Whether these
nuclei would be spherical or deformed depends strongly on the size
of the proton and neutron gaps (depending of the strength of 
tensor forces, for instance) and on the quadrupole 
collectivity that is at play.

Above $Z=20$, the $N=28$ shell gap is slightly decreasing from a
value of about 4.5~MeV at $Z=20$ to 4~MeV at $Z=26$. The evolution
of the neutron single-particle orbits $f_{7/2}$, $p_{3/2}$,
$p_{1/2}$ and $f_{5/2}$ have been determined from gathering
information on $(d,p)$ reactions in the $N=28$ isotones. Even if
only partial information exist, it was found that a strong
interaction between the protons $\pi f_{7/2}$ and neutrons  
$\nu f_{5/2}$, which can be due to a spin-isospin term, 
induces a reduction of the splitting $\nu
f_{7/2} - \nu f_{5/2}$ orbits. In $_{28}$Ni, the neutron $f_{5/2}$
has drop down to such an extent that it lies between the $p_{3/2}$
and $p_{1/2}$ orbits. This leads to the disappearance of the
$N=32$ shell closure between Ca and Ni, a topic which has been
discussed in Sect.\ref{bindenergyZ20}. Meanwhile the neutron
$p_{3/2} - p_{1/2}$ SO splitting seems to be constant.

\subsection{\it Evolution of the $Z=28$ shell closure\label{Z28}}

The $_{28}$Ni isotopic chain comprises 30 isotopes discovered so
far. At the two extremes of the proton- and neutron-rich sides,
the two doubly magic $^{48}_{28}$Ni$_{20}$~\cite{Blan00} and
$^{78}_{28}$Ni$_{50}$~\cite{Enge95,Hosn05} have remarkable
interest. The former is a prime candidate for the two-proton
radioactivity, whereas the second is an important waiting-point in
the path of the rapid-neutron capture process. So far a few
nuclei of $^{78}_{28}$Ni$_{50}$ have been produced worldwide.
Therefore its structure could not have been studies yet. It
therefore remains a wonderful challenge for the future.

From the neutron-deficient to neutron-rich nuclei, the ordering of
the neutrons shells is as follows: the $\nu f_{7/2}$ shell between
$^{48}_{28}$Ni$_{20}$ and $^{56}_{28}$Ni$_{28}$, the remaining
$fp$ shells up to $^{68}_{28}$Ni$_{40}$, and the $\nu g_{9/2}$
orbit up to $^{78}_{28}$Ni$_{50}$. The $Z=28$ shell gap is formed
between the occupied $\pi f_{7/2}$ and the valence $\pi p_{3/2}$
(or possibly $f_{5/2}$) orbits. The successive filling of the
neutron orbits could polarize the proton core by specific
proton-neutron interactions. Such effects can be revealed by the
studies of the binding energies of the proton orbits, would they
belong to the ground or excited configurations, and by the
evolution of the E($2^+$) and B(E2) values along the Ni isotopic
chain. The relative evolution of the $\pi f_{7/2}$ and $\pi
f_{5/2}$ spin-orbit partners will be examined to determine whether
tensor forces act to reduce the $Z=28$ shell closure as the
neutron $\nu g_{9/2}$ orbit is filled towards $^{78}$Ni.

\subsubsection{\it Evolution of the binding energies\label{evolgapZ28}}

\begin{figure}[h!]
\begin{minipage}[t]{9 cm}
\begin{center}
\epsfig{file=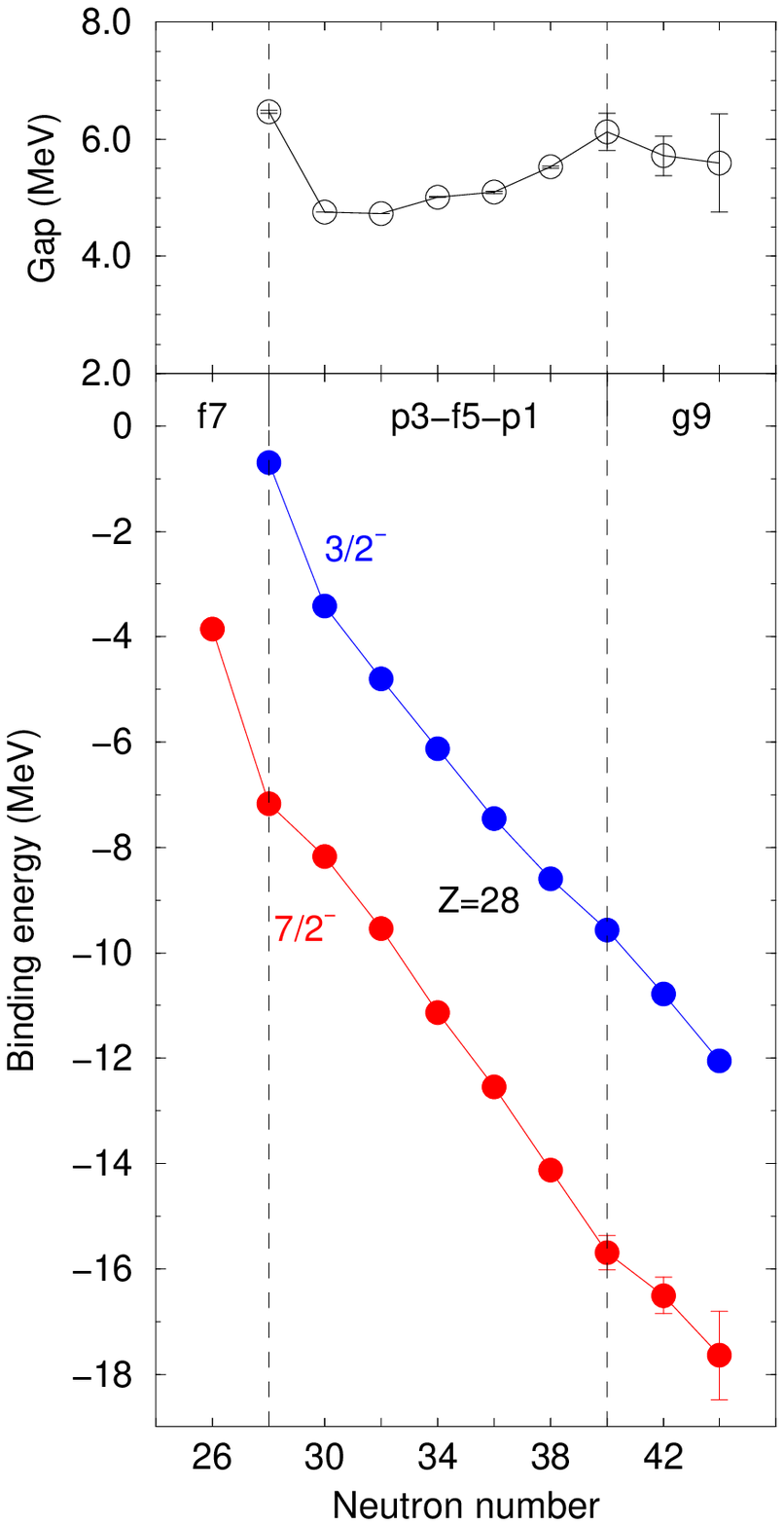,scale=0.5}
\end{center}
\end{minipage}\hfill
\begin{minipage}[t]{9 cm}
\epsfig{file=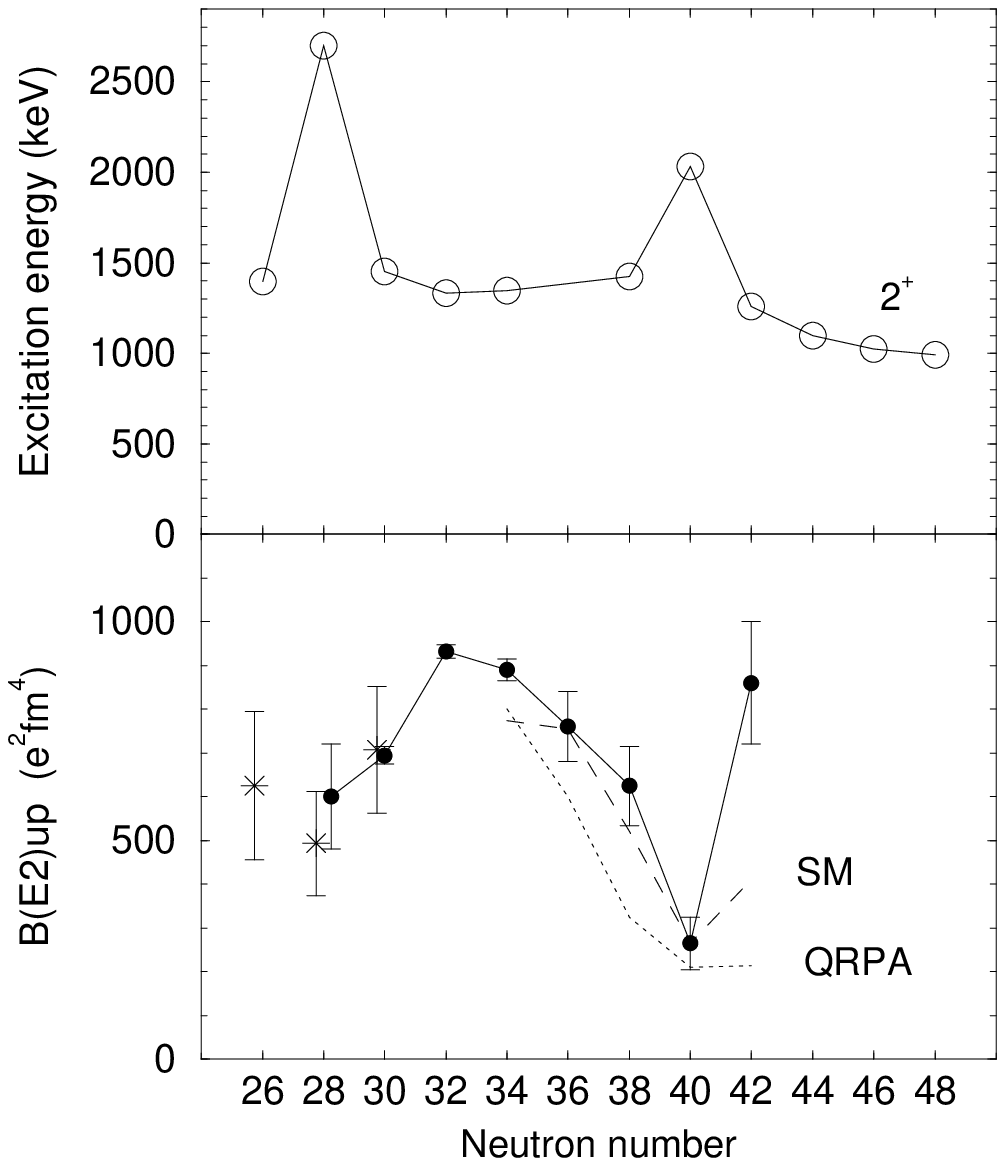,scale=0.6}
\end{minipage}
\begin{center}
\begin{minipage}[t]{16.5 cm}
\caption{{\bf Left}: Binding energies of the states located
just above and just below the $Z=28$ magic number and
difference of the binding energies of the two states surrounding
the gap at $Z=28$ (see Sect.~\ref{annex}). The neutron
orbitals which are getting filled as a function of increasing
$N$ are given in the middle of the figure.
{\bf Right}:
Experimental $E(2^+)$ and $B(E2; 0^+ \rightarrow
2^+)$ values in the $_{28}$Ni isotopic chain.
Values (drawn with filled circles) are taken from Ref.~\cite{Rama01}
for $^{56-68}$Ni  and from Ref.~\cite{Perr06} for
$^{70}$Ni.
The three additional data points (stars) for $^{54,56,58}$Ni
are from Ref.~\cite{Yurk04}. The B(E2) values calculated with the
shell model of Ref.~\cite{Sorl02} are drawn with dashed line,
those with the QRPA model of Ref.~\cite{Lang03} with dotted line.
\label{gapZ28}}
\end{minipage}
\end{center}
\end{figure}

The experimental evolution of the $Z=28$ shell gap, formed between
the $f_{7/2}$ and $p_{3/2}$ orbits is drawn in the left part of
Fig. \ref{gapZ28}, for $N=28$ to $N=44$. It has been obtained from
the  variation of the binding energies of the proton 7/2$^-$ and
3/2$^-$ states, reported in the same figure. Values of atomic masses 
come from the AME2003 table~\cite{Audi03}, to which new results 
for $^{70,72}$Ni and $^{73}$Cu nuclei~\cite{Raha07} have been added.
Singularities in the gap evolution are present for $N=28$
($^{56}$Ni), a self-conjugated doubly-magic nucleus (see
Sect.~\ref{annex}), and to a weaker extent for $N=40$ ($^{68}$Ni). In
average the $Z=28$ gap amounts to about 5 MeV. This is large
enough to maintain the spherical shape of all Ni isotopes, even
those located at mid-distance between $N=28$ and $N=40$.

Between $N=28$ and $N=40$ the variation of the proton binding
energies can be attributed  to the filling of the neutron $p_{3/2}$,
$f_{5/2}$ and $p_{1/2}$ orbits. As these orbits lie close in
energy,  the filling of the neutrons $fp$ orbits as a function of
$N$ is averaged between these three shells. Therefore the increase
of the binding energy of the 7/2$^-$ state -arising from the $\pi
f_{7/2}$ orbital- between $N=28$ and $N=40$ is mainly due to the
strongly attractive interaction between $f_{7/2}$ protons and
$f_{5/2}$ neutrons which have the same orbital momentum $\ell$=3
and opposite spin orientations. Other proton-neutron interactions
between the $\pi f_{7/2}$ and $\nu p$ orbits have weaker
intensity.

Above $N \simeq 40$, a weaker gain of the $\pi f_{7/2}$ binding
energy is found as the $\nu g_{9/2}$ orbital gets filled (see the
left part of Fig.~\ref{gapZ28}). This statement is valid with up
to about 2 $\sigma$ confidence level, an uncertainty which could
be reduced by measuring the atomic masses of the $_{27}$Co$_{N >
40}$ isotopes with a better accuracy. If pursued up to $N=50$,
this change of binding energy slopes induces a decrease of the
$Z=28$ gap -formed between the $f_{7/2}$ and $p_{3/2}$ orbits-
from $\simeq$6~MeV in $^{68}$Ni$_{40}$ to $\simeq$3.5~MeV in the
doubly-magic $^{78}$Ni$_{50}$ nucleus. In such a case, increased
proton excitations across the $Z=28$ gap -or core polarization-
would be expected.

\subsubsection {\it Trends of $2^+$ energies and B(E2)
values\label{enerprobaNi}}

Between the $N=28$ and $50$ shell closures the Ni isotopes have
been subject of extensive experimental and theoretical studies, as
can be found in Refs.
~\cite{Kenn01,Ishi00,Graw01,Oros01,Sorl02,Sawi03,Lang03,Hosm05,Mazz05,
Perr06}. The experimental results on the $2^+_1$ energies and
B(E2; $0^+\rightarrow2^+$) values are drawn in the right part of
Fig.~\ref{gapZ28}. The $2^+_1$ energies are maximum at the $N=28$
and $N=40$ (sub)shell closures and slightly decrease up to the
last known value of $^{76}$Ni. The curve formed by the B(E2)
values  varies in the opposite way, reaching a maximum at or near
mid-occupation by neutrons of the $fp$ shell ($N=34$) and
subsequently decreasing until $N=40$. Between $N_1=28$ and
$N_2=40$, the B(E2) curve follows the generalized seniority scheme
within the $fp$ shells. Within this approach~\cite{Talm77,Cast90}
which exploits the property of nucleons paired to J=0$^+$ in
nearby orbits, the B(E2) values of $^{56-68}$Ni follow a parabola
along the shell filling,
\begin{equation}\label{BE2_seniority}
B(E2; 0^+\rightarrow2^+_1) =  c(\pi fp,\nu fp) F(1-F)
\end{equation}
\noindent where $c(\pi fp,\nu fp)$ is representative of the
strength of the proton-neutron interaction in the $fp$ shells, and
$F$ is the fractional filling of the neutron shell which varies
between 0 at the beginning and 1 at the end of the shell. For a
nucleus containing $N$ neutrons between $N_1=28$ and $N_2=40$, F
expresses as $F=(N-N_1)/(N_2-N_1)$. As neutrons are chargeless
particles, they cannot contribute\emph{ directly} to the increase
of B(E2) but by inducing core-polarization. Therefore the height
of the parabola can be used to derive the strength of
neutron-induced proton-core excitations. The proton core
polarization occurs through quadrupole excitations via $(\pi
f_{7/2})^{-1}(\pi p_{3/2})^{+1}$
configurations~\cite{Kenn01,Sorl02}.

Deviation to the parabola of Eq.~\ref{BE2_seniority} exist at the
two extremes, $^{56}$Ni and $^{68}$Ni for which the B(E2) values
are slightly higher than expected with this simple model. For
$^{56}_{28}$Ni$_{28}$ this is due to the extended possibilities
for creating $2^+$ excitations across $N,Z=28$ in the $fp$ shell
for protons and neutrons~\cite{Krau94,Yana98} and a consequence of
a strong proton-neutron interaction characterizing the $N=Z$
nuclei~\cite{Otsu98,Pove99}. At $N=40$, the B(E2) value reaches a
minimum in the Ni chain. The low B(E2) of $^{68}$Ni$_{40}$, 3.2(7)
$W.u.$, is commonly assumed to be the signature for a doubly-magic
nucleus. This value is comparable to that of doubly magic nuclei,
$^{16}$O (3.3(3) $W.u.$), $^{40}$Ca (2.3(4) $W.u.$) and $^{48}$Ca
(1.6(5) $W.u.$). Such a minimum has been ascribed to the existence
of an $N=40$ sub-shell gap and to the lack of E2 excitations
between the $fp$ and the g$_{9/2}$ orbitals of different parity
value~\cite{Graw01,Sorl02,Lang03}. Without some amount of neutron
pair-scattering between the $fp$ and $g$ shells (or a superfluid
effect~\cite{Sorl02}) the B(E2) value would have even be weaker.

Beyond $N=40$ the B(E2) value rises steeply for
$^{70}$Ni$_{42}$, even though error bars are rather large. This
suggests an increased collectivity in the heavy Ni isotopes
induced by the interaction of protons in the $fp$ and neutrons in
the $g_{9/2}$ shell with a strength $c(\pi fp,\nu g)$ (cf. 
Eq.~\ref{BE2_seniority} with $N_1=40$ and $N_2=50$). This
enhanced collectivity goes in concert with a gradual reduction of
2$^+_1$ excitation energies from $^{70}$Ni$_{42}$ to
$^{76}$Ni$_{48}$ ~\cite{Mazz05} which cannot be understood solely
from pure neutron excitations within the $g_{9/2}$ shell. These
features can attributed to a reduction of the $Z=28$ gap, as
surmised already from the proton binding energy curves of the left
part of Fig.~\ref{gapZ28}. As a matter of comparison the steep
rise of the B(E2) between $^{68}$Ni$_{40}$ and $^{70}$Ni$_{42}$ could 
be compared to the more modest variations between
$^{56}$Ni$_{28}$~\cite{Krau94,Yana98} and
$^{54}$Ni$_{26}$ or $^{58}$Ni$_{30}$~\cite{Yurk04}.

A better insight on the polarization strength in the 2$^+_1$ state
of $^{70}$Ni can be obtained from the evolution of the B(E2;
$J\rightarrow J-2$) values along the 8$^+$, 6$^+$, 4$^+$, 2$^+$
components of the ($\nu g_{9/2}$)$^2$ multiplet. The 8$^+_1$ state
is expected to be almost only of neutron origin, as a 8$^+$ spin
value cannot be built solely with protons in the $fp$ orbitals.
The B(E2; $8^+_1 \rightarrow 6^+_1$) value would therefore
correspond to a reference value for the weakest core polarization.
Recently E2 transition strengths were measured for all the first
states in $^{70}$Ni~\cite{Lewi99,Mach03,Stan03,Perr06} , and a new
empirical T=1 effective interaction was derived in the pure
neutron $p,f_{5/2},g_{9/2}$ model space~\cite{Lise04}. In this
approach, the experimental B(E2; $8^+\rightarrow 6^+$) = 19(4),
B(E2; $6^+\rightarrow4^+$) = 43(1), and B(E2;$2^+ \rightarrow
0^+$)= 172(28)e$^2$fm$^4$ values are calculated as 17, 45 and 92
e$^2$fm$^4$ using an effective neutron charge $e_\nu = 1.2 e$. 
The good agreement
for the high-spin states breaks down for the $2^+_1 \rightarrow
0^+$ transition, which points to an enhanced proton-core
polarization at low excitation energy. This conclusion is at
variance with the QRPA~\cite{Lang03} and shell model
results~\cite{Sorl02,Lise05} (see the right part of
Fig.~\ref{gapZ28}) which predict that the B(E2; $0^+ \rightarrow
2^+_1$) strength in $^{70}$Ni predominantly corresponds to neutron
excitations, decoupled from the proton core. For the shell model
calculations of Refs.~\cite{Sorl02,Lise05}, the valence space that
was used may not be large enough to account for the observed
enhanced collectivity.

This strong polarization in the Ni isotopes beyond $N=40$ is
likely to be due to the\emph{ attractive} $\pi f_{5/2}-\nu g_{9/2}$
monopole interaction~\cite{Fran01,Graw04,Smir04,Lise04}, assigned
to the tensor force of the in-medium nucleon-nucleon
interaction~\cite{Otsu05}. This force is also predicted to act
through the\emph{ repulsive} $\pi f_{7/2}-\nu g_{9/2}$ interaction,
hereby explaining the weaker slope of the $\pi f_{7/2}$ binding
energy observed for $N>40$ in the left part of Fig.~\ref{gapZ28}.
On the whole, the tensor term $\pi f-\nu g_{9/2}$ would reduce
both the $Z=28$ gap and the $\pi f_{7/2}-\pi f_{5/2}$ splitting.
This will be discussed in the following section.

\subsubsection{\it Proton orbits above the $Z=28$ gap: Levels of the 
$_{29}$Cu isotopes\label{Cu}}

The variation of the binding energy of the proton orbits 
located\emph{ above} the $Z=28$ gap, i.e. $p_{3/2}$, $p_{1/2}$ and
$f_{5/2}$, can be inferred from experimental studies along the
$_{29}$Cu isotopic chain. They are depicted below as a function of
increasing neutron number. During the filling of the neutron $fp$ 
and $g$ shells the energies of these valence proton shells display 
large changes, as the $\pi f_{7/2}$ one evoked in Sect.~\ref{evolgapZ28}.

\begin{figure}[h!]
\begin{center}
\epsfig{file=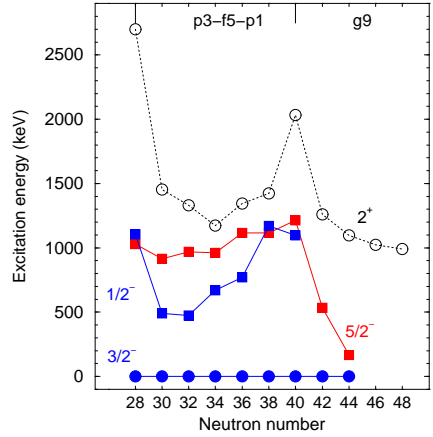,scale=0.55}
\begin{minipage}[t]{16.5 cm}
\caption{Evolution of the first states of the $_{29}$Cu
isotopes as a function of the neutron number. The energy of the
2$^+$ excitation of the Ni core is also drawn. The neutron
orbitals which are getting filled as a function of increasing $N$
are given on top of the figure.
\label{crossingCu}}
\end{minipage}
\end{center}
\end{figure}

The ground-state configuration of the $^{57-73}$Cu$_{28-44}$ isotopes
arises from the $\pi p_{3/2}$ orbital. Two excited states,
1/2$^-$ and 5/2$^-$, are located at excitation energies well below 
that
of the 2$^+$ state of the even-even Ni core. Their evolution
as a function of the neutron number is shown in
Fig.~\ref{crossingCu}.

The energy of the first $2^+$ of the $^{56}$Ni core is above
2.5~MeV. Therefore, the low-lying excited states measured in
$^{57}$Cu$_{28}$ likely correspond to the single-proton orbits
$\pi p_{1/2}$ and $\pi f_{5/2}$ rather than states originating
from the coupling $\pi p_{3/2} \otimes 2^+$. States in $^{57}$Cu
have been populated by the proton-capture reaction in inverse
kinematics, $p$($^{58}$Ni, $^{57}$Cu $\gamma$)$2n$ \cite{Zhou96}.
Photons of 1028(4), 1106(4) and 2398(10) keV have been measured in
coincidence with the $^{57}$Cu nuclei detected in a recoil
spectrometer. Using analogies with its mirror nucleus, $^{57}$Ni,
these $\gamma$-rays have been attributed to the direct decay of
three excited states (of spin (5/2$^-$), (1/2$^-$) and (5/2$^-$)
respectively) to the ground state \cite{Zhou96}. Only the two
first excited states, at 1028 and 1106 keV, are reported in
Fig.~\ref{crossingCu}. New measurements on $^{57}$Cu would be
welcome in order to confirm the energies and the spin values of
its excited states. Furthermore the determination of the
spectroscopic factors (SF) is essential to characterize the
orbitals $\pi p_{3/2}$, $\pi p_{1/2}$ and $\pi f_{5/2}$ for
$N=28$.

The $^{59-65}$Cu$_{30-36}$ isotopes have been studied using the
($^3$He,$d$) stripping reactions on stable Ni targets. Although a
lot of states with spin values 3/2$^-$, 1/2$^-$ and 5/2$^-$ have
been identified, the sum of their individual SF is systematically
lower than the maximum value expected for each orbital, meaning
that the single-particle energies lie at higher energy.
Nevertheless it is important to note that the three states drawn
in Fig. \ref{crossingCu} have the highest SF values, approximately
50 \% of the sum rule. Even though the 1/2$^-_1$ and 5/2$^-_1$
states have to be considered as a mixing of two components (the
individual configuration, $\pi p_{1/2}$ or $\pi f_{5/2}$, and the
coupling $\pi p_{3/2}$ $\otimes$ 2$^+$) their\emph{ relative}
evolution with respect to the 3/2$^-$ ground state will be used,
in the following, to figure out changes in the single-particle
energies of the $p_{3/2}$, $f_{5/2}$ and $p_{1/2}$ orbits. Between
$N=28$ and $N=40$, we can tentatively ascribe these changes to the
various proton-neutron interactions while neutrons occupy
successively the $p_{3/2}$, $f_{5/2}$ and $p_{1/2}$ orbits.

The increase of the 1/2$^-_1$ energy with respect to the ground
state observed for $32 \le N \le 38$, i.e. during the filling of
$\nu f_{5/2}$, is an interesting feature to start with. It may be
connected to a progressive increase of the energy spacing between
the two orbits, $\pi p_{1/2}$ and $\pi p_{3/2}$, since the 2$^+$
energy of the Ni core does not vary very much in this range of
neutron number (see Fig.~\ref{crossingCu}). The $\pi p_{3/2}-\pi
p_{1/2}$ splitting has increased by 695(20) keV with the
addition of 6 neutrons (see Fig.~\ref{crossingCu}), i.e. 116 keV
per neutron. Such an increase is related to the monopole difference
$V^{pn}_{p_{1/2}f_{5/2}}-V^{pn}_{p_{3/2}f_{5/2}}$ which contain a
difference of tensor terms of opposite signs, attractive
$\tilde{V}^{pn}_{p_{3/2}f_{5/2}}$ and repulsive
$\tilde{V}^{pn}_{p_{1/2}f_{5/2}}$. As the proton and neutron
orbits involved in these matrix elements are separated by two
units of orbital momentum ($\ell$=1 for the $p$ orbits and
$\ell$=3 for the $f$ ones), the tensor components could be
relatively weak.

The spacing between the $5/2^-_1$ and $3/2^-_1$ states evolves
smoothly from $N=28$ to $N=40$. This is suggestive of a subtle
balance between the two components of the $5/2^-_1$ state in the 
one hand, and between the relative evolution of the ESPE's due to the
proton-neutron monopoles involved there
(the tensor term is repulsive in 
$V^{pn}_{p_{3/2}p_{3/2}}$ and $V^{pn}_{f_{5/2}f_{5/2}}$, and
attractive in $V^{pn}_{p_{3/2}f_{5/2}}$) in the other hand.
The coupled component of the $5/2^-_1$ state is expected to be
reduced at $N=40$. Indeed its SF obtained from the 
$^{70}$Zn($d$, $^3$He) and ($t$, $\alpha$) pick-up reactions have 
very large values \cite{Zeid78,Ajze81}. Nevertheless as the 
peaks of interest are broader than the others, the authors have 
assumed that two levels in $^{69}$Cu are approximately equally 
populated, the spacing of the two unresolved states being less 
than 15~keV, each of them displaying a lower SF value.    

Above $N=40$, i.e. as the neutron $g_{9/2}$ orbit gets filled, the
situation changes dramatically. As shown in Fig.~\ref{crossingCu},
a strong decrease in energy of the 5/2$^-_1$ state is
observed in the $^{71,73}$Cu$_{42,44}$ nuclei from the
$\beta$-decay study of $^{71,73}$Ni~\cite{Fran98,Fran01}. 
Collective properties of the low-lying levels in the odd-A 
$^{67-73}$Cu have been recently investigated by Coulomb excitation 
with radioactive beams at REX-ISOLDE~\cite{Stef08}, where the values of
transition probabilities B(E2; 5/2$^-_1$ $\rightarrow$ 3/2$^-_{gs}$)
have been measured. A significant reduction of B(E2) is observed 
at $N=40$. It points to a transition from a rather collective 
5/2$^-_1$ state in $^{67}$Ni to a single-particle-like structure 
in $^{69-73}$Ni.    

\begin{figure}[h!]
\begin{minipage}{9 cm}
\begin{center}
\epsfig{file=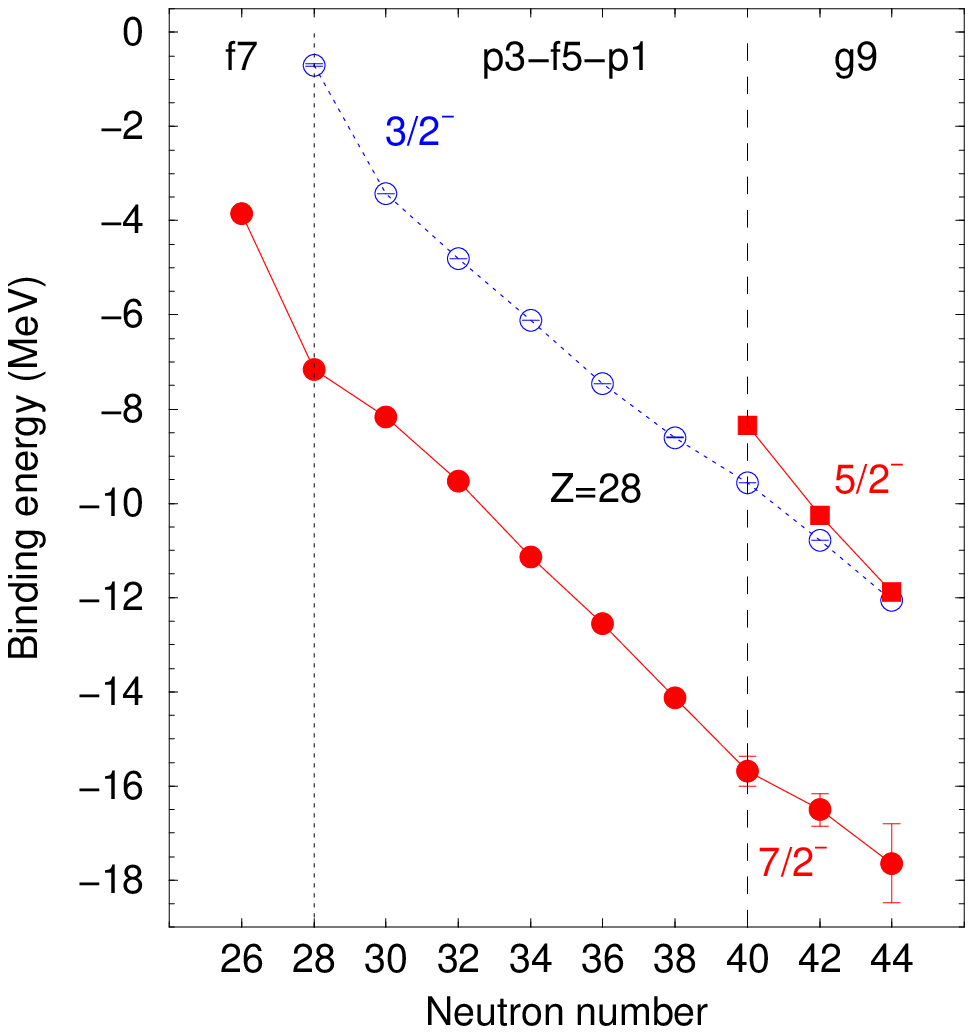,scale=0.5}
\end{center}
\end{minipage}\hfill
\begin{minipage}{9 cm}
\epsfig{file=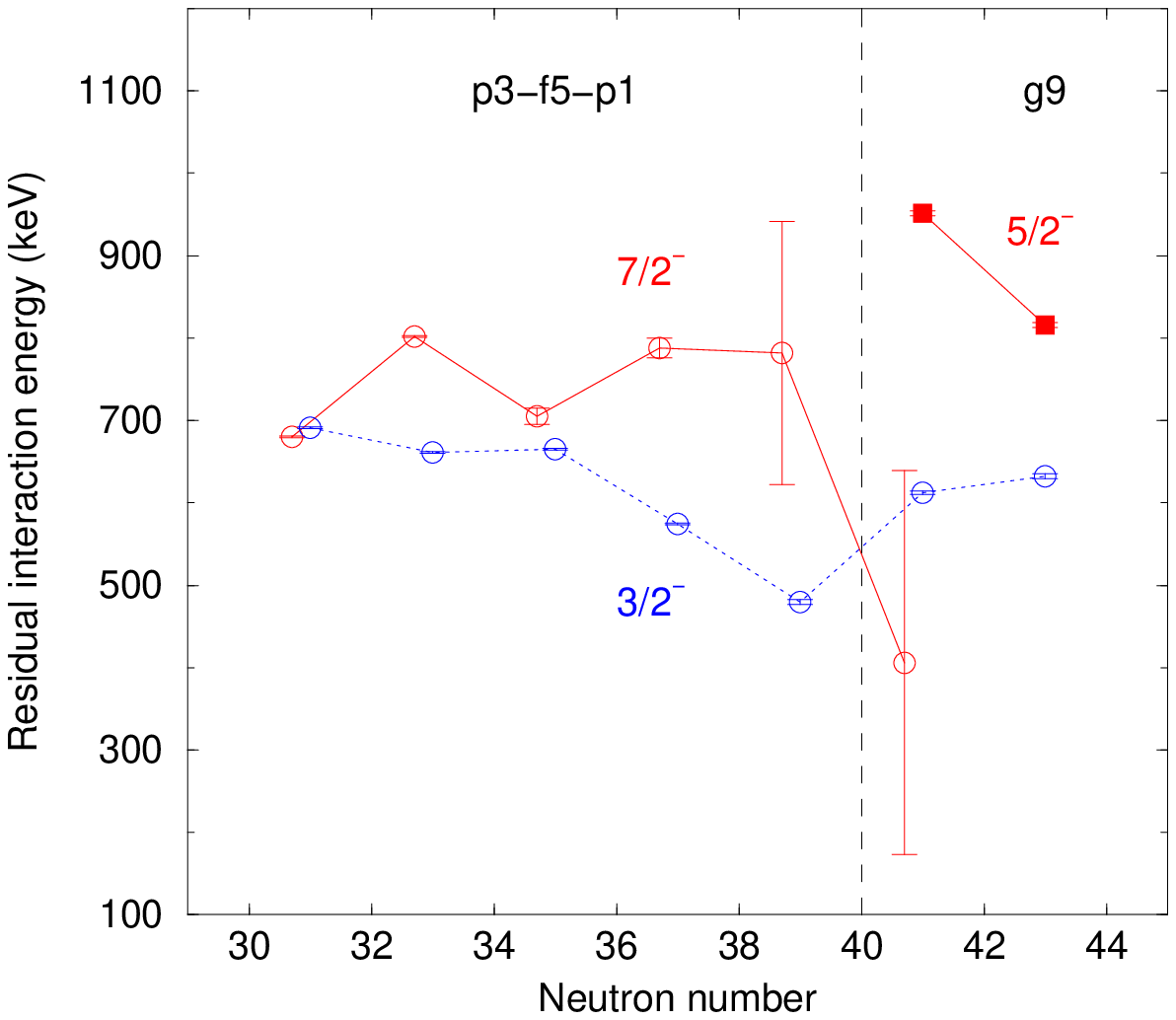,scale=0.5}
\end{minipage}
\begin{center}
\begin{minipage}[t]{16.5 cm}
\caption{{\bf Left}: Evolution the binding energies of the 7/2$^-$ and
5/2$^-$ states compared to those of the 3/2$^-$ states, on either
side of the $Z=28$ gap (see text).
{\bf Right}: Neutron-proton residual interactions
extracted from the slopes of the binding energies of the proton
states. The neutron orbitals which are getting filled as a function
of increasing $N$ are given on top of the figures.
\label{fig_pif7f5}}
\end{minipage}
\end{center}
\end{figure}

Assuming that the 5/2$^-_1$ state has a pure proton single-particle 
configuration
for $N=40-44$, the binding energy of the $\pi f_{5/2}$ level can
be computed and drawn together with the $\pi p_{3/2}$ and $\pi
f_{7/2}$ levels (see left part of Fig.~\ref{fig_pif7f5}). The
strong decrease of the $\pi f_{5/2} - \pi p_{3/2}$ energy spacing
arises in concert with the reduction of the $\pi f_{7/2}-\pi
p_{3/2}$ energy spacing. This leads to a reduction of the $\pi
f_{5/2}-\pi f_{7/2}$ spin-orbit splitting, which could be due to
the action of tensor forces between the protons $f$ and the
neutrons in the $g_{9/2}$ orbit.

As seen in the right part of Fig.~\ref{fig_pif7f5}, the residual
interactions $V^{pn}$ involving the $\pi f_{7/2}$ orbit, derived
simply from the evolution of its binding energy (as described in
Sect.~\ref{expmonopole}), have decreased from $\sim$ 750~keV
within the neutron $fp$ shells to $\sim$ 400~keV as soon as the
neutron $g$ orbit is filled, i.e. above $N=40$. There, the mean
intensity of the residual interaction for the 5/2$^-$ state -which
we assume to be of single-particle origin only- is larger
($\sim$ 900keV). However, it is worthwhile to remind that these
statements are not firm enough as the uncertainties on the atomic
masses of the Co isotopes around $N=40$ are relatively large and the 
spectroscopic factors of the 5/2$^-_1$ states in the heavy Cu isotopes 
are unknown.

The strong decrease of the energy difference between the 5/2$^-_1$
and 3/2$^-_1$ states (assumed to contain a significant part of the
$\pi f_{5/2}$ and $\pi p_{3/2}$ strengths) comes when the $\nu
g_{9/2}$ orbit is filled in the Cu isotopes. This situation is in
many aspect similar to the one arising between the 3/2$^+_1$ and
1/2$^+_1$ states (from the $\pi d_{3/2}$ and $\pi s_{1/2}$ orbits)
when the $\nu f_{7/2}$ orbit is filled in the P isotopes (see 
Fig.~\ref{ESPEsd} in Sect.~\ref{protonN28}). There, it is seen 
that the
slope of the energy difference E(1/2$^+_1$ - 3/2$^+_1$) (right
part of the Figure) is much steeper than that of the monopole
energy difference E(1/2$^+$ - 3/2$^+$) (left hand side of the same
Figure) when the neutron $f_{7/2}$ orbit starts to be filled after
$N=20$. The slope becomes much smoother later on. Deviation of the
E(1/2$^+_1$ - 3/2$^+_1$) energy difference to the monopole trend
E(1/2$^+$ - 3/2$^+$) can be assigned to correlations which are
maximum at mid-shell. These correlations arise from the coupling
between the single proton in the P nucleus and the $2^+$
excitations created by the neutrons in the Si chain. A similar
situation is expected in the Cu isotopic chain, in which
E(5/2$^-_1$ - 3/2$^-_1$) should vary steeply at the beginning of
the filling of the $g_{9/2}$ orbit and more smoothly when neutrons
are added to fill the shell to completion. Using this comparison,
it can be derived that a similar tensor force is at play in the
two cases. However the strength of the force is hitherto hard to
quantify in the Cu isotopic chain.\\

{\bf Theoretical results}\\
The neutron-rich Cu isotopes have been studied between $N=28$ and
$N=50$ within the framework of the shell-model approaches in
Refs~\cite{Smir04,Fran01,Lise05}. Their results are summarized
below.

The shell-model calculations of Smirnova et al.~\cite{Smir04} use
an effective interaction based on the G matrix
\cite{Hjor95,Nowa96}. The experimental energies of the first
3/2$^-$, 5/2$^-$ and 1/2$^-$ states in the Cu isotopic chain are
correctly reproduced by the model. The major discrepancy occurs at
$N=40$, where the calculated energy of the 5/2$^-_1$ state is
overestimated by about 600~keV. Beyond $N=40$, the 5/2$^-_1$ state
decreases with the same slope as observed experimentally. This
excess in energy is maintained for $N$ beyond 40 : the calculated
energy of the $5/2^-_1$ in $^{73}$Cu$_{44}$, $\simeq$~750~keV, is
larger than the measured value of 166~keV. Hence the crossing
between the 5/2$^-_1$ and 3/2$^-$ states is predicted at $N=50$ as
compared to $N=45$ extrapolated from the experimental data.
Concerning the SF values, the ground state 3/2$^-$ of the odd-A Cu
isotopes is found to be almost pure along the whole isotopic
chain, exhausting 70 \% - 100 \% of the single-proton $\pi
p_{3/2}$ state. On the other hand, the structure of the 5/2$^-_1$
state changes a lot as a function of the neutron number. While it
is rather pure single particle for $N=28$ or $N=50$, its
spectroscopic factor drops down to $\simeq$0 for $N=38$. This
suggests that at mid-distance between $N=28$ and $N=50$ this state
is very sensitive to correlations beyond the monopole
interactions. Experimental SF of the $5/2^-_1$ state in the Cu
chain determined up to $A=65$, are significantly larger that the
calculated ones. As mentioned by the authors of Ref.~\cite{Smir04}
this discrepancy points to a deficiency in the realistic
interaction of hitherto unknown origin. The 1/2$^-_1$ state has a
calculated SF value of about 40\% between $A=59$ and $A=77$.
Experimental values are available up to $A=65$. They are in
accordance with these calculations.

The excited levels of odd-A $^{69-79}$Cu$_{40-50}$ nuclei have
been also compared to results of realistic shell-model
calculations \cite{Fran01} using another residual interaction (the
so-called S3V set, from Ref.~\cite{S3V92}). In this work, the
ESPE's have been adjusted in order to reproduce the spectroscopy
of the Ni and Cu isotopes around $N=40$. Therefore the S3V set
applies to nuclei with $N \ge 40$ only. As in Ref.~\cite{Smir04}
the distance in energy between the $\pi p_{3/2}$ and $\pi f_{5/2}$
subshells decreases rapidly during the filling of the $\nu
g_{9/2}$ orbital. This was attributed to the strong attractive
monopole interaction $V^{pn}_{f_{5/2}g_{9/2}}$. It is worth
pointing out that the S3V interaction slightly overestimates the
2$^+_1$ energies of the neutron-rich $^{70-76}$Ni isotopes, as
shown in Fig.~4 of Ref.~\cite{Mazz05}.

The realistic interaction of Lisetskiy et al.~\cite{Lise04} has
been derived from a fit to existing experimental data for the Ni
isotopes from $A=57$ to $A=76$ and the $N=50$ isotones from
$^{79}$Cu to $^{100}$Sn, for neutrons and protons respectively. In
particular, its strength has been adjusted to produce the crossing
of the 3/2$^-_1$ and 5/2$^-_1$ states of Cu isotopes at $N \simeq
45$~\cite{Lise05}, as expected from the experimental results. This
interaction reproduces successfully the evolution of the 2$^+$
energies in the Ni isotopic chain up to $N=48$~\cite{Mazz05}. This
agreement indicates that the proton-neutron monopole matrix
elements involving the $g_{9/2}$ orbit would be well determined.
However the calculated E2 transition strength\footnote{using an effective
neutron charge $e_\nu=1.0 e$} of $^{70}$Ni, B(E2;2$^+_1$ $\rightarrow$ 
0$^+$) = 64$e^2fm^4$ \cite{Lise04} is smaller than the
measured one~\cite{Perr06}, B(E2)$\uparrow$=860(140)$e^2fm^4$,
giving B(E2)$\downarrow$=172(28)$e^2fm^4$. The model space that
has been used is not wide enough to account explicitely for the
full excitations to develop, in particular those across the $Z=28$
shell gap.

\subsubsection{\it Conclusion}

The Ni isotopic chain offers a variety of doubly-magic nuclei:\\
- the $^{48}_{28}$Ni$_{20}$, lying at the edge of particle
stability, which is the mirror of $^{48}_{20}$Ca$_{28}$,\\
- the $N=Z$
nucleus $^{56}_{28}$Ni$_{28}$,\\
- the "superfluid"
$^{68}_{28}$Ni$_{40}$ in which the neutron Harmonic Oscillator
shell number provides a very specific role with respect to parity
conservation,\\
- the $^{78}_{28}$Ni$_{50}$ nucleus which
is composed of two magic numbers $Z=28$ and $N=50$, both originating
from the spin-orbit interaction. \\
The structural evolution of the
Ni isotopes is related to that of the $Z=28$ shell gap, the size
of which varies according to proton-neutrons interactions. The
$Z=29$ Cu isotopic chain provides a mean to determine the strength
of the various effective interactions involved between protons and
neutrons in the $f, p, g$ shells and their decomposition in term
of central, spin-orbit, tensor components.

The evolution of the $Z=28$ shell closure, formed between the
proton $f_{7/2}$ and $p_{3/2}$ orbits, has been discussed using
the experimental and theoretical results of the Co, Ni and Cu
isotopic chains. The $Z=28$ gap slightly increases between $N=30$
to $N=40$ (from about 4.8 to 5.5~MeV) reaching a partial maximum
at $^{68}$Ni and a minimum of quadrupole collectivity there.
Adding neutrons above $N=40$ seems to provoke the decrease of the
$Z=28$ gap (see the left part of Fig.~\ref{gapZ28}), as the $\pi
f_{7/2}$ orbit gets less bound as compared to the $\pi p_{3/2}$
one. This picture could account for the increase of cross-shell
core excitations required to explain the B(E2) values of
$^{70}$Ni~\cite{Perr06}, as well as the neutron-rich Zn
isotopes (this is discussed in a next section, 
see Sect.~\ref{E2_B(E2)_N50}). 
In parallel,
following the trend of Fig.~\ref{crossingCu} one would expect that
the ground state of the heavy Cu isotopes become 5/2$^-$ instead
of 3/2$^-$. Would these states have a rather pure single particle
configuration, the $\pi f_{5/2}$ single-particle orbit decreases to
eventually cross the $\pi p_{3/2}$ orbit at a neutron number $N \simeq
45$. Then the $Z=28$ would become bound by the $\pi f_{7/2}$ and
$\pi f_{5/2}$ orbits.

As the increase of binding energy of the $\pi f_{5/2}$ orbit
occurs simultaneously to the decrease of binding energy for the
$\pi f_{7/2}$ ones, the filling of the $\nu g_{9/2}$ orbital
induces a reduction of the $\pi f_{7/2}- \pi f_{5/2}$ SO
splitting. This is qualitatively interpreted by the action of
tensor forces which are attractive (repulsive) between neutrons in
the $g_{9/2}$ and protons in the $f_{5/2}$ ($f_{7/2}$) orbits
(see Ref.~\cite{Otsu05} and Sect.~\ref{decompmonopol}). From 
these combined actions,
the $Z = 28$ gap in $^{78}$Ni could decrease down to 2.5~MeV only,
a value which is small enough to permit excitations across $Z=28$.  
We remind that the
present assumptions concerning the evolutions of the $\pi f_{5/2}$
and $\pi f_{7/2}$ orbits hold\emph{ mainly} on two facts, (i) the
evolution of the 5/2$^-_1$ state in the $_{29}$Cu isotopes, and (ii) 
the
determination of atomic masses for the $_{27}$Co isotopes. For the
first point the role of correlations should be quantified by
further experimental and theoretical works, whereas for the second
more precise atomic masses are required.

Noteworthy is the fact that the role of tensor forces that is
surmised here holds for several other regions of the chart of
nuclides, discussed in the next sections. 

%% file: texteNZ50_9avril.tex
\section{The magic number 50}

Various experimental facts indicating the particular stability of
shells of 50 neutrons or 50 protons have been listed sixty years
ago~\cite{Goep48}. For instance the large number of stable $N=50$
isotones (6) and $Z=50$ isotopes (10) has been stressed. At that
time, the magic number 28 was not put forward as the experimental
results were not conclusive enough. So the magic number 50 was the
first which did not have a natural description in terms of a
square well potential. This has been considered for a while as an
indication of the breakdown of the shell model for heavy nuclei,
until a strong spin-orbit coupling was assumed~\cite{Goep49,Haxe49}.
The magic number 50 originates from the spin-orbit part of the
nuclear interaction which lowers the energy of the $g_{9/2}$ orbit
from the $N=4$ major shell so that it is located close to the
orbits from the $N=3$ major shell. Its partner, the $g_{7/2}$
orbit, lies on the upper edge of the gap, immediately above or
below the $d_{5/2}$ orbit.

\subsection{\it Evolution of the $N=50$ shell closure}

Experimental information, even scarce, is available along the
$N=50$ isotonic chain from $^{78}_{28}$Ni to $^{100}_{~50}$Sn.
These two\emph{ a priori} doubly-magic nuclei attract physicists
attention, the former is very neutron rich ($N/Z \simeq 1.8$)
while the latter, located at the proton drip-line, is
self-conjugate. Both of them have been synthetized and identified
in-flight using the projectile fragmentation method. However, as
only a few nuclei of each have been produced worldwide their
excitation modes remain unknown and rely on the extrapolations of
the properties measured in their neighbors.

Possible evolution of the size of the $N=50$ shell gap between
$Z=28$ to $Z=50$ depends on proton-neutron interactions between
the $\pi fp$ and $\pi g_{9/2}$ shells and the neutron orbits which
form the $N=50$ gap, i.e. $\nu g_{9/2}$ below and $\nu d_{5/2}$ or
$\nu g_{7/2}$ above. As diverse interactions are involved, among
which some contain a tensor part, it is expected
that the $N=50$ gap will vary along the isotone chain. The
following sections intend to probe these evolutions through
various experimental methods: evolution of atomic masses, trends
of collective states, spectroscopy of high spin-states and
transfer reactions.

\subsubsection{\it Evolution of the binding energies\label{evolgapN50}}
\begin{figure}[h!]
\begin{minipage}{9.cm}
\begin{center}
\epsfig{file=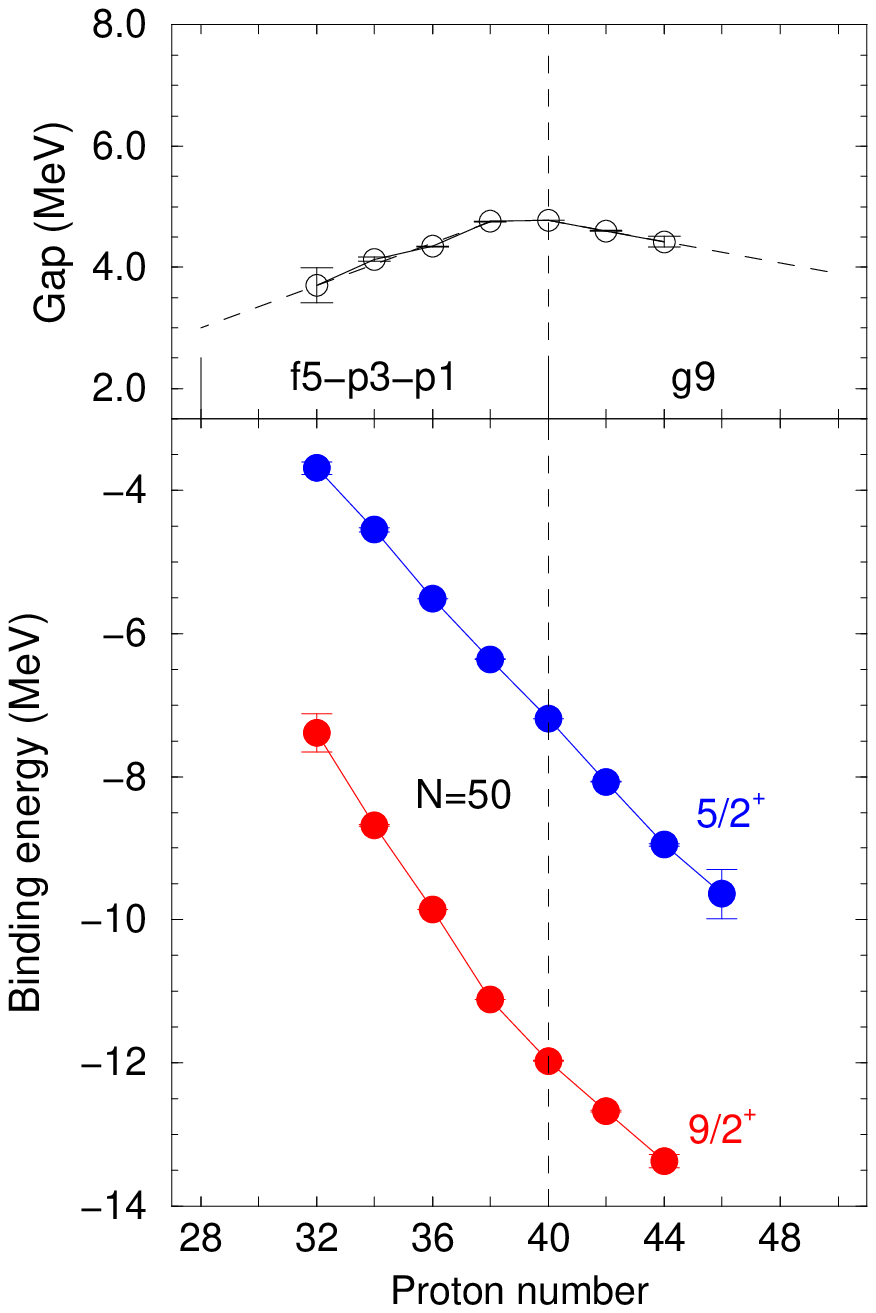,scale=0.55}
\end{center}
\end{minipage}\hfill
\begin{minipage}{9.cm}
\epsfig{file=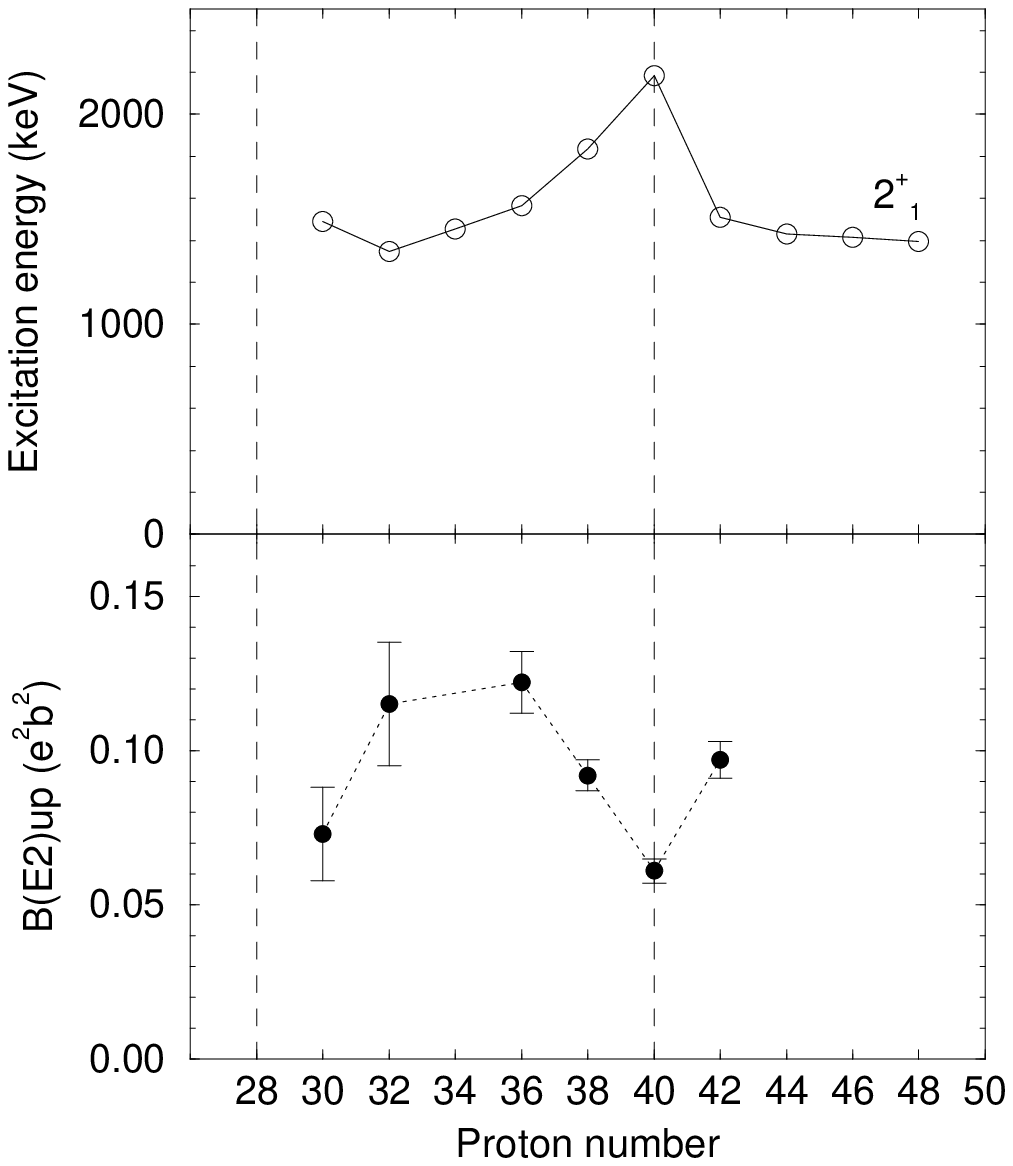,scale=0.55}
\end{minipage}
\begin{center}
\begin{minipage}[t]{16.5 cm}
\caption{{\bf Left}: Binding energies of the states $9/2^+$
($5/2^+$) located below (above) the $N=50$ magic number and
difference of the binding energies of these two states
surrounding the gap at $N=50$ (see Sect.~\ref{annex}).  The
proton orbitals which are getting filled as a function of
increasing $Z$ are given in the middle of the figure.
{\bf Right}:
Experimental $E(2^+)$ and $B(E2; 0^+ \rightarrow
2^+)$ values in the $N=50$ isotones. \label{gapN50}}
\end{minipage}
\end{center}
\end{figure}

The binding energies of the last neutron in the $N = 51,50$
isotones are drawn in the left part of Fig.~\ref{gapN50}. 
Their values are taken from the atomic mass table~\cite{Audi03}, to 
which the result for $^{83}$Ge has been added~\cite{Thom05}.
Whereas the curve corresponding to the 5/2$^+$ states shows a
quasi-linear behavior from $Z = 32$ to $Z = 46$, the one
corresponding to the 9/2$^+$ states displays a steeper slope for
$32 \le Z \le 38$ than for $Z \ge 40$. This change can be
related to the different interactions which come into play with
the progressive filling of the proton $f_{5/2}$, $p_{3/2}$, and
$p_{1/2}$ orbits for $Z~\le~40$, and $g_{9/2}$ for $Z~\ge~40$.

As shown in the left part of Fig.~\ref{gapN50} the size of the $N
= 50$ gap can be extrapolated to be 3.0(5)~MeV for $^{78}_{28}$Ni.
This extrapolation holds on the fact that similar proton-neutron
interactions are at play between $Z=38$ and $Z=28$ as protons are
equiprobably removed from the $f_{5/2}$ and $p_{3/2}$ orbits which
are very close in energy. A similar extrapolation of the $N=50$
gap can be also untertaken from $^{94}_{44}$Ru to $^{100}_{~50}$Sn
between which the $\pi g_{9/2}$ orbit is progressively filled. The
size of the $N=50$ gap would amount to 3.9(4)~MeV for 
$^{100}$Sn\footnote{This value does not contain the singularity
coming from the Wigner energy (see Sect.~\ref{annex}), such as
those discussed in previous sections for $^{40}$Ca or $^{56}$Ni.}.

The constancy of the slope of the binding energy for the $5/2^+$
states reveals that all the proton-neutron interactions involving
the $\nu d_{5/2}$ orbit have similar intensity, whatever the proton
orbits that are considered. On the other hand, the binding energy
curve of the $9/2^+$ states has an inflexion point at $Z=40$,
meaning that the filling of the $\pi g_{9/2}$ shell binds the $\nu
g_{9/2}$ orbit less than the $\pi fp$ shells do. In other words, the
matrix element $V^{pn}_{g_{9/2}g_{9/2}}$ is weaker than any
$V^{pn}_{(f,p) g_{9/2}}$ ones involving the proton $fp$ shells.
Among the latter, the $V^{pn}_{f_{5/2}g_{9/2}}$ matrix element
contains an attractive component from the tensor interaction, which
may enhance its strength with respect to other monopoles\footnote{
A similar matrix element is involved to account for the structural
evolution of the neutron-rich Cu isotopes (Sect.~\ref{Cu}). There
the binding of the $\pi f_{5/2}$ orbit increases with the filling of
the $\nu g_{9/2}$ one.}. 
On the contrary the
$V^{pn}_{g_{9/2}g_{9/2}}$ interaction (acting for $Z>40$) contains
a repulsive term arising from the tensor part of the nucleon-nucleon 
interaction.

\subsubsection{\it Trends of 2$^+$ and B(E2) values\label{E2_B(E2)_N50}}

As the $N=50$ shell gap is large, the 2$^+_1$ state in the $N=50$
isotonic chain is  expected to be mainly built on\emph{ proton}
excitations within the $fpg$ shells. The experimental 2$^+_1$
excitation energies and the B(E2; 0$^+$ $\rightarrow$ 2$^+_1$)
values shown in the right part of Fig.~\ref{gapN50} clearly point
to the persistence of the $N=50$ shell closure both towards the
heaviest and lightest isotones measured yet. As far as the B(E2)
measurement is concerned, the route to $^{100}_{~50}$Sn is longer
than to $^{78}_{28}$Ni. This is due to the recent works achieved
at the HRIBF (Oak Ridge) for $^{82}_{32}$Ge~\cite{Padi05} and the
Rex-Isolde facility for $^{80}_{30}$Zn~\cite{Vand07}. The
B(E2)$\uparrow$ value of $^{82}_{32}$Ge has been determined using
the Coulomb excitation of the radioactive beam produced from the
proton-induced fission of uranium, separated on-line and
post-accelerated to 220 MeV. The value of 0.115(20)~e$^2$b$^2$ is
close to the one of $^{86}_{36}$Kr. This situation is expected in
the framework of the generalized seniority scheme described in
Sect.~\ref{enerprobaNi} , where the B(E2) value is maximized at 
mid-shell (here at $Z=34$) and decays at each side symmetrically 
(see Eq.~\ref{BE2_seniority}). The
Coulomb excitation of the $^{80}_{30}$Zn nucleus was achieved at
the Rex-Isolde facility. The mass-separated radioactive Zn were
accumulated, cooled and bunched into a Penning trap. They were
subsequently brought to a high charge, separated in A/q and post
accelerated by the Rex linear accelerator to about 2.8 A.MeV. A
B(E2)$\uparrow$ value of 0.073(15)~$e^2b^2$ has been extracted for
$^{80}_{30}$Zn which, as shown in Fig.~\ref{gapN50}, is sensibly
weaker than that of $^{82}$Ge. Shell model calculations reproduce
rather well the experimental B(E2) results in the $N=50$ isotones
as well as in the Zn isotopes with two sets of interactions.
However, they require large proton $e_\pi$ effective charge. This
indicates a rather strong $Z=28$ core polarization~\cite{Vand07}.

\subsubsection{\it Evolution of the $N=50$ gap viewed from p-h states}

The size of a gap is related to the energy of the states arising
from the $1p-1h$ excitations across it. As the $N=50$ shell gap is
formed between the $g_{9/2}$ and $d_{5/2}$ orbits, the
corresponding $1p-1h$ states have a $(\nu
g_{9/2})^{-1}(\nu d_{5/2})^{+1}$ configuration. This gives rise to
a multiplet of six states with spin values $J$ ranging from 2 to
7. In $^{90}_{40}$Zr, all the members of this multiplet have been
selectively populated using the neutron pick-up reaction,
$^{91}$Zr$_{51}$($^3$He, $\alpha$)$^{90}$Zr$_{50}$ by requiring a
transferred angular momentum $\ell$=4.
\begin{figure}[h!]
\begin{minipage}{9.cm}
\begin{center}
\epsfig{file=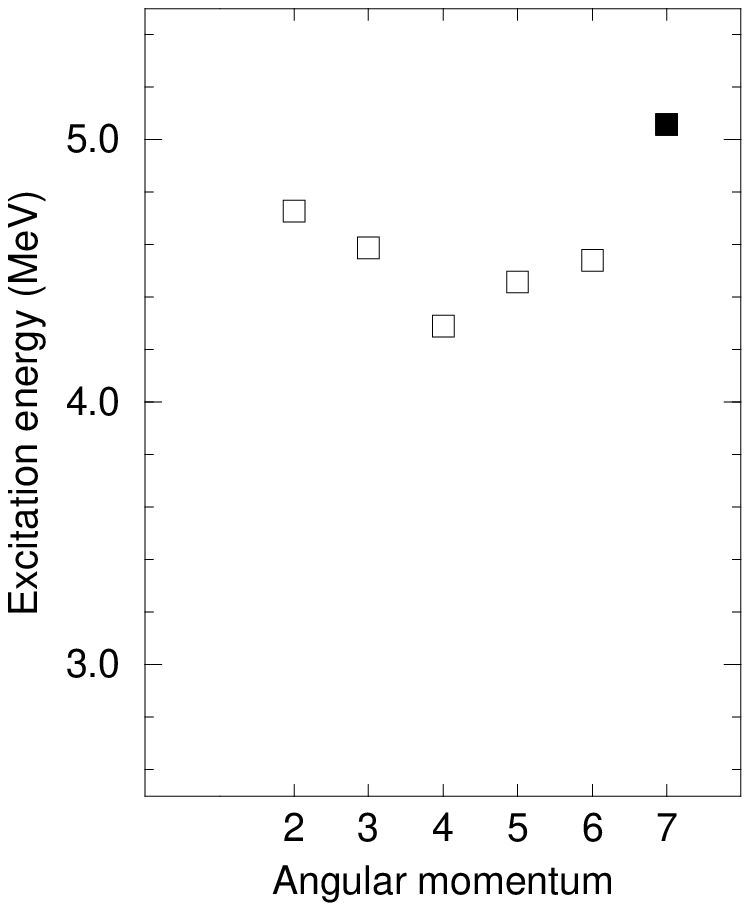,scale=0.6}
\end{center}
\end{minipage}\hfill
\begin{minipage}{9.cm}
\epsfig{file=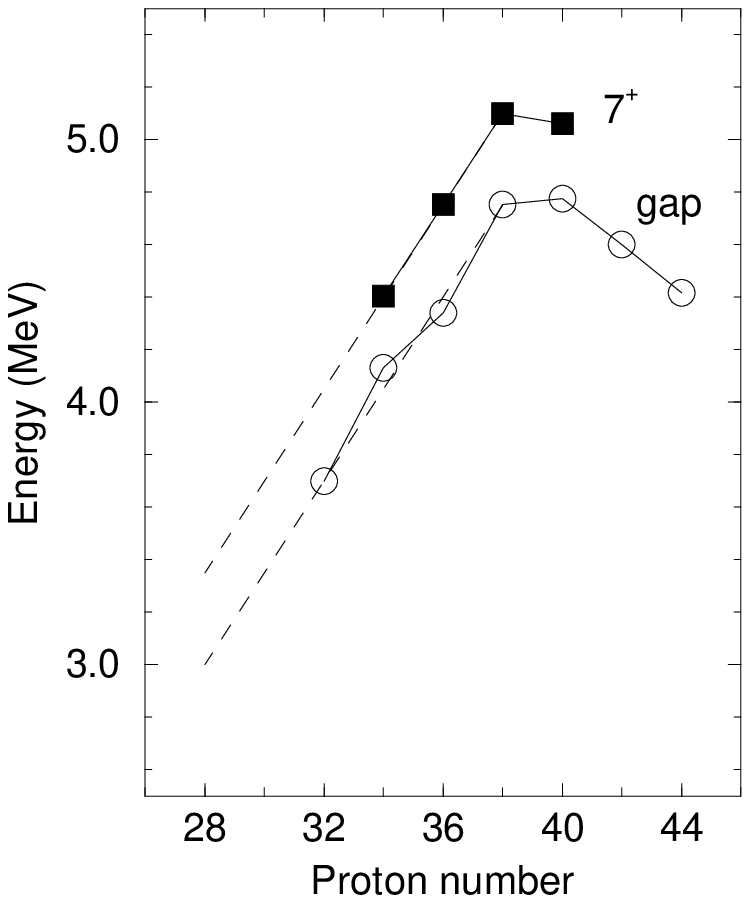,scale=0.6}
\end{minipage}
\begin{center}
\begin{minipage}[t]{16.5 cm}
\caption{ {\bf Left}: Excitation energies of the six states of the
$(\nu g_{9/2})^{-1}(\nu d_{5/2})^{+1}$ configuration measured in
$^{90}_{40}$Zr, as a function of their spin $J$. {\bf Right}:
Comparison of the energy of the 7$^+$ state from the $(\nu
g_{9/2})^{-1}(\nu d_{5/2})^{+1}$ configuration in the even-even
$N=50$ isotones (filled squares) and of the gap $N = 50$ (empty
circles, see the left part of Fig.~\ref{gapN50}).
\label{gapN50_2qp}}
\end{minipage}
\end{center}
\end{figure}
Their energies, reported in the left part of Fig. \ref{gapN50_2qp}
as a function of $J$, follow a parabola of positive curvature. The
energies of the two states having the extreme spin values, $J=2^+$
and $J=7^+$, are less lowered by the $p-h$ residual interaction, as
expected by the coupling rules reminded in Sect.~\ref{expmultiplet}.
These states have also been identified in $^{88}_{38}$Sr using the
neutron stripping reaction $^{87}$Sr$_{49}(d,p)$ in which a
transferred angular momentum $\ell$=2 was identified. Such a complete
identification of all the members of the $1p-1h$ multiplet could not
be obtained when unstable species are involved.

The route to access these states for neutron-rich unstable species
$A \sim 80-130$ is to use either fusion-fission reactions or
spontaneous fission of actinides. In these experiments the high
sensitivity of large arrays of Compton-suppressed Ge detectors is
needed to detect the large number of photons emitted by the $\simeq$
150 different nuclides produced simultaneously. The assignment of
$\gamma$-rays to a particular fragment requires an exceptional
selectivity. This is achieved by selecting, among the high-fold
data, those which contain several $\gamma$-rays originating from a
well identified fission partner.

Using the gamma-ray spectroscopy method from fusion-fission
reaction, the 7$^+$ yrast states of $^{86}_{36}$Kr and
$^{84}_{34}$Se have been recently identified, which enlarges the
knowledge of the evolution of the $(\nu g_{9/2})^{-1}(\nu
d_{5/2})^{+1}$ configuration in the light $N=50$
isotones~\cite{Prev04}. As seen in the right part of
Fig.~\ref{gapN50_2qp}, the energy of the 7$^+$ state follows
remarkably that of the gap, which confirms its decrease towards
$Z=28$. The energy of the 7$^+$ state is shifted by about +350~keV
with respect to the gap value. This difference comes from the
repulsive particle-hole interaction $(\nu g_{9/2}) ^{-1} (\nu
d_{5/2})^{+1}$ (see Sect.~\ref{expmultiplet}).

More recently, medium-spin states in $^{82}_{32}$Ge have been
studied from spontaneous fission of $^{248}$Cm \cite{Rzac07}. Two
new excited levels with spins and parities (5, 6$^+$) have been
proposed at 2930 and 3228 keV. At least one of them can be attributed
to the $(\nu g_{9/2})^{-1}(\nu d_{5/2})^{+1}$ configuration. Its
low energy, as compared to the heavier isotones, supports the
decrease of the $N=50$ shell gap.

Following the correlation between the $N=50$ gap and the energy of
the 7$^+$ state, one should find an hitherto unknown 7$^+$ state at
$\sim$ 4.0 MeV in $^{82}_{32}$Ge, $\sim$ 3.7 MeV in $^{80}_{30}$Zn,
and $\sim$ 3.4 MeV in $^{78}_{28}$Ni. As said earlier, this
extrapolation relies on the fact that the same $\pi p_{3/2}$ and
$\pi f_{5/2}$ orbits are involved  between $Z=38$ and $Z=28$ and
that they are closely packed in energy. In such a case the pairing
correlations smooth their occupancies, leading to a regular decrease
of the gap and of the $J=7^+$ energy as protons are removed.

Access to the neutron-deficient nuclei has been obtained by the use
of the fusion-evaporation technique. A 12$^+$ isomeric state has
been identified in the $^{98}_{48}$Cd isotope through the $\gamma$
transition at 4.207~MeV to the 8$^+$ component of the multiplet
arising from ($\pi g_{9/2})^{-2}$ configuration~\cite{Blaz04}. 
The composition of
this isomer is expected to be pure ($\pi g_{9/2})^{-2} (\nu
g_{9/2})^{-1} (\nu d_{5/2})^{+1}$, which involves the neutron
cross-shell excitation. Besides the repulsive interaction, 
$(\nu g_{9/2})^{-1}(\nu d_{5/2})^{+1}$ (such as that of the
neutron-rich isotones, as said previously) the 4.207~MeV 
energy also contains proton-neutron interactions,  
$(\pi g_{9/2})^{-1} (\nu d_{5/2})^{+1}$ (repulsive terms) and 
$(\pi g_{9/2})^{-1} (\nu g_{9/2})^{-1}$ (the 0$^+$ term is strongly 
attractive), which have to be determined beforehand in order to
get the $N=50$ neutron gap at $Z=48$.

\subsubsection{\it Neutron orbits above the $N=50$ gap: Levels of $N=51$
isotones\label{N51}}

As shown in Fig.~\ref{gapN50}, the $\nu d_{5/2}$ shell is located
just above the $N=50$ shell gap. The evolution in energy of the
two next shells, $\nu s_{1/2}$ and $\nu g_{7/2}$,  as a function
of increasing proton number can be derived from the study of the
$N=51$ isotones. This provides pertinent information on the
residual
proton-neutron interactions at work, as discussed now.\\

{\bf Experimental results}\\
Experimental results on the $N=51$ isotones cover a broad range of
masses, from $^{83}_{32}$Ge to $^{101}_{~50}$Sn. For stable nuclei
detailed spectroscopic information exist, whereas for neutron-rich
or neutron-deficient nuclei the information is more sparse. However
several results have been recently published on the $^{83}$Ge
\cite{Thom05,Perr06b}, $^{85}$Se \cite{Bard06,Porq07}, and $^{87}$Kr
\cite{Porq06} nuclei.  These add to the low-spin level scheme of
$^{85}$Se that was obtained about 15 years ago from the
$\beta$-decay of mass-separated $^{85}$As nuclei~\cite{Omtv91}.
As for the neutron-deficient part, first data on states of
$^{101}_{~50}$Sn have been newly obtained \cite{Kava07,Sewe07}.
All these results will be discussed in the following with the aim
of looking at the evolution of the neutron orbits as a function of
the valence protons.

The ground-state spin value of all $N=51$ isotones between $Z=36$
and $Z=44$ has been firmly established to be 5/2$^+$, whereas for
the others this $5/2^+$ assignment is compatible with the
spectroscopic properties known so far. Thus for all the proton
numbers considered here, the $\nu d_{5/2}$ sub-shell can be
considered as the first valence orbit above the $N=50$ gap. The
coupling of the $\nu d_{5/2}$ orbit to the first 2$^+$ excitation of
the proton core provides a multiplet of states with spin values
between 1/2$^+$ and 9/2$^+$. Therefore the 1/2$^+$ and 7/2$^+$
states originating from the $\nu s_{1/2}$ and $\nu g_{7/2}$ orbits
can be mixed to those obtained with the coupling to the core
excitation. The five states of the multiplet have been firmly
identified in $^{89}_{38}$Sr. Their centroid energy is similar to
that of the 2$^+$ state in $^{88}_{38}$Sr \cite{Perr06b}.

\begin{figure}[h!]
\begin{minipage}{10cm}
\begin{center}
\epsfig{file=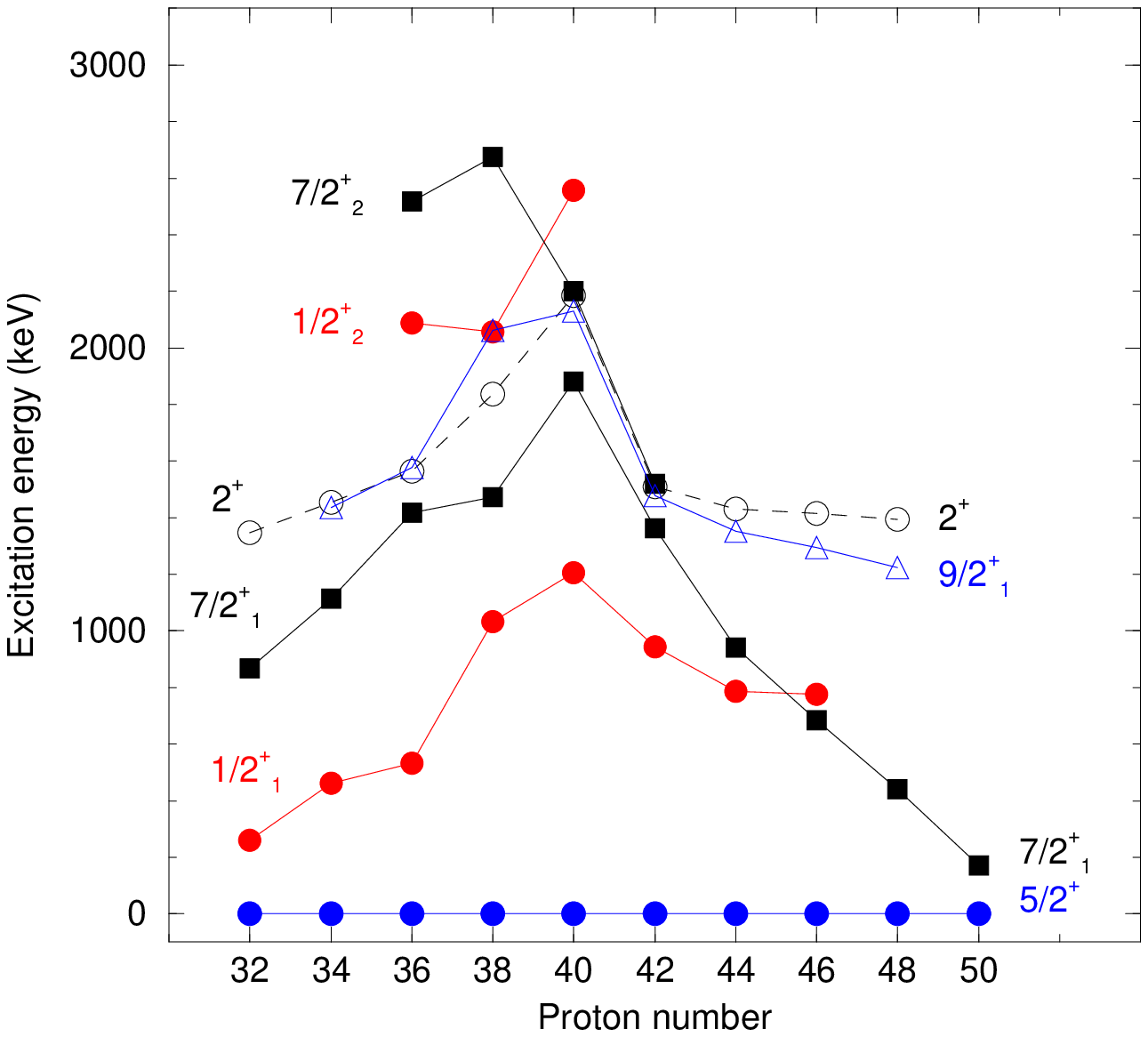,scale=0.6}
\end{center}
\end{minipage}
\begin{minipage}{9.cm}
\epsfig{file=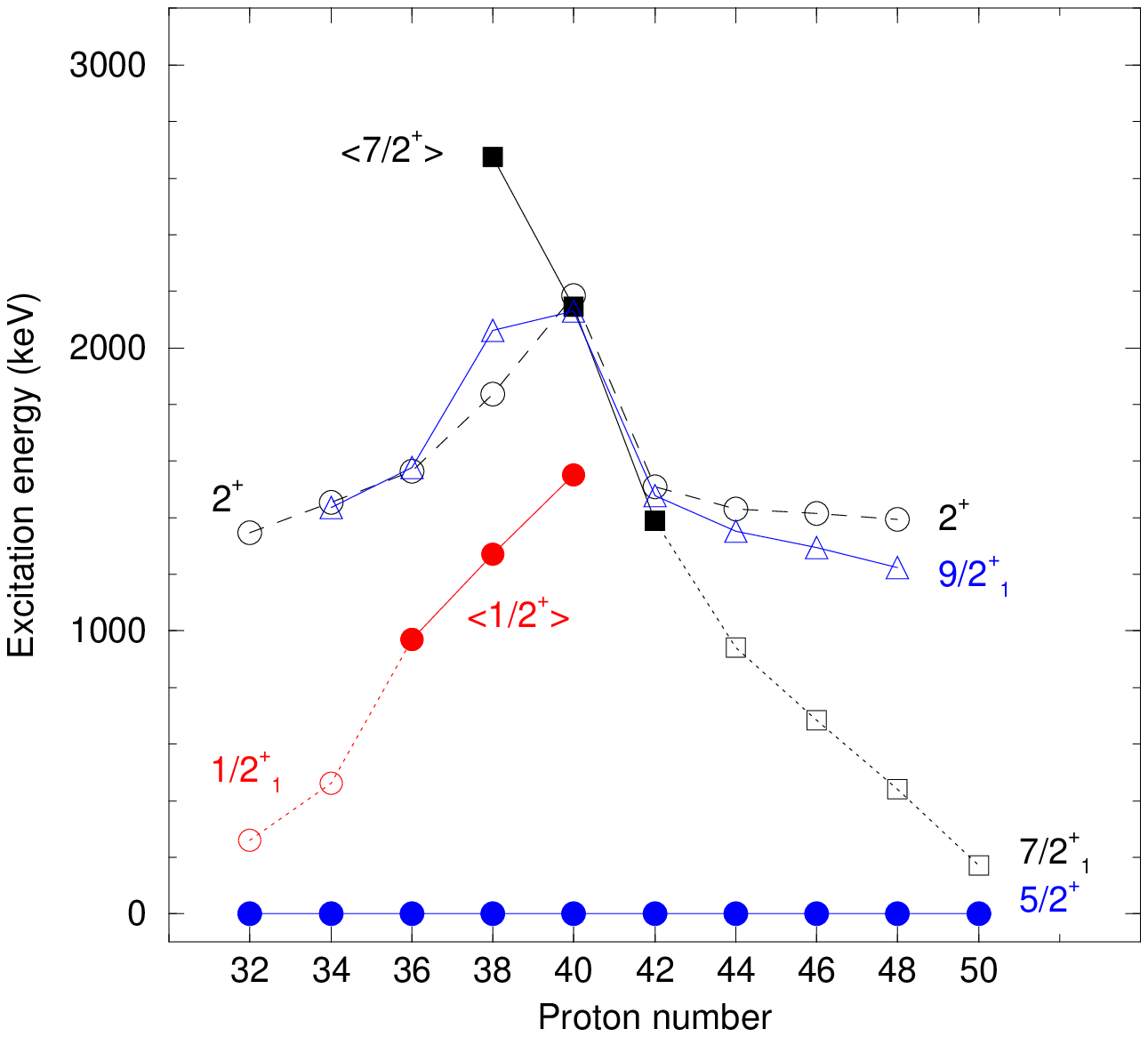,scale=0.6}
\end{minipage}
\begin{center}
\begin{minipage}[t]{16.5 cm}
\caption{{\bf Left}: First excited states observed in the $N=51$
isotones. The evolution of the 2$^+$ energy of the corresponding
core is also drawn. {\bf Right}: Possible evolution of the
$\nu s_{1/2}$ and $\nu g_{7/2}$ single-particle energies obtained
from (i) the average energy (filled symbols) of all the 1/2$^+$
and 7/2$^+$ states populated in the $(d,p)$ stripping reactions on
the stable targets, $^{86}_{36}$Kr, $^{88}_{38}$Sr, $^{90}_{40}$Zr,
and $^{92}_{42}$Mo, (ii) the energy of the 1/2$^+_1$ state (empty
circles) of the lightest isotones, (iii) the energy of the
7/2$^+_1$ state (empty squares) of the heaviest isotones.
\label{crossingN51}}
\end{minipage}
\end{center}
\end{figure}

The energies of the first excited states measured in the $N=51$
isotones are displayed in the left part of Fig.~\ref{crossingN51},
relative to that of the $5/2^+$ ground state. First of all, the
9/2$^+_1$ state is likely to have a rather pure coupled
configuration, $\nu d_{5/2} \otimes 2 ^+$,  as the $g_{9/2}$ orbit
is too deeply bound (by about 3.5~MeV) to give a fraction of
single-particle state at low energy. Such a configuration for the
9/2$^+_1$ state is confirmed by the fact that its energy follows
closely that of the 2$^+$ state. On the other hand, the energies of
the 1/2$^+_1$ and 7/2$^+_1$ states depart significantly from that of
the $2^+$ core at $Z <38$ and $Z
> 44$, respectively. This means that, for these proton numbers,
their composition is likely to be mostly of single-particle
origin, with a weaker amount of mixing with the coupled
configurations.

The $(d,p)$ transfer reactions on the stable $N=50$ targets help
to unfold the composition of both contributions, as the neutron
stripping to single-particle states is strongly enhanced as
compared to collective states. Several 7/2$^+$ and 1/2$^+$ states
have been populated
in $^{87}_{36}$Kr, $^{89}_{38}$Sr, $^{91}_{40}$Zr, and
$^{93}_{42}$Mo, their angular momenta being
assigned to $\ell = 4$ or $\ell = 0$, respectively. The 7/2$^+_1$
and 1/2$^+_1$ states have the largest spectroscopic factors. It
is therefore likely that the 7/2$^+_2$ and 1/2$^+_2$ states
displayed in the left part of Fig.~\ref{crossingN51} correspond
mainly to $\nu d_{5/2} \otimes 2^+$ configurations. The same holds
true for the 9/2$^+_1$ state which has not been populated in the
stripping reactions. This confirms the assumption made in
the previous paragraph.

Stripping $(d,p)$ reactions in inverse kinematics have been recently
performed with the radioactive $^{82}_{32}$Ge and $^{84}_{34}$Se
nuclei~\cite{Thom05,Bard06}. The two radioactive beams were produced
from the proton-induced fission of uranium, separated on-line, and
post-accelerated to 330 MeV and 378 MeV respectively. With such
low-energy beams, only the single-neutron states at low-energy could
be identified in $^{83}_{32}$Ge$_{51}$ and $^{85}_{34}$Se$_{51}$
through the detection of the protons in a highly-segmented silicon
detector array. Angular distributions of protons associated to the
population of their ground and first-excited states have been
obtained. They are consistent with $\ell$~=~2 and $\ell$~=~0
transferred momenta giving rise to 5/2$^+$ and 1/2$^+$ spin
assignements, respectively. Spectroscopic factors account for only
half of the single-particle strengths. The remaining part is likely
to be present at higher excitation energy. A similar result had been
already obtained in $^{87}$Kr, which displays around half of the
single-particle strengths, both for the $\nu d_{5/2}$ and the $\nu
s_{1/2}$ orbits.

The $\beta$ decay of $^{101}$Sn has been newly studied at the GSI
on-line mass separator, this very neutron-deficient isotope being
produced by using the $^{50}$Cr($^{58}$Ni, $\alpha$3n)
fusion-evaporation reaction~\cite{Kava07}. By comparing the
experimental energy spectrum of $\beta$-delayed protons to a
theoretical prediction, the 5/2$^+$ spin and parity assignments for
the $^{101}$Sn ground state have been confirmed. Moreover the
$^{46}$Ti($^{58}$Ni,3n) reaction was used to determine the first
excited state of $^{101}$Sn at the Argonne facility~\cite{Sewe07}.
Prompt $\gamma$-rays emitted by $^{101}$Sn have been identified
through correlations with $\beta$-delayed protons following its
decay, using the recoil-decay tagging (RDT) method. The
$\gamma$-rays were detected in a large array of Ge detectors
surrounding the target, whereas the products of the reaction were
separated from the unreacted beam and dispersed according to their
$M/Q$ ratio in the fragment mass analyzer. A new $\gamma$-ray line
with an energy of 171.7(6) keV has been assigned to $^{101}$Sn and
interpreted as the transition between the single-neutron $g_{7/2}$
and $d_{5/2}$ orbitals.

The right part of Fig.~\ref{crossingN51} displays the average energy
(filled symbols) of all the 1/2$^+$ and 7/2$^+$ states, weighted by
their SF values. These states were populated in the $(d,p)$
stripping reactions on the stable targets (using the same method as
in Sect. \ref{19K}). Added to these points the empty symbols display
the energy of the 1/2$^+_1$ and 7/2$^+_1$ states identified in the
lightest and heaviest $N=51$ isotones, respectively. The obvious
outcomes of this figure are (i) a reduction of the energy between
the $\nu d_{5/2}$ and $\nu g_{7/2}$ orbits when $Z$ increases, (ii)
a similar decrease of the energy spacing between the $\nu d_{5/2}$
and $\nu s_{1/2}$ orbits for low $Z$ values. Nevertheless the
excitation energy of the 1/2$^+_1$ (7/2$^+_1$) state measured in the
lightest (heaviest) isotones is not an absolute determination of the
single particle energy spacing because their wave functions may
contain an unknown fraction of mixing to the $\nu d_{5/2} \otimes
2^+$ configuration.

\begin{figure}[h!]
\begin{minipage}{10cm}
\begin{center}
\epsfig{file=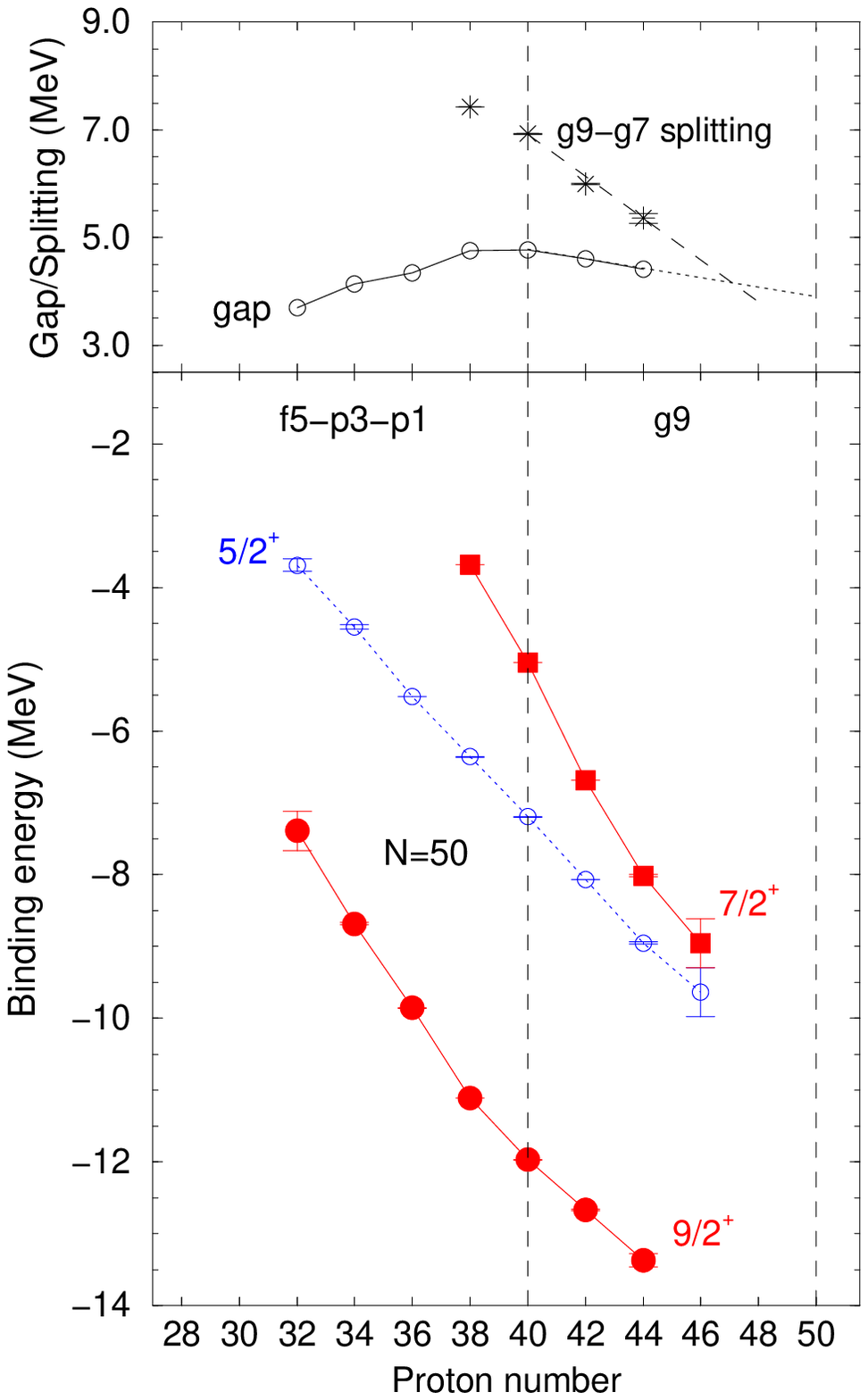,scale=0.5}
\end{center}
\end{minipage}\hfill
\begin{minipage}{8cm}
\epsfig{file=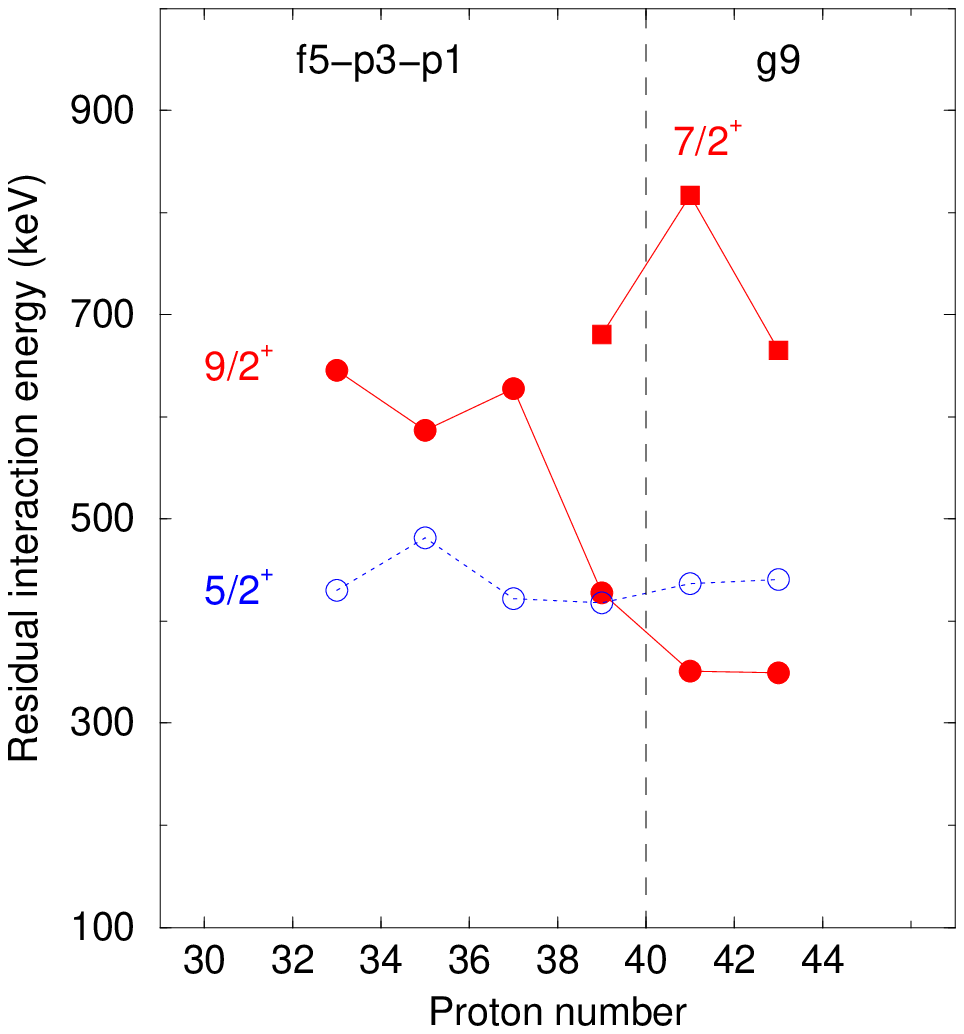,scale=0.5}
\end{minipage}
\begin{center}
\begin{minipage}[t]{16.5 cm}
\caption{{\bf Left}: Evolution of the binding energies of
the 9/2$^+$ and 7/2$^+$ states compared to those of the 5/2$^+$
states. The 9/2$^+$ and 5/2$^+$ states are the ground states of
the $N=49$ and $N=51$ isotones respectively, the 7/2$^+$ states
being those drawn in the right part of Fig.~\ref{crossingN51}.
Energy splitting between the 9/2$^+$ and 7/2$^+$ states (stars)
and the $N=50$ gap (empty circles).
{\bf Right}: Evolution of the proton-neutron residual
interactions extracted from the slopes of the binding energies
of the neutron states shown in the left part.
The proton orbitals which are getting filled as a function of
increasing $Z$ are given in the top of the figures.
\label{tenseurN50}}
\end{minipage}
\end{center}
\end{figure}

The item (i) can be ascribed to the attractive term of the
interaction between the $\pi g_{9/2}$ protons and the $\nu g_{7/2}$
neutrons, from the tensor term.
It is very strong in this particular configuration in which
protons and neutrons orbits with identical angular momentum and
opposite spin are involved, i.e. $ \pi g_\uparrow$ and $\nu
g_\downarrow$. Likewise the interaction between the $\pi g_{9/2}$
protons and the $\nu g_{9/2}$ neutrons, $V^{pn}_{g_\uparrow
g_\uparrow}$, displays a repulsive part, as already
mentioned in Sect.~\ref{evolgapN50}. These opposite behaviors
between the aligned $\nu g_{9/2}$ and anti-aligned $\nu g_{7/2}$
orbits during the filling of the $\pi g_{9/2}$ orbit can be seen in
the left part of Fig.~\ref{tenseurN50}.
The monopole interactions $V^{pn}$ involving the $\nu g_{9/2}$ and
$\nu g_{7/2}$ orbits can be tentatively derived from the evolution
of their binding energies as a function of the proton number (see
Sect.~\ref{expmonopole}). They are compared to the ones involving
the $\nu d_{5/2}$ orbit in the right part of Fig.~\ref{tenseurN50}.
From the constancy of the slope for the $\nu d_{5/2}$ orbit before
and after $Z=40$, one derives that the monopoles involving this
orbit have similar values of about 440~keV. For the $\nu g_{9/2}$
and $\nu g_{7/2}$ orbits the monopole matrix elements
$V^{pn}_{g9/2g9/2} \sim$ 350~keV and $V^{pn}_{g9/2g7/2} \sim$
750~keV can be extracted. The energy difference between the two
values is about 400~keV, which could be ascribed to the
tensor terms for these $\ell = 4$ orbits.

An extrapolation of the energy splitting between the 9/2$^+$ and
7/2$^+$ states can be made towards $Z=50$ to give a value of
3.0(3)~MeV for $Z=50$ as shown in Fig.~\ref{tenseurN50}. This value
is smaller than 3.9(5)~MeV, obtained from the extrapolated size of
the $\nu g_{9/2}$-$\nu d_{5/2}$ gap. This suggests that the $\nu
g_{7/2}$ orbit could cross or come close to the $\nu d_{5/2}$ one
around the $^{100}_{~50}$Sn nucleus. Experimentally, the 7/2$^+$
state remains about 170~keV above the 5/2$^+$ state at least up to
$Z=50$, as seen in Fig.~\ref{crossingN51} and surmised in
Ref.~\cite{Sewe07}. However both the size of
the shell gap and the SO splitting are reduced.\\

{\bf Theoretical results}\\
Shell-model calculations reproduce the decrease of the 7/2$^+_1$
energy of all the heavy isotones, except for the $^{99}_{48}$Cd
isotope~\cite{Dean04} in which the energy of the $\nu g_{7/2}$ orbit
has to be increased arbitrarily by 0.26 MeV in order to obtain a
good description of its high-spin states \cite{Lipo02a}. The same
values of single-particle energies and matrix elements
have been
then used to calculate the low-lying levels in $^{101}_{~50}$Sn,
unknown at that time. Its ground state has been predicted to be
5/2$^+$, with the 7/2$^+$ state lying $\sim$ 100~keV above. 
This is
in good agreement with the value of 172 keV energy newly measured in
$^{101}_{~50}$Sn~\cite{Sewe07}.

\begin{figure}[h]
\begin{center}
\epsfig{file=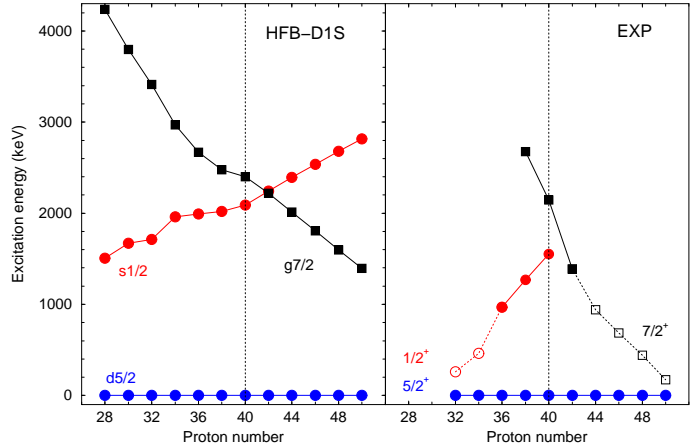,scale=0.5}
\begin{minipage}[t]{16.5 cm}
\caption{{\bf Left} Evolution of the $\nu s_{1/2}$ and $\nu g_{7/2}$
excitation energies in the N~=~51 isotones predicted within the
self-consistent approach using the D1S effective force of Gogny and
the blocking method (see text).
{\bf Right} Possible evolution of the 1/2$^+$ and 7/2$^+$ excitation
energies obtained from experimental
results (cf. the right part of Fig.~\ref{crossingN51})
\label{HFBD1S-N51}}
\end{minipage}
\end{center}
\end{figure}

We have carried out self-consistent calculations using the D1S Gogny
force and imposing the spherical symmetry, to obtain the first
excited states of the $N=51$ isotones. The blocking method was used,
imposing that the odd neutron is successively located in one
particular orbit, $d_{5/2}$, $s_{1/2}$, and $g_{7/2}$. The overall
change of excitation energy as a function of the proton number 
is\emph{ partly} accounted for as shown by comparing the left and 
right parts of Fig. \ref{HFBD1S-N51}.
The energy of the first excited state in the lightest
isotones (I$^\pi$=1/2$^+$), as well as in the heaviest ones
(I$^\pi$=7/2$^+$) are found at around 1.5 MeV. This is
significantly larger than the experimental values of 280(20)~keV
for $^{83}_{32}$Ge and 172~keV for $^{101}_{~50}$Sn. Moreover the
experimental slope of excitation energy and the reduction of the
$\pi g_{9/2} - g_{7/2}$ SO splitting are not reproduced by these
calculations. This leaves room for
additional effects of the nuclear interaction, such as tensor, 
to be added to the Gogny interaction. Nevertheless
one has to bear in mind that these experimental energies do not
directly correspond to those of the single-particle states $\nu
s_{1/2}$ and $\nu g_{7/2}$. Indeed the 1/2$^+$ and 7/2$^+$ states
states may comprise a certain mixing with $\nu d_{5/2} \otimes
2^+$ configuration. Therefore experimental pieces of information
has to be extended on the $N=51$ isotones to derive the\emph{ true}
single-particle energies at the two extremes of the valley of
stability. With this in hand, better comparisons between
experiment and theory could be made.

\subsubsection{\it Conclusion}
The $N=50$ shell gap is formed between the $g_{9/2}$ and the
$d_{5/2}$ or $g_{7/2}$ orbits. The evolution of this gap has been
studied over a wide range of masses thanks to the recent
experimental data obtained very far from stability. The largest
gap is found in nuclei close to the stability valley at $Z=40$. On
both side of this valley, towards $^{78}_{~28}$Ni and
$^{100}_{~50}$Sn, the gap decreases. These features have been
attributed to the specific proton-neutron interaction involved below
and above $Z=40$. 

Between $Z=28$ and $Z=38$ the attractive tensor
part of the $\pi f_{5/2} \nu g_{9/2}$ interaction could explain
the steep increase of the $\nu g_{9/2}$ binding energy as compared
to the $\nu d_{5/2}$ one during the filling of the $\pi f_{5/2}$
orbit. 

Above $Z=40$ the repulsive tensor 
component of the $\pi g_{9/2}- \nu g_{9/2}$ interaction reduces
the size of the $N=50$ shell gap. Simultaneously the strongly
attractive component of the $\pi g_{9/2}- \nu g_{7/2}$
interaction provokes a reduction of the $\nu g$ SO splitting and
the quasi-degeneracy of the $d_{5/2}$ and $g_{7/2}$ orbits at
$Z=50$. Altogether, significant reductions of the $N=50$ shell
evolution are found with hints of increasing core excitations.
Nevertheless, the present experimental results indicate that all
the $N=50$ isotones keep on behaving as spherical nuclei. Whether
this assertion would hold or not for $^{100}_{~50}$Sn and
$^{78}_{28}$Ni cannot be given today, as some experimental
information is still missing. The study of the two far edges of
the $N=50$ isotone chain offers a wonderful challenge for the
future. Thanks to the new worldwide facilities soon available,
they are close at hand.

\subsection{\it Evolution of the $Z=50$ shell closure\label{evolgapZ50}}
The $Z=50$ shell gap is formed between the proton $g_{9/2}$ and
$d_{5/2}$ or $g_{7/2}$ orbits. The $_{50}$Sn isotopic series
contains the two doubly-magic nuclei $^{100}$Sn$_{50}$ and
$^{132}$Sn$_{82}$. The synthesis of the self-conjugated $^{100}$Sn
isotope, located at the proton drip line, is a very difficult task
because of the very low cross sections. Thus no experimental data
on its excited states has been obtained up to now. On the other
hand, the $^{132}$Sn can be produced with a higher rate in the
asymmetric fission of actinides. Therefore properties of this
neutron-rich nucleus and its close neighbors are better
documented.

As for other shell gaps, possible variation of the proton 
single-particle energies could be caused by proton-neutron 
interactions. As the five neutron orbits involved between 
$N=50$ and $N=82$ are very close in energy, the
pairing correlation dilutes their respective occupancies and
smoothen possible changes of binding energies due to specific
proton-neutron interactions over several components.

The present section intend first to address the evolution of the
proton single-particle energies and collectivity around the Sn
isotopic chain. Afterwards the following paragraphs present
experimental data on the $_{51}$Sb isotopes obtained by transfer
reactions, followed by theoretical interpretations. As
experimental results disagree with respect to the single-particle
or mixed configuration of the proton states, the possibility to
extract the strength of the SO or tensor interactions directly
from experimental data is uncertain. This also sheds doubt on the
possibility to implement the tensor interaction there, as many
mean-field calculations tried without success.

\subsubsection{\it Binding energies\label{BEZ50}}

\begin{figure}[h!]
\begin{minipage}{9.cm}
\begin{center}
\epsfig{file=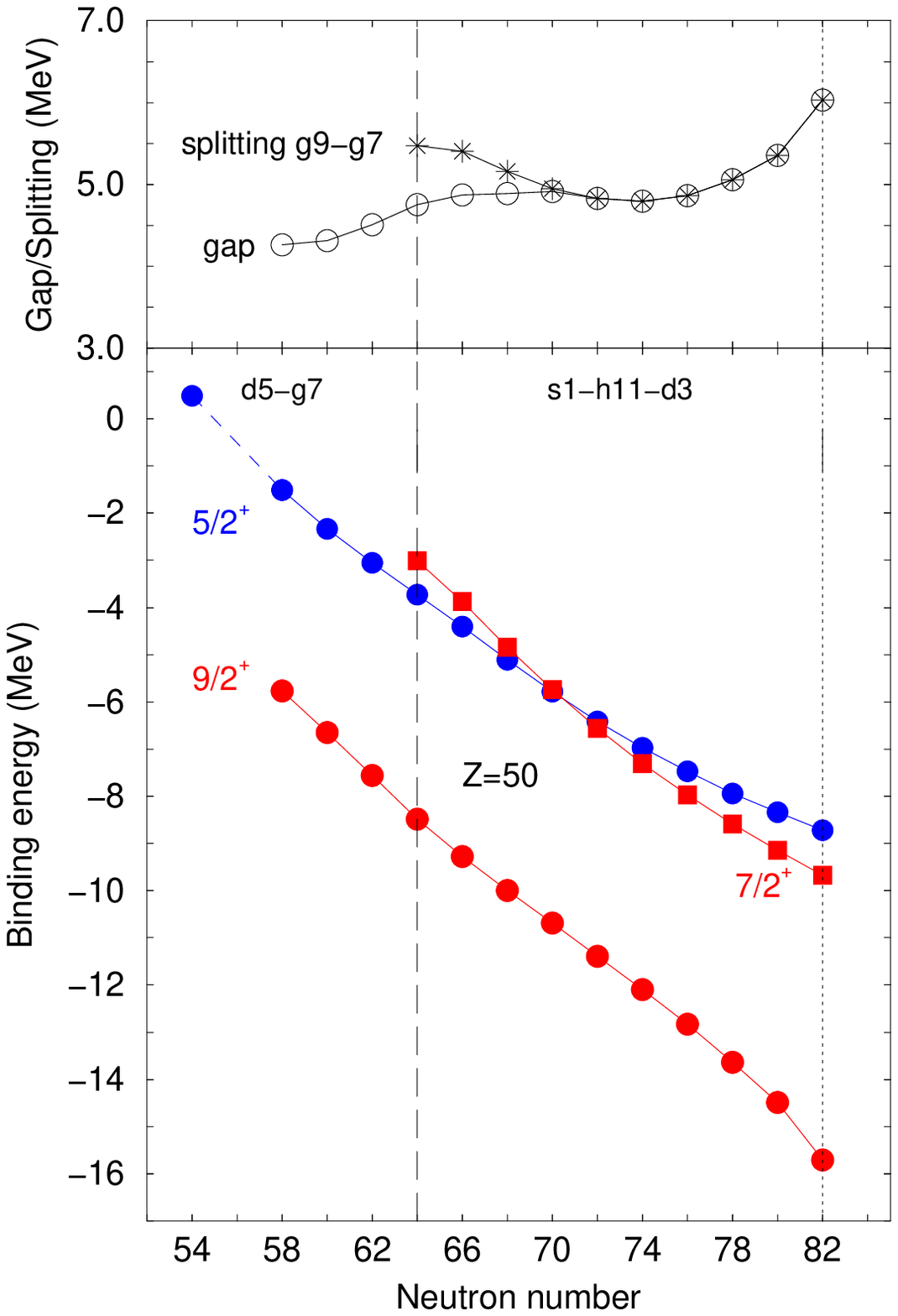,scale=0.5}
\end{center}
\end{minipage}\hfill
\begin{minipage}{9cm}
\epsfig{file=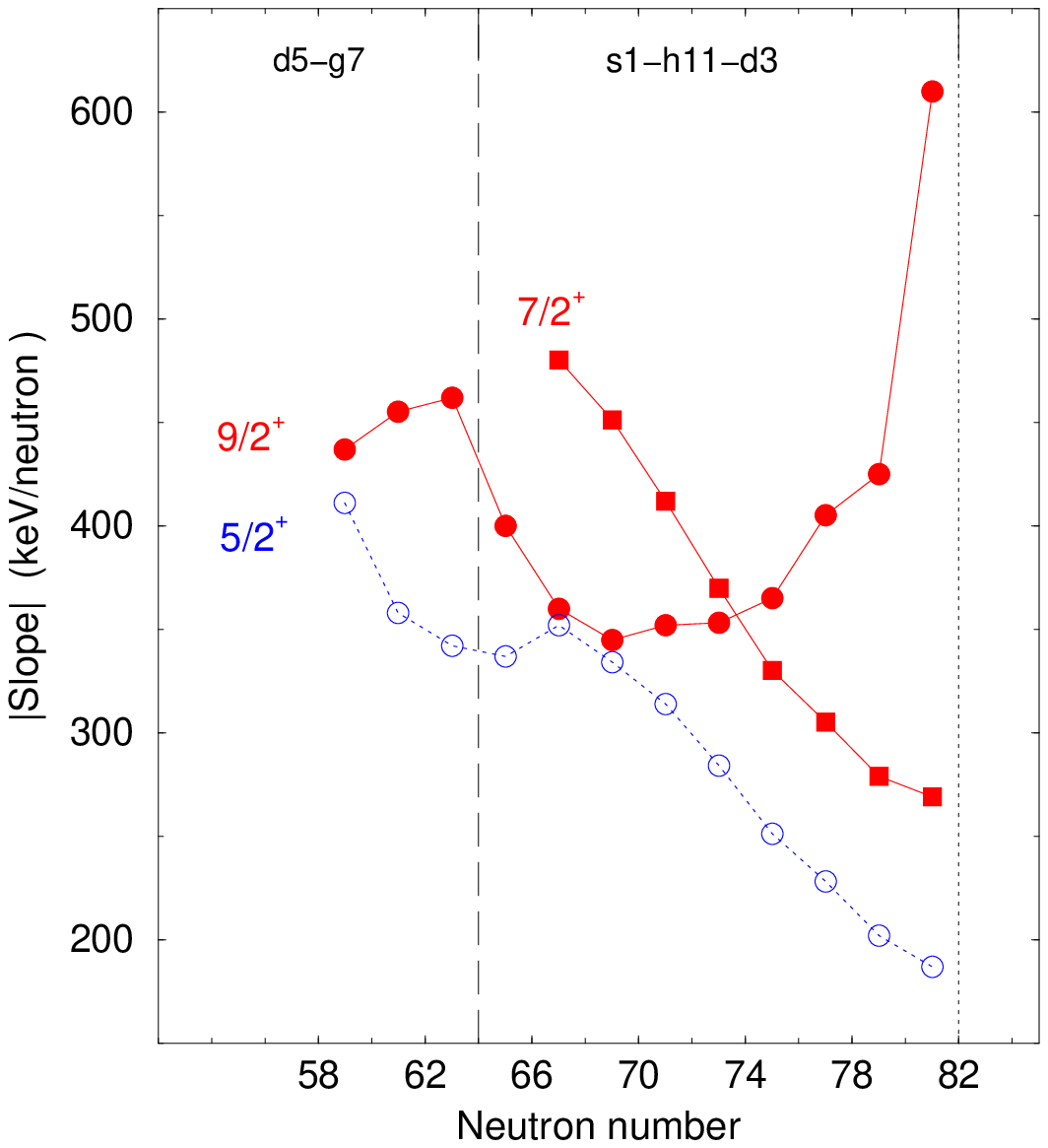, scale=0.5}
\end{minipage}
\begin{center}
\begin{minipage}[t]{16.5 cm}
\caption{{\bf Left}: Binding energies of the states located
just above and just below the $Z=50$ magic number and differences of the binding energies of the two states surrounding
the gap at $Z=50$ (empty circles) and of the binding energies of
the two $\ell=4$ spin-orbit partners (stars) (see
Sect.~\ref{annex}). The proton drip
line has been delineated at $^{105}$Sb$_{54}$ (being a proton
emitter, its proton binding energy is positive).
The neutron orbitals which are getting filled as a function of
increasing $N$ are given in the middle of the figure (see text).
{\bf Right}: Absolute values of the slopes of the experimental 
binding energies of the $9/2^+$,
$5/2^+$ and $7/2^+$ proton states bounding the $Z=50$ shell
closure, as a function of neutron number.
\label{gapZ50}}
\end{minipage}
\end{center}
\end{figure}

The evolution of the experimental binding energies of the proton
orbits located below and above the $Z = 50$ shell closure is shown
in the left part of Fig. \ref{gapZ50}. The slope of the binding
energy of the 9/2$^+$ state changes at the $N = 64$ subshell
closure : it is steeper when neutrons are filling the first
shells, $d_{5/2}$ and $g_{7/2}$, than when filling the next
$s_{1/2}$, $h_{11/2}$, and $d_{3/2}$ ones. This shell ordering is
reported in Fig. \ref{gapZ50}. It has been
established from the ground-state spin values of the odd-A Sn isotopes. As
mentioned earlier pairing correlations induce a simulataneous
filling of the neutron orbits located either below or above the
$N=64$ subshell. Their action is evidenced by the quasi-regular
variation of the binding energy curves before and after $N=64$.
Without pairing correlations, a sequence of various slopes of
binding energies would have been found, corresponding to the
progressive filling of individual orbits.

Above the $Z=50$ gap, the angular momentum of the\emph{ ground
state} of the odd-A $_{51}$Sb isotopes changes from I$^\pi$ =
5/2$^+$ to I$^\pi$ = 7/2$^+$ between $^{121}$Sb and $^{123}$Sb,
the two stable isotopes of antimony. Consequently the $Z=50$ gap
becomes progressively formed between the 9/2$^+$ and 7/2$^+$
states, which likely arise from the proton $g_{9/2}$ and $g_{7/2}$
orbits, respectively. If the 9/2$^+$ and 7/2$^+$ levels are pure
single-particle states, their energy spacing will reflect the
change of the spin-orbit splitting for the $\ell=4$ orbits. 
The left part of Fig.~\ref{gapZ50} shows both the evolution of the
$9/2^+-7/2^+$ splitting and that of the $Z=50$ gap. The latter 
evolves from 4.3~MeV ($N=58$) to 6~MeV ($N=82$).
Below $N=64$, the gap is reduced as the slope of the $\pi g_{9/2}$
binding energy is steeper than that of the $\pi d_{5/2}$ one 
(see the right part of Fig.~\ref{gapZ50}). This likely arises 
from the fact that the $V^{pn}_{d_{5/2}g_{7/2}}$ matrix element 
is weaker than $V^{pn}_{g_{9/2}g_{7/2}}$ one, as in the former, 
neither the orbital momenta nor the
number of nodes of the proton and neutron wave functions are
equal, contrary to the latter.

As done in the previous sections, we have calculated the slopes of the 
experimental binding energies of the states bounding the gap 
(see the right part of Fig.~\ref{gapZ50}). The results are very
different from the other cases, particularly above $N=64$ where 
the values changes continuously up to a factor $\simeq 2$. 
This hints for the presence of correlations beyond the pure 
monopole-induced variations. The relation~(\ref{shiftE}) 
(see Sect.~\ref{monopoleffect}) can no longer apply in this case,
the effects of the different monopole interactions involved 
there are obscured by changes of the cores. 

\subsubsection{\it Trends of first collective excitations of Sn isotopes}

\begin{figure}[h!]
\begin{minipage}{9.cm}
\begin{center}
\epsfig{file=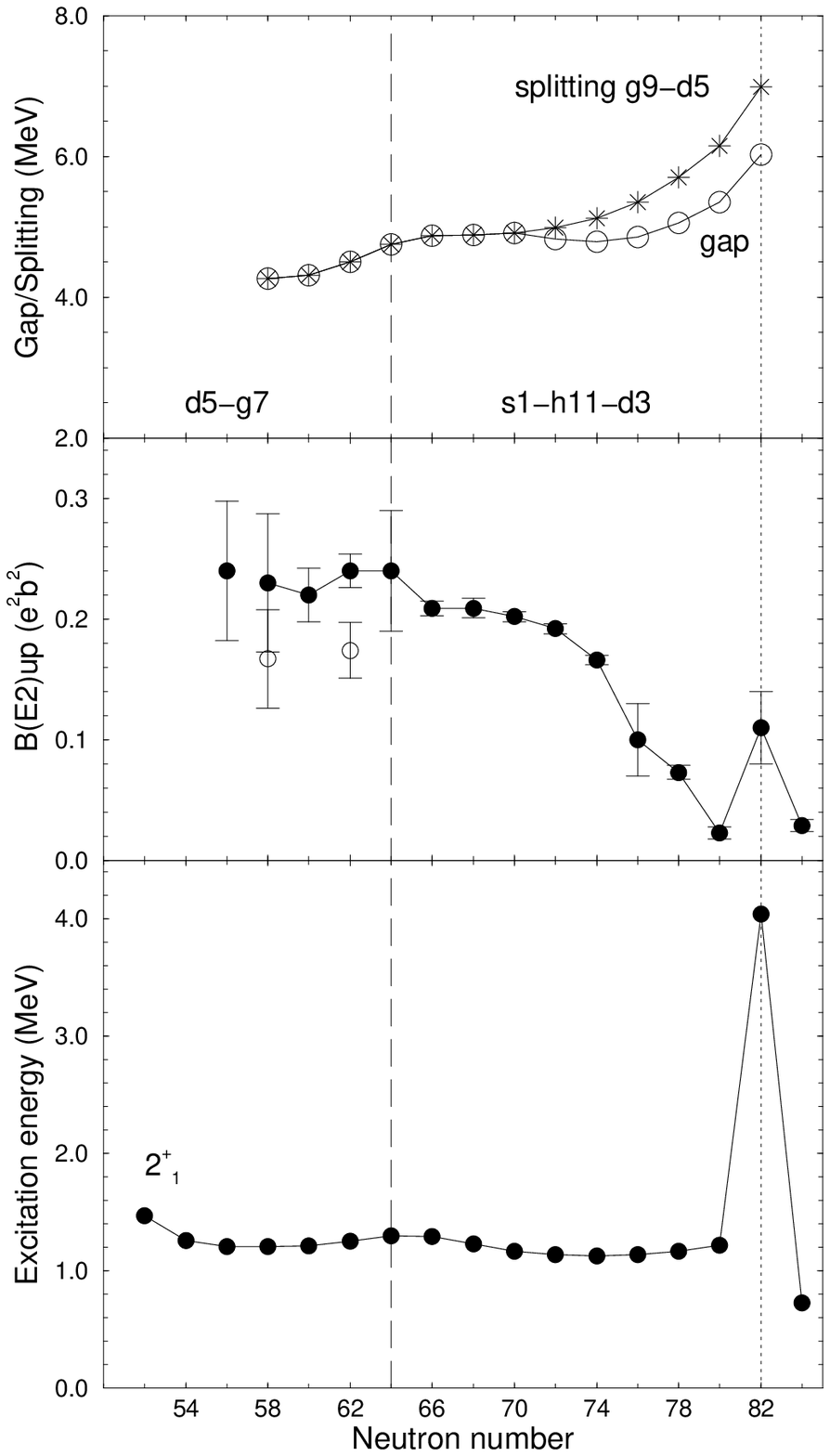,scale=0.5}
\end{center}
\end{minipage}\hfill
\begin{minipage}{9.cm}
\epsfig{file=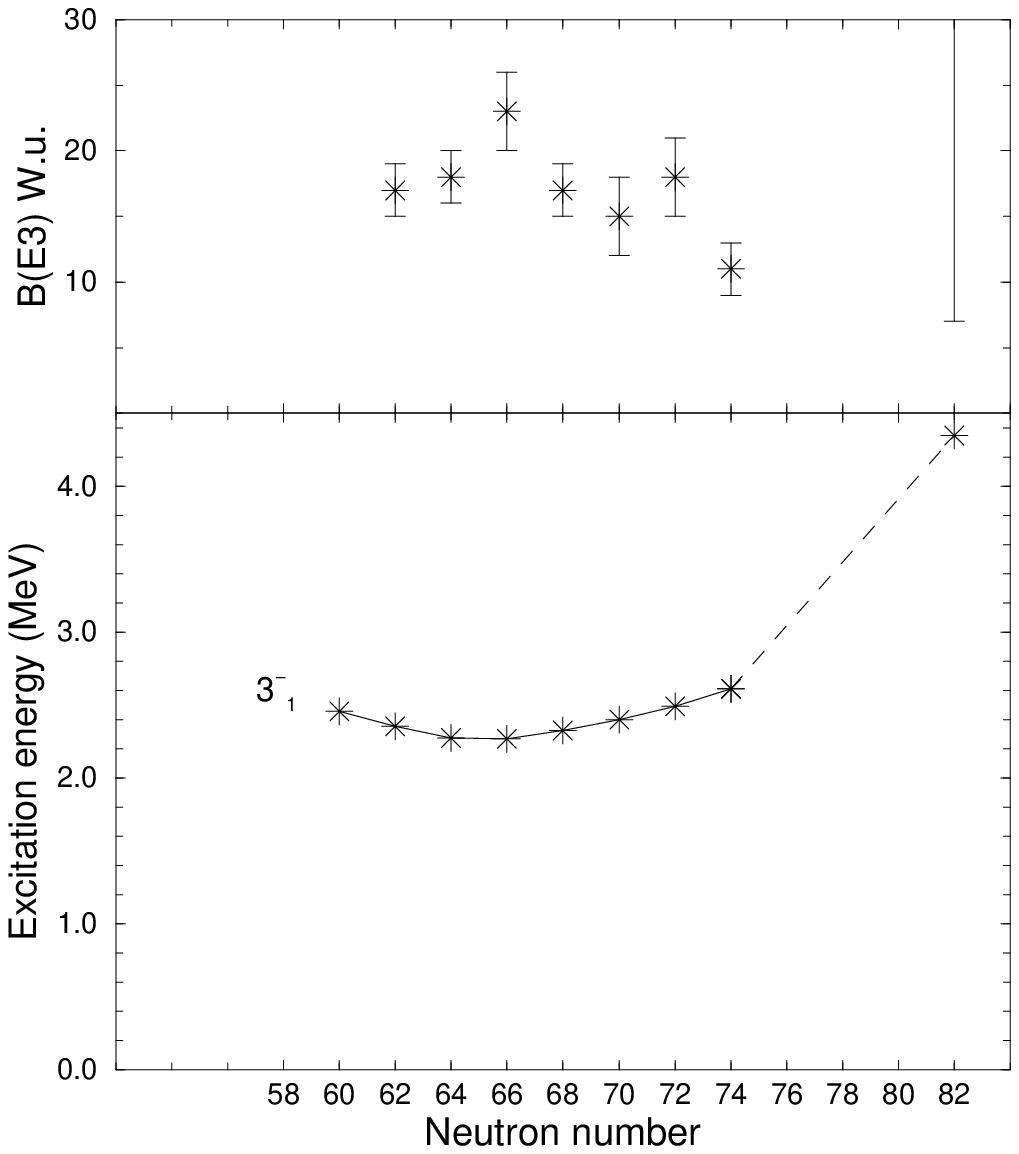,scale=0.5}
\end{minipage}
\begin{center}
\begin{minipage}[t]{16.5 cm}
\caption{{\bf Left} Experimental 2$^+_1$ energies (bottom) and
B(E2; 0$^+$ $\rightarrow$ 2$^+_1$) values (middle) in the Sn
isotopic chain. Differences of the
binding energies of the two states surrounding the gap at $Z=50$
and of the binding energies of the $\Delta \ell=2$ orbits (top).
The neutron orbitals which are getting filled as a function of
increasing $N$ are given in the middle of the figure (see text).
{\bf Right} Experimental 3$^-_1$ energies (bottom) and B(E3;
3$^-_1$ $\rightarrow$ 0$^+$) values (top) in the Sn isotopic
chain. \label{E2etE3_Sn}}
\end{minipage}
\end{center}
\end{figure}

As the $Z=50$ gap is almost constant and large over a wide mass
range, proton excitations across it require large excitation
energy. Consequently, apart from $^{132}$Sn, the first excited
$2^+$ levels of Sn at $\simeq$1.2~MeV (see Fig.~\ref{E2etE3_Sn})
are likely almost pure neutron states. The small rise and fall of
the state around $N=64$ witnesses the presence of a weak subshell
closure. Apart from this effect the almost constancy of the $2^+$
energies at each side of the $N=64$ subshell closure confirms that
pairing correlations acts to dilute shell occupancies and generate
$2^+$ states of similar configurations.

On the other hand, the drastic increase of the $2^+$ state for the
$^{132}_{~50}$Sn$_{82}$ nucleus arises from the fact that both
neutron and proton excitations across the $Z=50$ and $N=82$ shell
gaps require large energies. Important is to note also that the
$2^+$ state may not be of pure neutron origin anymore, but can
contain proton excitations as well.

Experimental B(E2)$\uparrow$ values shown in the left part of
Fig.~\ref{E2etE3_Sn} are taken from Ref.~\cite{Rama01} for the
stable Sn isotopes, whereas the filled data points for unstable
nuclei are obtained from Coulomb excitation using radioactive ion
beams~\cite{Vama07,Banu05,Cede07,Orce07,Radf05}. For $N > 64$ the B(E2)
value follows the parabolic trend of the generalized seniority
scheme described in Sect.~\ref{enerprobaNi}, with the sole exception for
the $N=82$ nucleus. There, as said earlier, the wave function of
the first $2^+$ state contains more proton excitations, which
enhances the B(E2) in a significant manner.

On the proton-rich side ($N < 64$), the B(E2) values exhibit a
very different pattern. Those drawn with filled symbols remain
almost constant, indicating an increase of collectivity towards
$^{100}_{~50}$Sn$_{50}$. This behavior goes in concert with the
slight decrease of the shell gap observed for $N < 64$ and shown
in the top left part of Fig.~\ref{E2etE3_Sn}. At first
consideration this corroborates an increase of collectivity due to
easier proton core excitations. However, the B(E2; 2$^+_1$
$\rightarrow$ 0$^+$) value of $^{112}$Sn has been
remeasured recently by using the the Doppler-shift
attenuation method~\cite{Orce07}. The new strength (drawn with empty symbol) is
significantly weaker than the one previously adopted. As the
former B(E2) value in $^{112}$Sn was used as normalization to
derive the B(E2) value in $^{108}$Sn~\cite{Banu05}, the value of
$^{108}$Sn should be downscaled accordingly. With these new values
for $^{108,112}$Sn shown with empty symbols, the B(E2) values in
the Sn isotopes display a more symmetric pattern at each side of
$N = 64$ mid-shell. Further experimental work has to be made to
reduce uncertainties and agree on mean B(E2) values.

On the theoretical point of view, the behaviour of the B(E2)
values along the isotopic chain of Sn has been analyzed up to
$^{130}$Sn in the framework of large-scale shell model
calculations\cite{Banu05}. Calculations well reproduce
experimental data on the neutron-rich side, whereas they agree or
are significantly underestimated below $N=64$ when comparing to
the empty or filled symbols in Fig.~\ref{E2etE3_Sn}, respectively.
As said earlier, the experimental situation has to be clarified
below $N=64$. On the shell model point of view, the evolution of
the $Z=50$ is an essential ingredient to consider to determine
B(E2) values. In the work of Ref.~\cite{Banu05} this gap is
increased below $N=64$ (towards $^{100}$Sn), whereas the
experimental trend shown in the top part of Fig.~\ref{E2etE3_Sn}
rather indicates a decrease.

The local increase of B(E2) at $^{132}$Sn was predicted in
Ref.~\cite{Tera02} using the quasiparticle random phase
approximation (QRPA) method. The wave function of the 2$^+_1$
states of tin isotopes is dominated by neutron excitations.
However, at $^{132}_{~50}$Sn$_{82}$,\emph{ both} proton and neutron
low-energy excitations are hindered due to the presence of shell
gaps.  Therefore the energy of the $2^+_1$ state suddenly
increases and its wave function have mixed components,\emph{ both}
from neutron and proton, the latter leading to the local increase
of the B(E2) value. After having passed the $N=82$ shell closure,
the neutron excitations take over again, a small B(E2) value being
measured in the $^{134}_{~50}$Sn$_{84}$ nucleus~\cite{Radf05}.

A 3$^-_1$ state is observed around 2.5~MeV excitation energy
throughout the Sn isotopic chain, as shown in the right part of
Fig.~\ref{E2etE3_Sn}. The large B(E3) values values of $\simeq
10-20~W.u.$ indicate that these states are collective. Their
configuration can be written in terms of neutron excitations
involving the negative parity intruder $h_{11/2}$ orbit and the
positive-parity ones $d_{5/2}$ and $g_{7/2}$.

In summary, the proximity of several neutron orbits with $\Delta
\ell=2$ and $\Delta \ell=3$ give rise to E2 and E3 collective
excitations in Sn isotopes. As these modes occur at relatively low
excitation energy the single-proton states above the $Z=50$ shell
closure could easily couple to them. Therefore they should be
considered carefully when determining the proton SPE in the
$_{51}$Sb isotopes.

\subsubsection{\it Proton orbits above the $Z=50$ gap: Levels of $_{51}$Sb
isotopes \label{levelSb}}

The present chapter deals with the evolution of the proton orbit
above the $Z=50$ shell gap as a function of increasing neutron
number. By virtue of the closed $Z=50$ core, the low-energy states
of the even-N $_{51}$Sb isotopes could be interpreted in terms of
single-proton states and their coupling to the $2^+_1$ and $3^-_1$
excitations of the Sn cores. In the following, we discuss on the
identification of the single-proton states located above the
$Z=50$ gap. The evolution of the proton $d_{5/2}$, $g_{7/2}$ and
$h_{11/2}$ orbits with respect to proton-neutron interactions is
discussed as well.\\

\begin{figure}[h!]
\begin{minipage}{9.cm}
\epsfig{file=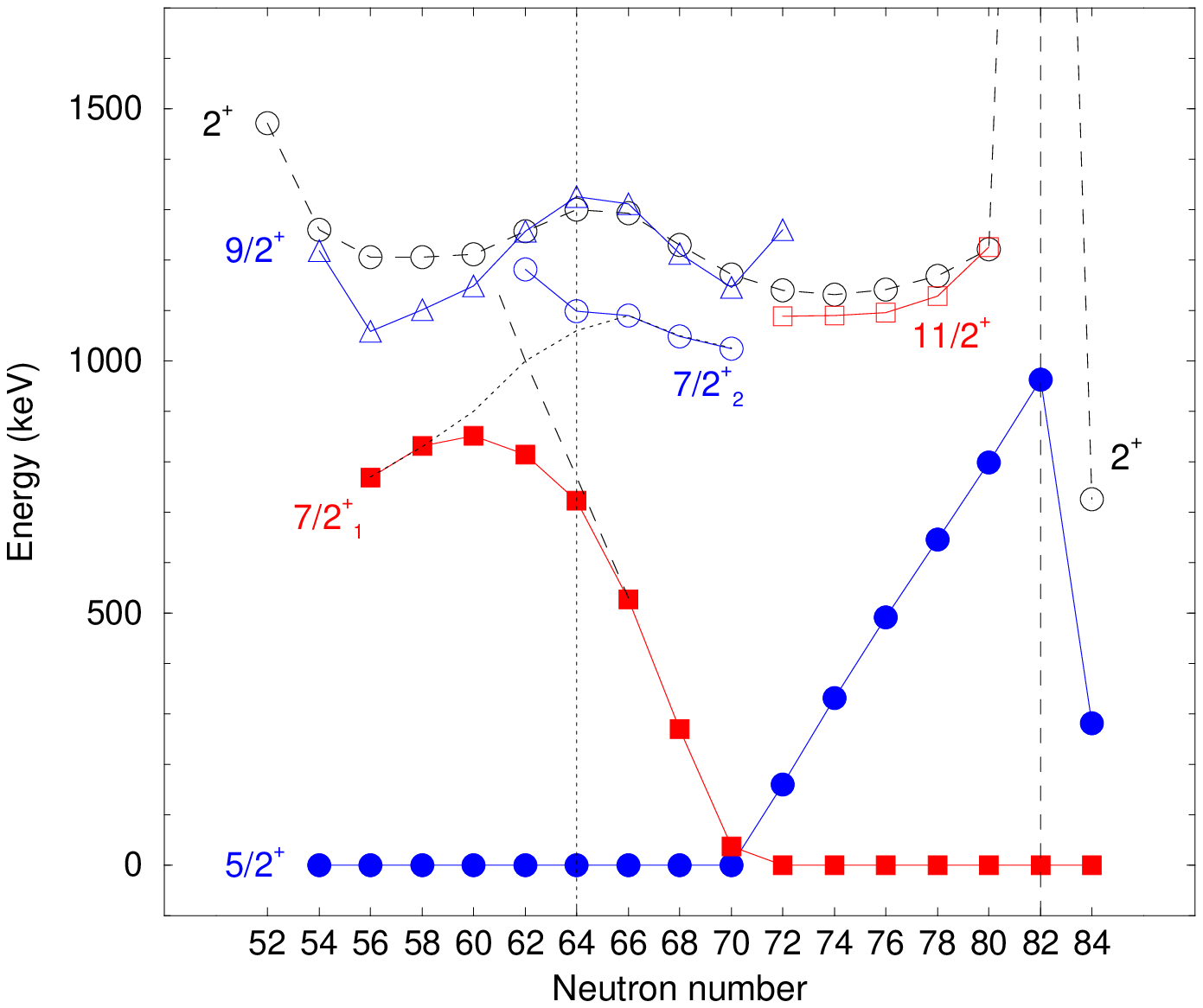,scale=0.55}
\end{minipage}\hfill
\begin{minipage}{9.cm}
\epsfig{file=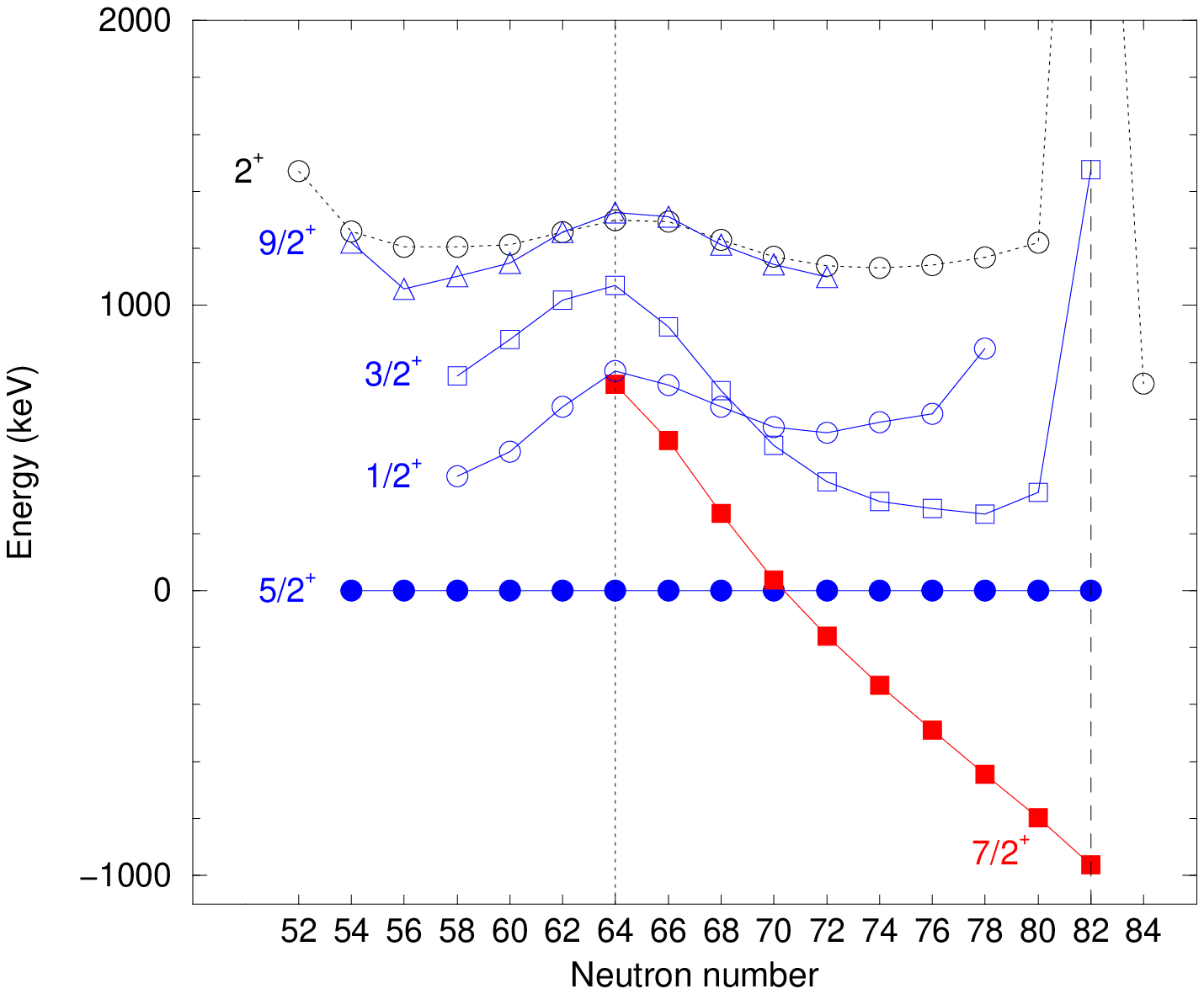,scale=0.55}
\end{minipage}
\begin{center}
\begin{minipage}[t]{16.5 cm}
\caption{{\bf Left}: Evolution of the first excited states with
$5/2^+ \leq I^\pi \leq 11/2^+$ in the odd-A
$_{51}$Sb isotopes as a function of the neutron number.
{\bf Right}: Evolution of the excitation energy of the 1/2$^+_1$,
3/2$^+_1$, 7/2$^+_1$ and 9/2$^+_1$ states relative to the 5/2$^+_1$ 
state in $^{105-133}$Sb$_{54-82}$.
The energy of
the 2$^+$ state of the Sn cores is also drawn (the 2$^+$ energy of
$^{132}$Sn is not shown, being well above the maximum scale).
\label{crossingSb}}
\end{minipage}
\end{center}
\end{figure}

{\bf Evolution of the $\pi d_{5/2}$ and $\pi g_{7/2}$ orbits}\\
The evolution of the first excited states with 
$5/2^+ \leq I^\pi \leq 11/2^+$ 
as well as the $2^+$ excitation of
the core nuclei are shown in the left part of
Fig.~\ref{crossingSb} for an increasing neutron number. The spin
value of the ground state of the odd-A Sb isotopes changes from
I$^\pi$ = 5/2$^+$  to I$^\pi$ = 7/2$^+$, between the two stable
isotopes of antimony, $^{121}$Sb$_{70}$ and $^{123}$Sb$_{72}$.
Taken at face value this indicates a crossing of the $\pi d_{5/2}$
and $\pi g_{7/2}$ orbits, if these states will be of
single-particle origin.

The coupling of the $2^+$ core excitation to these single-proton
states $d_{5/2}$ and $g_{7/2}$ also gives rise to 7/2$^+$ and
5/2$^+$ states. Therefore one needs to examin which levels
originate from (i) a core coupling, (ii) a mixing between these
two configurations or (iii) a single-particle state.

The most straighforward assignment for case (i) could be obtained
for the 9/2$^+$ and 11/2$^+$ states which have the maximum spin
values reached by the $2^+ \otimes d_{5/2}$ and $2^+ \otimes
g_{7/2}$ couplings, respectively. As supected the 9/2$^+$ and
11/2$^+$ states nicely follow the energy of the 2$^+$ excitation
of the Sn core~\cite{Lipo02b,Porq05a} in the light and heavy Sb
isotopes, respectively (see left part of Fig.~\ref{crossingSb}).

The behavior of the two 7/2$^+$ states between $N=62$ and $N=70$
is a good illustration for case (ii). Where a crossing of the
7/2$^+_1$ and 7/2$^+_2$ levels would have been expected at about
1~MeV, as indicated by the dashed and dotted lines in the left
part of Fig.~\ref{crossingSb}, they repel each others. The shift
in energy from the crossing point is similar for both state,
$\Delta$E = $\pm$~0.18~MeV, in agreement with a two-state mixing
process~\cite{Porq05a}. In the lightest isotopes the 7/2$^+_1$
state is a member of the multiplet from the $\pi d_{5/2} \otimes
2^+$ configuration, while in $^{111-115}$Sb$_{60-64}$, it has a
mixed wave function $\pi g_{7/2}$ and $\pi d_{5/2} \otimes 2^+$.

Above $N=66$ the 7/2$^+_1$ energy decreases quasi-linearly with
respect to the 5/2$^+$ one, whereas the 7/2$^+_2$ energy follows
that of the 2$^+$ of the Sn cores (therefore it corresponds mainly 
to a $\pi d_{5/2} \otimes 2^+$ configuration). These features 
indicate that
the amplitude of mixing is reduced above $N=66$. The 7/2$^+_1$
state becomes rather a single-particle state from the $g_{7/2}$
orbit (case (iii)). This is confirmed by the $^A$Sn($^3$He,d)
transfer reactions, in which the 7/2$^+_1$ state feeding amounts
to about 75\% of the $\pi g_{7/2}$ strength.
As already presented in Sect.~\ref{BEZ50}, the binding
energy curves of the 7/2$^+_1$ and 5/2$^+_1$ states are not linear
when increasing the neutron number (see Fig.~\ref{gapZ50}), which 
witnesses that some
correlations beyond the monopole trend are still present. As their
curvatures are similar, the origin of correlations is similar as
well.

Above the neutron magic number $N=82$, the energy of the first
5/2$^+$ excited state drops suddenly from 963 keV in $^{133}$Sb to
282 keV in $^{135}$Sb \cite{Korg01,Sher02,Sher05a}, as shown in
the left part of Fig. \ref{crossingSb}. This reverse behavior
between the 5/2$^+$ and 7/2$^+$ states triggered new studies. In
particular, using the fast time-delayed $\gamma \gamma$
coincidence technique, the half-life of the first excited state of
$^{135}$Sb at 282 keV has been measured to be T$_{1/2}$ = 6.0(7)
ns \cite{Korg05}. Assuming a pure M1 transition, this lifetime
implies a B(M1) of 2.9 10$^{-4}$ $\mu_N^2$, i.e. 1.6 10$^{-4}$ Wu.
This extremely small value was interpreted \cite{Korg05} to be due
to a transition between the two single-proton states, $\pi d_{5/2}
\rightarrow \pi g_{7/2}$. Such an M1 decay is hindered because of
the difference of angular momentum (2$\hbar$) between the $d$ and
$g$ orbits.

The sudden change between the energy of the $7/2^+_1$ and
$5/2^+_1$ states above $N=82$ has been recently studied in the
framework of the Shell-Model approach \cite{Cora05}.
Quantitatively, the two-body matrix elements involving the $\pi
g_{7/2}-\nu f_{7/2}$ and $\pi d_{5/2}-\nu f_{7/2}$ configurations
have been derived from the CD-Bonn nucleon-nucleon potential. The
resulting energy of the 5/2$^+_1$ state of $^{135}$Sb matches well
the experimental value. Moreover these new matrix elements give
rise to good results in the odd-odd $^{134}$Sb$_{83}$
nucleus~\cite{Cora06} and
in the even-even neighbors of $^{132}$Sn.

The trends of the 5/2$^+$ - 7/2$^+$ relative energy before and
after $N=82$ can be\emph{ qualitatively} interpreted in terms of
proton-neutron residual interaction and the overlap of the radial
wave functions. To a first order, it scales with the number of
nodes of the wave functions involved. The nodeless proton $\pi
g_{7/2}$ and neutron $h_{11/2}$ and $g_{7/2}$ wave functions have
large overlap. At the opposite the $\pi d_{5/2}$ wave function
has one node, making its overlap with the neutron $h_{11/2}$
and $g_{7/2}$ orbits weaker. Consequently, the absolute value of
$V^{pn}_{d_{5/2}h_{11/2}}$ is weaker than this of
$V^{pn}_{g_{7/2}h_{11/2}}$, leading to the steep decrease of the
7/2$^+$ state and the crossing with the 5/2$^+$ ones. Above $N=82$
the situation does change. As the first neutron orbit which comes
into play $f_{7/2}$ has one node, the arguments with respect to
the $\pi g_{7/2}$ and $\pi d_{5/2}$ orbits are reversed and the
5/2$^+$ - 7/2$^+$ spacing is brutally reduced.\\

The crossing of the $\pi d_{5/2}$ and $\pi g_{7/2}$ sub-shells at $N=70$
is well reproduced by the self-consistent calculations using the
D1S Gogny force \cite{Porq05b}.
Since the detailed spectroscopic study of all these antimony
nuclei (see the discussion in Ref.~\cite{Porq05a}) indicates that
the 5/2$^+_1$ and 7/2$^+_1$ states are associated to a spherical
shape, all the calculations have been performed imposing the
spherical symmetry.
\begin{figure}[h!]
\begin{center}
\epsfig{file=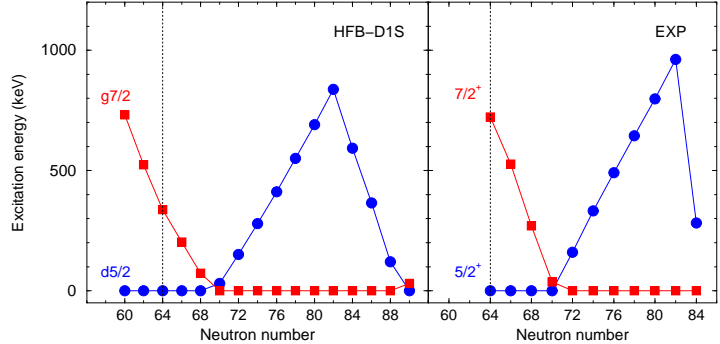,scale=0.5}
\begin{minipage}[t]{16.5 cm}
\caption{{\bf Left}: Evolution of the energies of the $\pi d_{5/2}$ and
$\pi g_{7/2}$ states of the odd-A Sb isotopes from N~=~60 to 90,
predicted in the HFB self-consistent calculations using the D1S
effective force of Gogny and the method of blocking, the spherical
symmetry being imposed (results published in ref. \cite{Porq05b}
for $N \le 82$). {\bf Right}: Evolution of the experimental
energies of the 5/2$^+_1$ and 7/2$^+_1$ states of the odd-A Sb
isotopes (part of Fig. \ref{crossingSb}). \label{SbHFBD1S}}
\end{minipage}
\end{center}
\end{figure}

The comparison between the theoretical results drawn in the left
part of Fig.~\ref{SbHFBD1S} and the experimental ones shown in the
right part indicates that (i) the crossing of the 5/2$^+$ and
7/2$^+$ states is found at the right neutron number, (ii) the 
decrease of the 5/2$^+$ state is found beyond the $N = 82$ shell 
closure. Therefore most of the underlying physics is taken into
account by the self-consistency of such a calculation, i.e. the
continuous change of the cores evidenced by the non-linear
variations of the binding energies (see 
Fig.~\ref{gapZ50}). However, as for the results obtained in the $N=51$
isotones (see Fig.~\ref{HFBD1S-N51}), there is room for additional effects
of the nuclear interaction, as the predicted variations are not as
large as the observed ones.\\

{\bf The $\pi s_{1/2}$ and $\pi d_{3/2}$ orbits}\\
The  $\pi s_{1/2}$ and $\pi d_{3/2}$ orbits are located above the
$Z=50$ gap. They should lead to 1/2$^+$ and 3/2$^+$
single-particle states in the Sb isotopes. However their exact
location is not trivial, as the $\pi d_{5/2} \otimes 2^+$ coupling
also gives rise to 1/2$^+$ and 3/2$^+$ states. From several
stripping or pick-up reactions, numerous states with I$^\pi$ =
1/2$^+$ and 3/2$^+$ (populated with $\ell$~=~0 and 2 respectively)
have been identified in $^{113-129}$Sb$_{62-78}$. As their 
spectroscopic factors are weak their wave functions are likely
mixed. By looking at the right part of Fig.~\ref{crossingSb}, it
is seen that the 1/2$^+_1$ and 3/2$^+_1$ states follow the trend
of the 2$^+$ of the Sn core, but in a different manner before and
after $N=64$. This can be ascribed to a change of the
configuration of the 2$^+$ of the Sn core. Even though its\emph{ energy} 
does not vary significantly, its\emph{ composition}
changes as a function of the neutron number. Below $N=64$ the $\nu
d_{5/2}$ and $\nu g_{7/2}$ orbits are strongly involved to form a
2$^+$. Above $N=64$ it is the turn of the $\nu h_{11/2}$ orbit.
Therefore the behaviors of the 1/2$^+_1$ and 3/2$^+_1$ states of 
Sb isotopes are driven by\emph{ different combinations} of 
residual proton-neutron interactions.\\

\begin{figure}[h!]
\begin{minipage}[t]{9cm}
\begin{center}
\epsfig{file=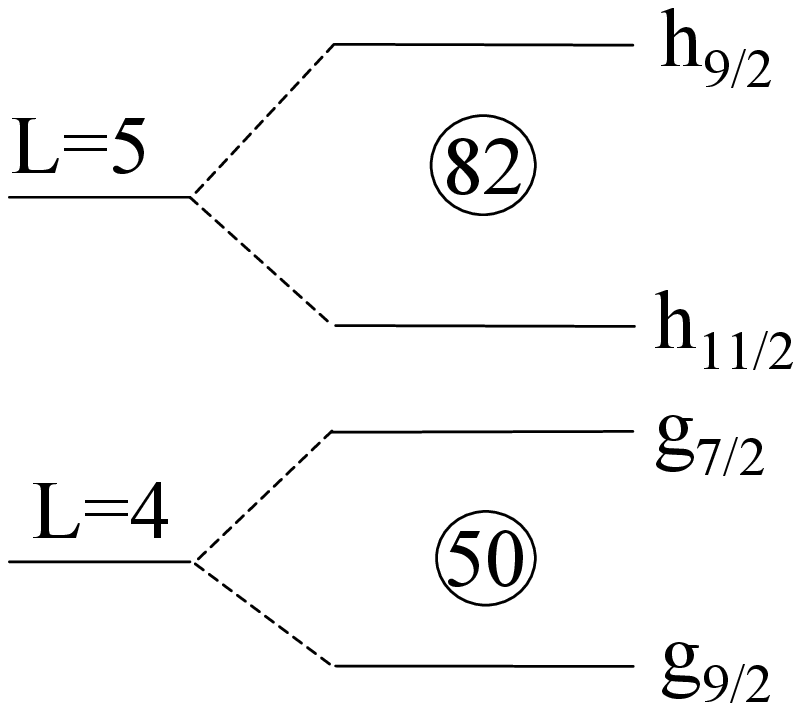, scale=0.7}
\end{center}
\end{minipage}
\begin{minipage}[t]{9cm}
\begin{center}
\epsfig{file=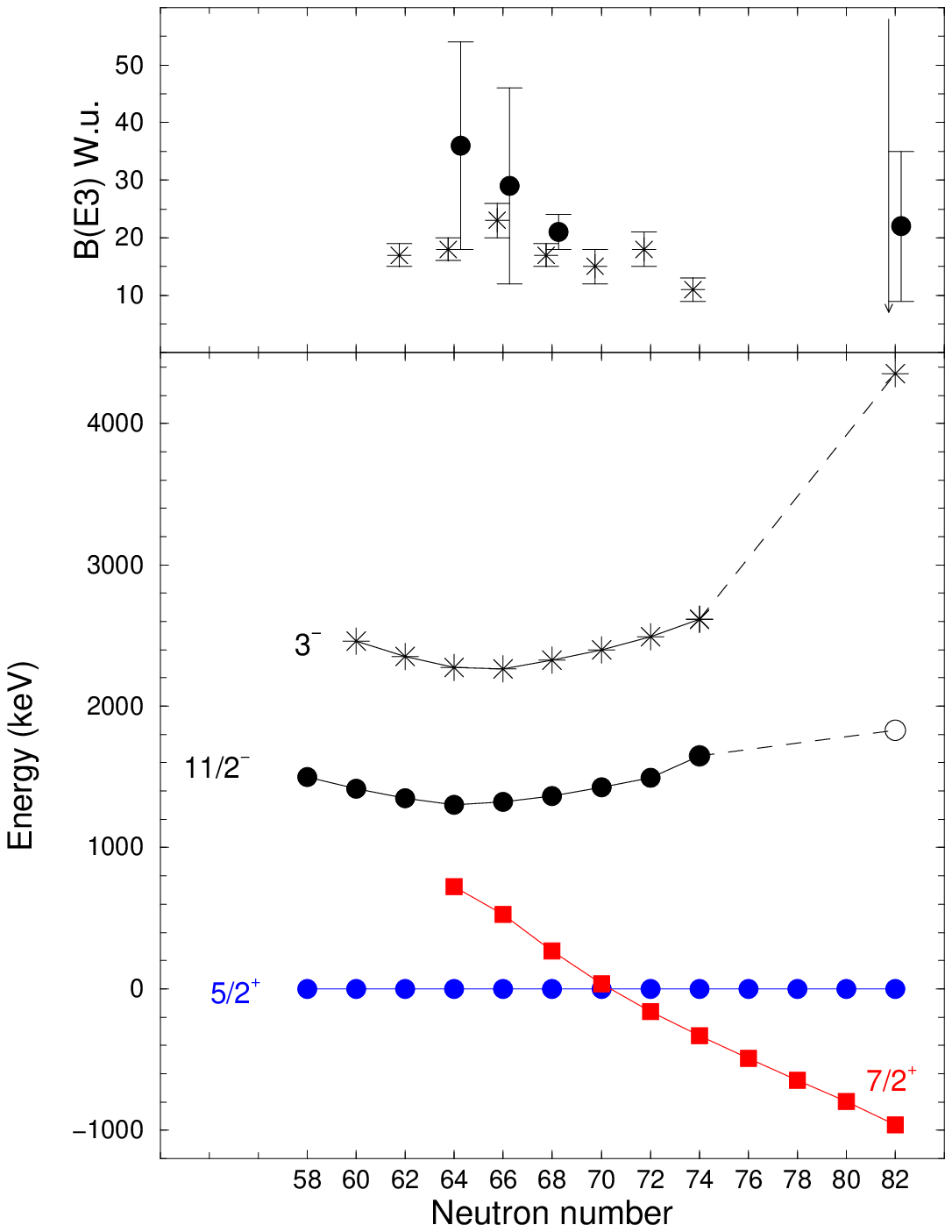,scale=0.6}
\end{center}
\end{minipage}
\begin{center}
\begin{minipage}[t]{16.5 cm}
\caption{ \textbf{Left} : Schematic view of the effect of the SO
interaction on the proton $g$ and $h$ splittings. Any reduction of
this interaction enhances the $g_{7/2} - h_{11/2}$ spacing.
\textbf{Right}, {\bf Bottom}: Evolution of the excitation energy
of the 11/2$^-_1$ state relative to the 5/2$^+_1$ state in the
$^{109-125}$Sb, and of the 3$^-$ octupole state of the
corresponding $^{A-1}$Sn cores. {\bf Top}: Probability of reduced
transition B(E3), in Weisskopf units, for the 11/2$^-$
$\rightarrow$ 5/2$^+$ transition  in the odd-A Sb isotopes
(circles) and the 3$^-$ $\rightarrow$ 0$^+$ transition in the
even-even Sn cores (stars). \label{octu_SbSn}}
\end{minipage}
\end{center}
\end{figure}

{\bf The $\pi h_{11/2}$ orbit}\\
The $h_{11/2}$ orbit, with an angular momentum $\ell=5$, arises
from the Harmonic Oscillator shell $N=5$. The strength of the
spin-orbit interaction is large enough to lower this $j= \ell
+1/2$ state into the preceeding $N = 4$ major shell in which the
$g_{7/2}$ orbit ($\ell$=4 and $j= \ell -1/2$) is located. Any
variation of the SO interaction for the $\ell$=5 or/and $\ell$=4
orbits would be magnified when looking at the energy difference
between the states assigned to these proton $h_{11/2}$ and
$g_{7/2}$ orbits, as shown in the left part of
Fig.~\ref{octu_SbSn}. This method was proposed recently in
Ref.~\cite{Schi04} to derive the variation of the SO interaction
as a function of the neutron number, using experimental results
obtained in the Sb isotopic chain.
The discovery of the 11/2$^-$ states in the
$^{113-125}$Sb$_{62-74}$ isotopes was made in the seventies using
the one-nucleon transfer ($^3$He,$d$) reaction on
isotopically-enriched even Sn targets \cite{Conj68}. An $\ell$~=~5
state, with a spectroscopic factor of about 0.5, was identified
around 1.4 MeV above the 5/2$^+$ state for all isotopes. Later on,
a more complete spectroscopic information has been obtained for
$^{117}$Sb$_{66}$ using the $^{116}$Sn($^4$He,$t$) reaction at 80
MeV~\cite{Gale85}. The spectroscopic factors of the first excited
states are in agreement with the ones of Ref.~\cite{Conj68}, 0.64
for the 7/2$^+_1$ state and 0.42 for the 11/2$^-_1$ state.
Moreover several states with $\ell=4,5$ were identified between
1.5 and 5.0~MeV, pointing to a large fragmentation of both the $\pi
g_{7/2}$ and $\pi h_{11/2}$ single-particle strengths. One can
therefore conclude from this work that the 11/2$^-_1$ level has
not to be assigned to a pure $\pi h_{11/2}$ single-particle state,
regardless the complex analysis of the cross sections to extract
the spectroscopic factors.

To get a better insight on the single-particle nature of the
11/2$^-_1$ state in the $^{A+1}_{~~51}$Sb isotopes, one can
compare the evolution of its energy and lifetime to that of the
$3^-$ state in the $^A_{50}$Sn core nuclei. From these comparisons
one would appreciate whether the wave function of the 11/2$^-_1$
state contains some amount of $\pi d_{5/2} \otimes 3^-$ coupling.
As said before, the $3^-$ state of the Sn isotopes is provided by neutron excitations involving
mainly the $\nu g_{7/2}$ or $\nu d_{5/2}$ orbits with the $\nu
h_{11/2}$ one. As shown in the bottom part of
Fig.~\ref{octu_SbSn}, the excitation energy of the 11/2$^-_1$
state above the 5/2$^+_1$ state follows exactly that of the $3^-$ state. The lifetimes of
the 11/2$^-_1$ states in the $^{113-119}$Sb$_{62-68}$ isotopes
were determined through their direct decay to the 5/2$^+$ ground
state. From these lifetimes, B(E3; 11/2$^-$ $\rightarrow$ 5/2$^+$)
transition probabilities are extracted and compared to the B(E3;
3$^-$ $\rightarrow$ 0$^+$) values of the core nuclei in the top
part of Fig. \ref{octu_SbSn}. The B(E3) values of the Sb isotopes
are close to those measured in the Sn cores. On the basis of the
energy evolution of the 11/2$^-_1$ with respect to the $3^-$
state, the B(E3) systematics, the former determination of SF in
Ref.~\cite{Conj68,Gale85} and the fragmentation of the $\pi
h_{11/2}$ strength measured in $^{117}$Sb~\cite{Gale85} one can
reasonably deduce that the 11/2$^-_1$ state is not of pure
single-particle origin. It may therefore be hazardeous to use 
the\emph{ absolute} evolution of the 11/2$^-_1$ state with respect to
the 7/2$^+_1$ state to derive the variation of the
$\pi h_{11/2}-\pi g_{7/2}$ splitting. For instance, shell model 
calculation has been undertaken in the $^{114}$Sn 
nucleus~\cite{Brow06}. It shows that by including proper correlations, the binding energy of the
11/2$^-_1$ and 7/2$^+_1$ states may change by about 1~MeV 
and 0.5~MeV, respectively.

The recent results obtained by Schiffer et al.~\cite{Schi04} are
clearly at variance with the aforementioned conclusions. In this
work the spectroscopic factors of the 7/2$^+_1$ and 11/2$^-_1$
states in $^{113-125}$Sb$_{62-74}$ have been  determined from a
new set of experimental data on the $^A$Sn ($^4$He,$t$)$^{A+1}$Sb
transfer reactions, with $A$ extending from $112$ to $124$. By
using a beam of $\alpha$ particles at 40 MeV, the proton pick-up
with angular momenta $\ell=4,5$ is kinetically well matched giving
rise to relatively large cross sections. Tritons were momentum
analyzed in an Enge split-pole magnetic spectrograph. The measured
normalized spectroscopic factors of the 7/2$^+_1$ and 11/2$^-_1$
states were found to be close to 1 for the whole set of nuclei,
which was given as the signature of a single-particle character of
the 7/2$^+_1$ and 11/2$^-_1$ states. Their energy spacing varies
from 576 keV at $N=64$ to 1890 keV at $N=74$ (see the bottom part
of Fig.~\ref{octu_SbSn}).
By adding 10 neutrons this amounts to an increase of the energy
spacing $\pi g_{7/2} - \pi h_{11/2}$ by about 1.4~MeV, and a
reduction of the relative SO splittings involving the $\ell=4$
and $\ell=5$ orbits by the same value~\cite{Schi04}.\\

Several mean field calculations, relativistic or
not~\cite{Otsu06,Brow06,Colo07,Lesi07} tried to reproduce these
latter experimental data by adding tensor terms in their
interaction. This procedure required a subsequent readjustment of
parameters of these effective forces, which were established from
other nuclear properties. Whatever the effective force in
use, the gap between the two proton orbits $g_{7/2} - h_{11/2}$ is
predicted to widen from $N=64$ to $N=82$, i.e. during the filling
of the $\nu h_{11/2}$ orbit, the tensor term giving both a
repulsive effect on the excited $\pi h_{11/2}$ orbit and an
attractive one on the low-lying $\pi g_{7/2}$ orbit. Nevertheless
the\emph{ absolute} distances between the two single-proton
energies is very far from the experimental one given in
Ref.~\cite{Schi04} (see Fig.~\ref{octu_SbSn}). These discrepancies,
on the one hand between old and new experimental data, and on the
other hand between new experimental data and theory, should
encourage other physicists to delve into this subject. In
particular a proper unfolding of correlations should be undertaken
throughout the Sb isotopic chain to derive the variation of the
single-particle energies. With this in hand, a more precise
discussion on the effect of the tensor interaction could be
envisaged.

At $N=82$, the 11/2$^-_1$ energy of $^{133}$Sb$_{82}$ does not
follow the rise in energy of the $3^-$ state of its Sn core, as shown in
Fig. \ref{octu_SbSn}. This state could be assigned as a pure 
$\pi h_{11/2}$ single particle. However, the
B(E3;~11/2$^-$~$\rightarrow$~5/2$^+$) value of 22(13)~$W.u.$
extracted from its lifetime is, within the error bars, too large 
to correspond to the transition between two single-proton states
($\le$ 4~$W.u.$ in the neighboring nuclei). It is closer to
the collective octupole strength of the $^{132}_{~50}$Sn$_{82}$
core, B(E3)~$>$~7~$W.u.$. Therefore the configuration of the
11/2$^-_1$ state in $^{133}$Sb is still likely to be mixed, 
$\pi h_{11/2}$ single particle and 
$\pi d_{5/2} \otimes 3^-$ coupling.
Transfer reactions using radioactive beams around $^{132}$Sn would
give us this important information in a near future.

\subsubsection{\it Conclusion}

The $Z=50$ shell closure has been analyzed in a large range, 
between the two doubly
magic nuclei $^{100}$Sn at $N=50$ and $^{132}$Sn at $N=82$.
Numerous recent experimental data such as transfer, Coulomb
excitation, decay spectroscopy, have been considered to obtain a
global understanding of the evolution of this proton shell gap and
on the proton orbits located above it. These aimed in particular
at determining the resistance of the $Z=50$ gap against
excitations and the role of tensor forces.

To start with, the Sn nuclei exhibit $2^+$ and $3^-$ exitations
are about 1.2 and 2.5~MeV, respectively. These states are composed
of neutron excitations involving the $d_{5/2}$, $g_{7/2}$ and
$h_{11/2}$ orbits. As they are present at relatively low energy,
the proton single-particle states can easily couple to these
modes. Therefore the variation of the 5/2$^+$, 7/2$^+$ and
11/2$^-$ states observed in the $_{51}$Sb isotopes cannot be 
directly attributed to that of the $\pi d_{5/2}$,
$\pi g_{7/2}$ and $\pi h_{11/2}$ single-particle states, 
respectively.

From the trends derived from the proton binding energies, 
it can be inferred that the size of
the $Z=50$ remains large above $N=64$. This accounts well for the
doubly magic character of $^{132}$Sn. On the other hand this gap
is decreasing below $N=64$ (towards $^{100}$Sn), a fact that could
favor core excitation and increase collectivity. Experimental
B(E2) values, even if not all agreeing, globally point to this
collectivity enhancement for the lightest Sn isotopes.

Several tensor interactions were expected to
play a role in this isotopic chain. Two of them were discussed in
the text. First the $\pi g_{9/2} - \nu g_{7/2}$ interaction was
invoked to explain the reduction of the $Z=50$ gap for the
lightest Sn isotopes. Second the crossing of the 5/2$^+$ and
7/2$^+$ states (likely to originate partly from the $\pi d_{5/2}$
and $\pi g_{7/2}$ orbits) was indebted to the filling of the
$\nu h_{11/2}$ shell. However, as the filling of neutron
orbits between $N=50$ and $N=82$ is diluted by the effect of
pairing correlations and spread over several orbits, the effects of
specific proton-neutron interactions are considerably smoothened.
Therefore a proper unfolding of the various kind correlations
involved -pairing, quadrupole and octupole- should be undertaken
to draw decisive conclusion about the effect of these
interactions.

These isotopic chains contain many potential interests on both
sides of the valley of stability that have been recently put
forward. Many theoretical models have started exploiting and
interpreting experimental results. This will undoubtedly continue
in the future with further discoveries to come, as for the study
of $^{100}$Sn.

%% file: texteNZ82_10avril.tex
\section{The magic number 82}
This magic number appears after the complete filling of the $N=4$
major shell and the $h_{11/2}$ intruder orbit. As shown below, the
orbits bounding the $N=82$ neutron and the $Z=82$ proton gaps are
not the same. This occurs partly because the Coulomb field -not
negligible anymore - distorts the mean potential felt by the
protons, changing the order of the levels accordingly. Also,
very different $\pi \nu$ interactions come into play to modify the
binding energies of the orbits involved,
because of the large differences in the respective nucleon numbers
(the $N=82$ isotones with $Z \simeq 50-70$ as compared to the
$_{82}$Pb isotopes with $N \simeq 104-130$).

\subsection{\it Evolution of the  $N=82$ shell closure\label{evolgapN82}}

This shell closure is formed between the $h_{11/2}$ (or the nearby
$s_{1/2}$ and $d_{3/2}$) and the $f_{7/2}$ orbit. The $N=82$
closure is so strong that seven $N=82$ isotones with $54 \leq Z
\leq 62$ are stable. The $N=82$ chain contains one known
doubly-magic nucleus, $^{132}_{~50}$Sn. For the heavy $N=82$
isotones, the $Z=64$ proton gap is large enough to provide a
doubly-closed character to $^{146}_{~64}$Gd, its first excited
level is a 3$^-$ state, as for the doubly-magic nucleus
$^{208}$Pb. The lighter $N=82$ isotones
have been much less extensively studied. Potentially another magic
nucleus, $^{122}_{~40}$Zr, could be found. It is formed by the
combination of the $Z=40$ shell closure, the effect of which has
been evidenced in the $^{90,96}$Zr nuclei, and by the $N=82$ one
evidenced for $^{132}_{~50}$Sn. The persistence of the $N=82$
closure below $Z = 50$ is an open question which is in
particular essential for modelling the explosive neutron-capture
nucleosynthesis processes.

In the following, experimental results on the neutron binding
energies of states surrounding the gap as well as the trends of
collective states are presented to shed light on the $N=82$
shell evolution as a function of the proton number. Afterwards the
action of some terms of the $NN$ forces to reduce the $\nu h_{11/2}
- \nu h_{9/2}$ SO splitting will be demonstrated and quantified.
Finally the motivations for studying the lightest $N=82$ isotones
($Z < 50$) will be emphasized in the frameworks of the $NN$
interactions and astrophysics.

\subsubsection{\it Evolution of the binding energies}

\begin{figure}[h!]
\begin{minipage}{9.cm}
\begin{center}
\epsfig{file=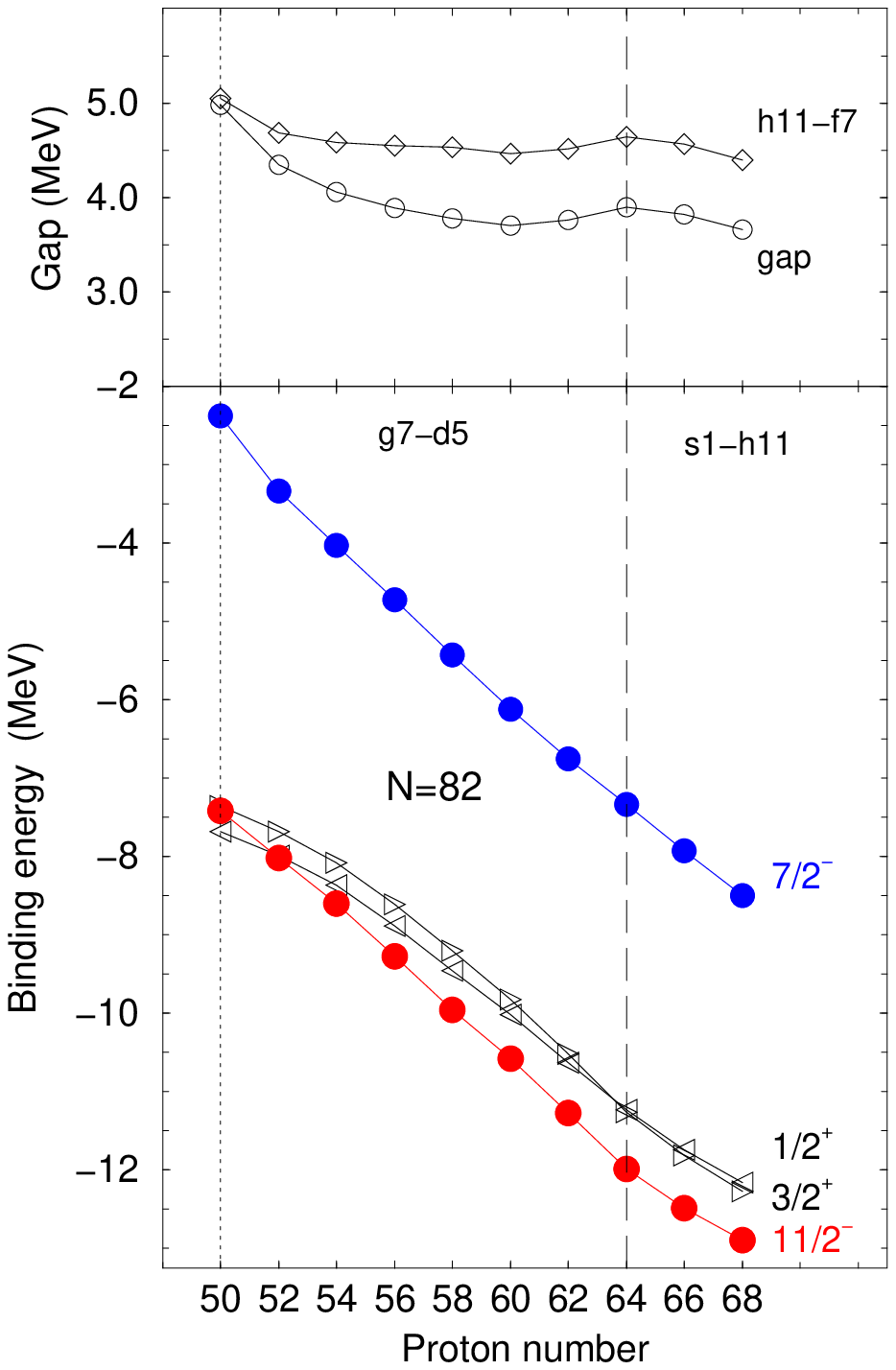,scale=0.65}
\end{center}
\end{minipage}\hfill
\begin{minipage}{9.cm}
\epsfig{file=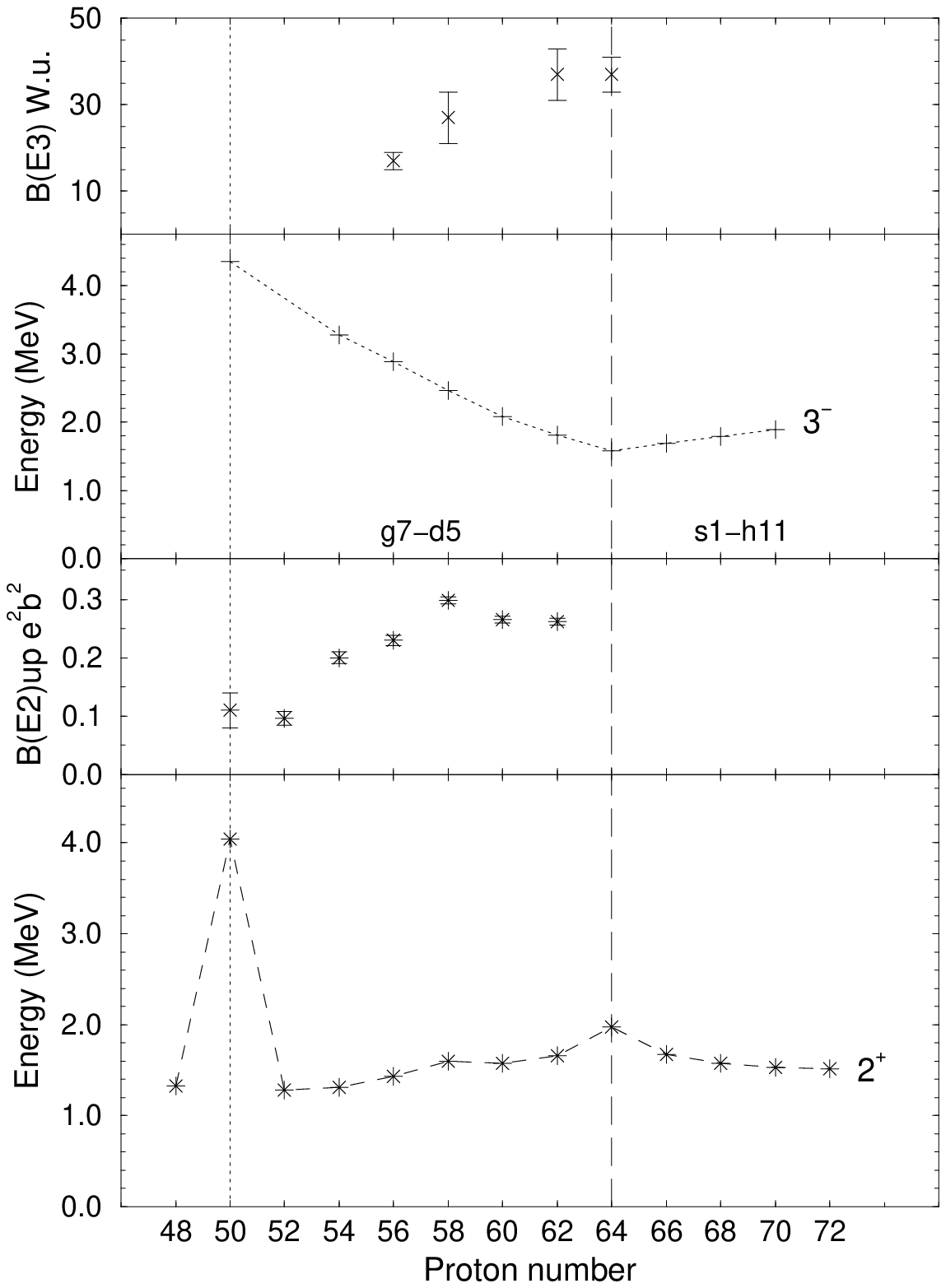,scale=0.6}
\end{minipage}
\begin{center}
\begin{minipage}[t]{16.5 cm}
\caption{{\bf  Left}: Binding energies of the states located below
and above the $N=82$ magic number and differences of the binding
energies of the two states surrounding the gap at $N=82$ (labelled
as 'gap') and of the 11/2$^-$ and 7/2$^-$ states (labelled as
'h11-f7')(see Sect.~\ref{annex}). The proton orbitals which are
getting filled
as a function of increasing $Z$ are given in the middle of the
figure. {\bf Right}: Experimental 2$^+_1$ energies and
B(E2)$\uparrow$ values (bottom), 3$^-_1$ energies and B(E3) values
(top) in the $N=82$ chain. B(E2) values of $^{132}$Sn and $^{134}$Xe come
from Refs.~\cite{Radf02,Jako02} respectively.\label{gapN82}}
\end{minipage}
\end{center}
\end{figure}

Figure~\ref{gapN82} displays the experimental binding energies of
states around the $N=82$ shell closure as a function of proton
number, ranging from $Z=50$ ($^{132}$Sn) to $Z=68$ ($^{150}$Er).
They are taken from the atomic mass table~\cite{Audi03}, to
which the very new results for $^{131-133}$Sn isotopes~\cite{Dwor08} have
been added.

While the heavy-mass part is close the proton-drip line, there is
a wide region to explore in the neutron-rich nuclei below $Z=50$.

The $N=81$ isotones exhibit three states below 800 keV excitation
energy, with spin values 1/2$^+$, 3/2$^+$, and 11/2$^-$
respectively. As indicated from neutron pick-up reactions using
all available stable targets ($Z=54-62$, $N=82$) these states
correspond to neutron holes in the three orbits located below the
$N=82$ gap, namely $\nu s_{1/2}$, $\nu d_{3/2}$ and $\nu
h_{11/2}$. Their spectroscopic factors are rather large, greater
than 0.7. For $^{131}_{~50}$Sn, the excitation energy of its
long-lived isomeric state 11/2$^-$ has been determined using the
$\beta$-decay of $^{131}_{~49}$In. It is proposed to lie 65.1 keV
above the 3/2$^+$ ground state~\cite{Foge04}.

The linear variation of the binding energy of the 11/2$^-$ state ,
and henceforth of the
$\nu h_{11/2}$ orbit, indicates that the filling of the proton
$g_{7/2}$ and $d_{5/2}$ orbits occur simultaneously from $Z=50$ to
$Z=64$. A successive filling of these two orbits would give rise
to a change of slope at $N=58$. Therefore its actual variation
results from the combined effects of two proton-neutron
interactions, $\pi g_{7/2}-\nu h_{11/2}$ and $\pi d_{5/2}-\nu
h_{11/2}$. The former monopole matrix element should dominate over
the latter, as discussed already in Sect.~\ref{levelSb}.

The binding energy for the 1/2$^+$ and 3/2$^+$ states display a
more rounded curve as protons are added above $Z=50$ (see the left
part of Fig.~\ref{gapN82}). There the $1/2^+$ and $3/2^+$ states
are likely influenced by correlations beyond the monopole
interactions.

On the theoretical point of view, whatever the mean field
potentials in use (such as those obtained from several
parameterizations of the Skyrme force, BSk1, SLy6, and SkI3, the
D1S Gogny force, or the relativistic mean-field forces NL3 and
NL-Z2), the $\nu h_{11/2}$ orbit is predicted to lie at about
1~MeV above any other shells as the $\nu d_{3/2}$ and $\nu
s_{1/2}$ ones~\cite{Bend03}. The ground state of all the $N=81$
isotones is therefore predicted to have a $\nu h_{11/2}^{-1}$ hole
state configuration, at variance with experimental results.

The $N=82$ gap, of about 4~MeV for $Z>54$, increases for lower $Z$
values (see left part of Fig.~\ref{gapN82}). This happens as the
slope of the 3/2$^+$ binding energy is reduced. This variation can
be attributed to a progressive change of configuration of the
3/2$^+$ state, which becomes of more single-particle origin
towards $Z=50$. From this trend, it follows that the $N=82$
gap is large, in particular for the $^{132}$Sn nucleus. Combined
with a
large $Z=50$ gap, $^{132}$Sn has a well established doubly-magic
structure. This statement is confirmed from its collective
properties described in the following section.

Below $Z=50$, the properties of the $N=82$ gap cannot be
extrapolated directly as the change of binding energy of the
neutron orbits surrounding the gap will be dictated by the yet
unknown residual interactions involving the $\pi g_{9/2}$ orbit.
First $\beta$- and $\gamma$-spectroscopic decay studies
of $^{130}_{~48}$Cd are in favor of an $N=82$ shell
quenching~\cite{Dill03}. This assumption was derived from the
rather high $Q_\beta$ value of $^{130}_{~48}$Cd which compares well
with results from mass models including a shell quenching. A direct
determination of a possible reduction of the $N=82$ shell gap when
the $\pi g_{9/2}$ orbit is being emptied needs at least accurate
atomic masses of $^{129,130,131}_{~~~~~~~~48}$Cd isotopes.

\subsubsection{\it Trends of the first collective
excitations of the $N=82$ isotones}

The experimental energies and reduced transition probabilities for
the first 2$^+$ and 3$^-$ states of the $N=82$ isotones are
reported in the right part of Fig.~\ref{gapN82}. Neutrons
excitations at $N=82$ can hardly produce a low-energy $2^+$ state
as the spacing between orbits $h_{11/2}$ and $f_{7/2}$ which can
generate quadrupole excitations is large (see the curve labelled
'h11-f7' in the left hand side of Fig.~\ref{gapN82}). Therefore
the configuration of the 2$^+_1$ state in the $N=82$ isotones is
mainly built with protons from the $\pi g_{9/2}$ orbit  below
$Z=50$, from the $\pi g_{7/2}$ and  $\pi d_{5/2}$ orbits between
$Z=52$ and $Z=64$, and from the $\pi h_{11/2}$ orbit above $Z=64$.

The $^{132}_{~50}$Sn$_{82}$ nucleus exhibits all the
charateristics of a doubly-magic nucleus, i.e. a high energy for
the first excited states and a weak transition probability
B(E2). Apart from this spectacular behavior of $^{132}_{~50}$Sn,
the energy of the $2^+$ state varies smoothly along the whole
$N=82$ isotonic chain. A small increase is observed at $Z=64$,
which is due to the sub-shell closure between the $\pi
g_{7/2}/d_{5/2}$ and $\pi h_{11/2}$ orbits.

Below $Z=50$, the $\pi g_{9/2}$ orbit is likely to be involved in
turn. Two experiments have been carried out at the GSI
facility to search for the cascade of $\gamma$-rays de-exciting
the $8^+$ isomeric state expected from the $(\pi g_{9/2})^{-2}$
configuration of $^{130}_{~48}$Cd$_{82}$, as in
$^{98}_{48}$Cd$_{50}$. The isomer has been populated both in the
fragmentation of a $^{136}$Xe beam as well as in projectile fission
of $^{238}$U~\cite{Gors07}. The energy of the $2^+ \rightarrow 0^+$
transition, displayed in the bottom right part of
Fig.~\ref{gapN82}, is found at 1.325~MeV, an energy close to
that of the valence mirror nucleus $^{98}_{48}$Cd$_{50}$
(1.395~MeV).

The $N=82$ isotones exhibit a collective 3$^-$ state with large
transition probabilities $B(E3)=20-40~W.u.$. When present at low
excitation energy, this state originates from $p-h$ excitations
between the positive-parity\emph{ proton} orbits located below
$Z=64$ ($g_{7/2}$ and $d_{5/2}$)  and the negative-parity
$h_{11/2}$ orbit above $Z=64$. At the opposite neutrons can hardly
generate a $3^-$ state across the $N=82$ shell gap. The B(E3)
value is maximum at $^{146}_{~64}$Gd, where the energy of this
3$^-$ state is minimum. This is attributed to the fact that the
number of particle and holes to generate the $3^-$ state is
maximized at $Z=64$, which is approximately at the mid-distance
between $Z=50$ and $Z=82$.

As the $2^+$ and 3$^-$ states of the $N=82$ nuclei have relatively
small excitation energy, the single-neutron particle strengths in
the neighboring odd-$N$ will be fragmented due to couplings with
these collective states. This feature should be kept in mind in
the following paragraphs.

\subsubsection{\it Neutron orbits above the $N=82$ gap: Levels of
the $N=83$ isotones\label{N83}}

The low-energy orbits in the $N=83$ isotones arise from the $N=5$
major shell. The first one, $\nu f_{7/2}$, is the ground state of
all the $N=83$ isotones which have been identified so far from $Z
= 50$ to $Z = 70$. The coupling of the $\nu f_{7/2}$
single-neutron state to the first 2$^+$ and 3$^-$ excitations of
the core gives rise to two multiplet of states with spin values
ranging from 1/2$^-$ to 11/2$^-$ and 1/2$^+$ to 13/2$^+$,
respectively. Single-neutron states coming from the $\nu p_{3/2}$
and $\nu h_{9/2}$ orbits, as well as from the $\nu i_{13/2}$
intruder one are expected to be present in addition to states
provided by couplings to core excitations. Therefore mixing
between the single-particle and coupled states are expected and
cannot be neglected.

\begin{figure}[h!]
\begin{center}
\epsfig{file=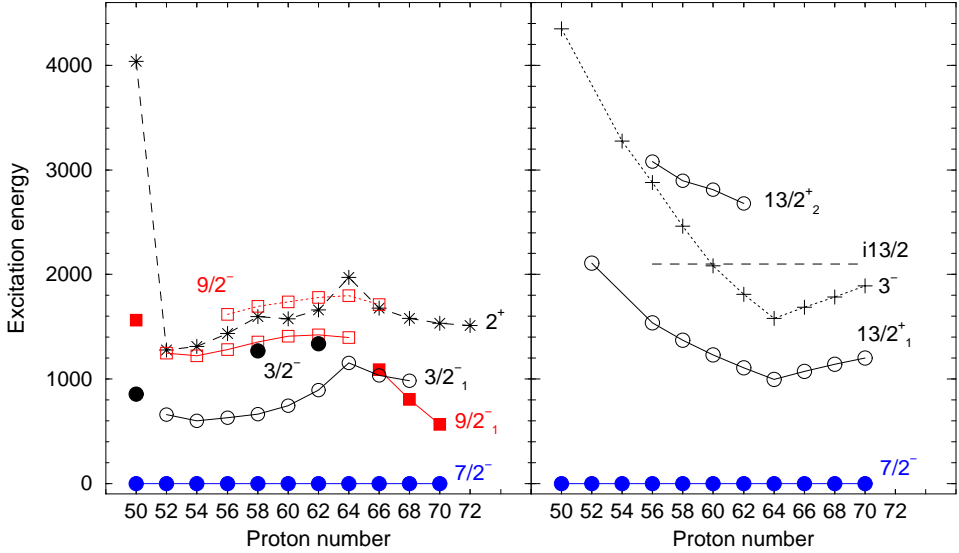,scale=0.7}
\begin{minipage}[t]{16.5 cm}
\caption{First excited states observed in the $N=83$ isotones: The
full symbols correspond to the states having a large
single-particle character, the empty symbols those containing a
large collective component (see text). The evolution of the 2$^+$
and 3$^-$ energies of the corresponding core are also drawn, as
well as the $\nu i_{13/2}$ energy assumed in the two-state mixing
calculation \cite{Piip90} (see text). \label{crossingN83}}
\end{minipage}
\end{center}
\end{figure}

The experimental energies of the first excited states in the $N=83$
isotones are given in Fig. \ref{crossingN83}, those with negative
(positive) parity being gathered in the left (right) part. The
quadrupole (octupole) collective excitations of the core nuclei are
also reported in the left (right) part of this figure. Until
recently, the 13/2$^+$ excited state in the $N=83$ isotones had not
been identified below $^{139}_{~56}$Ba. The neutron transfer
reactions, $^{9}$Be($^{134}$Te,$^{8}$Be) and
$^{13}$C($^{134}$Te,$^{12}$C), were studied with $^{134}$Te
radioactive beams at energies slightly above
the Coulomb barrier in order to determine the energy of
the 13/2$^+$ level (2109 keV) in $^{135}_{~52}$Te \cite{Radf05}.

For $Z<68$, the first excited state of the $N=83$ isotones
has I$^\pi$ = 3/2$^-$ (see
the left part of Fig.~\ref{crossingN83}). Its energy
follows the 2$^+$ excitation energy of the core. Moreover its
normalized spectroscopic factor, measured by neutron-transfer
$(d,p)$ reactions from $_{54}$Xe to $_{62}$Sm,  amounts to $\simeq
0.5$. This confirms that this first state has not a pure
single-neutron $\nu p_{3/2}$ configuration. In $^{141}_{~58}$Ce
and $^{145}_{~62}$Sm other 3/2$^-$ states have been identified at
higher excitation energy, which brings the sum of the
spectroscopic factors close to one. The energy of the $\nu
p_{3/2}$ orbit, reported with a full symbol, was calculated by
averaging the energies weighted by the corresponding spectroscopic
factors. As the 2$^+$ state of $^{132}$Sn is very high, one can
speculate that the 3/2$^-$ state of $^{133}$Sn is mainly of
single-neutron character. Therefore a full symbol has also been
included for this nucleus in Fig.~\ref{crossingN83}. Using these
three results, one would speculate that the spacing in energy
between the $\nu f_{7/2}$ and $\nu p_{3/2}$ orbits is slowly
increasing from Z = 50 to Z = 62, i.e. during the filling of the
$\pi g_{7/2}$ and $\pi d_{5/2}$ orbits. Nevertheless this argument
is weak by many aspects. For instance, the assignment of the first
excited state of $^{133}$Sn at 853 keV has to be confirmed, being
based from a single experiment with low statistics~\cite{Hoff96}.

One or two states with I$^\pi$ = 9/2$^-$ have been identified, as
shown in the left part of Fig.~\ref{crossingN83}. In every isotone
but $^{133}$Sn and those with Z $\ge$ 66, the energy of the
9/2$^-_1$ is slightly lower than the 2$^+$ excitation of the core.
In parallel the energy of the 9/2$^-_2$  is slightly higher. The
presence of two $9/2^-$ states at each side of the $2^+$ one
points to a mixing of wave functions, $\nu h_{9/2}$ and $\nu
f_{7/2}$ $\otimes$ 2$^+$. Furthermore, the spectroscopic factors
measured by neutron-transfer $(d,p)$ reactions from $_{54}$Xe to
$_{62}$Sm, confirms that none of these 9/2$^-$ states has the
single-neutron $\nu h_{9/2}$ configuration.
The large mixing of the two components is also proved in
$^{147}_{~64}$Gd, as its two 9/2$^-$ states are strongly populated
(log$ft$=4.3 and 4.8) in the favored $\beta$-decay of the
$^{147}_{~65}$Tb isomeric state, $\pi h_{11/2} \rightarrow \nu
h_{9/2}$. On the other hand, only one Gamow-Teller branch is
observed  in the heavier masses, meaning that the 9/2$^-_1$ states
of $^{149}_{~66}$Dy, $^{151}_{~68}$Er, and $^{153}_{~70}$Yb
progressively become more of single-particle character. Beyond $Z
=66$ the energy of the 9/2$^-_1$ state decreases, whereas the
2$^+$ excitation remains almost constant. This likely implies a
strong decrease of the energy spacing between the two orbits, $\nu
f_{7/2}$ and $\nu h_{9/2}$ during the filling of the $\pi
h_{11/2}$ orbit. We will come back to this important point in the
next section (Sect.~\ref{nuh9h11}) when discussing about the
proton-neutron interactions.

As regards the $\nu i_{13/2}$ orbit in the $N=83$ isotones, the
evolution of its energy has been recently discussed
in terms of change of the nuclear spin-orbit interaction with increasing
neutron excess~\cite{Schi04}, similarly to the evolution of the $\pi h_{11/2}$
orbit in the $Z=51$ isotopes (see Sect. \ref{levelSb}). In this
work, the first excited state with I$^\pi$ = 13/2$^+$, identified
in every $N=83$ isotone, has been assumed to have the
single-neutron $\nu i_{13/2}$ configuration. The location of the
$\nu i_{13/2}$ single-particle energy was debated already in
Ref.~\cite{Piip90}, in which configurations of the high-spin
states of nuclei close to $^{146}_{~64}$Gd$_{82}$ were assigned.
Noticeable is the fact that the excitation energy of the
13/2$^+_1$ state of the $N=83$ isotones closely follows that of
the 3$^-$ collective state measured in the corresponding N = 82
cores, displaying in concert a minimum for $Z=64$ (as shown in the
right part of Fig.~\ref{crossingN83}). By converting the measured
half-life of the 13/2$^+_1$ level into B(E3) values, it is found
that the B(E3) values are much larger than expected for a $\nu
i_{13/2}$ $\rightarrow$ $\nu f_{7/2}$ single-neutron transition.
These two sets of energies and lifetimes have been analyzed with a
schematic two-state mixing calculation in Ref.~\cite{Piip90}.
Using a \emph{ constant} value of 2.1~MeV (indicated by dashed line
in the right part of Fig.~\ref{crossingN83}) for the $\nu
i_{13/2}$ single-particle energy from $Z=56$ to $Z=70$, the
measured energies of the 13/2$^+_1$ and 13/2$^+_2$ states (known
for $Z=56-62$), as well as the B(E3, 13/2$^+_1$ $\rightarrow$
7/2$^-$) values are quite well reproduced~\cite{Piip90}. Contrary
to what is said in Ref.~\cite{Schi04}, this result rather
indicates no change of the binding energy of the $\nu i_{13/2}$
orbit for increasing Z values, but a fragmentation of the $\nu
i_{13/2}$ force occurring from the coupling with the $3^-$ state. A
similar conclusion was obtained in Sect.~\ref{levelSb} with
respect to a possible evolution of the $\pi h_{11/2}$ orbit within
the $Z=50-82$ major shell.

\subsubsection{\it Reverse behaviors of the $\nu h_{11/2}$ and
$\nu h_{9/2}$ orbits for $Z>64$\label{nuh9h11}}

The present paragraph intends to show a plausible evolution of the
$\nu h_{11/2} - \nu h_{9/2}$ SO splitting as the proton orbit $\pi
h_{11/2}$ is progressively filled.

\begin{figure}[h!]
\begin{minipage}{9.cm}
\begin{center}
\epsfig{file=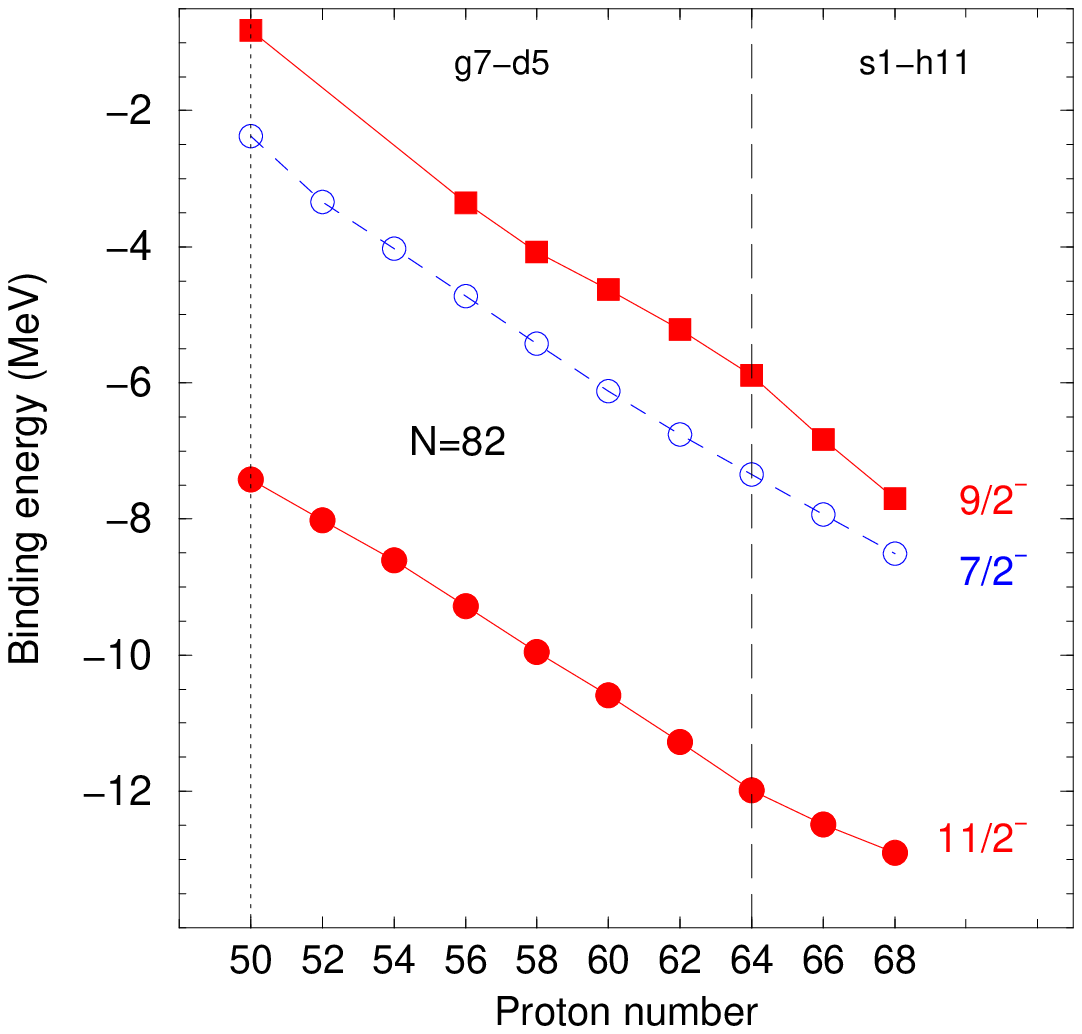,scale=0.6}
\end{center}
\end{minipage}\hfill
\begin{minipage}{9.cm}
\epsfig{file=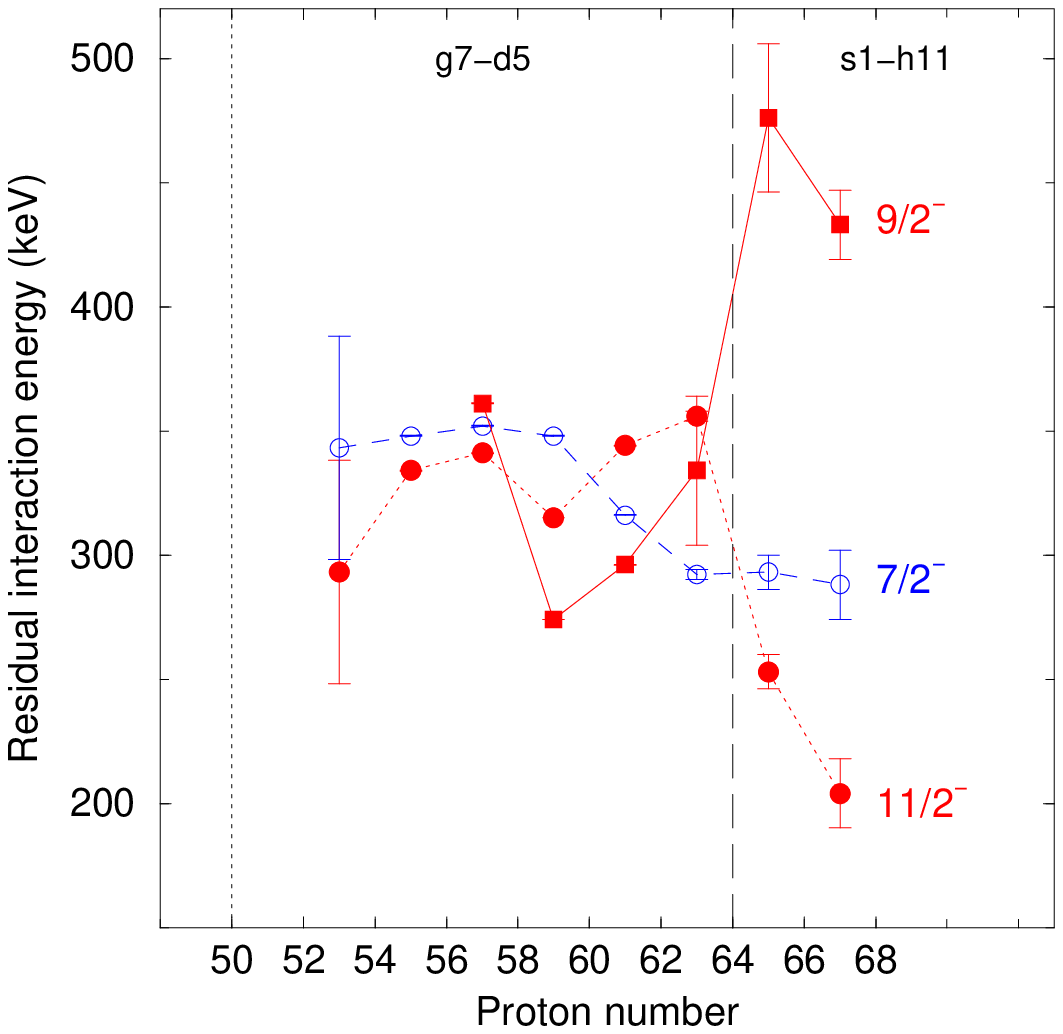,scale=0.6}
\end{minipage}
\begin{center}
\begin{minipage}[t]{16.5 cm}
\caption{{\bf Left} Evolution of the binding energies of the
11/2$^-$ and 9/2$^-$ states compared to those of the 7/2$^-$
states, around the $N=82$ gap. {\bf Right} Evolution of the
neutron-proton residual interactions (absolute values) extracted
from the slopes of
the binding energies of the three neutron states. The proton
orbitals which are getting filled as a function of increasing $Z$
are given on the top each figure. \label{fig_nuh9h11}}
\end{minipage}
\end{center}
\end{figure}
The left-hand side of Fig.~\ref{fig_nuh9h11} displays the binding
energies of the 9/2$^-$ and the 7/2$^-$ states of the $N=83$
isotones, as well as the binding energies of the 11/2$^-$ level of
the $N=81$ isotones. For $Z=56-64$ the energy of the 9/2$^-$ has
been averaged from the two 9/2$^-$ states, weighted by their
spectroscopic factors. Beyond $Z=64$, it was discussed in the
previous section that the $9/2^-$ become more of single particle
character. The binding energies of these three states display a
quasi-linear behavior in the $Z=50-64$ interval, while steeper
(smoother) slopes are observed beyond $Z = 64$ for the 9/2$^-$
(11/2$^-$) states. These changes of slopes are directly connected
to the the new neutron-proton residual interactions
that come into play beyond $Z =64$. Using the method explained in
Sect.~\ref{expmonopole} the corresponding monopole interactions have
been derived from the binding energy slopes and are drawn in the
right part of Fig.~\ref{fig_nuh9h11}. When the protons are filling
the $\pi g_{7/2}$ and $\pi d_{5/2}$ orbits, their residual
interactions with $\nu h_{11/2}$, $\nu f_{7/2}$, or $\nu h_{9/2}$
have almost the same values, between -300 and -350 keV.
On the other hand, as soon as the $\pi h_{11/2}$ orbit starts to
be filled, the $\nu h_{11/2}$ and $\nu h_{9/2}$ orbits exhibit
reverse behaviors, the $V^{pn}_{h_{11/2}h_{9/2}}$ matrix element
($\sim$ -450 keV) is much more attractive than the
$V^{pn}_{h_{11/2}h_{11/2}}$ one ($\sim$ -200 keV). This change of
effective interaction can be ascribed to the tensor force.

\subsubsection{\it Study of the $N=82$ shell closure below $Z=50$,
Astrophysical implications}

As mentioned along the previous chapters of the present work, the
properties of $NN$ interactions, such as those of the tensor
forces, can significantly alter or reinforce the strength of a
shell closure. In addition, the change of the mean field potential
of the nucleus, in particular the increased surface diffuseness
for large N/Z, weakly bound systems, also modifies the energy of
the orbits. Taken together a profound change of the spacing and
ordering of levels can occur when moving from stability to the
drip line, following for instance a given isotonic chain. The
present section shows how these structural changes impact on the
development of the rapid-neutron capture nucleosynthesis. We refer
the reader for instance to Refs.~\cite{Pfei01,Krat07} for recent
review papers.

After a brief introduction on the role of this process and the
possible astrophysical site where it prevails, the key 'integral'
parameters (binding energy, $\beta$-decay lifetime and neutron
capture cross section) to its development will be listed.  Nuclear
structure evolution can modify these parameters significantly,
changing for instance the location and duration of the r process.
Enlighted by the previous chapters on the evolutions of shell
closures, expected structural evolutions at $N=82$ will be commented
 and put in an astrophysical perspective.\\

{\bf Introduction}\\
Approximately half of the elements beyond Fe are produced via
neutron captures on very short time scales in neutron-rich
environments. Despite its importance, the exact site(s) where the
r-process(es) occurs still remain one of the greatest mystery in
science. The most frequently suggested astrophysical environments
are high-entropy ejecta from type II supernovae (SN) and
neutron-star mergers. The shock-heated He or C outer layer of type
II SN could provide a moderate neutron flux (a weak r-process)
through $^{13}$C($\alpha$,n) reactions which could account for
several isotopic anomalies found in pre-solar grains. The key for
understanding the r-process(es) resides in a close interaction
between astronomy, cosmochemistry, astrophysical modeling of
explosive scenarios and nuclear physics.

\begin{figure}[h!]
\begin{center}
\begin{minipage}[t]{16 cm}
\epsfig{file=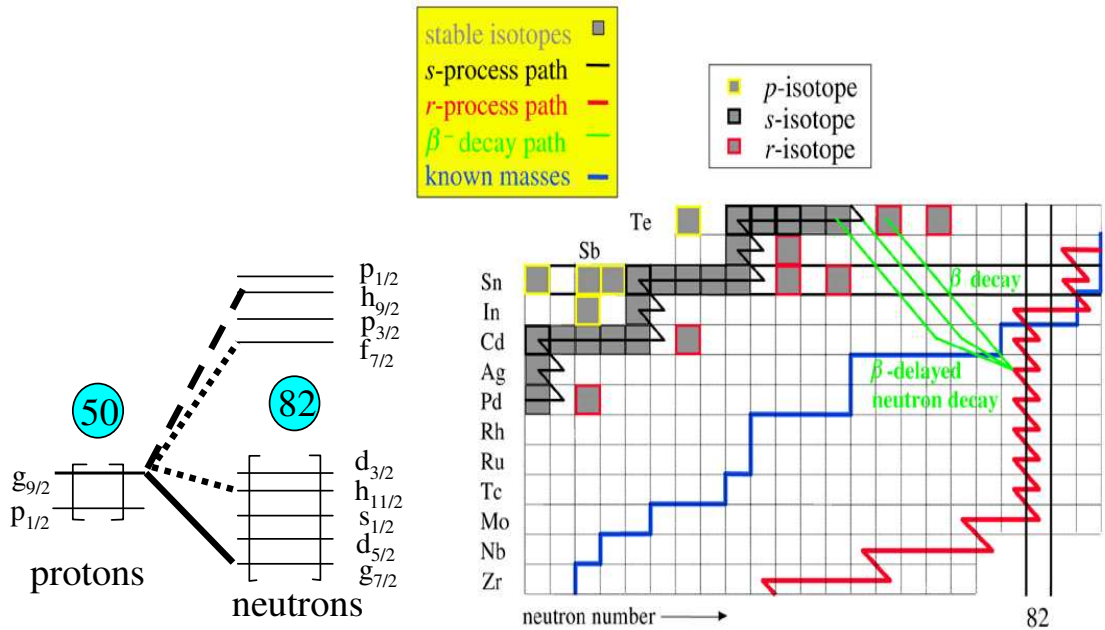,scale=0.8}
\end{minipage}
\end{center}
\begin{center}
\begin{minipage}[t]{16.5 cm}
\caption{\textbf{Left} : Schematic view of the proton-neutron
interactions which play a role for the evolution of the $N=82$
shell gap and for the properties of the $N \simeq 82$ nuclei as
proton are removed from the $g_{9/2}$ orbit. Interactions
represented with dotted (dashed) lines are expected to contain
repulsive (attractive) tensor terms. The interaction between the
$\pi g_{9/2}$ and $\nu g_{7/2}$, indicated by a full line, is
expected to be the largest among all others involved.
\textbf{Right}: Schematic representation of the location of the
r-process path at the $N=82$ shell closure (by courtesy of D.
Lunney). Present experimental knowledge on the atomic masses is
indicated by a thick line. As far as $\beta$-decay lifetimes are
concerned, measurements extend further from stability, reaching in
particular the $^{130}$Cd \cite{Dill03} and $^{129}$Ag
waiting-point nuclei~\cite{Dill03,Kaut98}. \label{rproc}}
\end{minipage}
\end{center}
\end{figure}

In very hot and neutron-dense environments, the r-process develops
through neutron captures on very short timescales until reaching
nuclei with sufficiently small binding energy ($S_n \simeq $
2-3~MeV). Typically this occurs at the major shell closures, such
as the $N=82$ one. There, the rate of captures is balanced by
that of photodisintegration induced by the ambient photon bath of
the exploding star. The process is stalled at these so-called
waiting-point nuclei until $\beta$-decays occur, depleting the
nucleosynthesis to the upper $Z$ isotopic chain where subsequent
neutron captures could occur. After successive $\beta$-decays and
neutron captures at the $N=82$ closed shell, the process is
progressively driven closer to stability where $\beta$-decay
lifetimes ($T_{1/2}$) become longer (see the zig-zag line along
the $N=82$ shell in the right part of Fig.~\ref{rproc}).
There, around the Sn isotopic chain, neutron captures are expected
to be shorter that photodisintegration and $\beta$-decay rates,
driving the r-process nucleosynthesis towards the next shell
closure. At the end of the r-process, radioactive progenitors
decay back to stability via $\beta$ or $\beta$-delayed neutron
emission towards the valley of stability. The accumulation of
r-elements which form the r-peaks at masses $A\simeq 80$, $\simeq
130$ and $\simeq 195$ on the abundance curve of the elements is a
direct imprint of the existence of waiting point progenitors far
from stability. The location, height and shape of the $A=130$ peak
could be traced back from the neutron separation energies ($S_n$),
the half-lives ($T_{1/2}$), the neutron delayed emission
probability ($P_n$), and the neutron-capture cross section
($\sigma_n$) of the nuclei located at the $N=82$ shell closure.
These parameters are imbricated in the structural evolution of the
nuclei as depicted in the
following paragraph.\\

{\bf Structural evolution}\\
The evolution of the $N=82$ shell gap south to the doubly magic
nucleus $^{132}_{50}$Sn$_{82}$ is ruled by the combined effects of
the proton-neutron interactions and by the progressive proximity
of the continuum states. As from $Z=50$ to $Z=40$ solely the
proton orbit $g_{9/2}$ orbital is progressively emptied, all the
proton-neutron monopole interactions which intervene in the
evolution of nuclear structure between $^{132}_{50}$Sn$_{82}$ and
the close-to-drip-line nucleus $^{122}_{40}$Zr$_{82}$, involve the
same proton orbit. As shown schematically in the left part of
Fig.~\ref{rproc}, the differential values of these monopoles could
modify the ordering of the neutron orbits, the size of the $N=82$
gap and subsequently the $S_n$ values as protons are removed.
Related to this, the tensor interactions (through the $\pi g_{9/2}
- \nu h_{11/2}$ or $\pi g_{9/2} - \nu f_{7/2}$ monopoles) are
likely to play important roles.

Likewise, the strength of the tensor term (through the $\pi
g_{9/2} - \nu g_{7/2}$ monopole) influences strongly the
$\beta$-decay lifetimes of the $N=82$ nuclei. The reason for so is
that the $\beta$-decay of $N \simeq 82$ nuclei south to $Z=50$
proceeds mainly through the Gamow-Teller (GT) transition $\nu
g_{7/2} \rightarrow \pi g_{9/2}$ transition, as described in
Ref.~\cite{Dill03}. As protons are progressively removed from the
$g_{9/2}$ orbit, the neutron $g_{7/2}$ orbit becomes gradually
less deeply bound due to the missing proton-neutron interactions
$\pi g_{9/2} - \nu g_{7/2}$. Consequently, the aforementioned GT
transition will occur at gradually smaller excitation energy $E^*$
in the daughter nuclei, leading to a drastic shortening of
$\beta$-decay lifetimes $T_{1/2}$, which scales with $(Q_\beta
-E^*)^{5}$.

Added to the properties of the $NN$ interactions, self-consistent
mean-field calculations which encompass the treatment of continuum
states show a quenching of the $N=82$ shell gap when approaching
the neutron drip-line~\cite{Doba94}. This is primarily caused by
the lowering of the low-$j$ orbits relative to those of high-$j$
values. The low-$j$ neutron orbits, such as the $\nu p_{1/2}$ and
$\nu p_{3/2}$ ones, may progressively become the valence states
immediately above the $N=82$ gap. Such effects were ascribed to
the fact that, close to the drip-line, the low-$j$ neutron orbits
of the continuum interact strongly with bound states, whereas the
interaction with the high-$j$ ones is much less effective.
Qualitatively, the nuclear mean field close to the drip-line could
be mimicked by a Nilsson potential without the $\ell ^2$
term~\cite{Doba94}. At exploding star temperatures of typically
$10^9~K$ (or neutron energies of about 100~keV) neutrons can
hardly overcome large centrifugal barriers created by high $\ell$
orbits. Therefore, the presence of low-$j$ or low-$\ell$ neutron
valence orbits at low excitation energy would enhance the neutron
capture cross-sections $\sigma_n$ by several orders of magnitudes,
shortening the neutron capture time at the waiting point nuclei
accordingly.\\

{\bf Astrophysical consequences of changes in the
$S_n$, $T_{1/2}$ and $\sigma_n$ values}\\
The influence of the $S_n$ values at the $N=82$ shell closure to
the r-abundance peaks has been emphasized in
Refs.~\cite{Chen95,Pfei01}. Using masses from "unquenched" models
(as FRDM or ETFSI-1), one obtains a local increase of the $S_n$
values around $N=70$ and a very abrupt  drop immediately after
$N=82$. They originate from a predicted strong quadrupole
deformation around $^{110}$Zr and a strong shell closure at
$N=82$, respectively. As a result, neutron captures are stalled
for a while around $A=110$, $N=70$ and are subsequently directly
driven to the closed shell $N=82$. This leaves few r-progenitors
in between these two regions, leading to a significant trough in
the fit of the abundance curve of the elements at $ A \simeq
120$~\footnote{Noteworthy is the fact that if $^{110}$Zr was a
doubly magic, a sudden drop of $S_n$ value at $N=70$ will be
found, leading to the build up of r-process elements at $A \simeq
110$.}. On the other hand, the "quenched" mass models (as ETFSI-Q
or HFB/SkP) bring back r-progenitors before reaching the $N=82$
shell closure, thus filling the trough at $A \simeq 120$ in closer
agreement with the solar r-abundance curve. The recent
determination of the $Q_\beta$-value in the $\beta$-decay of the
$^{130}$Cd waiting-point nucleus accredits this
weakening~\cite{Dill03}, which should be confirmed by further
investigations. From these arguments, it follows that the
determination of the $S_n$ values at the $N=82$ shell closure is
essential. So far theoretical models diverge soon after departing
from the last measured nucleus, as discussed for instance in
Ref.~\cite{Lunn03}, meaning that the underlying physics which
could modify the binding energies of the orbits is poorly known
and far from being consensual.

The influence of the $\beta$-decay lifetimes of neutron-rich
nuclei around $N=82$ on the abundance curve of the r-elements has
been discussed extensively for instance in Ref.~\cite{Pfei01}. When
short-lived nuclei are paving the r-process path, the abundance
peaks of the r-elements are barely formed and a total duration of
the explosive process of few milliseconds could be long enough to
reach the heaviest elements of the chart of nuclides. At the
opposite, long-lived nuclei located in the r-process path are the
major genitors of stable r-nuclei in the universe (after series of
$\beta$ or $\beta$-neutron decays to stability) and are
responsible for the existence of significant peaks in the
abundance curve of the elements. Therefore accurate half-life
predictions or measurements are required along the r- process both
to explain the shape of observed peaks but also to constrain the
duration of the r-process.

Neutron capture cross-sections are often calculated in the
framework of the statistical Hauser-Feschbach model, which assumes
the presence of a high density of states above the neutron
separation energy with various spin and parity values. The use of
this statistical approach is not appropriate to calculate neutron
capture cross sections for neutron-rich at closed shell, since the
neutron-separation energy is small and the level density low. In
such cases, the main contribution is obtained from direct captures
on few bound states of low $\ell$ states, mainly through $s$ or
$p$ waves, or/and by resonance capture slightly above the
neutron-energy threshold. The fact that the neutron $p$ orbits
become the few first valence states is therefore extremely
important. This feature is discussed in Ref.~\cite{Raus98} for the
$^{133}$Sn nucleus. As the direct capture cross section proceeds
mainly through the same $\nu p_{3/2}$ and $\nu p_{1/2}$ states
from $Z=50$ down to $Z=40$, the evolution of their energy and
spectroscopic factors determines how neutron-capture cross
sections will change along the $N=82$ shell closure. It tells to
what extent the r-process will be blocked at this shell closure
when the temperature of the exploding system decreases. Since the
direct determination of neutron-capture cross sections on very
unstable nuclei $A$ is technically not feasible, their
determination should be provided by models, constrained by
spectroscopic information (such as energy, spin, spectroscopic
factors of the bound and unbound levels in the $A+1$ nucleus)
obtained by $(d,p)$ transfer reactions. Such pionneering studies
have been undertaken for the $N=28$ shell closure in the
neutron-rich $^{48}$Ca and $^{46}$Ar nuclei where similar $p$
states come into play in the neutron-capture cross
sections~\cite{Kapp85,Krau96,Gaud06b,Sorl03b}.

To summarize this astrophysical part, experimental and theoretical
achievements aiming at a comprehensive understanding of the
evolution of the $N=82$ shell closure will provide significant
insight to the understanding of the r-process nucleosynthesis, as
these topics are intimately entwined.

\subsubsection{\it Conclusion}

On the basis of neutron binding energies, energies and reduced
transition probabilities of the $2^+$ and $3^-$ states, it is
derived that the $N=82$ shell gap is large enough to maintain a
spherical configuration for the $N=82$ isotones. The
$^{132}_{~50}$Sn nucleus behaves as a doubly-magic nucleus.
Between $Z=50$ and $Z=64$ many different proton-neutron
interactions act to modify the neutron binding energies of the
$N=81$ and $N=83$ nuclei. Owing to the large pairing correlations
between neighboring proton orbits such as $g_{7/2}$, $d_{5/2}$,
$s_{1/2}$ and $h_{11/2}$, this change of neutron binding energies
owes to the action of several interactions that cannot be
identified individually. After having passed $Z=64$ the filling of
the proton $h_{11/2}$ orbit induces a reduction of the $\nu
h_{11/2}- \nu h_{9/2}$ SO splitting, which can be attributed to
the tensor terms of the $NN$ interactions.

The study of the lightest isotones $Z < 50$ is in its infancy as
this requires the most powerfull facilities which are coming up
soon. However hints for weakening of the $N=82$ shell gap were
proposed from $\beta$-decay studies. From $Z = 50$ to $Z = 40$,
the same proton orbit $g_{9/2}$ is involved. Consequently
proton-neutron monopoles involving this orbit play decisive roles
to modify the $S_n$, $T_{1/2}$ and $\sigma_n$ values, which are
all important parameters for the r-process nucleosynthesis. As
theoretical calculations give contradictory results at the present
time on a possible quenching of the $N=82$ shell gap, some key
measurements are prerequisite to improve extrapolations far from
the valley of stability.

\subsection{\it Evolution of the  $Z=82$ shell closure\label{evolgapZ82}}

\subsubsection{\it Evolution of the binding energies}

\begin{figure}[h!]
\begin{minipage}{9.cm}
\begin{center}
\epsfig{file=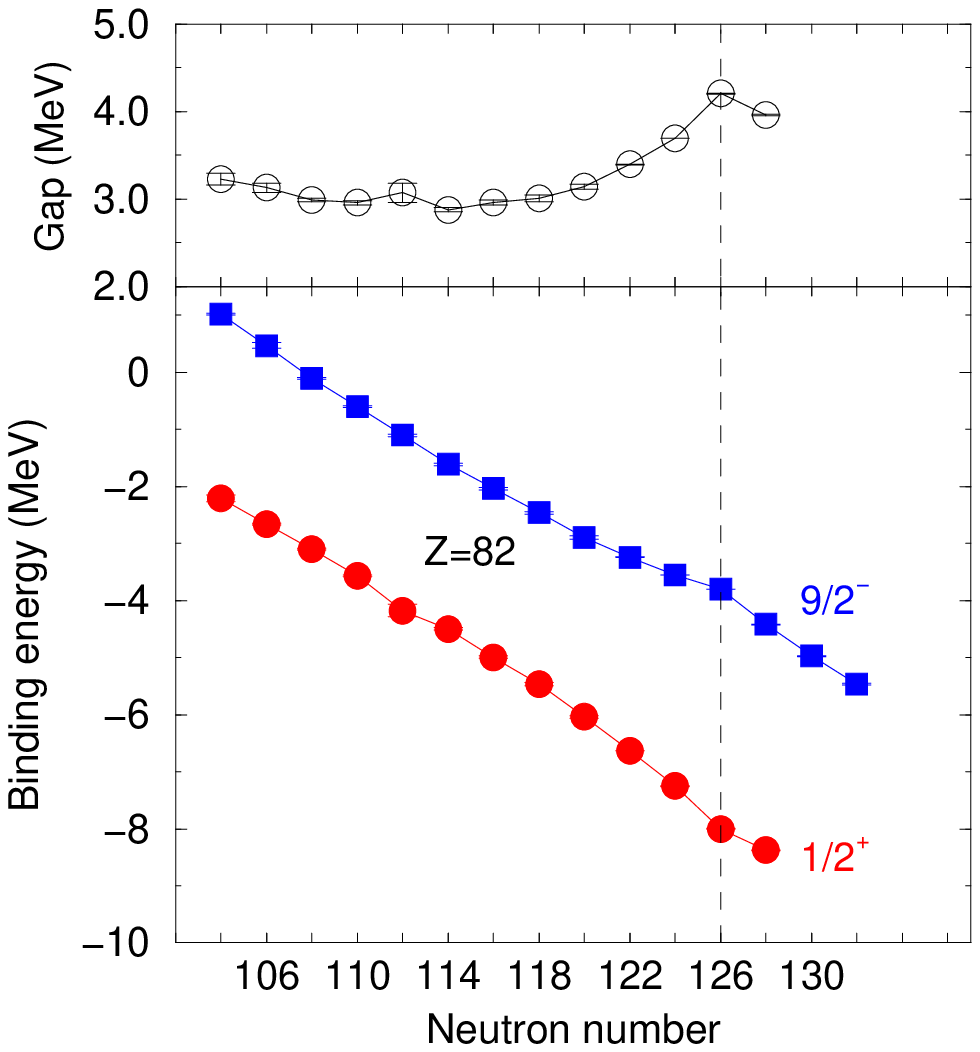,scale=0.6}
\end{center}
\end{minipage}\hfill
\begin{minipage}{9.cm}
\epsfig{file=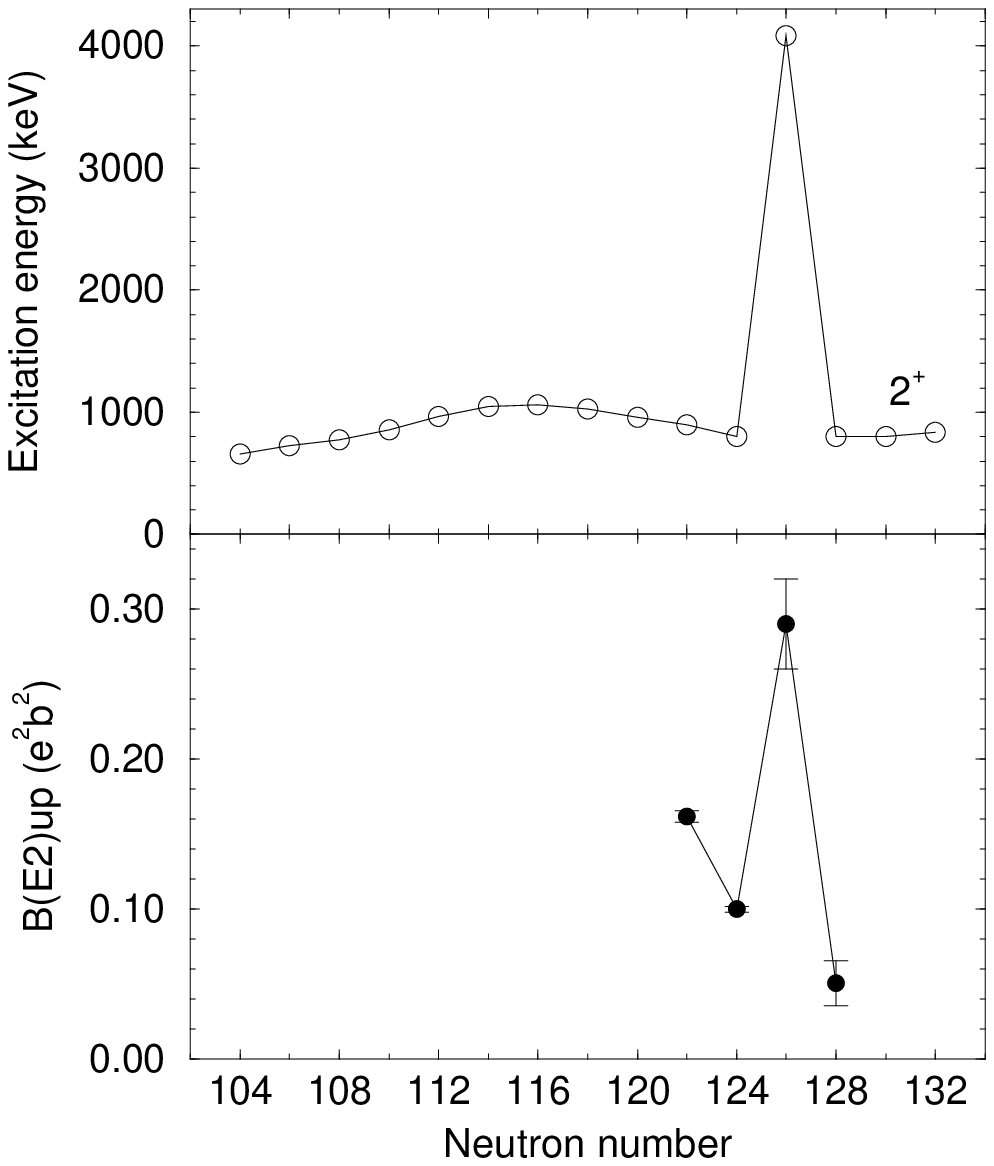,scale=0.55}
\end{minipage}
\begin{center}
\begin{minipage}[t]{16.5 cm}
\caption{{\bf Left}: Binding energies of the states located just
above and just below the $Z=82$ magic number, for $104<N<132$
and difference of the binding energies of the two states
surrounding the gap at $Z=82$ (see Sect.~\ref{annex}).
{\bf Right}: Experimental B(E2; 0$^+$ $\rightarrow$ 2$^+_1$)
values and 2$^+_1$ energies in the $_{82}$Pb isotopes.
\label{gapZ82}}
\end{minipage}
\end{center}
\end{figure}

The $Z=82$ magic number is surrounded by the $\pi s_{1/2}$
(bottom) and the $\pi h_{9/2}$ (top) orbits, giving rise to a
1/2$^+$ ground state for the $_{81}$Tl isotopes and a 9/2$^-$
ground state for the $_{83}$Bi isotopes, respectively.
The binding energies of the last proton of the $Z=83$ and $Z=82$
isotopes are drawn in the left part of Fig.~\ref{gapZ82}. The results
on the lightest isotopes are not taken into account, because of
the numerous shape coexistences identified at low
excitation energy in the nuclei having $N \le 104$
\cite{Heyd83,Andr00}. In even-even Pb nuclei close to $N=104$, three
different nuclear shapes are predicted to coexist at low excitation
energy: a\emph{ spherical} ground
state associated with the $Z=82$ shell closure, an\emph{ oblate} $2p-2h$
proton configuration with a moderate deformation
($\beta_2 \simeq -0.15$), and a more strongly deformed\emph{ prolate}
configuration ($\beta_2 \simeq +0.25$) associated with an excitation
of four or even six protons ($4p-4h$ or $6p-6h$) from the core. These
are the assignments of the three 0$^+$ states which are the lowest
three levels observed in the experimental spectrum of
$^{186}$Pb$_{104}$~\cite{Andr00}.

Apart from the singularities coming from the extra binding energy
of the $^{208}$Pb doubly-magic nucleus (see Sect.~\ref{annex}),
the two curves are very smooth, their slopes remain almost
constant along the whole neutron range. This is attributed to the
fact that neutron orbits are close in energy, which dilutes
effects of specific $\pi \nu$ interactions.

The size of the $Z=82$ spherical gap, $\sim$ 3 MeV, is not
large enough to overcome the development of shape coexistences,
as evidenced by the intruder states lying at low excitation energy
in the three isotopic chains, $_{81}$Tl, Pb, and $_{83}$Bi.
Unlike all the ground states of the
Tl and Bi isotopes which keep enough
single-particle character, the excited states being too mixed
with collective excitations cannot be used for determining the
evolution of the other proton orbits lying below or above the
$Z=82$ gap.

Speculations were made about a quenching of the $Z=82$ gap far off
the $\beta$-stability line, in the neutron-deficient side, in
order to explain the variation of the two-proton separation
energy gaps, $\Delta S_{2p}$ around $Z=82$~\cite{Novi02}.
However the computation of these $\Delta S_{2p}$ values involves
the atomic masses of neutron-deficient $_{80}$Hg and $_{84}$Po
isotopes which have
deformed ground states. Hence the $\Delta S_{2p}$ values no longer
provide a measurement of the size of the spherical $Z=82$ gap.
Self-consistent mean-field models~\cite{Bend02} account well for the
observed evolution of $\Delta S_{2p}$, as the ground states of
Hg and Po isotopes are predicted to be deformed.

\subsubsection{\it Trends in $E(2^+)$ and B(E2)}

Except for the doubly-magic nucleus $^{208}_{~82}$Pb$_{126}$ for
which a value as large as 4~MeV is found, the
$2^+$ energies for $104 \le N \le 132$ display a smooth variation
with a partial maximum at
around $N=114$ (see right part of Fig.~\ref{gapZ82}). The nearly
constancy of the $2^+$ energy is due to the closeness of the neutron
orbits involved to generate quadrupole excitations. The situation
changes dramatically in the doubly-magic nucleus
$^{208}_{~82}$Pb$_{126}$ which also displays a large B(E2) value.
This is the same
situation as the  $^{132}_{~50}$Sn$_{82}$ one (see
Fig.~\ref{E2etE3_Sn}), where the $2^+_1$ state also acquires some proton
excitations leading to a local increase of the
B(E2), a quantity which is mainly sensitive to excitation of
charged particles.

\subsubsection {\it Conclusion}
The $Z=82$ gap is not strong enough to maintain spherical shapes
for the neighboring isotopic series, at variance to all the cases
studied in the preceding sections. The density of neutron and
proton states increases as a function of the nuclear masses,
meaning that the mean distance between the orbits is decreasing,
as well as the gap values. Only the doubly-magic nucleus
$^{208}_{~82}$Pb$_{126}$ unambiguously shows a spherical behavior.
At mid-major shell ($N \simeq 104$), the $_{82}$Pb isotopes are
typical examples of different shapes coexisting at very low
excitation energy, as quadrupole correlations play a major role.
A similar situation could be foreseen in the very neutron-rich
isotopes, as soon as the number of neutrons above the $N=126$
magic number is large enough.

%% file: texteN126_25fev.tex
\section{The magic number N=126\label{126}}
The magic number 126 
is due to the spin-orbit coupling which lowers the energy 
of the $i_{13/2}$ orbit ($N = \ell = 6$, parallel spin) which 
intrudes into the lower-shell states ($N=5$). As a result, 
the magic number 126 is bound by the $g_{9/2}$ orbit 
(the lower one of the $N=6$ major shell, that having $\ell=4$ 
and a parallel spin) and by the $p_{1/2}$ orbit (the 
upper one of the $N=5$ major shell, that having the lowest 
value of the orbital momentum and an anti-parallel spin). 

\begin{figure}[h!]
\begin{minipage}{9.cm}
\begin{center}
\epsfig{file=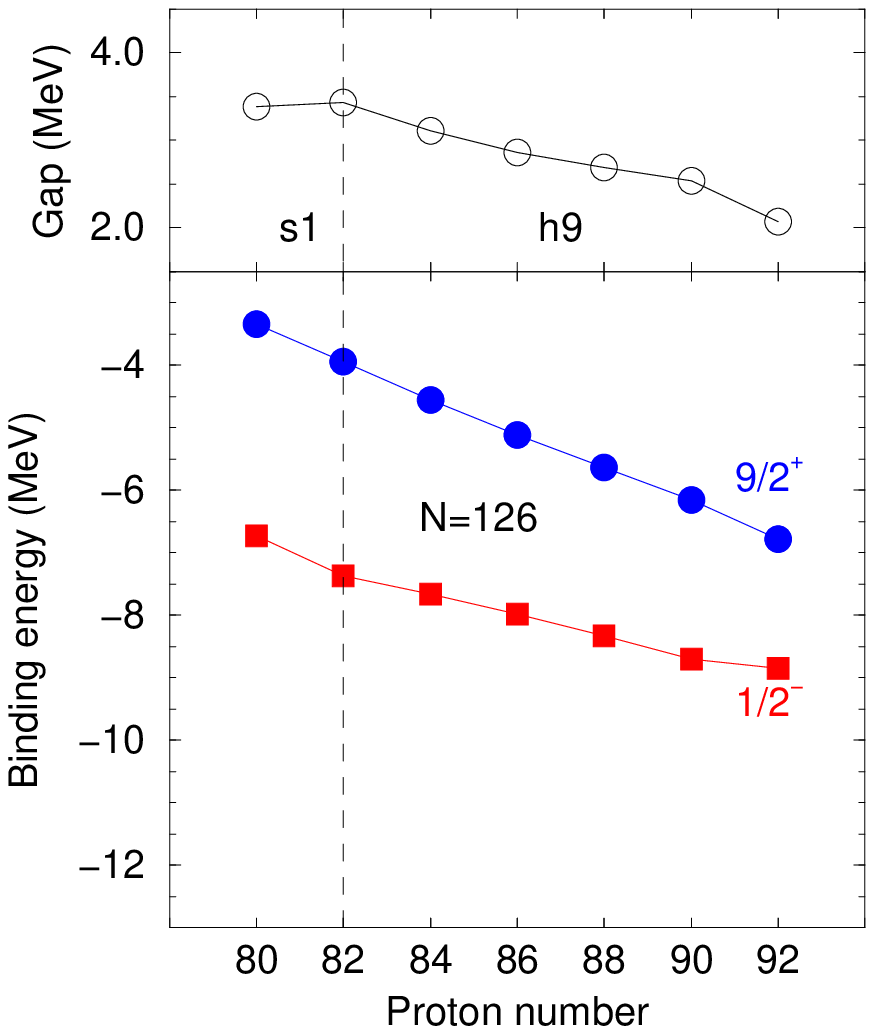,scale=0.55}
\end{center}
\end{minipage}\hfill
\begin{minipage}{9.cm}
\epsfig{file=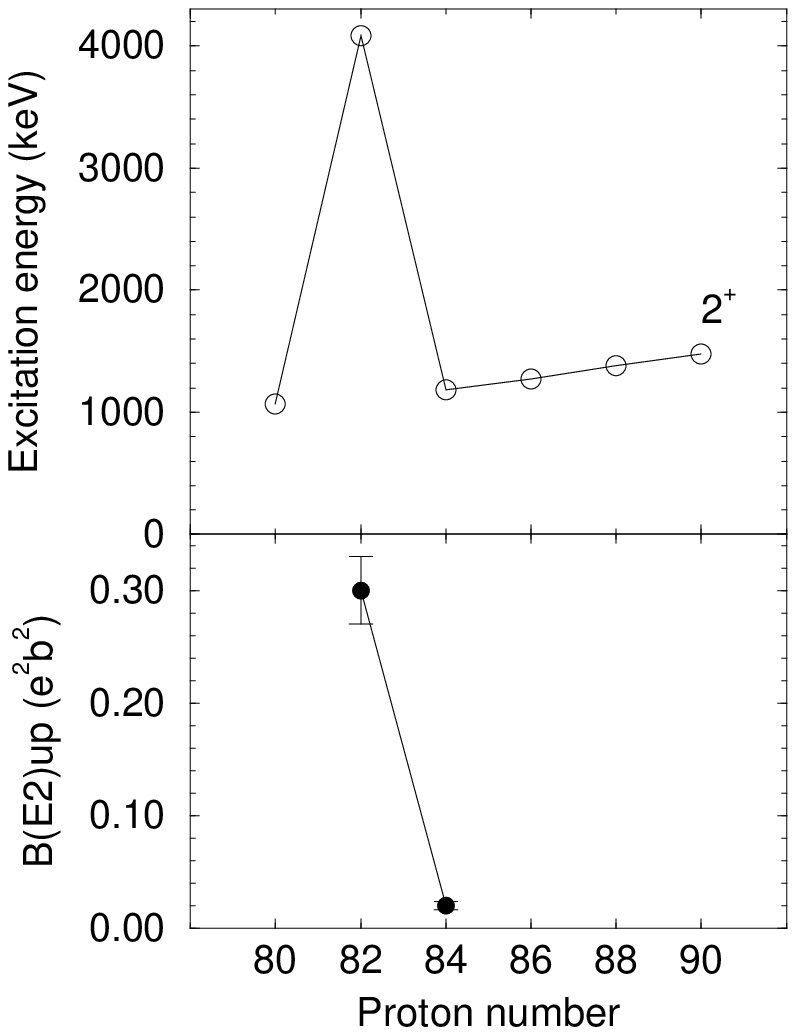,scale=0.55}
\end{minipage}
\begin{center}
\begin{minipage}[t]{16.5 cm}
\caption{{\bf Left}: Binding energies of the states $1/2^-$
($9/2^+$) located below (above) the $N=126$ magic number and 
difference of the binding energies of these two states 
surrounding the gap at $N=126$ (see Sect.~\ref{annex}). 
{\bf Right}:
Experimental $E(2^+)$ and $B(E2; 0^+ \rightarrow
2^+)$ values in the $N=126$ isotones. 
\label{N126}} 
\end{minipage}
\end{center}
\end{figure}

The binding energies of the last neutron in the $N=127$ and $N=126$
isotones are drawn in the left part 
of Fig.~\ref{N126}. 
The curves do not show any abrupt change. Starting 
around 3.5 MeV, the value of the $N=126$ gap seems to collapse for 
the largest values of Z (see the top of 
Fig.~\ref{N126}): It amounts to only 2 MeV for $Z=92$. Hence 
we cannot exclude that the ground states of the $N=126$ isotones 
with $Z>92$, which are never been synthesized up to now, 
are deformed. 

The 2$^+_1$ energies and $B(E2; 0^+ \rightarrow 2^+)$ values 
measured so far in the $N=126$ isotones are drawn 
in the right part of Fig.~\ref{N126}. Except the case of 
the doubly magic $^{208}$Pb showing a very large value, all the 
other energies are close to 1.2 MeV. For $84<Z<90$, 
the valence protons being mainly in one particular orbit, 
$\pi h_{9/2}$, this can be understood in terms of seniority 
scheme. Besides the 2$^+_1$ state, the low-lying part of 
the level schemes of these isotones display the same structure 
with the 4$^+$, 6$^+$, and 8$^+$ states coming from the 
$(\pi h_{9/2})^2$ configuration. This interpretation has been 
confirmed by large-scale shell model calculations~\cite{Caur03}. 
On the other 
hand, the $B(E2; 0^+ \rightarrow 2^+)$ value of $^{210}_{~84}$Po 
measured by the $(d,d')$ inelastic scattering is very puzzling. 
The experimental result is more than a factor of 6 less than the
SM predictions, while the calculations describe very well all 
the E2 strengths measured in the upper part of its 
$(\pi h_{9/2})^2$ structure.

In conclusion, the experimental knowledge of the $N=126$ isotonic 
chain remain very scarce at the present time, particularly 
on the nuclei belonging to the r-path involved in the 
nucleosynthesis of heavy masses.

%% file: texteLastZ_9avril.tex
\section{The super-heavy elements: What is the next proton shell
closure?}
It has long been known that in absence of shell
effects, any super-heavy element would have vanishing barrier 
against fission and a spontaneous fission half-life of the order 
of 10$^{-22}$~sec. An extrapolation of nuclear shell structure to
super-heavy nuclei has been done forty years ago~\cite{Myer66}. The
possible existence of "islands of stability" is already mentioned 
in that publication. The hypothetic presence of magic numbers beyond 
the end of the periodic table was analyzed, it would give  an extra 
binding energy of the corresponding nucleus leading to a barrier against
fission around 9~MeV, which is enough to stabilize the otherwise
highly fissile nucleus. Even at that time, it was suggested that shapes 
other than spherical might be stabilized by special shell structure.
Several years later, the nuclear potential
energy surfaces were systematically studied~\cite{Nils69} using the 
liquid-drop energy and shell corrections. The latter were 
evaluated by the Strutinsky method which, being not restricted to 
the spherical shape, allows for the map of a whole range of deformation.
For $102 < Z < 114$ and $N < 176$, all the nuclei appear to
have deformed ground states as, even though the liquid drop would 
favour spherical shapes, shell corrections are enough to create a
deformed minimum in the energy surface. On the other hand, the ground
states of nuclei with $N > 176$ remain spherical whereas secondary
minima are found at large deformation, which correspond to the
fission isomers discovered in the 1960s. These secondary deformed minima 
are also due to shell effects{\footnote{Such secondary deformed minima
have been identified many years later in 
the whole chart of nuclides, giving rise to the superdeformed 
states~\cite{Sing02}}. 
During the last 25 years, self-consistent microscopic calculations have been
largely developed, giving similar conclusions. A review of theoretical
studies on the structure of super-heavy elements have been recently 
published~\cite{Sobi07}, with a particular emphasis on the region of 
deformed nuclei. In this section, we only report on the
candidates for \emph{spherical} proton magic number next to $Z=82$. 

As shown in the previous sections, the amplitude of\emph{ spherical} 
shell effects
decreases as the size of the system size is increased. Moreover
the occurence of a spherical proton shell closure is linked to
proton-neutron interaction. It becomes therefore difficult to find
well pronounced gaps for super-heavy elements. The proton shell
closure beyond $Z=82$ was surmised to be $Z=114$ for 
many years, due
to the large spin-orbit splitting calculated for the $5f$ shell.
This prediction was common to various models in which the
spin-orbit interaction was of similar strength, such as for the
Woods-Saxon or some potentials derived from HF
calculations~\cite{Nils95}.

More recently the shell structure of superheavy nuclei has been
investigated within various parametrizations of relativistic and
nonrelativistic nuclear mean-fied models~\cite{Bend99}. Spherical
doubly-magic superheavy nuclei are found at $^{298}_{184}$114,
$^{292}_{172}$120, or $^{310}_{184}$126 depending on the
parametrization. The $Z=114$ shell closure is predicted only by
forces which overestimates the proton spin-orbit splitting in
$^{208}$Pb, which makes the existence of a large shell gap at
$Z=114$ very questionable. The $Z=120$ and $N=172$ shell closures
are predicted by the relativistic models and some Skyrme
interactions~\cite{Bend99}. These self-consistent microscopic
calculations find a\emph{ central depression} in the nuclear density
distribution which generates a wine-bottle shaped nucleonic
potential. This peculiar shape brings different 
magic numbers from
those obtained with flat-bottom potential wells. The central
depression is a consequence of the different density distributions
of the single-particle states : high-$j$ orbits are located near
the surface and low-$j$ orbits near the center. As high-$j$ proton
orbits are pushed to large radii by the Coulomb interaction, the
depression at the interior of the nucleus is enhanced.

When the nucleon numbers are increased beyond $Z=120$ and $N=172$,
the occupation of low-$j$ orbits removes this central depression.
This creates new shell gaps, corresponding to a flat density
distribution in the central part of the nucleus \cite{Afan05}.
From a very simple extrapolation of nuclear shell structure to
superheavy nuclei, $Z=126$ would be the next spherical shell
closure. Within the relativistic mean-field theory, spherical
shell gaps appear  at $Z=126$ and $N=184$.

The "cold fusion" reaction has been successfully used during the
last two past decades to synthesize new elements with
$Z=107-112$ using $_{82}$Pb and $_{83}$Bi targets~\cite{Hofm00}.
At GSI the heaviest element obtained so far 
($Z=112$, $N=165$) has been
produced by the $^{70}$Zn +$^{208}$Pb reaction with a cross
section as low as $1$ pb. More recently, one
event has been assigned to the decay chain of the new element 
 $^{278}113$ which has been synthesized in the cold fusion 
 $^{70}$Zn+$^{209}$Bi at RIKEN~\cite{Mor04}
As reaction cross section are expected to plummet as $Z$ is increased, 
this method will
probably not be suitable to reach the proton magic numbers $Z=120$
and $Z=126$.

The situation has been recently improved by using more
neutron-rich actinides isotopes as targets and less massive nuclei
as projectiles, the compound nuclei being produced in "hot fusion"
reactions. After positive results obtained on the synthesis of the
new isotopes, $^{283}$112 and $^{287}$114, the 116 and 118
elements have been produced for the first time~\cite{Ogan06}.
Three $\alpha$-decay chains have been attributed to the even-even
$^{294}$118$_{176}$ isotope, which was produced with a cross
section of 0.5 pb in the 3n-evaporation channel of the $^{48}$Ca
+$^{249}$Cf fusion. On the face of it, these experimental results
cannot confirm or  reject  the theoretical predictions for the next proton
magic number. Similar sets of results are expected for neighboring
nuclei to learn about nuclear shell closures.

As the variation of the spin-orbit splittings - and their isospin
dependence both from the SO potential through the modification of the
nucleon density and from the tensor terms of the $\pi \nu$ monopole 
interactions- is the leading factor for determining the location of
the island of stability of the superheavy elements, a global
understanding of the $NN$ interaction along the chart of nuclides is
required. New studies foreseen for lower-Z exotic nuclei would
bring a complementary information into the present quest. 

%% file: texteConclusions_10avril.tex
\section{Conclusions and Outlooks\label{outlooks}}

Every localized fermionic object, such as atomic cluster,
molecule, atom or nucleus, displays shell structure. The non uniform
energy distribution of the shells gives rise to shell gaps. In
atomic nuclei it has thorough consequences, among which : (i) The
stable magic nuclei -which exhibit large shell gaps between the
last occupied and the first empty orbit- are the most abundant on
earth, and the radioactive ones are expected to be the main
'survivors' in explosive stellar environments such as in X-ray
bursts and supernovae, (ii) The closed-shell nuclei are the major
building blocks which serve for a global description of all
existing nuclei.

The concept of a magic number has remained a solid pillar for a
long time, until the study of nuclei with extreme N/Z ratios has
revealed their fragility. Guided by various spectroscopic
information (e.g. binding energies of orbits on each side of the
gap, trends of excitation energy, collective properties of states,
spectroscopic factors,...) the present work intended to review the
vulnerability of the major-shell closures, and the emergence of
new ones. An exploratory 'journey' throughout the chart of
nuclides has therefore been undertaken to adress these evolutions,
which were tentatively ascribed to specific properties of the $NN$
force, particularly the tensor term which strongly acts in
proton-neutron configurations, either between orbitals having the
same orbital momenta $\Delta \ell =0$ or $\Delta \ell =1$.
Nevertheless the single-particle states often easily couple to
collective states since contrary to molecular systems, there is no
clean cut between single-particle excitation, vibrational and
rotational states in the atomic nucleus. Therefore the present
statements on the observed structural evolution in terms of
fundamental properties of the nuclear force 
remain\emph{ semi-quantitative}, as a proper unfolding with collective
motions has still to be achieved in most of the cases.

Two major kinds of shell gaps exist in atomic nuclei and their
ability to resist to erosion depends on their origin, size and
composition. The shell gaps $8$, $20$ and $40$ are provided by the
mean field description of the nuclei using a Harmonic Oscillator
potential, which is a consequence of the saturation of the nuclear
interaction. In heavier systems, these shell gaps are replaced by
those created by the strong spin-orbit interaction, which
originates from the spin-dependence of the nuclear interaction.

\begin{figure}[h!]
\begin{minipage}{5.7cm}
\begin{center}
\epsfig{file=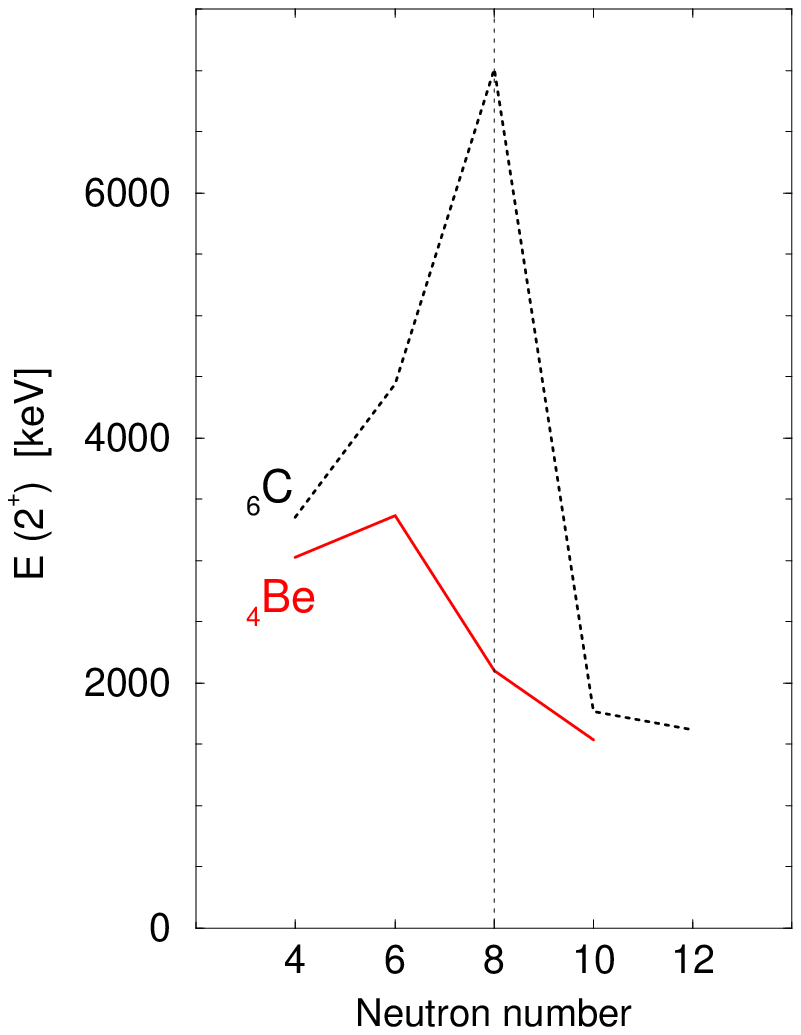,scale=0.6}
\end{center}
\end{minipage}\hfill
\begin{minipage}{5.7cm}
\begin{center}
\epsfig{file=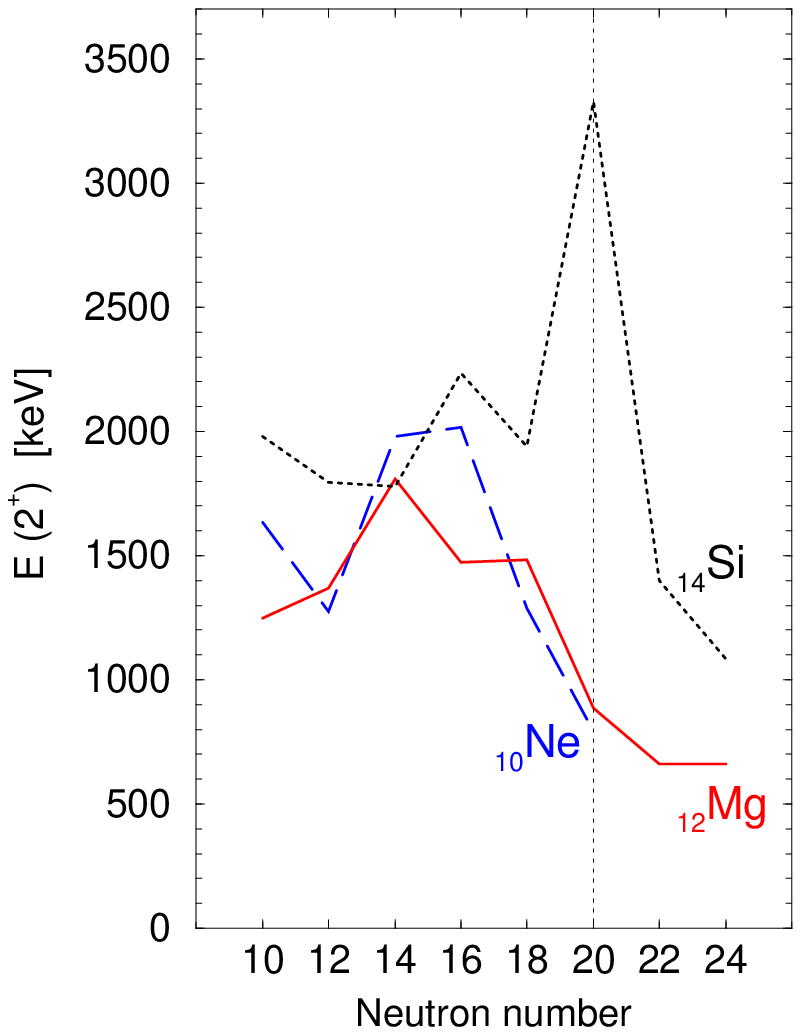,scale=0.6}
\end{center}
\end{minipage}\hfill
\begin{minipage}{5.7cm}
\begin{center}
\epsfig{file=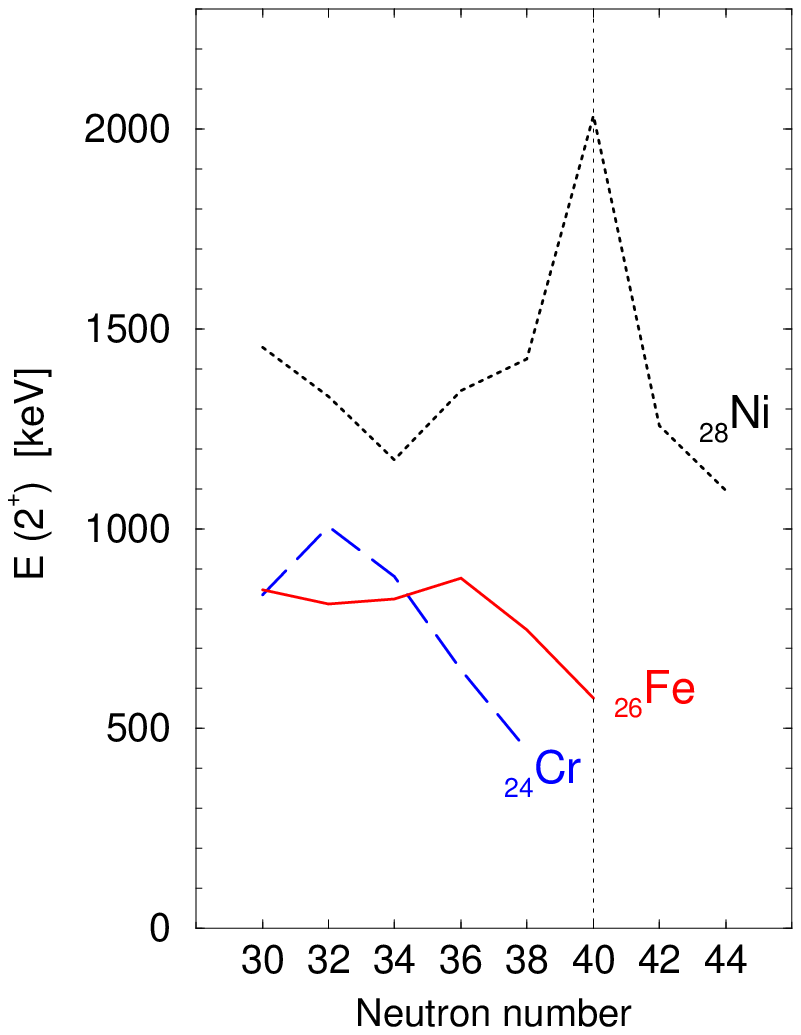,scale=0.6}
\end{center}
\end{minipage}
\begin{center}
\begin{minipage}[t]{16.5 cm}
\caption{Experimental E(2$^+_1$) values for several isotopic
chains, as a function of the neutron numbers where the
$N=8$ (right), $N=20$ (middle) and $N=40$ (right) shell closures
are crossed.  \label{deuxplus_HOgap}}
\end{minipage}
\end{center}
\end{figure}

The HO shell gaps are formed between two orbits of opposite
parity. Consequently excitation which preserve parity, such as
quadrupole ones, should occur through a minimum of one pair of
particle excitations, e.g. ($2p-2h$), across the gap. This
specific property gives rise to a relatively enhanced rigidity
with respect to collective excitations, unless the size of the gap
is reduced or/and many particle excitations are permitted, as in
semi-magic nuclei. It is remarkable that these two effects thrive
in the three known cases of HO shells in a similar and sudden
manner. Starting from the very rigid
$^{14}_{~6}$C$_{8}$, $^{34}_{14}$Si$_{20}$ and
$^{68}_{28}$Ni$_{40}$ nuclei the removal of only two protons
provokes a sudden deformation of the $^{12}_{~4}$Be$_{8}$,
$^{32}_{12}$Mg$_{20}$ and $^{66}_{26}$Fe$_{40}$ nuclei,
respectively, in which the intruder deformed states have the
lowest energy. This is illustrated in 
Fig.~\ref{deuxplus_HOgap} by the large changes in
the energy of the 2$^+_1$ state  for $N=8$, $N=20$ and $N=40$.

\begin{figure}[h!]
\begin{center}
\epsfig{file=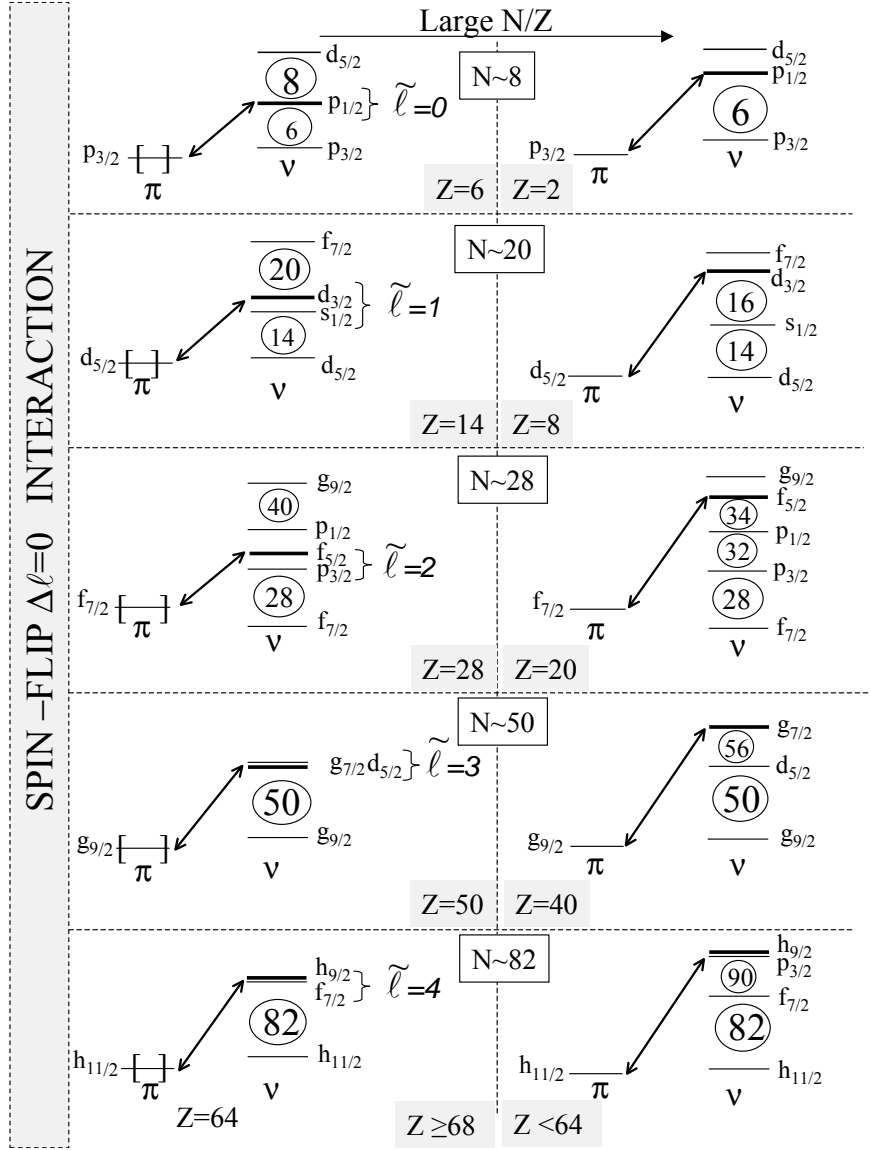,scale=0.6}
\begin{minipage}[t]{16.5 cm}
\caption{The qualitative effect of the spin-flip $\Delta \ell=0$
proton-neutron interaction is shown for different regions of the
chart of nuclides. By comparing the left and right hand sides of
the Figure, the removal of protons induces a change in the spacing
and possibly ordering of the neutron states. When this interaction
is missing, the major Harmonic Oscillator shell gaps $N=8, 20, 40$
are reduced to the benefit of new subshell gaps at $N=6, 16, 32$.
Additionally the degeneracy of the pseudo-spin doublets $\tilde
\ell=1,2,3,4$ is lifted. \label{spinflipdl0}}
\end{minipage}
\end{center}
\end{figure}

Noteworthy is the fact
that this passage from a rigid configuration with a normal filling
of the orbits to the so-called island of inversion is triggered by
the same phenomenon : the sudden release of a strong \emph{ 
spin-flip $\Delta \ell =0$} proton-neutron force. For the three
rigid nuclei mentioned above, the forces that came into play to
produce large $N=8$, $N=20$ and $N=40$ gaps are $\pi p_{3/2} -
\nu p_{1/2}$, $\pi d_{5/2} - \nu d_{3/2}$ and $\pi f_{7/2} - \nu
f_{5/2}$, respectively (see the first three rows of
Fig.~\ref{spinflipdl0}).
As soon as protons are removed from the $p_{3/2}$, $d_{5/2}$ or
$f_{7/2}$ orbits, the corresponding neutron orbits ($p_{1/2}$,
$d_{3/2}$ or $f_{5/2}$) become less bound. Thus the neutron gaps
$N=8$, $N=20$ or $N=40$ are reduced significantly and become more
vulnerable to excitations. In parallel new shell gaps $N=6$,
$N=16$, or $N=32,34$ are created below, as shown in the right part
of Fig.~\ref{spinflipdl0}.

\begin{figure}[h!]
\begin{center}
\epsfig{file=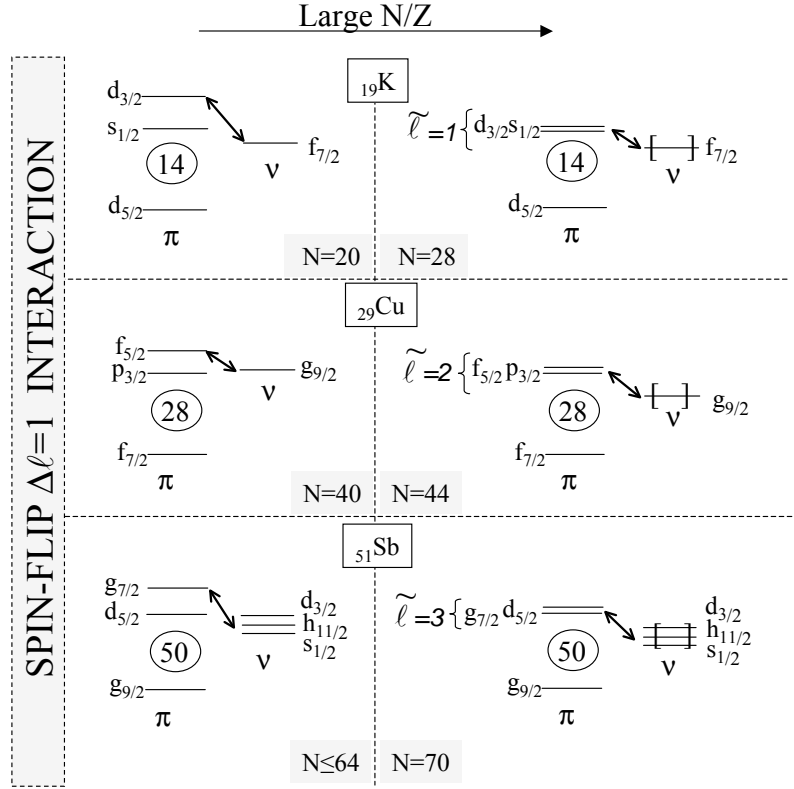,scale=0.6}
\begin{minipage}[t]{16.5 cm}
\caption{The spin-flip $\Delta \ell=1$ proton-neutron interaction
acts in a similar manner in different regions of the chart of
nuclides, as shown in the K, Cu and Sb isotopic chains. By
comparing the left and right hand sides of the Figure, the
addition of neutrons induces a shrink in the spacing and possibly
ordering of the proton states forming the doublets labelled with
$\tilde \ell=1,2,3$ values. When this interaction operates, the
spin-orbit shell gaps $Z=14, 28, 50$ are reduced.
\label{spinflipdl1}}
\end{minipage}
\end{center}
\end{figure}

\begin{figure}[h!]
\begin{center}
\epsfig{file=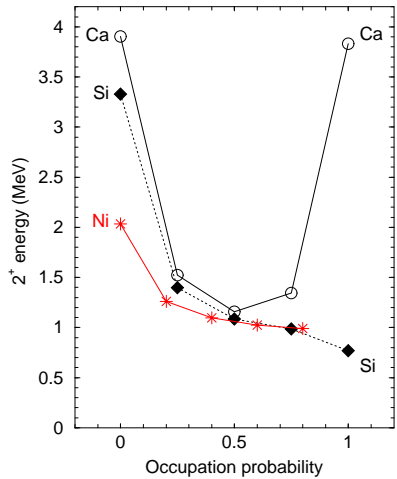,scale=0.6}
\begin{minipage}[t]{16.5 cm}
\caption{Evolution of the energy of the $2^+_1$ states in the
$_{14}$Si, $_{20}$Ca, and $_{28}$Ni isotopic chains during
the filling of the high-$j$ neutron shell (see text). The occupation 
probability is calculated on the assumption of spherical shapes.
The doubly-magic nuclei 
$^{34}$Si$_{20}$, $^{40}$Ca$_{20}$, and  $^{56}$Ni$_{28}$ are present 
in the left-hand side of the Figure. On the right-hand side, 
it is the turn of the $^{42}$Si$_{28}$, $^{48}$Ca$_{28}$, and 
$^{78}$Ni$_{50}$ nuclei.
\label{deuxplusCaSiNi}}
\end{minipage}
\end{center}
\end{figure}

The SO-like magic numbers, such as $14$, $28$ and $50$, have
different properties. As shown in Fig.~\ref{spinflipdl1} for the
K, Cu and Sb isotopic chains, these magic numbers are defined
between the two SO partners (having aligned and anti-aligned spin
and angular momenta), in between which another orbit with $\ell -
2$ is present. 
In these examples the spin-aligned orbits
$d_{5/2}$,$f_{7/2}$ and $g_{9/2}$ at located the bottom of the
shell gaps $Z=14$, $Z=28$ and $Z=50$, respectively. The couples of
valence orbits above the shell gaps are ($s_{1/2}, d_{3/2}$),
($p_{3/2}, f_{5/2}$), and ($d_{5/2}, g_{7/2}$), respectively. The
\emph{ spin-flip $\Delta \ell$=1} proton-neutron force acts to
reduce the SO splittings and shrink the space between the two
valence states (see comparison of the proton orbits on the left-
and right-hand sides of Fig.~\ref{spinflipdl1}). This situation
seems to be generic in the whole chart of nuclides. This reduction
could trigger quadrupole collectivity -and the development of
deformation- between the occupied and valence states. This
jeopardizes the rigidity of the spherical shell gap, as in the
case of the neutron-rich nucleus $^{42}_{14}$Si$_{28}$. The very
low energy of its first excited state indicates the erosion of
both the $Z=14$ and $N=28$ shell closures by the action of the
mutual proton and neutron spin-flip $\Delta \ell =1$ forces ($\pi
d_{5/2}$ acts to reduce the neutron $f_{7/2} - f_{5/2}$ SO
splitting and $\nu f_{7/2}$ to shrink the proton $d_{5/2} -
d_{3/2}$ one). The next doubly-magic nucleus formed with
two SO shell gaps at $Z=28$ and $N=50$, is $^{78}_{28}$Ni$_{50}$. It
is being thought to behave as a spherical rigid nucleus. Whether
it would develop a permanent quadrupole ground-state deformation
or not, depends on both the size of the $Z=28$ and $N=50$ gaps and
on the strength of the quadrupole correlations. As regards the
evolution of the shell gaps, significant reductions are foreseen
from the experimental results obtained so far while approaching
$^{78}$Ni. Complementary to what was discussed in the different
sections of this review, it is instructive to look at the 
evolution of the $2^+$ energies in the Si, Ca, and Ni isotopic 
chains during the filling of the high-$j$ neutron shell involved
($f_{7/2}$ for Si and Ca, and $g_{9/2}$ for the Ni isotopes). Starting at
$^{34}$Si, $^{40}$Ca, and $^{56}$Ni which exhibit large $2^+$
energies, the three chains behave similarly up to the mid
occupancy of the shell, i.e. at $^{38}$Si, $^{44}$Ca, and $^{62}$Ni 
(see Fig.~\ref{deuxplusCaSiNi}).  
Then the $2^+$ energy of the Si isotopes does not follow the
parabola curve as for the Ca isotopes, but strongly decreases. 
The actual trend of the $2^+$ energy in the Ni isotopic
chain, immediately before reaching $^{78}$Ni, looks very close to the
one in the Si isotopes. This might indicate that $^{78}$Ni would
behave like $^{42}$Si, rather than $^{48}$Ca. This feature is very
challenging to look at in the future. For heavier nuclei, the
combination of the magic numbers of SO-origin gives rise to
spherical rigid nuclei, such as $^{132}_{~50}$Sn$_{82}$ and
$^{208}_{~82}$Pb$_{126}$.

As shown in the various examples presented in
Figs.~\ref{spinflipdl0} and \ref{spinflipdl1}, the 'spin-flip
$\Delta \ell=0$' and 'spin-flip $\Delta \ell=1$' $\pi \nu$ forces
play decisive roles in the evolution of shell gaps, would they be
of HO or SO origin. They also remarkably bring the
quasi-degeneracy of two states separated by two units of orbital
momentum (i.e. $\ell$ and $\ell +2$). These states belong to the
so-called 'pseudo-spin' doublet, having pseudo-orbital angular
momentum $\tilde {\ell} =\ell +1$ and and pseudo-spin 
$\tilde {s}$, such as $j=\tilde {\ell} \pm \tilde {s}$. 
For instance, the proton doublet with $\tilde {\ell} =1$,
formed with the $\pi s_{1/2}$ ($\ell=0$) and $\pi
d_{3/2}$($\ell=2$) states, become degenerate when the $\nu
f_{7/2}$ orbit is full, e.g. in $^{47}$K$_{28}$. Same holds true
for the other proton $\tilde {\ell} = 2,3$ orbits which become
degenerate while the neutron $g_{9/2}$, $h_{11/2}$ orbits are
filled, respectively (see Fig.~\ref{spinflipdl1}). Similarly the
neutron $\tilde {\ell} =2,3,4$ doublets are present while the
proton $d_{5/2}$, $f_{7/2}$, $g_{9/2}$, $h_{11/2}$ orbits are
filled, respectively, as shown in Fig.~\ref{spinflipdl0}. This
pseudo-spin symmetry has been shown by J. N.
Ginnochio~\cite{Gino97} to occur naturally in a relativistic
mean-field description of single-particle states. Gathering the
empirical results of Figs.~\ref{spinflipdl0} and \ref{spinflipdl1}
it is found that the breaking of this symmetry is linked to the
spin-isospin dependence of the nuclear interaction (the spin-flip
$\Delta \ell =0,1$ terms). Effective interactions which are not 
taking into account these effects will never reach a satisfactory 
level of predictive power.

The present review has addressed the evolution of shell closures
for nuclei studied so far. Fascinating enough is to remind that
the spin-dependence of the nuclear force governs the
existence of most of the magic nuclei. Despite many decades of
research, the role of the various components of this force, and in
particular the tensor one, is still veiled. With the future
radioactive ion beam facilities worldwide and the increasing
detector and spectrometer sensitivities, some of the scenarios
suggested here on the evolution of shells and on the underlying 
nuclear forces at play will be confirmed or refuted. In parallel 
the way to link the 'bare' $NN$ force to the in-medium 
forces, would they be in the valley of stability or very weakly 
bound, is being searched for. These experimental and theoretical 
progresses are to be led together. This is essential to build up
a consistent description of all kind of nuclear systems.

%% file: texteAnnex_9avril.tex
\section{Annex: Binding energy of the last nucleon in  
nuclei close to the semi-magic nuclei\label{annex}} 
The spherical nuclei having nucleon numbers close to magic numbers
can be considered as systems of {\emph independent} nucleons moving
in a common potential well. The eigenvalues of the corresponding
Hamiltonian give the single-nucleon spectra.  
From an experimental point of view, the binding energies of
such nuclei can be used to determine the separation energies of 
nucleons, that are supposed to be close to the single-particle
energies we are looking for. For that, we have to assume that the 
semi-magic core remains closed when adding or removing one nucleon,
i.e. there are no rearrangement or collective excitation.
  
The energy $\epsilon_b$ of the orbit located {\emph below} a proton 
gap is equal to the inverse of the proton separation energy 
$S_p(Z_{magic})$ of the semi-magic nucleus having $Z_{magic}$ 
protons, i.e.
$\epsilon_b=-S_p(Z_{magic})$, where $S_p(Z_{magic})$ is the
difference between the binding energies,
$BE(Z_{magic},N)-BE(Z_{magic}-1,N)$. Similarly, the energy
$\epsilon_a$ of the orbit situated {\emph above}  the proton gap is 
equal to the inverse of the proton separation energy of the 
nucleus having $Z_{magic}+1$ protons, i.e. 
$\epsilon_a=-S_p(Z_{magic}+1)$. 
The distance between these two orbits is representative of the 
gap in energy  between the last-occupied and the
first-valence orbit, $E_{gap}(Z_{magic},N) = S_p(Z_{magic},N)-
S_p(Z_{magic}+1,N)$. As said before, the use of this relation has 
to be limited to spherical nuclei, for which the $\epsilon_{a}$ 
and $\epsilon_{b}$ values are close to the single-particle energy 
of the first valence and last occupied orbitals, respectively. 
Note that the
mean field potential, which is an average of nucleon-nucleon 
interactions inside a nucleus, is slightly changing between the 
nuclei with $Z_{magic}$ and $Z_{magic}+1$ which are used here to 
define the size of the gap, $E_{gap}(Z_{magic},N)$. Therefore, the 
intrinsic value of
$E_{gap}(Z_{magic})$ is approximate. Rather, its evolution 
along an isotopic chain is meaningful to look at the modification
of shell closures. 

The same procedure applies to the evolution of
gaps $E_{gap}(Z,N_{magic})$ formed by between occupied and valence
neutron single-particle orbitals. In addition to these definitions 
and words
of caution, several other considerations apply to the binding
energy curves that are used extensively throughout the present
work :

\begin{itemize}
\item  Only the {\emph measured} values reported in the AME2003
atomic mass evaluation~\cite{Audi03} have been used, not the
extrapolated ones. Added to these are the recent atomic mass
measurements of $^{23}$N and $^{24}$O~\cite{Jura07}, $^{47}$Ar~\cite{Gaud06}, 
$^{70,72}$Ni and $^{73}$Cu~\cite{Raha07}, $^{83}$Ge~\cite{Thom05}, 
and $^{131-133}$Sn~\cite{Dwor08}. 
\item The uncertainties on the
neutron (proton) separation energies are shown in the figures,
except when they are smaller than the size of the symbols. 
\item We restrict the plots to data corresponding to {\emph spherical} 
nuclei, whose shapes are confirmed by the properties of their 
first excited states. In the case of deformed nuclei,
the nucleon separation energies, that is the energy required to
remove the least bound particle from the nucleus, are the results
of a complex situation, the two involved nuclei even having 
different deformations.
\item The curves are labeled by the
$j^\pi$ values of the states of the odd-A nuclei and not by the
quantum numbers ($\ell ,j, \pi)$ of the expected single-particle
orbits. This choice is governed by the fact that the spectroscopic
factor of the first $j^\pi$ state does not always exhaust the sum
rule arising from the $j^\pi$ orbit. Such cases are discussed in the
sections following the results on the binding energies.
\item Most of the 
figures only display the binding energies of the two states which
surround the magic number that is considered. When two orbits are
crossing each others, would it be above or below the shell gap,
both are displayed in the figures.
\item Many results involving doubly-magic nuclei display  
singularities in the curves. One has to notice
that the so-called Wigner term which gives an additional binding
to nuclei with neutrons and protons occupying the same shell-model
orbitals, is the largest when $N=Z$. As a consequence,
every separation energy calculated using a doubly-magic $N=Z$ 
nucleus displays a sudden variation at this number of nucleons,
resulting in a very large value of the shell gap. Chasman~\cite{Chas07} 
has recently proposed a method to calculate values of shell gap 
associated to the $N=Z$ nuclei by combining the binding energies of 
various nuclei surrounding the self-conjugated isotopes, in such a 
way they are free from the Wigner energy contribution. The obtained
values are now very close to those predicted in mean-field
approaches.
\item The small effects possibly observed for the $N \not= Z$ 
doubly-magic nuclei ($^{48}$Ca, $^{132}$Sn or $^{208}$Pb) show
evidence of the limits of this method. Even though all the 
nuclei chosen for the plots shown in this paper have a spherical 
shape attested by their excited states, various correlations can 
give more stability to the odd-A nuclei than to the semi-magic 
core, leading to a larger binding energy of its ground state.
Therefore for nucleon number between two magic numbers, the 
value of the separation energy is not inevitably the  
the energy of the orbit in interest. Nevertheless any sharp 
variation remains relevant for the analysis of the single-particle 
behavior.
\end{itemize}